# High-Throughput Transition-State Searches in Zeolite Nanopores

Pau Ferri-Vicedo, Alexander J. Hoffman, Avni Singhal, Rafael Gómez-Bombarelli*

Department of Materials Science and Engineering, Massachusetts Institute of Technology, Cambridge, MA 02139, USA

E-mail: rafagb@mit.edu

## ABSTRACT

Zeolites are important for industrial catalytic processes involving organic molecules. Understanding molecular reaction mechanisms within the confined nanoporous environment can guide the selection of pore topologies, material compositions, and process conditions to maximize activity and selectivity. However, experimental mechanistic studies are time- and resource-intensive, and traditional molecular simulations rely heavily on expert intuition and hand manipulation of chemical structures, resulting in poor scalability.

Here, we present an automated computational pipeline for locating transition states (TS) in nanopores and exploring reaction energy landscapes of complex organic transformations in pores. Starting from the molecular structure of potential reactant and products, the Pore Transition State finder (PoTS) locates gas-phase transition states using density functional theory (DFT), docks them in favorable orientations near active sites in nanopores, and leverages the gas-phase reaction mode to seed condensed-phase DFT calculations using the dimer method. The approach sidesteps tedious manipulations, increases the success rate of TS searches, and eliminates the need for long path-following calculations.

This work presents the largest ensemble of zeolite-confined transition states computed at the DFT level to date, enabling rigorous analysis of mechanistic trends across frameworks, reactions, and reactant types. We demonstrate the applicability of PoTS by analyzing 644 individual reaction steps for transalkylation of diethylbenzene in BOG, IWV, UTL and FAU zeolites, and in skeletal isomerization of 162 individual reaction steps in BEA, FER, FAU, MFI and MOR zeolites finding good experimental agreement in both cases. Lastly, we propose a path to address the limitations we observe regarding unsuccessful TS searches and insufficient theory in other reactions, like alkene cracking.

## INTRODUCTION

Understanding catalytic reactivity is fundamental to advancing materials science and drug discovery, yet experimental investigations, from synthesis to reactor optimization, remain time-consuming and costly. Currently, computational methods are widely applied for modeling potential energy surfaces (PES)[1] and elucidating reaction mechanisms through transition state (TS)[2] identification, linking computed free energy barriers to reaction rates via transition state theory.[3,4] Searching for a TS involves locating first-order saddle points on the PES. These points, defined by a unique negative curvature along the reaction coordinate, are computationally demanding to locate even in the gas phase, often requiring manual, intuition-driven approaches with limited success.[5,6,7] State-of-the-art TS identification methods are classified as either double-ended or single-ended.[8,9] Double-ended methods, such as the Nudged Elastic Band (NEB),[10,11] generate a complete reaction path using both reactant and product structures,[12] but they incur significant computational cost due to the iterative optimization of many path images and are prone to failure if the initial guess path differs from the true one. Single-ended methods like the Dimer method[13,14] and eigenvector-following[15] require only an initial guess, reducing the computational effort, but are even more critically reliant on the accuracy of the initial structure and the chosen reaction coordinate. The complexity increases in heterogeneous catalysis, where periodic boundary conditions introduce further challenges. In such systems, the PES is affected by both local interactions and the collective effects of an infinite array of periodic images,[16] leading to dependencies on cell geometry, long-range interactions,[17] and nonlocal couplings absent in isolated molecules. This complexity necessitates specialized summation techniques and rigorous numerical convergence to accurately capture the intricate energy landscapes of catalytic reactions.[18]

Heterogeneous catalysts offer unique opportunities for reaction control. A prime example is provided by zeolites, microporous materials composed of substituted $SiO_4$ tetrahedra that feature uniform pore structures of various sizes and topologies. These pores enable shape-selective adsorption and catalysis across a broad spectrum of chemical reactions.[19] Substitution of $SiO_4$ with $AlO_4^-$ generates localized negative charges that, when balanced by protons or metal cations, create Brønsted or Lewis acid sites, respectively, thereby enhancing catalytic activity.[20,21] This confinement not only stabilizes specific transition states but also favors particular intermediates or facilitates unique rearrangements that differ markedly from those observed in bulk phases.[22,23] Understanding reaction kinetics in zeolites is critical because the rates of adsorption, reaction, and desorption directly influence turnover frequency and product selectivity.[24,25] However, the rigid and confined zeolite environment poses significant challenges for TS identification,[26] as subtle, orientation-dependent noncovalent interactions can trap manually generated TSs in local minima, preventing optimal framework stabilization. Although systematic reorientation approaches have been explored by Hibbits et al.[27,28] they have been limited to a narrow set of TSs within specifically chosen zeolite frameworks and still require a manually generated initial guess from a user, thereby restricting the broader applicability of the approach.

We introduce a novel computational approach to address these challenges that bridges gas-phase TS identification with the periodic environments of zeolites. In the PoTS method, gas-phase TS structures are automatically located with molecular DFT and placed into zeolite frameworks close to active sites. These guess geometries then seed high-throughput, single-ended TS searches via the Dimer method, exploiting the availability of gas-phase reactive modes. By eliminating the need for a prior, computationally expensive NEB calculation, our approach reduces computational cost and significantly enhances TS identification success in intricate catalytic systems (**Figure 1a**).

The pipeline was evaluated on three acid zeolite-catalyzed reactions: diethylbenzene transalkylation, cycloalkene isomerization, and alkene cracking (**Figure 1c**). Diethylbenzene transalkylation is an industrially relevant process catalyzed by Brønsted-acid zeolites that converts less desirable diethylbenzene byproducts to ethylbenzene, a crucial precursor for styrene and polystyrene. This reaction improves the efficiency and increases the overall yield of industrial ethylbenzene production.[31,32] This reaction is predominantly favored in zeolites with bidimensional pore systems defined by 12-membered rings (12MR), which are constructed from 12 tetrahedrally coordinated T-sites.[33] Accordingly, we targeted frameworks such as BOG (12 × 10), FAU (12 × 12 × 12), IWV (12 × 12), and UTL (14 × 12).

Next, we examined cycloalkene isomerization reactions in which ethylcyclohexene is protonated to form ethylcyclohexyl cations that serve as crucial intermediates in the conversion of biomass, natural gas, and crude oil into fuels and value-added chemicals.[34,35] Finally, we explored alkene cracking, a process essential for producing fuels and light olefins, particularly in the context of renewable feedstocks and shale gas valorization.[36,37] To elucidate the structure-activity relationships governing these reactions, we selected zeolite frameworks, FAU (12 × 12 × 12), BEA (12×12×12), FER (10), MFI (12×12), and MOR (12), which are used in multiple industrial catalytic processes due to their diverse and well-defined pore architectures (**Figure 1b**).[38,39] Each framework was modeled with one unique Brønsted acid site, and topology-specific models were generated to evaluate how acid site positioning influences TS identification and reaction outcomes (see **Method** and **S2**).

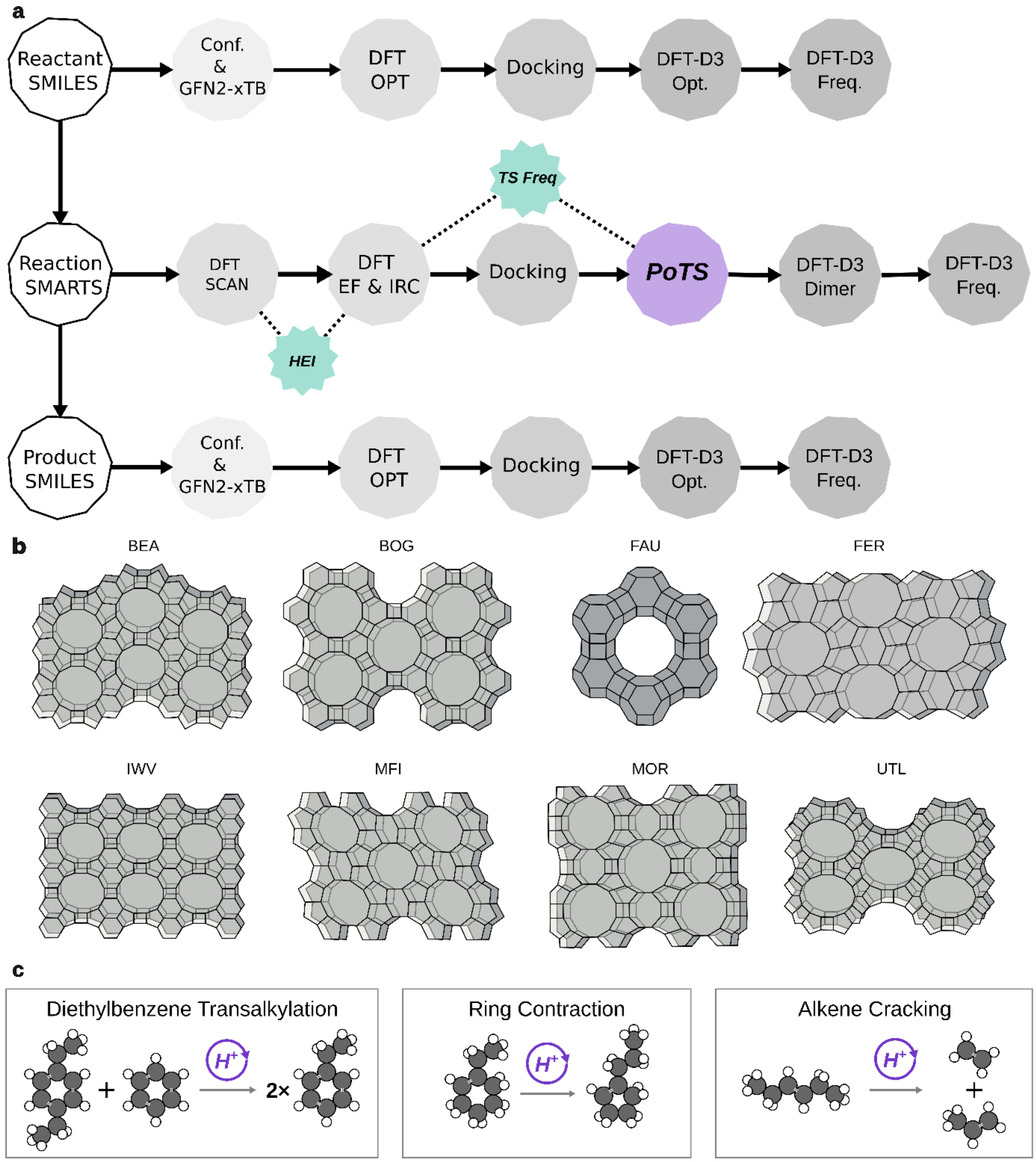

**Figure 1.** a) Schematic overview of the computational workflow from SMILES-based conformer generation and reaction enumeration to transition-state (TS) structures in zeolites via PoTS. The pipeline begins with reaction enumeration (SMARTS) between reactant and product SMILES, followed by conformer generation (ETKDGv3 + GFN2-xTB) and gas-phase DFT optimization (M06-2X/def2-SVP). TS identification proceeds through a relaxed scan to locate the highest energy image (HEI), then eigenvector following (EF) and intrinsic reaction coordinate (IRC) calculations. The resulting structures are docked into zeolites using a modified Monte Carlo algorithm that applies a cation–acid site distance threshold. Finally, the PoTS methodology aligns the gas-phase TS imaginary vibrational modes to the docked pose in the periodic system, after which periodic DFT (PBE-D3) optimizations, Dimer calculations, and frequency calculations provide the free energies of reactants, TS, and products under confinement. b) Zeolites studied in this work, ordered alphabetically, BEA (12x12x12), BOG (12x10), FAU (12x12x12), FER (10), IWV (12x12), MFI (10x10x10), MOR (12), UTL (14x12). c) Scheme for the main reactions studied, diethylbenzene transalkylation, ring contraction and alkene cracking.

# RESULTS

## Diethylbenzene Transalkylation

Ethylbenzene (EB) production requires benzene alkylation with ethene and generates diethylbenzene (DEB) as a byproduct, which is recycled via transalkylation to improve yield. DEB transalkylation follows two mechanisms. In the alkyl-transfer pathway,[40] DEB dealkylation produces EB and a surface-bound ethoxy species that reacts with benzene but can also lead to over-ethylation or ethene formation, which reduces selectivity (**Figure 2a**). In the diaryl-mediated pathway, $DEB^+$ cations undergo transalkylation through hydrogen transfer (**Figure 2b**). The initial $DEB^+$ formation is kinetically challenging, though extra-framework Al may assist this step.[41] Once formed, $DEB^+$ enters a catalytic cycle where transalkylation proceeds through $I1^+$ and $I4^+$ intermediates via either a direct (TS1) or multistep (TS2, TS3, TS4) HS path. A similar reaction with DEB instead of benzene leads to disproportionation, yielding EB and triethylbenzene (TEB), reducing EB yield (depicted as red substituents in **Figure 2b**). Zeolites that favor the diaryl cycle while suppressing disproportionation are the most effective EB catalysts;[42] our study focuses on analyzing the diaryl mechanism inside the BOG, IWV, UTL, and FAU topologies (see **Method** and **Section S1** of the **Supporting Information (SI)**).

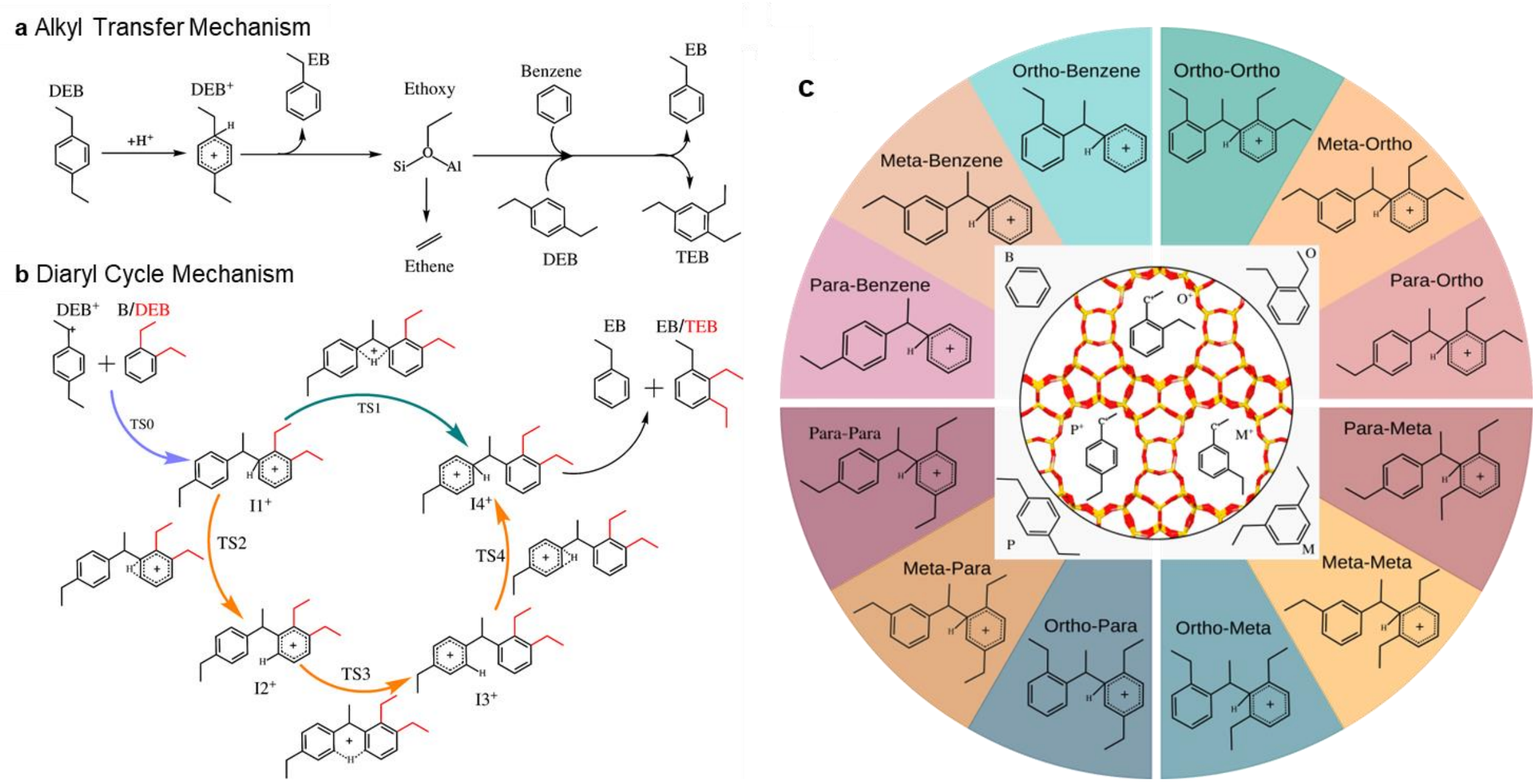

**Figure 2. a** Diethylbenzene transalkylation mechanisms: **a** Alkyl transfer mechanism via the zeolite framework. **b** Diaryl mechanism proceeding through TS0 (purple) followed by a direct pathway via TS1 (teal) or multistep pathway via TS2, TS3, and TS4 (orange). In **a2** black sticks represent $DEB^+$ transalkylation with benzene intermediates, while black plus red sticks indicate disproportionation intermediates from the $DEB^+$ + DEB reaction. **c** All $I1^+$ isomer pairs considered for transalkylation and disproportionation. Blue shades denote $I1^+$ intermediates derived from ortho-$DEB^+$, orange from meta-$DEB^+$, and burgundy from para-$DEB^+$.

All diaryl combinations in DEB transalkylation and disproportionation were systematically explored. In transalkylation, each $DEB^+$ isomer (*ortho*, *meta*, and *para*) reacts with benzene, while disproportionation involves each potential $DEB^+$ isomer reacting with another neutral DEB, resulting in nine distinct $DEB^+$–DEB pairs (**Figure 2c**). For each of the 12 resulting $I1^+$ intermediates, one direct pathway to $I4^+$ was identified along with multiple multistep routes, dictated by substituent arrangement and resulting in 42 unique energy paths. Overall, this exploration yielded 92 unique TS across the twelve $I1^+$ intermediates (see **S2.1**).

Using PoTS, we successfully identified 560 internal proton-transfer TS for DEB transalkylation at the DFT-D3 level, a tenfold increase compared to the 48 TS analyzed in our previous work[42], which were restricted exclusively to para-DEB$^+$ isomers. Despite this substantial advancement, the 84 bimolecular diaryl TSs (TS0 in **Figure 2b**) remained elusive (see **The Scope of Static DFT and PoTS Performance)**. Nonetheless, the gas-phase DFT results highlight that the key mechanistic pathways are sufficiently represented by these TS, as the rate-determining steps consistently involve proton transfers between aromatic rings (see **S2.2**).

The 42 distinct reaction pathways identified for each zeolite and active site by our automated workflow, depicted in **Figure 3**, enable a comprehensive analysis of the diaryl mechanism. Moreover, the free energy barriers at the experimental temperature 513 K (20–160 kJ/mol, with disproportionation barriers exceeding 200 kJ/mol in BOG) align with earlier theoretical findings for this reaction in more limited studies.[42]

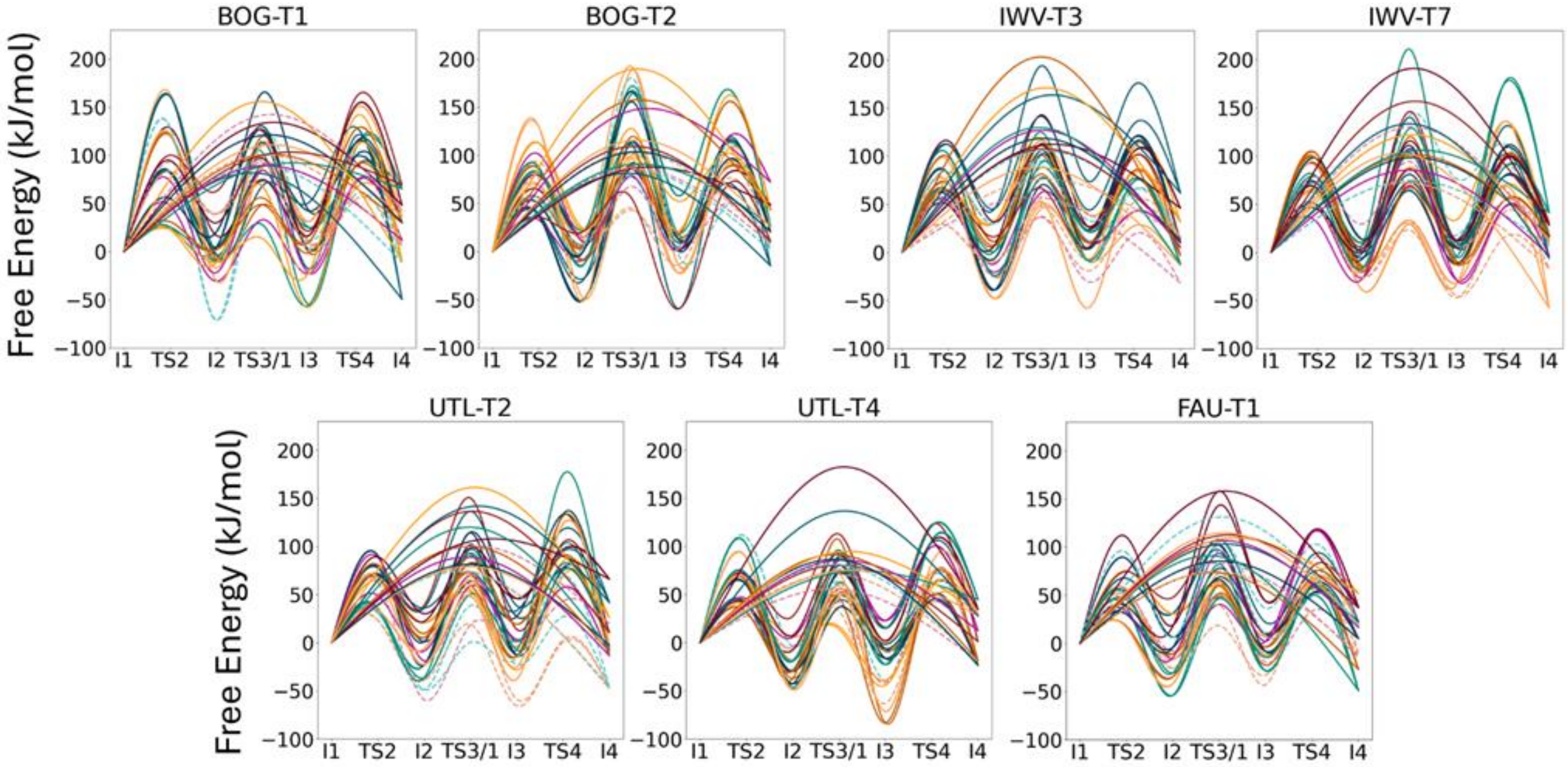

**Figure 3.** The 42 mechanistic pathways identified per zeolite system, comprising 12 direct and 30 multistep paths, with colors matching **Figure 2b**. Blue shades denote I1$^+$ intermediates derived from ortho-DEB$^+$, orange from meta-DEB$^+$, and burgundy from para-DEB$^+$. Dashed paths correspond to transalkylation; solid paths correspond to disproportionation.

Reaction pathways were systematically evaluated for each zeolite and active site at the isomer-specific level. For each isomer pair, the activation barrier of the direct mechanism (TS1) was compared with the MEP of the matching multistep route (TS2, TS3, TS4), defined as the lowest rate-determining step (RDS) (see **S2.3**). The pathway with the lowest overall barrier is expected to be indicative for feasibility of the diaryl cycle. Finally, isomer-specific minimum energy pathway (MEP) were examined across reactions (columns in **Figure 4**) and zeolite active sites (rows in **Figure 4**), with the lowest energy path highlighted with a black box and a gold star.

The lowest RDS barriers for transalkylation reveal zeolite-dependent trends (**Figure 4**). IWV with IWV-T3 (intersection with 8.54 $Å^3$ volume) and IWV-T7 (channel 6.9×6.2 $Å^2$ cross-section) exhibiting low barriers for *para*-DEB$^+$ + benzene (57 kJ/mol, direct) and meta-DEB + benzene (60 kJ/mol, multistep) transalkylations, highlighting an effective 12MR-2D topology. UTL similarly shows low barriers (56–58 kJ/mol) in its 14MR channels (UTL-T4, 9.5×7.1 Å), but higher barriers (66 kJ/mol) at the 14×12MR intersection (UTL-T2, 9.3 $Å^3$). FAU-T1 cavity, 11.24$Å^3$, closely mirrors IWV for para-DEB$^+$ (57 kJ/mol), but its ortho and meta isomers exhibit substantially higher barriers (94 and 75 kJ/mol). By contrast, BOG is least favorable (76–93 kJ/mol), as its narrow 10MR channels, 5.8×5.5 Å, favor the alkyl transfer mechanism and the 12×10 intersections, 8.03 Å$^3$, and monodimensional 12MR channels, 7.0x7.0 Å, are too constrained for the diaryl cycle to proceed. These observations reflect tight confinement effects for BOG (see **S2.4**) and align with our previous experimental findings,[42] who reported a low transalkylation rate of 349 $mol_{EB}/(mol_{acid}{\cdot}h)$, an ethene yield of 4.8%, and a TEB yield of 4.4% in BOG. These data confirm the participation of alkyl transfer mechanism on the overall reaction. Therefore, we exclude BOG from further disproportionation analysis, which involves even bulkier diarylic intermediates.

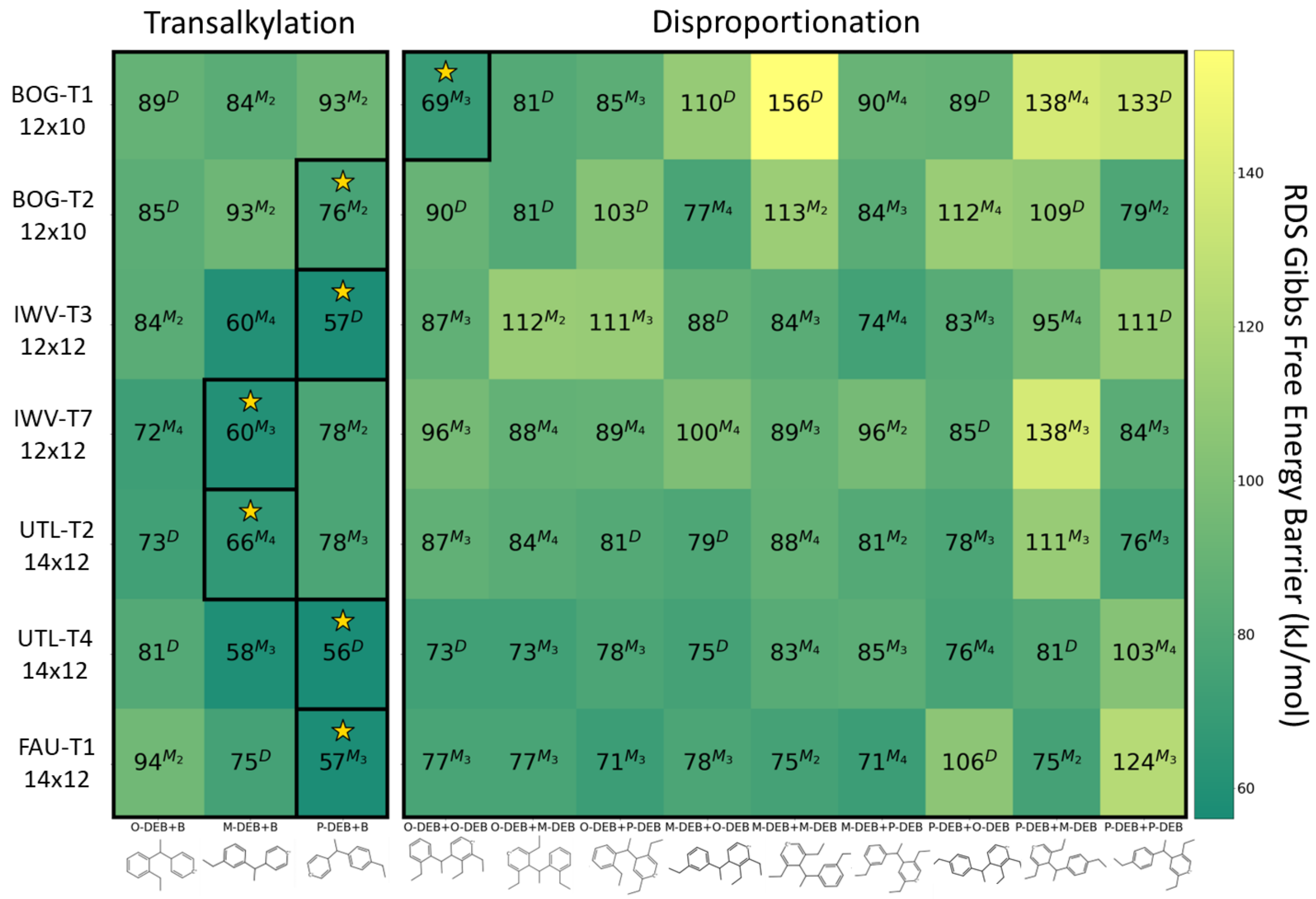

**Figure 4**. Heat map showing the free energy barriers at 513 K for the rate-determining step (RDS) of each DEB reaction across the investigated zeolite frameworks. Superscript D denotes the direct pathway (via TS1), while M indicates a multistep pathway; M2 identifies TS2 as the RDS, M3 identifies TS3, and M4 identifies TS4. Each row corresponds to a specific zeolite active site, with transalkylation (left) and disproportionation (right) barriers shown side by side. Complete energy profiles for each framework are presented in **S2.3**, full dataset provided in **Table S17–18**.

Disproportionation shows a distinct energy landscape, with all zeolites displaying higher RDS barriers than for transalkylation. With BOG excluded from the disproportionation analysis, the highlighted barrier for BOG-T1 (69 kJ/mol) appears artificial, reflecting limited stabilization of bulky diaryl reactants relative to smaller transalkylation species. In IWV, nearly all RDS exceed 83 kJ/mol due poor accommodation of bulky intermediates. Experimental data from past work confirm IWV's strong selectivity for transalkylation with the highest transalkylation rate observed (1926 $mol_{EB}/(mol_{acid}{\cdot}h)$, 94.1% EB, 0.5% TEB).[46] UTL displays consistently lower disproportionation barriers, particularly in its wide 14MR channels, (UTL-T4, 9.5x7.1 Å), where five pathways fall within 73–78 kJ/mol. Notably, 8 of the 9 lowest barriers across all zeolites occur in UTL. Scaled pathways (see **S2.5**) indicate that UTL-T2, located at the 14×12 MR intersection (9.3 $Å^3$), provides the most thermodynamically favorable routes. There results highlight UTL's capacity to accommodate bulky intermediates via its large 2D intersections and channels, consistent with experimental data showing non-existent ethene yield (0.0%) and enhanced TEB production that reduces UTL transalkylation rate when compared to IWV's (1599 $mol_{EB}/(mol_{acid}\cdot h)$, 85.7% EB, 12.0% TEB).[46] FAU shows moderate disproportionation activity, with 7 of 9 RDS falling between 71–78 kJ/mol. Its large 11.2 $Å^3$ cavity facilitates bulky intermediate accommodation, though its loose confinement corresponds to a lower transalkylation ratio observed in previous studies (1075 $mol_{EB}/(mol_{acid}\cdot h)$).[46] estimates of confinement effects confirm that UTL and FAU can accommodate bulky intermediates by providing up to ~20 $Å^2$ free pore area (see **S2.4**), while BOG and IWV impose stronger confinement over disproportionation diaryl intermediates. See **Figure 5** for a depiction of the RDS TS structures identified.

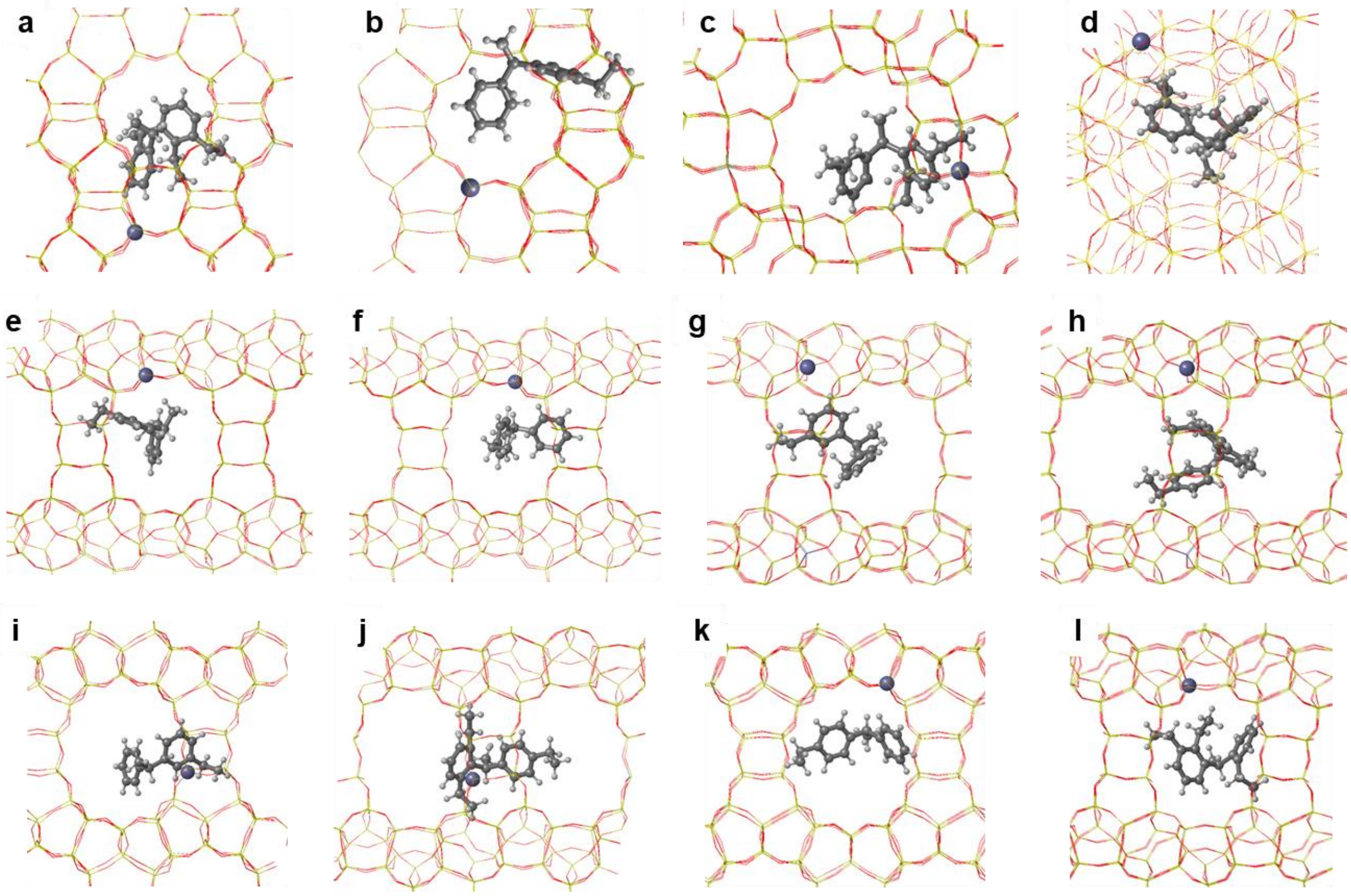

**Figure 5.** TS structures obtained through the PoTS pipeline. Corresponding barriers at T = 513.15 K on **Figure 4** to the TS structures shown in this figure are **a** BOG-T1, TS3, ortho-ortho disproportionation, 69 kJ/mol **b** BOG-T2, TS2, para-benzene transalkylation, 76 kJ/mol **c** FAU-T1, TS3, meta-para disproportionation 71kJ/mol **d** FAU-T1, TS3, meta-ortho disproportionation 78 kJ/mol **e** IWV-T3, TS4, meta-benzene transalkylation, 60 kJ/mol **f** IWV-T3, TS1, para-benzene transalkylation, 57 kJ/mol **g** IWV-T7, TS3, meta-benzene transalkylation, 57 kJ/mol **h** IWV-T7, TS3, para-meta disproportionation, 138 kJ/mol **i** UTL-T2, TS4, meta-benzene transalkylation, 66 kJ/mol **j** UTL-T2, TS3, para-meta disproportionation, 111 kJ/mol **k** UTL-T4, TS1, para-benzene transalkylation, 56 kJ/mol **l** UTL-T4, TS1, ortho-ortho disproportionation, 73 kJ/mol.

Experimental confirmation of DEB isomer behavior in these reactions is challenging, as isomerization within the zeolite can obscure the actual isomer distribution. As a proof of concept of PoTS capabilities, our dataset enabled an isomeric comparison of transalkylation and disproportionation mechanisms, detailed in **S2.5**. All pathways were rescaled to each zeolite's minimum intermediate I1, focusing on thermodynamically favored cases ($\Delta G_{I4\text{–}I1} < 0$). For transalkylation and disproportionation, *meta* and *para* pathways were found to be consistently more stable across all frameworks than *ortho*-$DEB^+$. Coupled with the higher RDS barriers observed in **Figure 4** for this isomer, these findings identify ortho-$DEB^+$ as the least favoured $DEB^+$. These trends match gas-phase results (**S2.1**). While gas-phase calculations favor multistep mechanisms, zeolite confinement stabilizes some direct TS1 routes, 22 of 84 RDS in **Figure 4**, though multistep paths generally remain preferred due to smoother six-membered TS geometries.

## Cyclohexene Skeletal Isomerization

Biomass-derived naphtenes have been proposed as a renewable alternative to fossil fuels, with zeolite-catalyzed hydroisomerization optimizing biofuel production.[34,43] Zeolites enable naphthenes selective skeletal isomerization by protonating a paraffin π bond to form a carbocation, which, if not already generated during the protonation, typically rearranges to a tertiary carbon through HS. This intermediate enables a range of conversion pathways, including cracking, and ring contraction (RC) isomerization through a protonated cyclopropane-like (PCP) structure.[44,45]

This section focuses on RC TSs to further showcase PoTS capabilities. Specifically, we examine three ethylcyclohexyl carbocations ($ECH1^+$, $ECH2^+$, and $ECH3^+$) formed by protonation of 1-ethylcyclohex-1-ene and 1-ethylcyclohex-2-ene, focusing on their rearrangements via HS1 and HS2 and different RC pathways (TSA, TSB, TSC, and TSD), as illustrated in **Figure 6a** and **S3.1**. Additionally, we explore how methyl substitution in 1-ethyl-3-methylcyclohexene ($EMCH1^+$) and 1-ethyl-(3,5)-dimethylcyclohexene ($EDMCH1^+$) influences these mechanistic pathways (highlighted with red and blue sticks in **Figure 6a**). Zeolites from the "Big Five" group were selected for their industrial relevance, stability, and diverse active sites. These include FER-T1 (10×8), MOR-T2 (12×8), BEA-T1a/b/T2 (12×12×12), and MFI-T4/T10/T11 (12×12×12) (see **Method**). FAU-T1 was used as in **Diethylbenzene Transalkylation**.

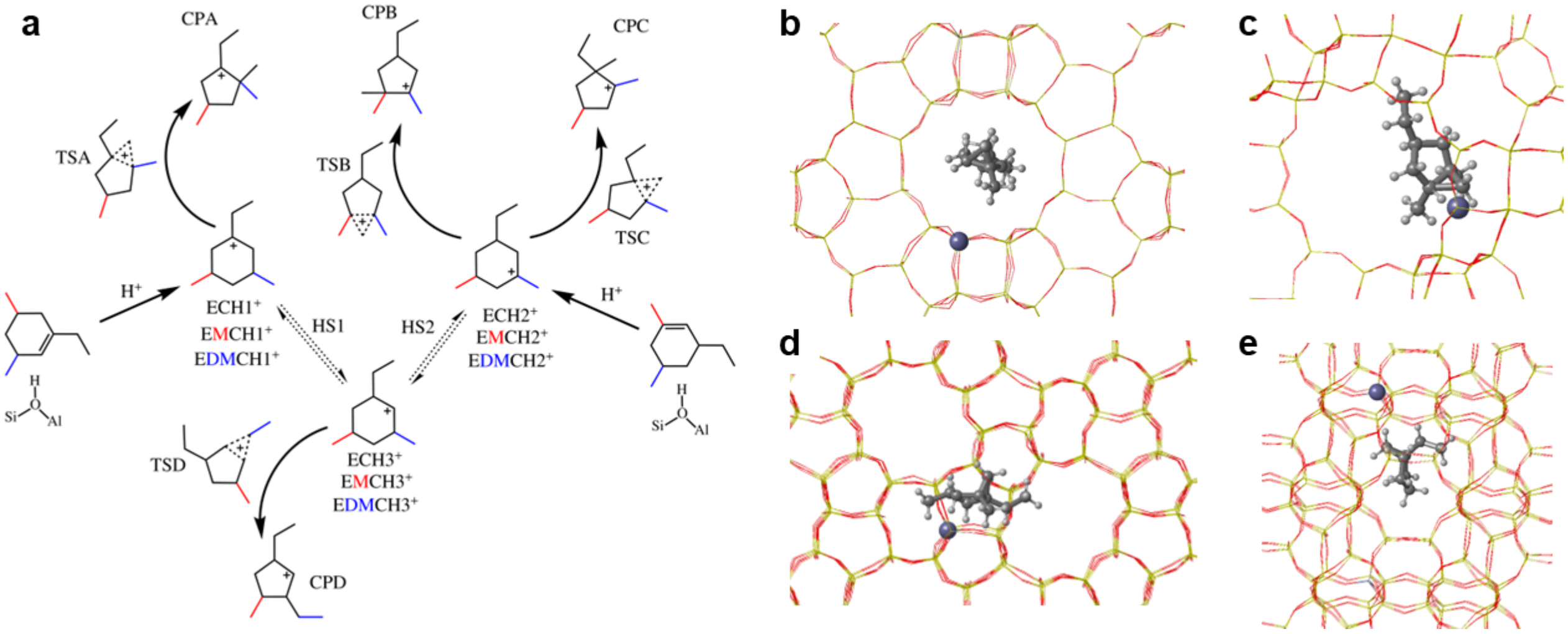

**Figure 6. a** Ring contraction skeletal isomerizations through TSA, TSB, TSC and TSD, studied for 1-ethyl-1-cyclohexyl ($ECH^+$), 1-ethyl-3-methyl-cyclohexyl ($EMCH^+$) (blue sticks), and 1-ethyl-(3,5)-cyclohexyl ($EDMCH^+$) (blue and red sticks). TS structures obtained through the PoTS pipeline for ring contraction: **b** BEA-T2 $EMCH^+$ TSA, **c** FAU-T1 $EDMCH^+$ TSB, **d** MFI-T4 $EDMCH^+$ TSC, and **e** MFI-T11 $ECH^+$ TSA.

The gas-phase study in **S3.2** indicates that although the secondary ethylcyclohexyl carbocation ($ECH3^+$) has lower ring-contraction barriers than its tertiary counterparts ($ECH1^+$ and $ECH2^+$), it is intrinsically less stable and thus short-lived. In zeolites, this instability is amplified, as secondary carbocations rapidly rearrange into more stable tertiary forms via hydrogen or methyl shifts. This means that in DFT calculations, these metastable species cannot be "isolated," as the isomerization happens so quickly that the optimizer converges directly to the thermodynamic product unless artificial constraints are imposed. Thus, although RC TS were located, secondary-carbocation pathways were excluded from kinetic analysis due to the lack of well-defined reactant states and unreliable barrier estimates. Moreover, no HS1 or HS2 (see **Figure 6a**) were identified. Nevertheless, the converged secondary carbocation RC structures are provided in **S3.3**.

We computed Gibbs free energy barriers at 550.15 K, based on prior experimental and theoretical studies.[44,45] For RC via TSA of $ECH1^+$ (**Table 1**), our results align with those of Gutierrez-Acebo et al.,[45] who reported barriers of 59–107 kJ/mol at 550 K using PBE-D2. Comparing topologies, 10MR channel zeolites generally yield lower barriers, such as FER-T1 (5.4x4.2 Å channels) (112 kJ/mol) and especially MFI-T11 (6.4 $Å^3$ intersection) (100 kJ/mol). BEA, despite their larger, 12MR 3D pore systems, 5.9x5.9 Å channels and 6.59 $Å^3$ intersections, also show relatively low barriers (T1a: 118; T1b: 121; T2: 119 kJ/mol). In contrast, frameworks with wider cavities or straight 12MR channels, such as FAU-T1 (11.2 $Å^3$ cavity) (143 kJ/mol) and MOR-T2 (7.0x6.5 Å channel) (147 kJ/mol), exhibit higher barriers due to weaker TS stabilization from looser confinement.

**Table 1.** Gibbs Free Energies (ΔG, $kJ/mol^{-1}$) at 550.15 K for ring contraction of ethyl-substituted cyclohexyl cations in FAU, FER, MOR, BEA, and MFI zeolite frameworks. $\Delta G_{TS-R}$ represents the Gibbs free energy difference between TS and reactant (R). $\Delta G_{P-R}$ represents Gibbs free energy difference between product (P) and R.

| Path | | FAU T1 | FER T1 | MOR T2 | BEA T1a | BEA T1b | BEA T2 | MFI T4 | MFI T10 | MFI T11 |
|---|---|---|---|---|---|---|---|---|---|---|
| **1-Ethyl-Cyclohexyl Cation Ring Contraction** | | | | | | | | | | |
| **A ($ECH1^+$)** | $\Delta G_{TS-R}$ | 143 | 112 | 147 | 118 | 121 | 119 | 140 | 115 | 100 |
| | $\Delta G_{P-R}$ | 5 | -19 | -5 | 23 | 11 | -3 | 19 | -22 | -11 |
| **1-Ethyl-3-MethylCyclohexyl Cations Isomers Ring Contraction** | | | | | | | | | | |
| **A ($EMCH1^+$)** | $\Delta G_{TS-R}$ | 115 | 159 | 137 | 121 | 124 | 118 | 161 | 111 | 111 |
| | $\Delta G_{P-R}$ | -21 | 56 | -4 | 4 | 17 | 19 | 93 | -23 | 9 |
| **B ($EMCH2^+$)** | $\Delta G_{TS-R}$ | 106 | 176 | 167 | 103 | 102 | 95 | 150 | 121 | 124 |
| | $\Delta G_{P-R}$ | -3 | 5 | -21 | -4 | 0 | -11 | 43 | -14 | 51 |
| **C ($EMCH2^+$)** | $\Delta G_{TS-R}$ | 127 | 175 | 150 | 143 | 142 | 138 | 142 | 120 | 121 |
| | $\Delta G_{P-R}$ | -7 | 98 | -3 | 68 | -2 | 37 | 26 | -4 | 38 |
| **1-Ethyl-3,5-DimethylCyclohexyl Cations Isomers Ring Contraction** | | | | | | | | | | |
| **A ($EDMCH1^+$)** | $\Delta G_{TS-R}$ | 124 | 225 | 124 | 125 | 132 | 134 | 35 | 120 | 141 |
| | $\Delta G_{P-R}$ | 10 | 30 | 15 | 8 | 7 | 16 | -92 | -36 | -20 |
| **B ($EDMCH2^+$)** | $\Delta G_{TS-R}$ | 90 | 57 | 161 | 101 | 96 | 100 | 53 | 129 | 136 |
| | $\Delta G_{P-R}$ | 0 | -28 | 37 | -16 | -3 | 4 | -84 | 67 | -25 |
| **C ($EDMCH2^+$)** | $\Delta G_{TS-R}$ | 119 | 105 | 162 | 147 | 159 | 154 | 97 | 121 | 151 |
| | $\Delta G_{P-R}$ | 6 | -69 | 12 | -8 | 70 | 8 | -80 | 26 | 33 |

Methyl substitution on the cyclohexyl ring allows us to probe its effect on RC barriers (TSA, TSB, TSC) in $EMCH^+$ and $EDMCH^+$ cations (**Figure 6a**). One-dimensional zeolites with straight channels can generally be excluded due to either prohibitively high barriers (e.g., 225 kJ/mol for TSA in $EDMCH1^+$ in FER-T1; 167 kJ/mol for TSB in $EMCH1^+$ in MOR-T2) or artificially low ones (e.g., 57 kJ/mol for TSB in $EDMCH2^+$ in FER), caused by reactant destabilization as bulky intermediates and TSs outgrow the channel. A similar distortion is seen in MFI-T4, where Al sits in the narrow sinusoidal 5.1x5.5 Å channel, leading to irregular barriers compared to MFI-T10 and MFI-T11 (Al at intersections), which yield more consistent results (see **S3.4**). As molecular size increases, RC catalysis becomes restricted to bulkier cavities like FAU and multidimensional 12MR frameworks such as BEA. These frameworks consistently yield lower TSB barriers, 106 kJ/mol and 90 kJ/mol for $EMCH2^+$ and $EDMCH2^+$ in FAU, and 95–103 kJ/mol and 96–101 kJ/mol in BEA, due to the absence of the bulky ethyl group present in TSA and TSC (**Figure 6**). Additionally, TSB products tend to have near-neutral or favorable $\Delta G_{P-R}$ values, unlike TSA or TSC in BEA and FAU, making TSB the most favorable RC pathway for $EMCH2^+$ and $EDMCH2^+$.

Coupling these theoretical trends with experimental data from Cerqueira et al.[44] on methylcyclohexane hydroisomerization in FAU, BEA, and MFI reveals clear selectivity differences. MFI shows a cracking-to-isomerization ratio exceeding 17.1, while BEA and FAU exhibit much lower values (2.94–3.94 and 1.45–2.03, respectively). These results rule out MFI as a viable RC catalyst, despite low RC barriers in MFI-T10 and MFI-T11, its narrow 10MR (5.1x5.5 and 5.3x5.6) pores favor cracking over RC. A similar behaviour should be expected from FER, 10MR 5.4x4.2 Å. In contrast, while BEA and FAU display comparable RC barriers, FAU's looser confinement more effectively destabilizes cracking TSs, reducing its cracking selectivity. Thus, wide-pore zeolites like FAU may offer a better balance for promoting isomerization over cracking, even without significantly lower RC barriers.

Finally, we acknowledge that catalytic cracking is an important competing reaction to RC that is not included in this section, but cracking was found to be not amenable to PoTS (see below).

## Alkene Cracking

Zeolite-catalyzed alkene cracking is essential for fuel production and increasing light olefin yields from $C_4$–$C_8$ alkenes, despite its high energy requirements.[36,37] The industrial process entails oligomerization, isomerization, hydrogen transfer, and cracking, with β-scission of carbenium ions as the dominant mechanism. This β-scission produces free alkenes and smaller reactive ions, while rapid isomerization via hydrogen and methyl shifts forms carbenium ions with varying stabilities.[46,47]

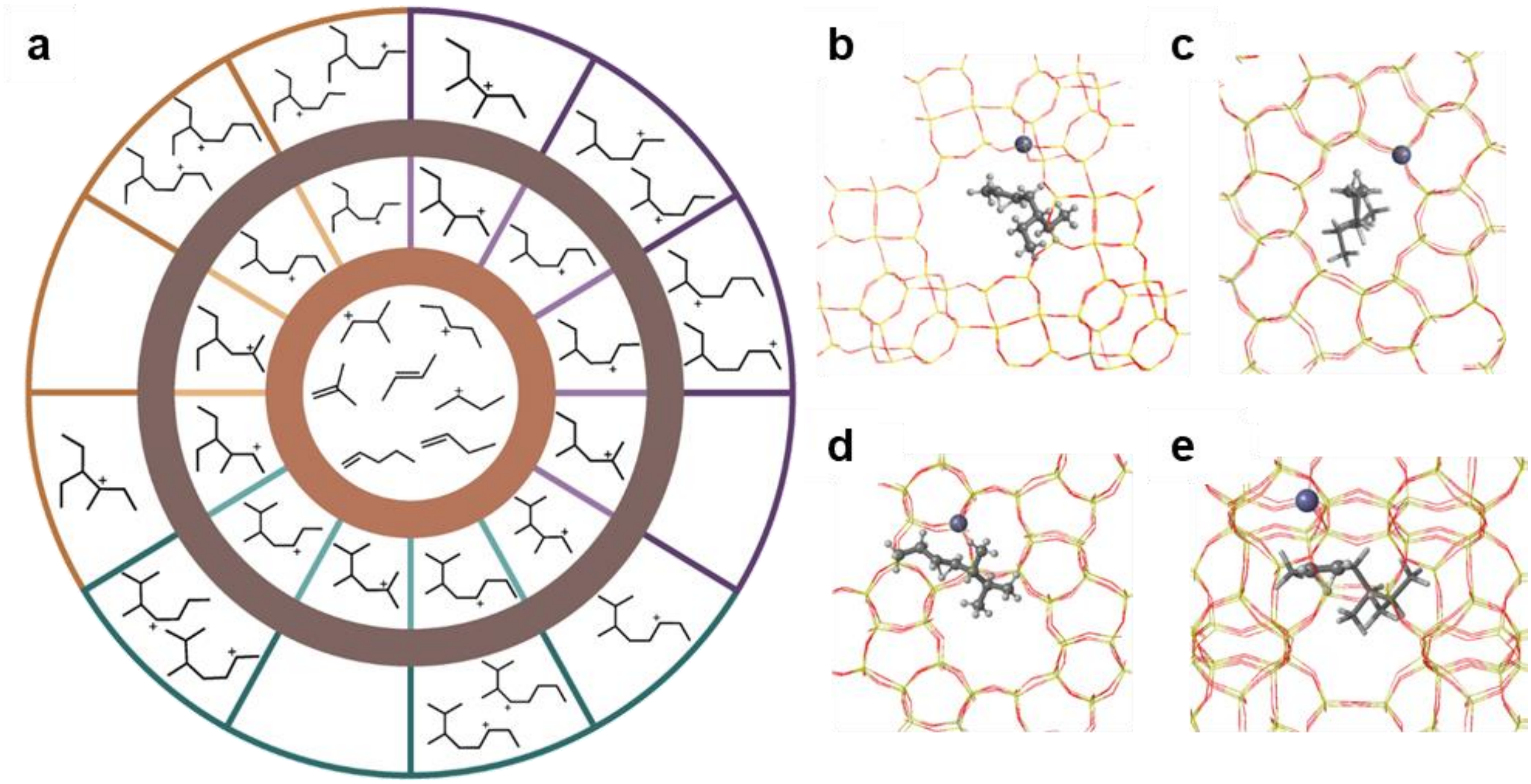

**Figure 7. a** Circular schematic illustrating the cracking and hydrogen-shift reactions examined in this study. The light-brown ring highlights the key carbocations, 2-butyl, 3-pentyl, and 2-methyl-3-butyl, that react with neutral alkenes, 1-butene, 2-butene, isobutene, and 1-pentene. The dark-brown ring encloses the resultant $C_8$–$C_{10}$ molecules; purple, teal, and beige sectors denote products from 2-butyl, 2-methyl-3-butyl, and 3-pentyl carbocations, respectively. The outermost ring depicts the additional molecules produced via secondary-to-secondary carbocation hydrogen shifts investigated in this work. Hydrogen shift TS structures found for **b** FAU-T1 CC[CH+]CC(CC)CC **c** MFI-T11 CCCC[CH+]C(CC)CC **d** MFI-T4 CC[CH+]CC(C)C(C)C **e** MFI-T11 CC[CH+]CC(C)C(C)C.

In this section, we examine β-scission and internal HS in alkene cracking by pairing 1-butene, 2-butene, isobutene, or 1-pentene with 2-butyl, 3-pentyl, or 3-methyl-2-butyl carbocations, yielding twelve $C_8$–$C_{10}$ species and ten distinct secondary-to-secondary hydrogen shifts (HS) pathways (**Figure 7a** and **S4**). We use the same set of zeolites described in **Alkene Cracking**. Zeolites from the "Big Five" group were selected for their industrial relevance, stability, and diverse active sites. These include FER-T1 (10×8), MOR-T2 (12×8), BEA-T1a/b/T2 (12×12×12), and MFI-T4/T10/T11 (12×12×12) (see **Method**). FAU-T1 was used as in **Diethylbenzene Transalkylation**. Bimolecular minima and TS structures were generated using the algorithm described in **Method**.

Because of limitations of static DFT discussed below in **The Scope of Static DFT and PoTS Performance**, our PoTS pipeline identified only four HS, three in MFI and one in FAU (**Figure 7b**), and no bimolecular cracking TS converged. In the FAU cavity, the free energy barrier is 50 kJ/mol, whereas in MFI it ranges from 85 to 137 kJ/mol. This stark contrast arises because FAU's wider pores accommodate easily the long $C_8$–$C_{10}$ chains, while MFI's narrower channels, particularly in the MFI-T4 with Al in the sinusoidal channels, constrain the longer $C_8$–$C_{10}$ chains, increasing the barrier, (**see S4**).

## The Scope of Static DFT and PoTS Performance

Through this work, the PoTS pipeline was applied to a total 1004 DFT-D3 TS searches in 15 distinct active site positions across nine zeolite frameworks, successfully converging 672 of them. To our knowledge, this constitutes the most extensive DFT-level dataset of zeolite-confined TS structures reported to date. However, two clear accuracy limits emerged: difficulties in locating bimolecular TS, and challenges associated with identifying HS TSs in some carbocation rearrangements. Identifying bimolecular TS is challenging because they are very dynamic. Static DFT struggles to capture how subtle structural changes in transient molecular configurations affect energies.[48] Similarly, HS reactions present a very flat PES, complicating the identification of valid negative-curvature TS, especially with GGA-based functionals such as PBE-D3, which also suffer from delocalization errors and incomplete hydrogen-bonding treatment.[49,50]

The large dataset of successfully converged TS highlights the robustness and broad applicability of the PoTS pipeline, delivering comprehensive mechanistic insights. Furthermore, the scale and reproducibility of the simulations indicate the boundaries of transition state theory, which relies on singular, static geometries to explain chemical reactivity, compared with other approaches that better include dynamic effects; and the choice of exchange-correlation functionals to calculate energies and forces (and Hessians) on the PES at the DFT level of theory. Therefore, this work benchmarks where complementary approaches using dynamic TS searches or hybrid functionals are needed.[51]

## Conclusions

We have developed an automated computational pipeline that bridges gas-phase reaction mechanism exploration with periodic DFT-D3 transition state identification within zeolite frameworks. This integrated approach employs a modified Monte Carlo docking algorithm and the PoTS method, which together enable the generation of realistic starting geometries and the seamless alignment of gas-phase TS structures to their confined counterparts. By leveraging single-ended DFT searches via the Dimer method, our pipeline circumvents the need for computationally expensive double-ended methods, facilitating high-throughput TS identification.

Applied to zeolite-catalyzed transalkylation of diethylbenzene, cycloalkene skeletal isomerization, and alkene cracking, PoTS proved robust, identifying 672 TS structures out of 1004 attempts across nine zeolite frameworks. Notably, our approach facilitated detailed mechanistic insights into diethylbenzene transalkylation pathways, identifying IWV as a particularly effective catalyst due to its favorable pore dimensions. The pipeline also highlighted the significant effects of zeolite confinement on reaction energetics.

For cyclohexene skeletal isomerization, PoTS identified key ring-contraction pathways and revealed the limitations imposed by pore topology. One-dimensional zeolites like FER and MOR, as well as narrow 10MR channels in MFI, were found unsuitable for RC catalysis. In contrast, multidimensional 12MR frameworks such as BEA and FAU consistently provided lower barriers and allowed stable TS formation. Experimental cracking-to-isomerization ratios further support these trends, showing significantly reduced cracking in BEA and FAU compared to MFI. These findings highlight FAU's looser confinement as a promising feature for suppressing cracking while maintaining viable RC reactivity.

The resulting PoTS dataset represents the largest DFT-level exploration of zeolite-confined transition states to date, enabling both mechanistic discovery and benchmarking of TS search methodologies under confinement. Overall, our study establishes a robust framework for mechanistic exploration in zeolite catalysis, enhancing our understanding of reaction pathways and defining the operational limits of current computational methodologies.

# Method

## Pore Transition State Finder Pipeline (PoTS)

**Conformer generation:** Simplified Molecular Input Line Entry System (SMILES)[52] strings for the molecules were generated in house according to the reactions under study. Conformers were generated using the distance-geometry approach ETKDGv35[53] and relaxed with the MMFF94[54] force field as implemented in RDKit,[55] after explicit enumeration of all stereoisomers for each molecule. Followed by optimization with GFN2-xTB.[56]
Bimolecular reaction poses were generated using an algorithm that positions the reactive indices of the reactants in proximity. The process begins by centering each reactant at its center of mass and applying a random rotation to randomize their orientations. To position the reactants near their reactive indices, the algorithm identifies a random point near each selected index. It does so by generating a random normalized vector, and evaluating potential crowding by projecting the vector onto the displacement vector between the index and nearby points. If the projection is less than the minimum distance (0.75Å) , the algorithm tries a new direction. Once a suitable direction is found, the vector is scaled to a random distance within a pre-defined range of 0.75Å to 2Å from the index. The algorithm then computes a rotation matrix (aligner) to align the two points. This matrix is applied to rotate and translate the reactants, ensuring their reactive sites are positioned appropriately for interaction.

**DFT gas-phase:** The two lowest energy conformers were further refined using Density Functional Theory (DFT) in ORCA[57] with the M06-2X functional[58] and the def2-SVP basis set.[59] Reaction enumeration was performed using the SMILES Arbitrary Target Specification (SMARTS) language to encode reactive sites.[60] The positions of all C and H atoms in the cationic intermediates and transition states (TS) were fully optimized with eigenvector-following.[61] Constraints on the bond angles and distances were applied to guide the reaction mechanism and prevent undesired interactions when needed. All stationary points were characterized by means of harmonic frequency calculations at the same theory level. Intrinsic reaction coordinate (IRC) calculations were performed to confirm the transition state found through generation of the mass-weighted minimum energy path from the TS.[62]

**Docking:** Generation of TS-zeolite and intermediate-zeolite poses was performed using a new modification in the Monte Carlo (MC) docking algorithm implemented in the VOID package.[63] Only one molecule was loaded inside each zeolite framework either with its minimum or with its TS geometry. The modified Monte Carlo docking algorithm employs two reward functions to ensure realistic placement of the docked molecule within the zeolite framework. The first reward function guarantees that the cationic atom in the docked structure is positioned within a 3.5 Å threshold from the acidic site of the zeolite, specifically one of the four oxygen atoms bonded to aluminum. The second reward function enforces a steric constraint, ensuring that no atom from the docked molecule is closer than 1.5 Å to any atom in the zeolite framework. Both conditions must be satisfied for the docking configuration to be accepted. The combined reward function R is given by:

$$eq.1) \quad R = \{(1\ if\ any\ d_{cat} \leq\ t_{cat}\ and\ all\ d_{col} \geq\ t_{col})\ 0\ otherwise\}$$

Where $d_{cat}$ represents the distance between the cationic atom and the nearest acidic oxygen atom in the framework, $d_{col}$ represents each distance between each docked molecule atom and each framework atom. And $t_{cat}$ and $t_{col}$ are the cationic and steric constraint thresholds mentioned previously.

The modified MC docking algorithm ensured more realistic catalytic poses as outputs of this process. **Figure M1 a)** illustrates the kind of poses a random MC algorithm might produce when the distance between the cationic part of the molecule and the anionic part of the framework is not monitored, ending up in too large distances. Meanwhile, the poses generated by the modified MC algorithm rewarding the distance between the guest cation and the framework anion are more suited to study chemical reactivity, see **Figure M1 b).**

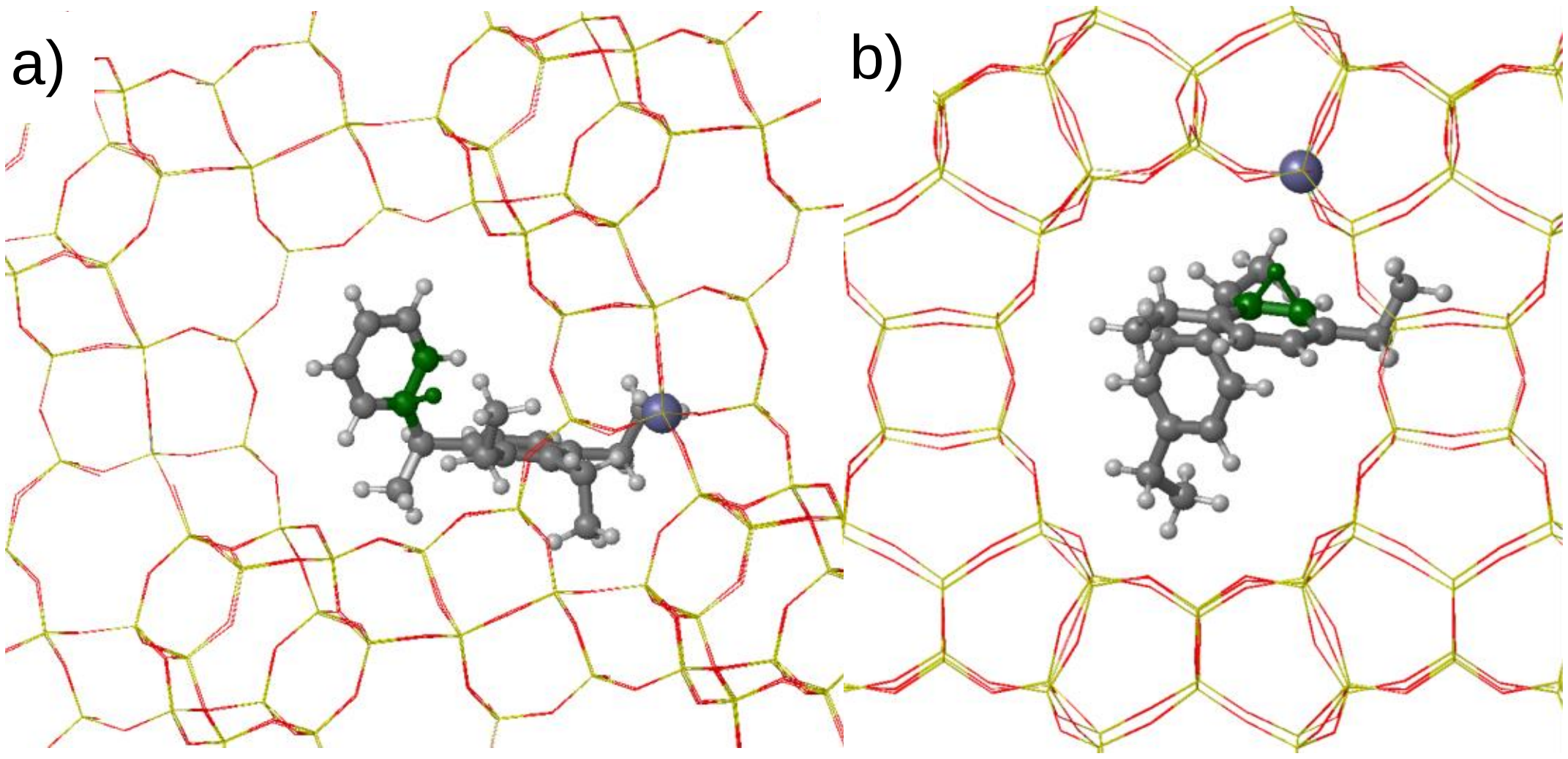

**Figure M1**. **a)** Docked pose obtained using the random MC algorithm, where the cationic center (green) in the guest molecule is positioned 8.20 Å from the zeolite acid site. **b)** Docked pose obtained using the modified MC algorithm, reducing the cationic center–acid site distance to 2.85 Å.

A public version for the code used to produce the TS-zeolite and Intermediate-zeolite poses presented in this work can be found on the Learning Matter group GitHub repository https://github.com/learningmatter-mit/VOID

**PoTS, Gas to Zeolite Vibrational Mode Adaptation:** First, the coordinates and vibrational displacements associated with the imaginary frequency of the gas-phase TS are adjusted to fit within the periodic unit cell. Then, the gas-phase TS geometry with its counterpart in the periodic system are aligned, the coordinates of the docked molecule $rdock$ are identified and isolated from the periodic structure. A rotation matrix $R$ is generated by translating both the gas-phase TS and the docked molecule to a common origin. The covariance matrix $H$ between the two geometries is computed as:

$$eq.2)\quad H = \sum_i (r_{gas,i} - c_{gas})(r_{dock,i} - c_{dock})^T$$

where $c_{gas}$ and $c_{dock}$ are the centroids of the gas-phase and docked geometries respectively. The rotation matrix $R$ is then obtained using Singular Value Decomposition (SVD) of $H$:

$$eq.3)\quad H = U\sum V^T$$

$$eq.4)\quad R = VU^T$$

Next, the vibrational displacement vectors $v_{frac}$ from the gas-phase TS are rotated using this matrix to align them with the docked geometry:

$$eq.5)\quad v_{rot} = R \cdot v_{frac}$$

This step ensures that the directions of the vibrational modes are properly aligned with the docked geometry in the periodic system.

After aligning, the atomic coordinates and vibrational modes are sorted alphabetically by element to match the format used in periodic calculations. The final adjusted vibrational modes $v_{adjusted}$ are then compiled into a MODECAR file for Dimer calculations in VASP. For framework atoms like Si, O, and Al, small dummy displacement vectors $v_{dummy}$ are used to focus the vibrational analysis on the adsorbed molecule:

$$eq.6)\quad v_{dummy} = [0.001, 0.001, 0.001]$$

This process ensures that the vibrational modes are transferred appropriately from the gas-phase to the periodic system, preserving their physical relevance in the solid-state environment.

A public version for the code used to adapt the gas-phase TS vibrational frequencies to zeolite unit cells and docked poses rotations in this work can be found on the Learning Matter GitHub repository https://github.com/learningmatter-mit/PoTS

**DFT in Periodic Solids:** Periodic DFT calculations were performed using the Perdew-Burke-Ernzerhof (PBE) exchange-correlation functional within the generalized gradient approach (GGA)[64,65], as implemented in the Vienna Ab initio Simulation Package (VASP) code.[66] The valence density was expanded in a plane wave basis set with a kinetic energy cutoff of 600 eV, and the effect of the core electrons in the valence density was taken into account by means of the projected augmented wave (PAW) formalism.[67] Integration in the reciprocal space was carried out at the Γ k-point of the Brillouin zone. Dispersion corrections to the energies were evaluated using the D3 Grimme's method.[68,69] Electronic energies were converged to $10^{-6}$ eV and geometries were optimized until forces on atoms were smaller than 0.01 eV/Å. Transition states were obtained using the DIMER algorithm.[70,71] During geometry optimizations, the positions of all atoms in the system were allowed to relax without restrictions.

## Zeolites

As this study focuses on chemical reactivity, we introduced an acid site in each framework by substituting a Si atom with an Al atom, forming an $AlO4^{-}$ unit to compensate for the positive charge introduced by the reactive cation. The T-site positions in each framework were chosen based on the intrinsic stability of an empty cell with a background negative charge at the PBE-D3 theory level. Subsequently, the final positioning was determined according to the specific topology of each framework. Images and a more detailed description for each model can be seen in **S1 Models** section.

### BEA

In this work, we used a chiral polymorph-A BEA model featuring a three-dimensional network of 12-membered-ring (12MR) channels (5.6–7.3 Å), described by a tetragonal unit cell (a = b = 12.593 Å, c = 26.495 Å, α = β = γ = 90°) with 64 Si and 128 O atoms. The BEA framework contains nine T-sites, with T1 and T2 being the most stable crystallographic positions for single Al substitution. Because the 12MR channels exhibit minimal differences between T1 and T2, we constructed three Si/Al=63 models to capture all relevant topological variations: two with Al in T1 or T2 along distinct 12MR channels, and one with Al placed at a T1 site near the 12MR intersection.

### BOG

The Boggsite (BOG) framework features 12MR channels (7.0 Å aperture) running along the a-axis, each with offset 10MR windows (5.8 × 5.2 Å) branching into channels parallel to the b-axis. For modeling, we used a smaller primitive unit cell obtained via a transformation matrix that halves the original volume, yielding triclinic symmetry (P1) with parameters a = 12.812 Å, b = c = 16.791 Å, α = 74°, β = 68°, γ = 68°, and containing 48 Si and 96 O atoms. BOG has six T-sites, with T1 and T2 being the most stable for single Al placement. Because BOG presents a bidimensional 12MR × 10MR channel system and T1/T2 differ by only ~1 kJ/mol, we created two Si/Al=47 models placing Al in a 12×10MR intersection at distinct positions within the cell, capturing the main topological variations.

### FAU

Faujasite (FAU) consists of sodalite cages connected in a cubic arrangement via six-membered double rings (d6R), forming intersecting 12MR channels (7.4 Å aperture) along [111]. We modeled FAU using a smaller primitive unit cell (one-quarter the original volume) obtained by the transformation matrix 1/2a+1/2c, 1/2a+1/2b, 1/2b+1/2c, which has trigonal symmetry (a = b = c = 17.215 Å, α = β = γ = 60°) and contains 48 Si and 96 O atoms. Because FAU has a single T-site, only one Si/Al=47 model was used, placing Al in that unique site.

### FER

Ferrierite (FER) is a multipore zeolite with an orthorhombic structure formed by a two-dimensional system of 8MR and 10MR ring openings. The 10MR channels run along [0 0 1] with a 5.4 × 4.2 Å aperture, while the 8MR channels run along [0 1 0] with a 4.8 × 3.5 Å aperture, forming the FER cage accessible only through 8MR windows. The orthorhombic FER unit cell (a = 18.816 Å, b = 14.217 Å, c = 7.493 Å; α = β = γ = 90°) contains 36 Si and 72 O atoms, and was expanded to a 1 × 1 × 3 supercell (108 Si, 216 O) for modeling, increasing c to 22.455 Å. FER has four T-sites, with T1 being the most stable for single Al substitution. As its 10MR channel is monodimensional, different Al placements are not expected to dramatically affect reactivity, so we created a single Si/Al=107 model by placing Al at T1.

## IWV

ITQ-27 (IWV) features a two-dimensional pore system of 12MR channels that intersect at a 120° angle, forming a 14MR pore. We modeled IWV by creating a smaller primitive unit cell (one-quarter the original volume) via the transformation matrix −c, −(b+c)/2, −(a+c)/2, yielding triclinic symmetry (P1) with lattice parameters a = 14.544 Å, b = 15.784 Å, c = 19.055 Å, α = 48°, β = 54°, and γ = 78°, and containing 38 Si and 76 O atoms. IWV has seven T-sites, with T7 and T3 being the most stable for single Al substitution and differing by just 2 kJ/mol. Consequently, two Si/Al=37 models were built to capture its key topological features: one placing Al at T3, oriented solely toward the 12MR channels, and another placing Al at T7 in the 14MR intersection.

## MFI

ZSM-5 (MFI) has two 10MR channels: a two-dimensional sinusoidal channel (5.1 × 5.5 Å) along the a-axis ([100] and [101] facets) and a one-dimensional straight channel (5.3 × 5.6 Å) along the b-axis ([010] facets), intersecting at about 9 Å. We modeled MFI with an orthorhombic unit cell (a = 20.348 Å, b = 19.865 Å, c = 13.366 Å; α = β = γ = 90°) containing 96 Si and 192 O atoms. Of the 12 distinct T-sites in MFI, T4 is the most stable for single Al substitution, though T3 and T7 are slightly more stable but less suitable for catalytic positioning. Because MFI presents a three-dimensional 10×10×10MR network, we created three Si/Al=95 models, each placing Al in a different T-site relevant to reactivity: T4 (in the sinusoidal channel), T11 (in the straight channel), and T10 (at the channel intersection).

## MOR

The Mordenite (MOR) framework is built from 5-membered tetrahedral rings (mor composite units) linked into chains along c, which join via 4-rings to form puckered sheets with 8MR openings parallel to (010). These sheets connect further via 4-rings to generate 12MR channels along [001], while the 8MR of successive sheets do not align to create channels along [010]. MOR has an orthorhombic unit cell (a = 18.204 Å, b = 20.426 Å, c = 7.501 Å; α = β = γ = 90°) containing 48 Si and 96 O atoms; we used a 1 × 1 × 3 supercell (144 Si, 288 O) that extends c to 22.504 Å. Among four T-sites (Table S7), T2 is the most stable for single Al substitution. Since MOR has a monodimensional 12MR channel, different Al placements are not expected to significantly affect reactivity, so we created a single Si/Al=143 model with Al at T2.

## UTL

Zeolite UTL features a two-dimensional pore system with perpendicular 12MR (8.5 × 5.5 Å) and 14MR (9.5 × 7.1 Å) channels. It was modeled using a monoclinic unit cell (a = 29.160 Å, b = 14.098 Å, c = 12.180 Å; α = γ = 90°, β = 105°) containing 76 Si and 152 O atoms. Of the 12 T-sites in UTL, T2 is the most stable for single Al substitution, so one model (Si/Al=75) placed Al at T2 in the 14×12 channel intersection. To investigate how the 14MR and 12MR channels and intersections influence ethylbenzene transalkylation, a second model located Al at T4, identified as the second most stable T-site, in the 14MR channel.

## Data and Code Availability

All data supporting the findings of this study are publicly accessible on Zenodo at https://zenodo.org/records/15200092. The computational code-base described in the methods section for building zeolite models and running TS searches with Dimer method on VASP reported in the work are all available free of charge under the MIT license: https://github.com/learningmatter-mit/VOID & https://github.com/learningmatter-mit/PoTS

## Acknowledgements

The authors acknowledge the MIT SuperCloud and Lincoln Laboratory Supercomputing Center for providing HPC resources that have contributed to the research results reported within this paper. This research used resources of the National Energy Research Scientific Computing Center (NERSC), a Department of Energy Office of Science User Facility using NERSC awards BES-ERCAP-m4604 and BES-ERCAP-m4866. P.F.V. thanks Fundacion Ramon Areces for his postdoctoral fellowship. A.S. also acknowledges support from the National Defense Science and Engineering Graduate Fellowship. We thank Mingrou Xie (MIT) for helpful discussions and suggestions. P.F.V., A.H. and R.G.-B. acknowledges financial support from Deshpande Center and ExxonMobil Technology and Engineering Company.

Supplementary Information

# Scaling Transition-State Searches in Zeolite Nanopores

Pau Ferri-Vicedo, Alexander J. Hoffman, Avni Singhal, Rafael Gómez-Bombarelli*

Department of Materials Science and Engineering, Massachusetts Institute of Technology, Cambridge, MA 02139, USA

E-mail: rafagb@mit.edu

## Table of Contents

# S1 Zeolite Modeling

All unit cells presented were obtained after full relaxation of the experimental lattice parameters downloaded from the International Zeolite Association (IZA) database[1] using PBE-D3 calculations as detailed in **Method** on the main work. To achieve full relaxation of lattice parameters, atomic positions, cell shape and cell volume were considered as degrees of freedom for calculations. Results are summarized in **Table S1**.

**Table S1.** Crystallographic data for the zeolite frameworks used in this study, including IZA code, cell type, and unit cell parameters (a, b, c in Å; α, β, γ in °). The final column lists the number of Si (T) atoms per unit cell.

| IZA code | Cell Type | Unit Cell Parameters | | | | | | T atoms |
|---|---|---|---|---|---|---|---|---|
| | | a | b | c | α | β | γ | |
| BEA | Tetragonal | 12.593 | 23.593 | 26.495 | 90 | 90 | 90 | 64 |
| BOG | Triclinic | 12.812 | 16.791 | 16.791 | 74 | 68 | 68 | 48 |
| FAU | Trigonal | 17.215 | 17.215 | 17.215 | 60 | 60 | 60 | 48 |
| FER | Orthorhombic | 18.816 | 14.217 | 22.455 | 90 | 90 | 90 | 108 |
| IWV | Triclinic | 14.544 | 15.784 | 19.055 | 48 | 54 | 78 | 38 |
| MFI | Orthorhombic | 20.348 | 19.865 | 13.366 | 90 | 90 | 90 | 96 |
| MOR | Orthorhombic | 18.204 | 20.426 | 22.504 | 90 | 90 | 90 | 144 |
| UTL | Monoclinic | 29.16 | 14.098 | 12.180 | 90 | 105 | 90 | 76 |

## S2.1. BEA

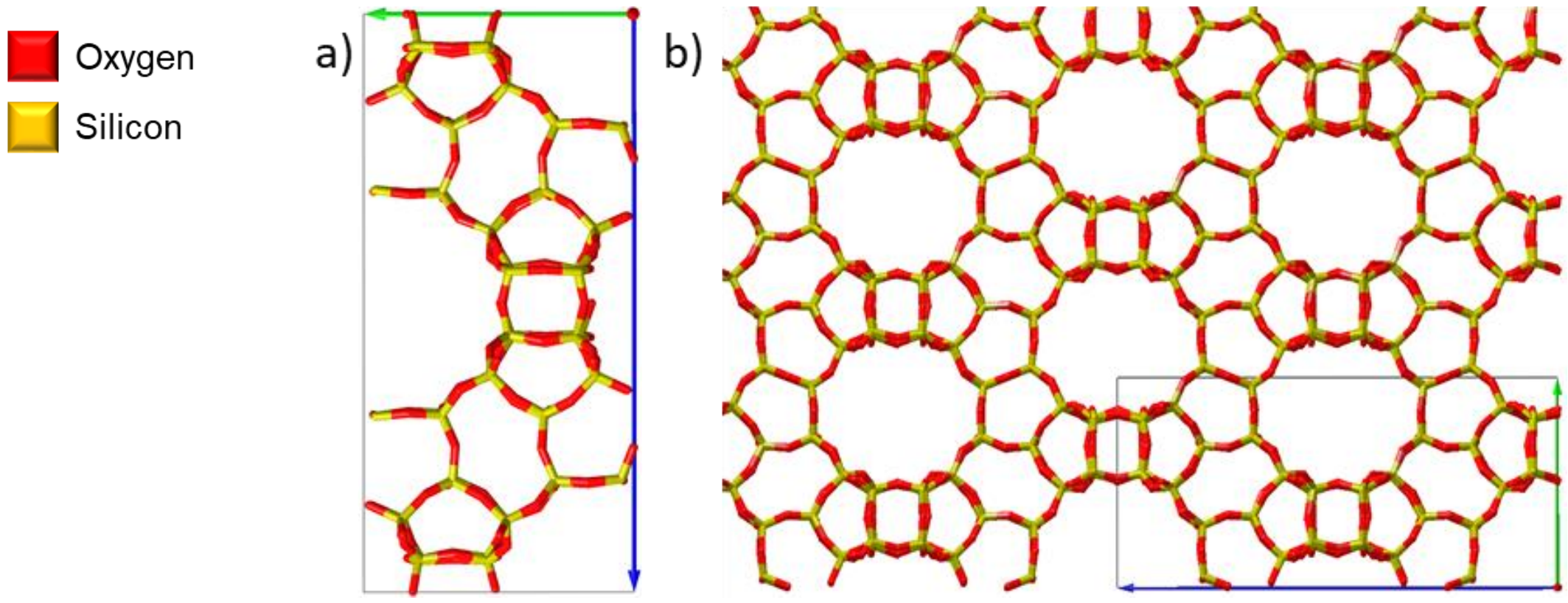

**Figure S2**. a) BEA unit cell. b) BEA structure expanded.

Zeolite beta (BEA) is a typical stacking-fault intergrowth structure of several distinct but closely related polymorphs. These polymorphs are built from an identical centrosymmetric layer by different stacking sequence. In this work we have used a pure model of the chiral polymorph-A which with an irregular three-dimensional channel system with 12-membered (12MR) ring apertures of 5.6–7.3 Å and a high external surface area.
Thus, BEA structure was modeled by means of a tetragonal unit cell with lattice parameters a = b = 12.593 Å, c = 26.495 Å, α = β = γ = 90°, containing 64 Si atoms and 128 O atoms.

## S2.2. BOG

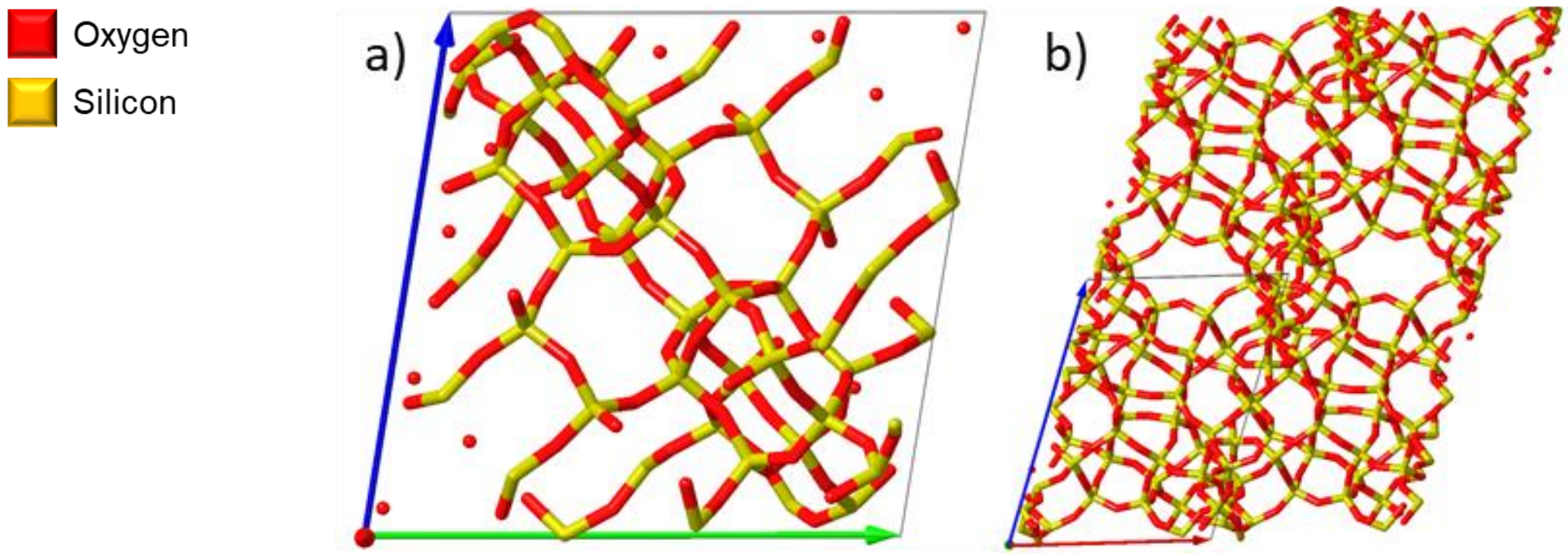

**Figure S3**. a) BOG primitive unit cell. b) BOG structure expanded.

The Boggsite (BOG) framework has wide channels parallel to the *a*-axis confined by 12MR with a ring aperture of 7.0 Å. Each 12MR ring channel along the *a*-axis has offset ten-membered (10MR) ring windows into the left and right channels parallel to the *b*-axis with a ring aperture of 5.8 x 5.2 Å.
Thus, BOG was modeled using a smaller primitive reduced unit cell obtained through the transformation matrix: c,-1/2a-1/2b+1/2c,1/2a-1/2b+1/2c that has a 2 times smaller volume. This unit cell presents triclinic symmetry (P1) and lattice parameters a = 12.812 Å, b = c = 16.791 Å, α = 74, β = 68 and γ = 68 containing 48 Si atoms and 96 O atoms

## S2.3. FAU

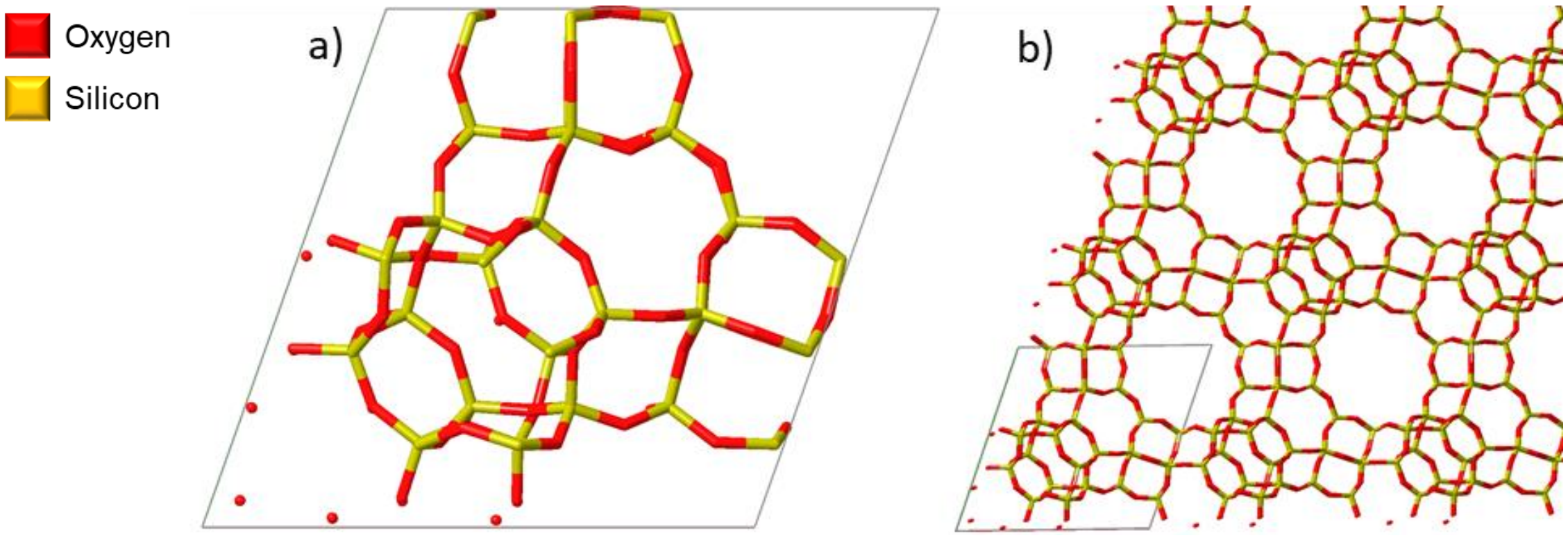

**Figure S4**. a) FAU primitive unit cell. b) FAU structure expanded.

The Faujasite (FAU) structure consists of sodalite cages connected in a cubic manner over six-membered double rings (d6R). Then, 12MR intersecting channels are formed parallel to [111] with a 7.4 Å aperture. Thus, FAU was modeled using a smaller primitive reduced unit cell obtained through the transformation matrix: 1/2a+1/2c,1/2a+1/2b,1/2b+1/2c that has a 4 times smaller volume. This unit cell presents trigonal symmetry and lattice parameters a = b = c = 17.215 Å, α = β = γ = 60 containing 48 Si and 96 O atoms.

## S2.4. FER

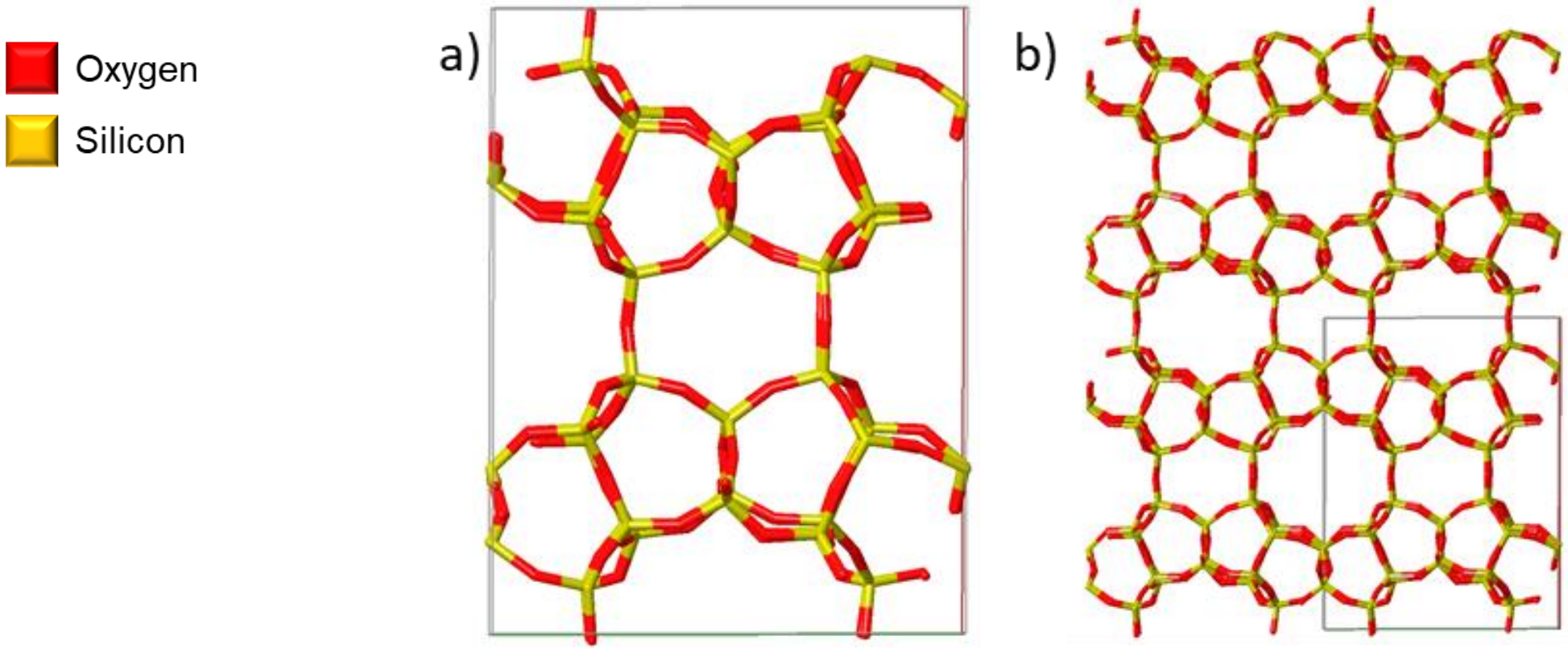

**Figure S5**. a) FER unit cell. b) FER structure expanded.

Ferrierite (FER) is a multipore zeolite with an orthorhombic crystallographic structure, formed by a two-dimensional system of straight channels interconnected perpendicularly, with 8-membered (8MR) and 10MR ring openings. The 10 MR pores, parallel to the c-axis [0 0 1], are 5.4 x 4.2 Å, and those of the 8 MR pores, parallel to the b-axis [0 1 0], are 4.8 x 3.5 Å. Along the 8 MR channels, a cavity known as the FER cage is formed, which is only accessible through 8 MR windows.

Thus, FER presents an orthorhombic unit cell, lattice parameters a = 18.816, b = 14.217 and c = 7.493 Å, α = β = γ = 90.0°, containing 36 Si atoms and 72 O atoms. To model the FER structure, we used a larger 1 × 1 × 3 supercell containing 108 Si atoms and 216 O atoms, increasing lattice parameter c = 22.455 Å.

## S2.5. IWV

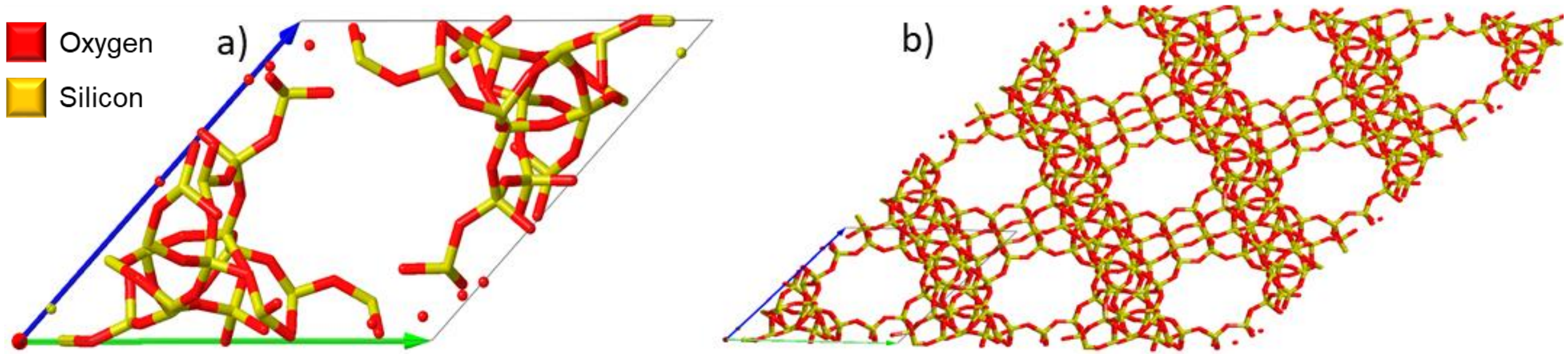

**Figure S6**. a) IWV primitive reduced unit cell. b) IWV structure expanded.

ITQ-27 (IWV) has a two-dimensional pore system bounded by 12MR that lead to internal cross-sections with a 120-degree angle containing a 14-membered ring (14MR).

Thus, IWV structure was modeled using a smaller primitive reduced unit cell obtained through the transformation matrix: -c,-1/2b-1/2c,-1/2a-1/2c that has a 4 times smaller volume. This unit cell presents triclinic symmetry (P1) and lattice parameters a = 14.544 Å, b = 15.784 Å, c= 19.055 Å, α = 48, β = 54 and γ = 78 containing 38 Si atoms and 76 O atoms.

## S2.6. MFI

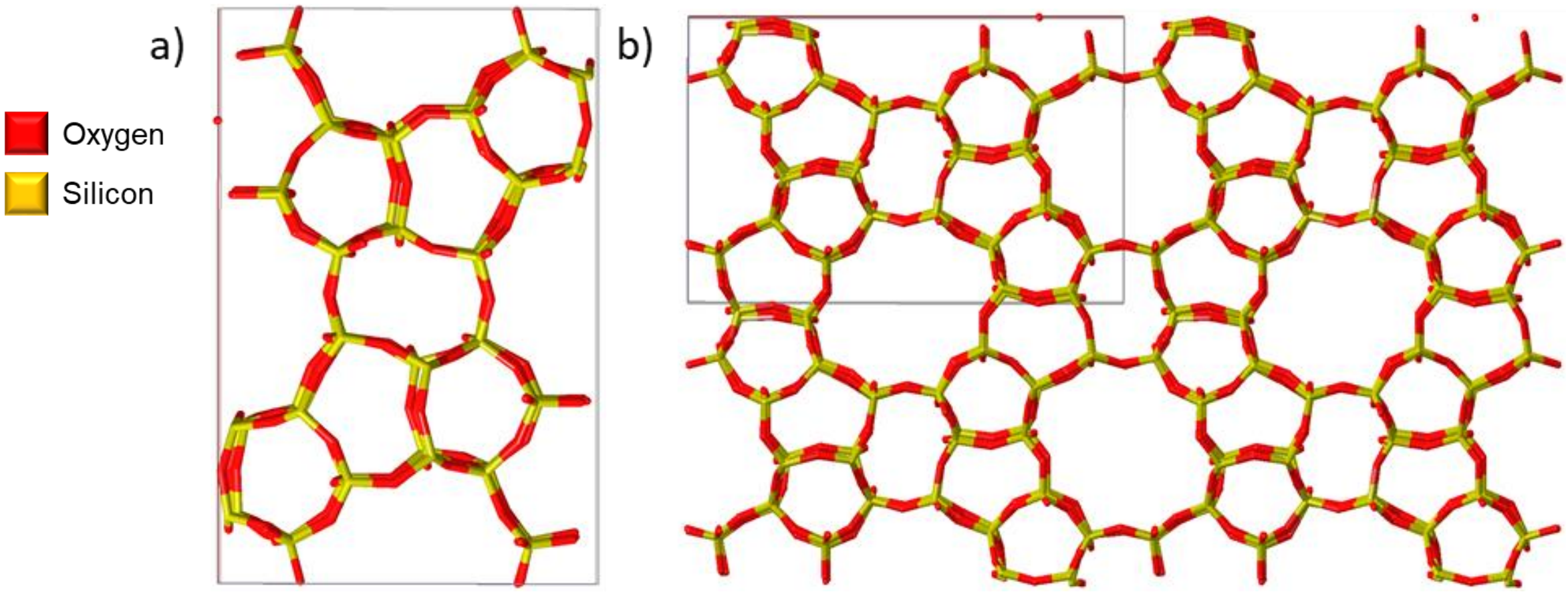

**Figure S7**. a) MFI unit cell. b) MFI structure expanded.

ZSM-5 zeolite (MFI topology) consists of two-dimensional sinusoidal channels (5.1 × 5.5 Å) and one-dimensional straight channels (5.3 × 5.6 Å), including two 10-MR micropore structure. The straight channels follow the direction of the b-axis with exposed [010] facets, while sinusoidal channels are parallel to the a-axis with exposed [100] and [101] facets. The critical dimensions at the intersection of the straight and sinusoidal channels are close to 9 Å.

Thus, MFI framework, was modeled by means of a orthorhombic unit cell with lattice parameters a = 20.348, b = 19.865, and c = 13.366 Å, α = β = γ = 90°, and contains 96 Si atoms and 192 O atoms.

## S2.7. MOR

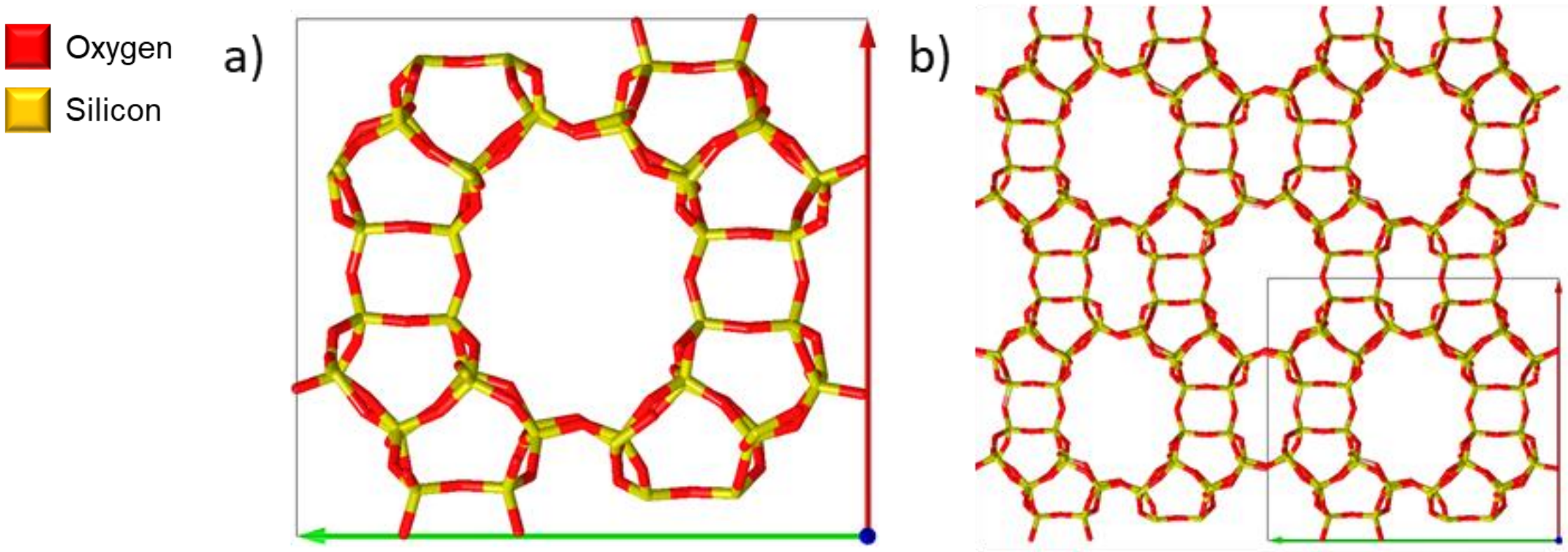

**Figure S8**. a) MOR unit cell. b) MOR structure expanded.

The topology of the Mordenite (MOR) framework is characterized by 5-member tetrahedral rings, which are part of the mor composite building unit. These building units are linked by edge-sharing into chains along c, which are in turn linked together by 4-rings to form a puckered sheet perforated with 8-ring holes. These are oriented parallel to (010). Linking these sheets together with 4-rings, 12MR are formed parallel to [001]. The 8MR of successive sheets do not align to make channels parallel to [010].

Thus, MOR presents an orthorhombic unit cell with lattice parameters a = 18.204, b = 20.426, and c = 7.501 Å, α = β = γ = 90.0°, containing 48 Si atoms and 96 O atoms. To model the MOR structure, we used a larger 1 × 1 × 3 supercell containing 144 Si atoms and 288 O atoms, increasing lattice parameter c = 22.504.

## S2.8. UTL

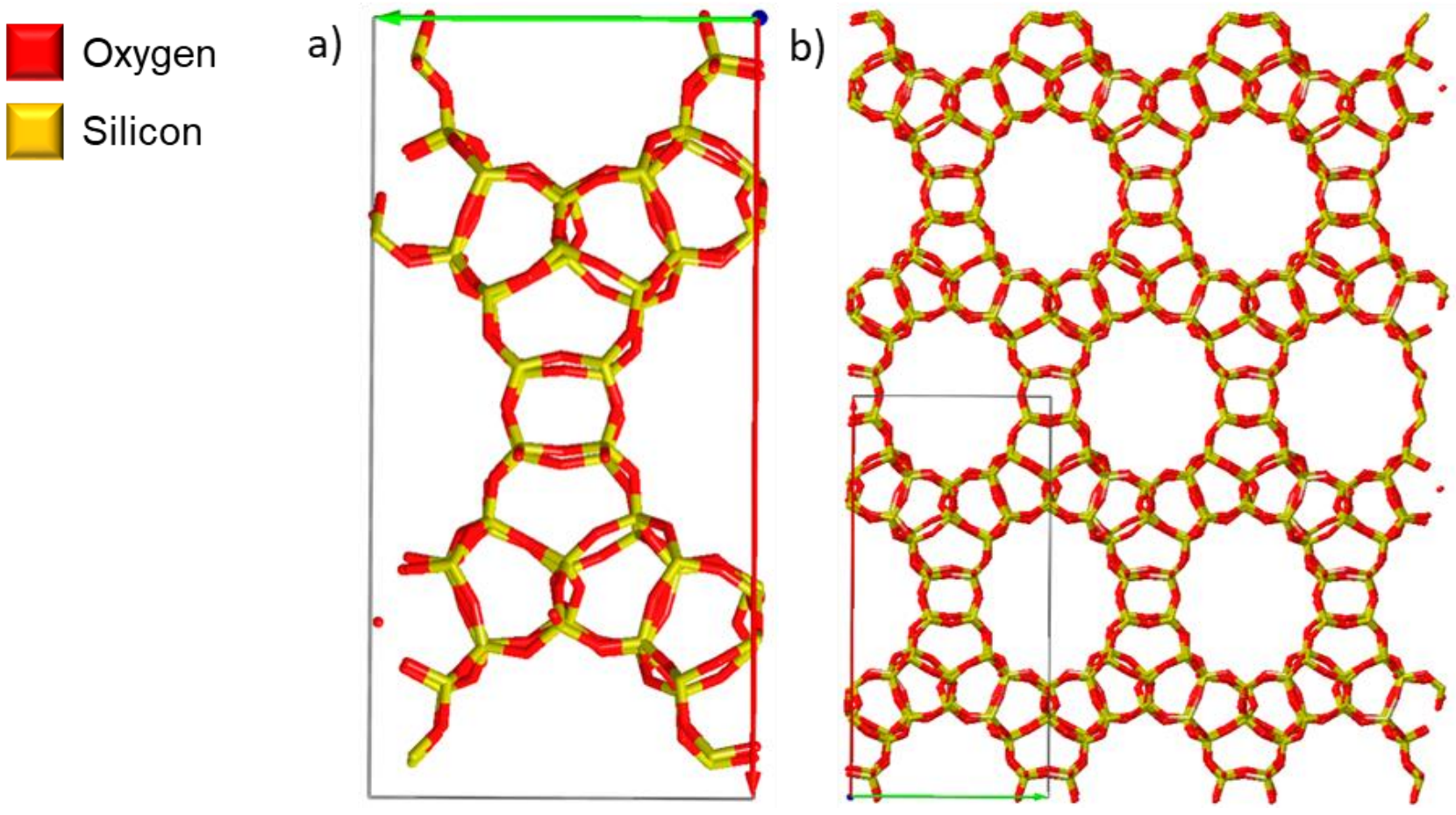

**Figure S9**. a) UTL unit cell. b) UTL structure expanded.

Zeolite UTL possesses two-dimensional pore system with perpendicular 12- and 14-ring channels of diameter 8.5 × 5.5 and 9.5 × 7.1 Å, respectively.

Thus, UTL framework was modeled by means of a monoclinic unit cell with lattice parameters a = 29.160, b = 14.098, and c = 12.180 Å, α = γ = 90°, β = 105° and contains 76 Si atoms and 152 O atoms.

## S2.9. Al Positioning

As this study focuses on chemical reactivity, we introduced an acid site in each framework by substituting a Si atom with an Al atom, forming an $AlO4^{-}$ unit to compensate for the positive charge introduced by the reactive cation. The T-site positions in each framework were chosen based on the intrinsic stability of an empty cell with a background negative charge at the PBE-D3 theory level, detailed at section **Method** in the main work. Subsequently, the final positioning was determined according to the specific topology of each framework.

### S2.9.1. BEA

BEA framework presents 9 different T-sites, its relative stability is shown at **Table S2** with T1 and T2 being its most stable crystallographic positions for a single Al positioning.

**Table S2.** Relative intrinsic stabilities (in kJ/mol) of each crystallographic T-site in the BEA framework, referenced to T2 = 0.0 kJ/mol.

| T-site | T1 | T2 | T3 | T4 | T5 | T6 | T7 | T8 | T9 |
|---|---|---|---|---|---|---|---|---|---|
| Erel (kJ/mol) | 0.1 | 0.0 | 12 | 15 | 13 | 11 | 13 | 11 | 11 |

As BEA framework presents a tridimensional 12MR channel topology and the differences between T1 and T2 are non-existent, we created three different BEA models with Si/Al=63 that encompass all BEA topological specifities. We placed one Al in T1 and in T2 oriented to different straight 12MR channels in order to account for possible differences arising from Al positioning, see **Figure S10 a)** and **b)**. Moreover, in order to acquire deeper understanding about the BEA topology we created a third model with an Al atom placed in a T1 site on the 12MR intersection between, see **Figure S10 c)**.

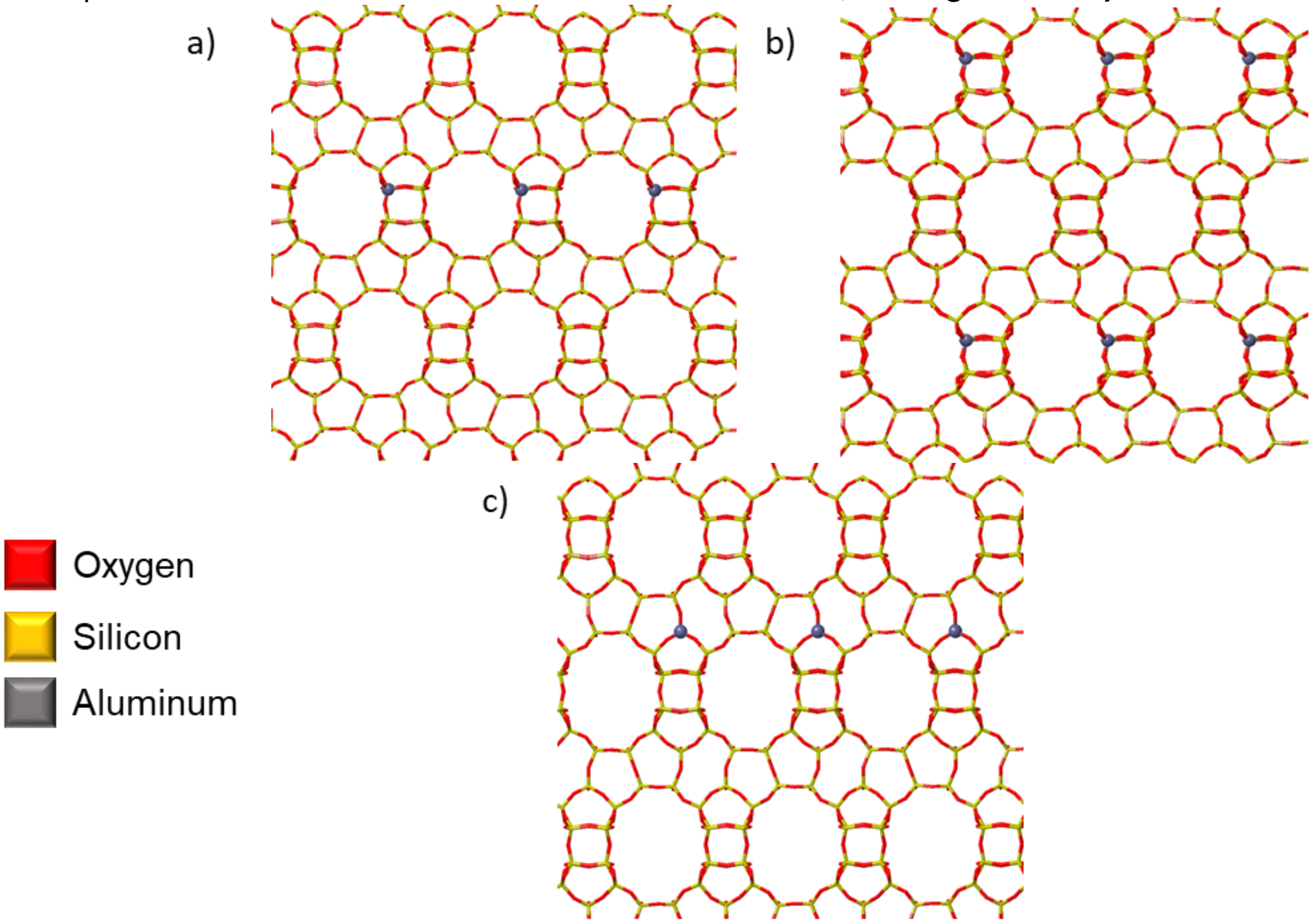

**Figure S10**.a) BEA-T1 12MR channel b) BEA-T2 12MR channel c) BEA-T1 12MR intersection models.

## S2.9.2. BOG

BOG framework presents 6 different T-sites, its relative stability is shown at **Table S3** with T1 and T2 being its most stable crystallographic positions for a single Al positioning.

**Table S3.** Relative intrinsic stabilities (in kJ/mol) of each crystallographic T-site in the BOG framework, referenced to T2 = 0.0 kJ/mol.

| T-site | T1 | T2 | T3 | T4 | T5 | T6 |
|---|---|---|---|---|---|---|
| Erel (kJ/mol) | 1 | 0 | 12 | 6 | 7 | 2 |

As BOG framework presents a bidimensional 12MR x 10MR channel topology and the differences between T1 and T2 are just 1 kj/mol. We created two different BOG models with Si/Al=47 that encompass all BOG topological specifities, in this case both models present an Al atom in a 12x10MR intersection but in different positions around the unit cell, see **Figure S11.**

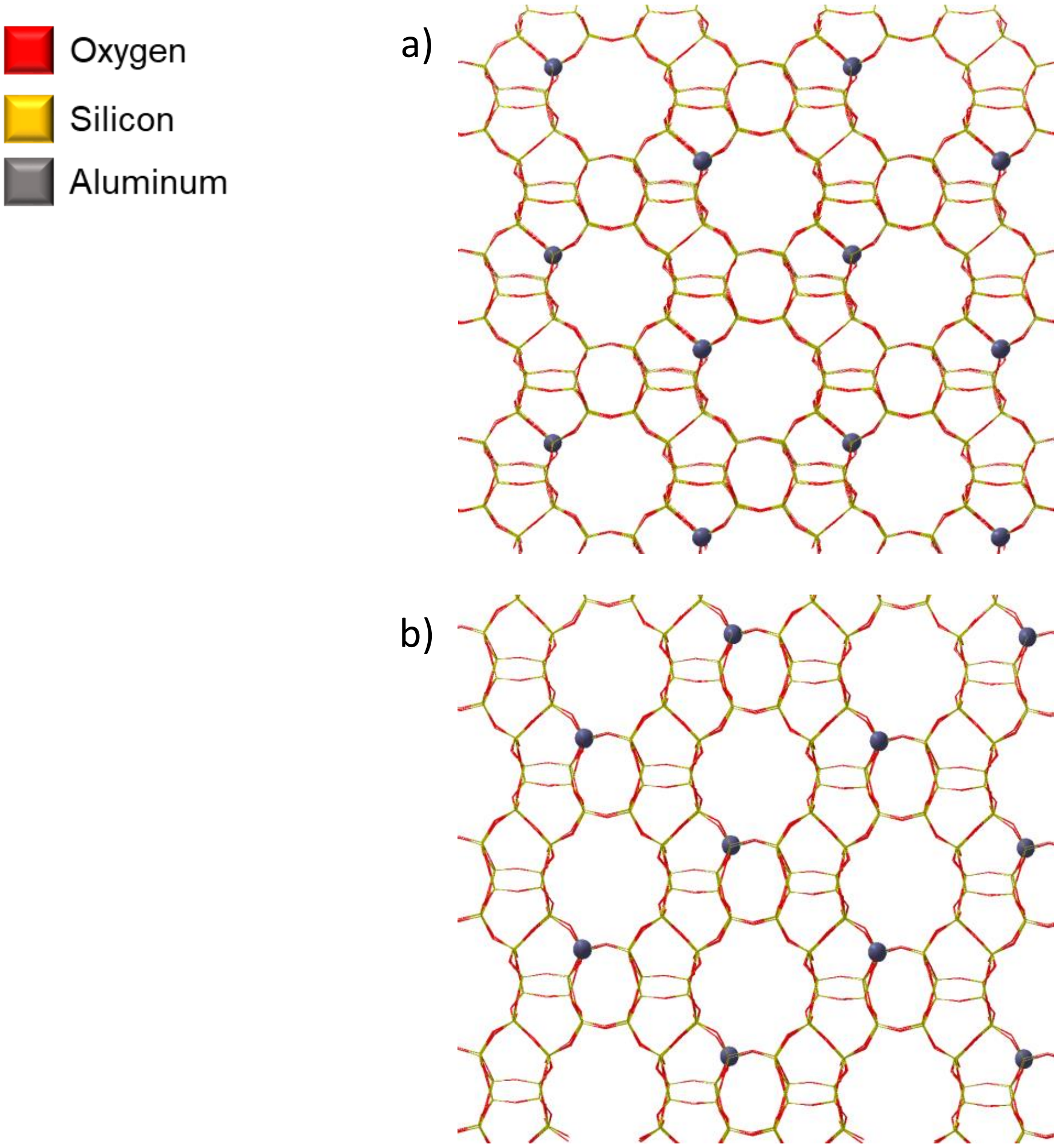

**Figure S11**. a) BOG-T1 b) BOG-T2 12x10 MR intersection models.

## S2.9.3. FAU

FAU framework only presents one unique T-site, then, only one model with Si/Al=47 was used with an aluminium atom placed as shown in **Figure S12**.

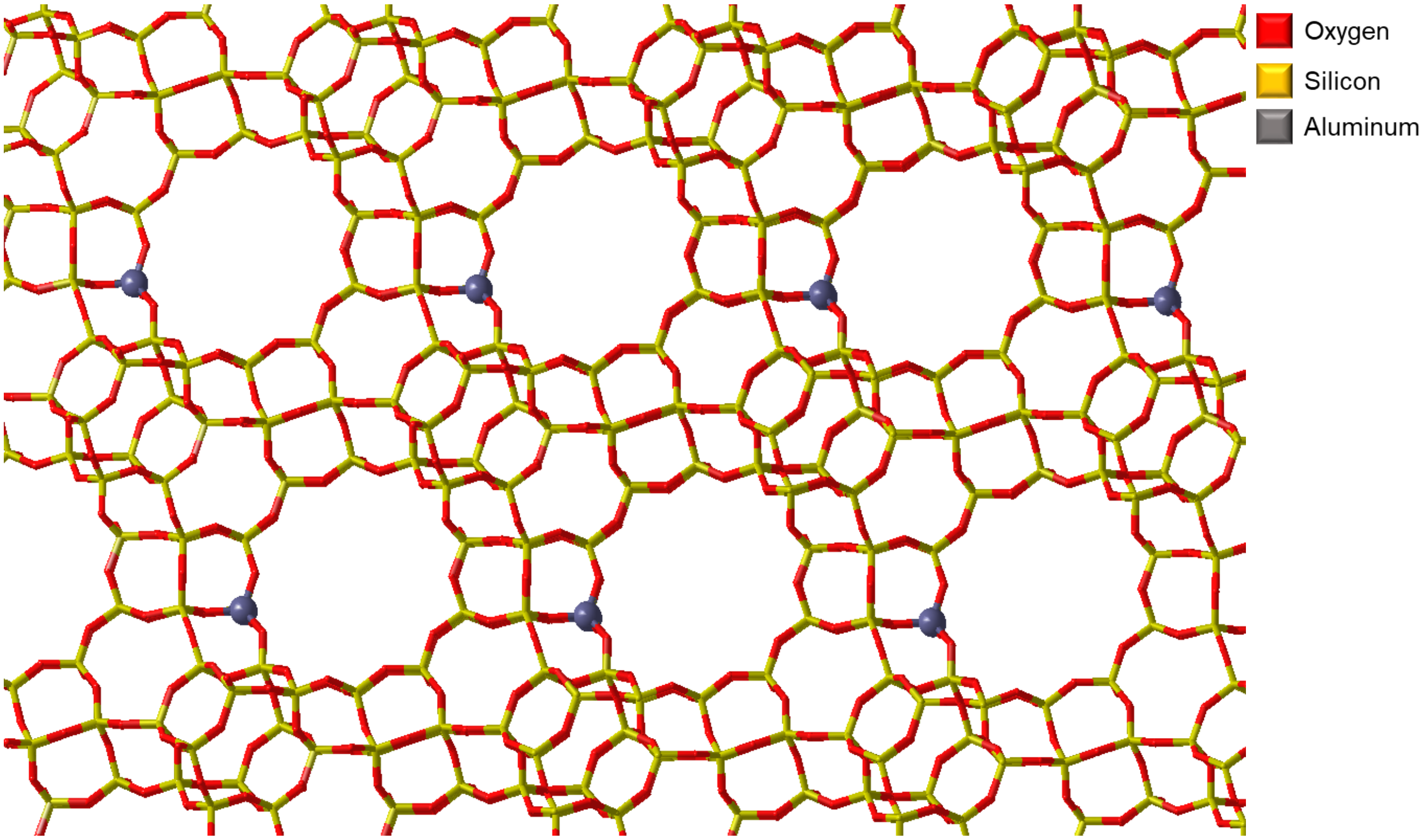

**Figure S12**. FAU-T1 model.

## S2.9.4. FER

FER framework presents 4 different T-sites, its relative stability is shown at **Table S4** with T1 being its most stable crystallographic position for a single Al positioning.

**Table S4.** Relative intrinsic stabilities (in kJ/mol) of each crystallographic T-site in the FER framework, referenced to T1 = 0.0 kJ/mol.

| T-site | T1 | T2 | T3 | T4 |
|---|---|---|---|---|
| Erel (kJ/mol) | 0 | 5 | 16 | 20 |

As FER framework presents a monodimensional 10 MR channel, differences in Al positioning along this channel are not expected to bring dramatic changes to reactivity. Hence, we created a unique FER model with Si/Al=107 with Al sitting at a T1 crystallographic site, as depicted in **Figure S13**.

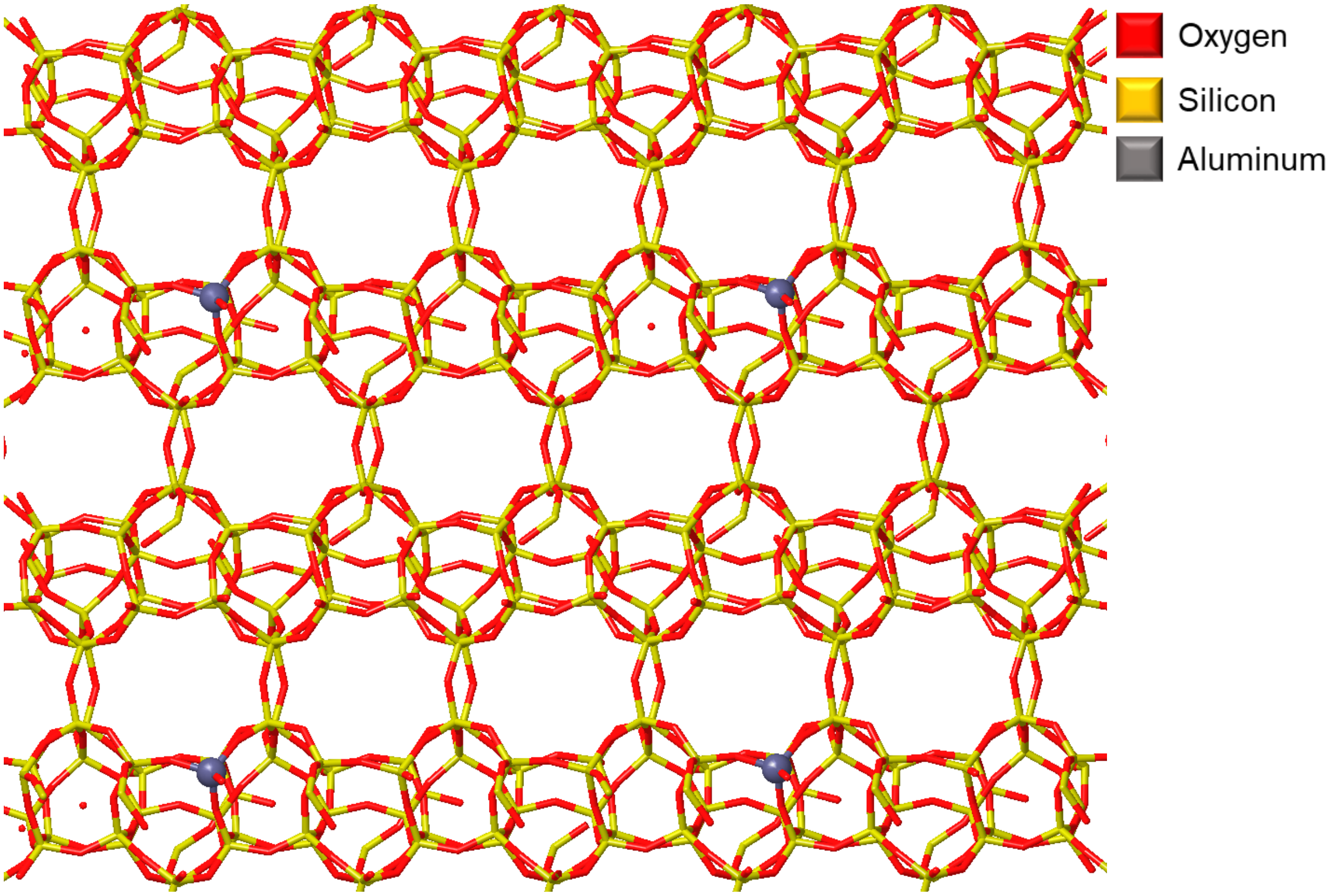

**Figure S13**. FER-T1 model.

## S2.9.5. IWV

IWV framework presents 7 different T-sites, its relative stability is shown at **Table S5** with T7 and T3 being its most stable crystallographic positions for a single Al positioning.

**Table S5.** Relative intrinsic stabilities (in kJ/mol) of each crystallographic T-site in the IWV framework, referenced to T7 = 0.0 kJ/mol.

| T-site | T1 | T2 | T3 | T4 | T5 | T6 | T7 |
|---|---|---|---|---|---|---|---|
| Erel (kJ/mol) | 36 | 37 | 2 | 31 | 10 | 49 | 0 |

As IWV framework presents a binodimensional 12 MR channel topology and a 14MR pore intersection and the differences between T3 and T7 are just 2 kj/mol. We created two different IWV with Si/Al=37 models that encompass all IWV topological specifities, one with the Al atom placed at a T7 pointing exclusively towards the 12MR channels, see **Figure S14 a)**, and a second one with the Al atom placed at T3 inside the 12x12MR pore intersection, see **Figure S14 b)**.

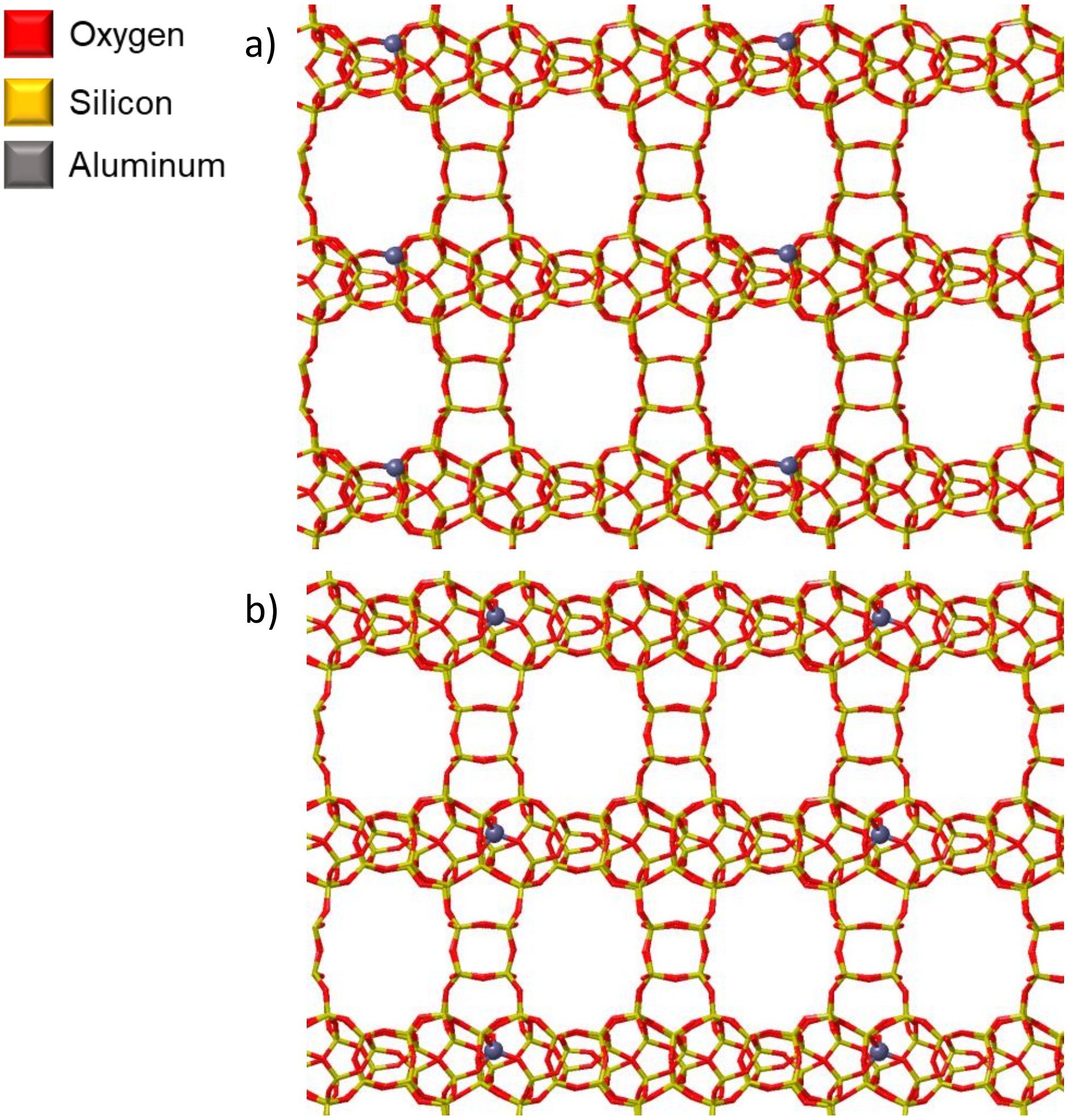

**Figure S14**. a) IWV-T3 at the 12MR channel b) IWV-T7 at the 14MR pore intersection models.

## S2.9.6. MFI

MFI framework presents 12 different T-sites, its relative stability is shown at **Table S6** with T4 being its most stable crystallographic positions for a single Al positioning.

**Table S6.** Relative intrinsic stabilities (in kJ/mol) of each crystallographic T-site in the MFI framework, referenced to T4 = 0.0 kJ/mol.

| T-site | T1 | T2 | T3 | T4 | T5 | T6 | T7 | T8 | T9 | T10 | T11 | T12 |
|---|---|---|---|---|---|---|---|---|---|---|---|---|
| Erel (kJ/mol) | 18 | 16 | 6 | 0 | 12 | 27 | 6 | 18 | 22 | 8 | 10 | 13 |

As MFI framework presents a tridimensional 10x10x10MR channel topology with interconnected straight and sinusoidal channels, we aimed to create three different MFI models with Si/Al=95 that capture all the MFI topological features combining intrinsic stability for Al positioning and availability of the T-sites at the different channels and intersections. First, we created one MFI model with Al placed at a T4 found exclusively at the sinusoidal 10MR channel, see **Figure S15 a)**. Second, a similar model was created placing Al at a T11 found on the straight 10MR channel, see **Figure S15 b)**. Finally, we created an MFI model with Al placed at T10 and found in the intersection between sinusoidal and straight 10MR channels, see **Figure S15 c)**. Although the T3 and T7 T-sites are slightly more stable, they were discarded due to the inability to place Al in chemically relevant positions, as was possible with the T10 and T11 models.

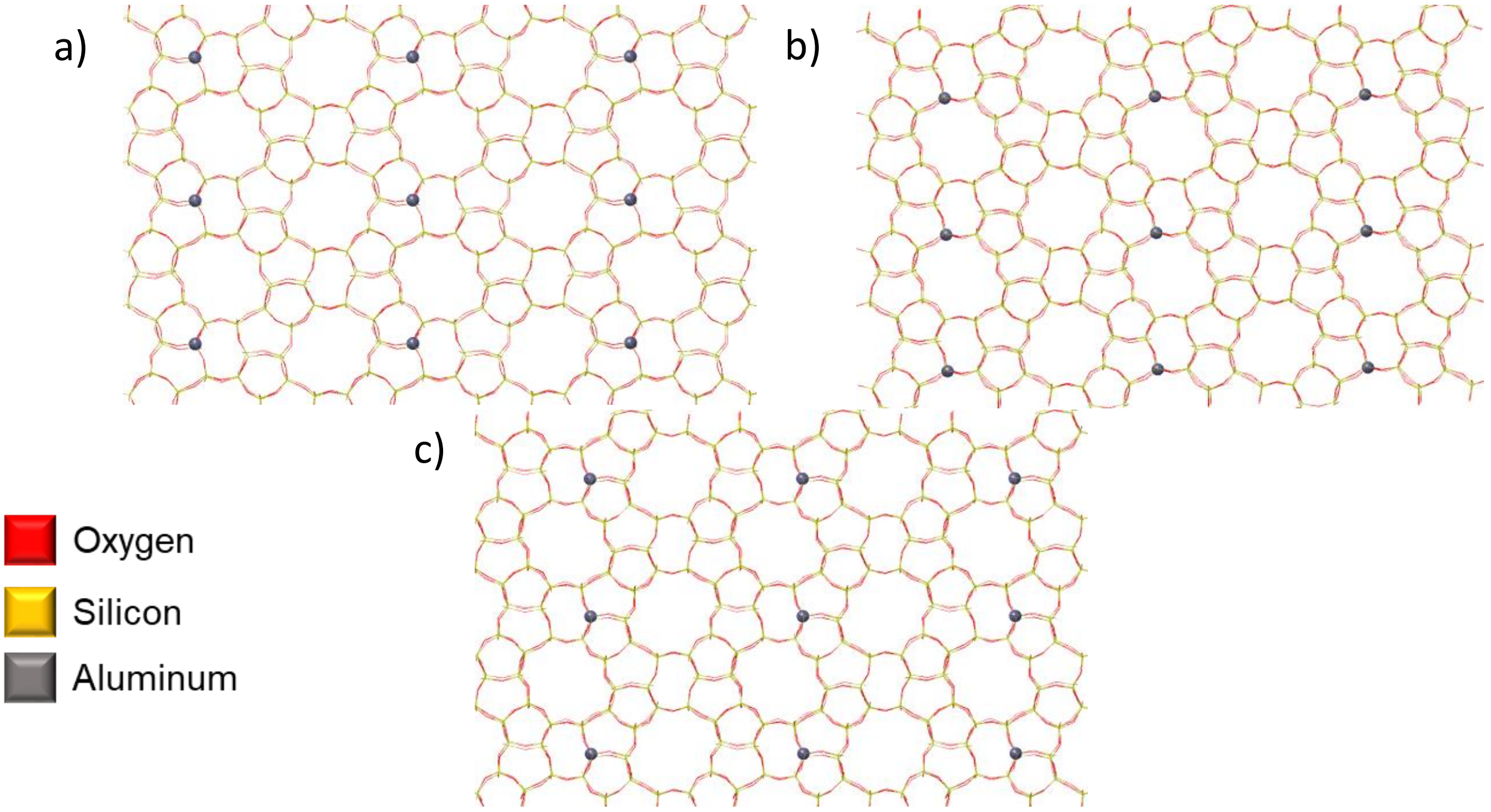

**Figure S15**. a) MFI-T4 sinusoidal, b) MFI-T11 straight c) MFI-T10 10x10MR intersection models.

## S2.9.7. MOR

MOR framework presents 4 different T-sites, its relative stability is shown at **Table S7** with T2 being its most stable crystallographic position for a single Al positioning.

**Table S7.** Relative intrinsic stabilities (in kJ/mol) of each crystallographic T-site in the MOR framework, referenced to T2 = 0.0 kJ/mol.

| T-site | T1 | T2 | T3 | T4 |
|---|---|---|---|---|
| Erel (kJ/mol) | 7 | 0 | 1 | 10 |

As MOR framework presents a monodimensional 12 MR channel, differences in Al positioning along this channel are not expected to bring dramatic changes to reactivity. Hence, we created a unique MOR model with Si/Al=143 with Al sitting at a T2 crystallographic site, as depicted in Figure S16.

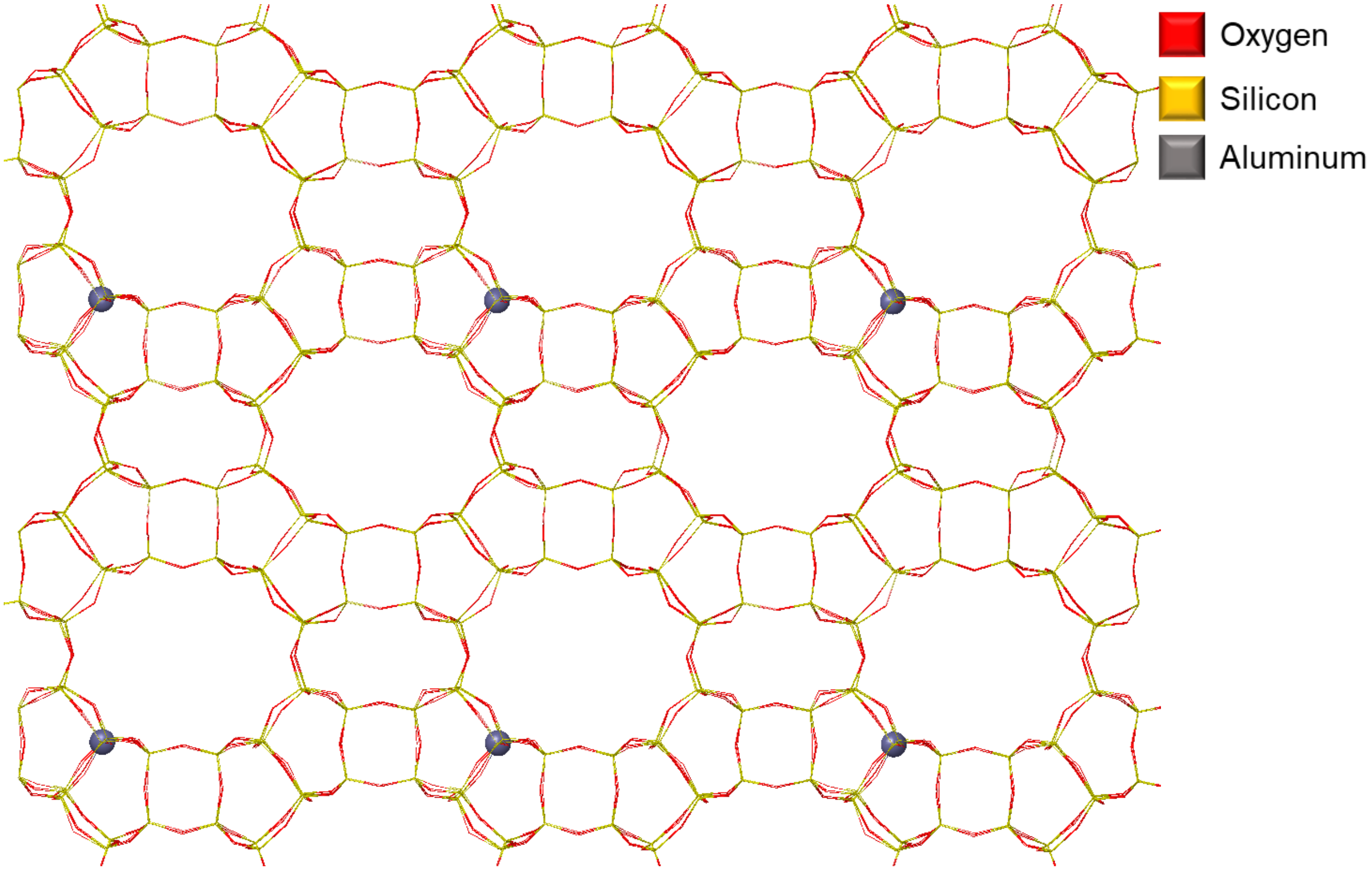

**Figure S16.** MOR-T2 model.

## S2.9.8. UTL

UTL framework presents 12 different T-sites, its relative stability is shown at Table S8 with T2 being its most stable crystallographic position for a single Al positioning.

**Table S8.** Relative intrinsic stabilities (in kJ/mol) of each crystallographic T-site in the UTL framework, referenced to T2 = 0.0 kJ/mol.

| T-site | T1 | T2 | T3 | T4 | T5 | T6 | T7 | T8 | T9 | T10 | T11 | T12 |
|---|---|---|---|---|---|---|---|---|---|---|---|---|
| Erel (kJ/mol) | 26 | 0 | 21 | 16 | 31 | 22 | 32 | 41 | 38 | 40 | 29 | 42 |

Hence, we created a UTL model with Si/Al=75 with an Al atom placed at T2 in the 14x12 channel intersection, see Figure 17 a). Moreover, in order to study how the topological differences between UTL 14MR and 12MR channels and intersections affect ethylbenzene transalkylation reactivity, we created another model for UTL with an Al atom placed at T4 in the 14MR channel because it is the second most stable T-site position, see Figure 17 b).

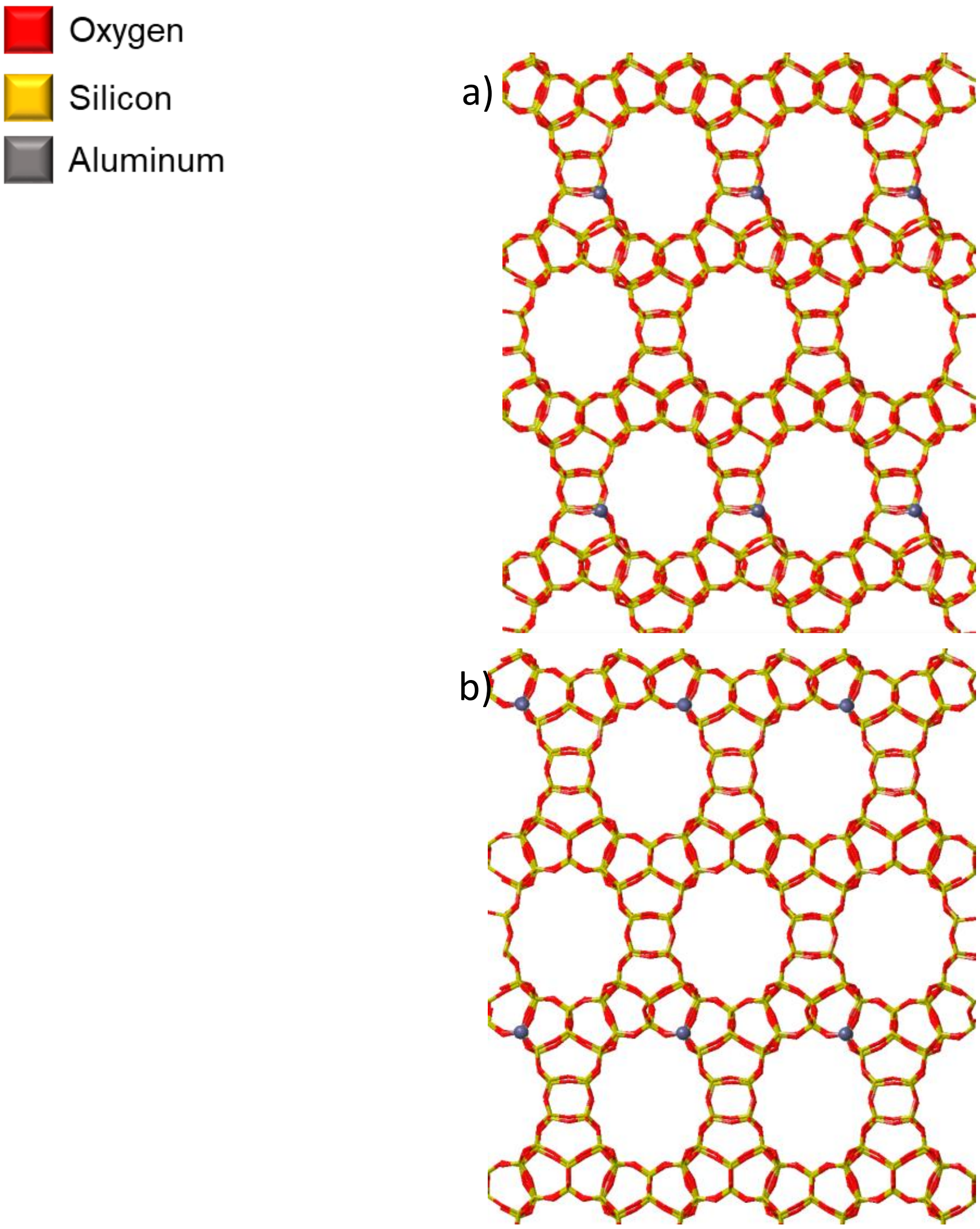

**Figure S17.** a) UTL-T2 at the 14x12 intersection b) UTL-T4 at the 14MR channel models.

# S2 Diethylbenzene Transalkylation Results

## S2.1 Reactions Studied

This section presents a detailed breakdown of all individual reaction steps investigated in this work for diethylbenzene transalkylation and disproportionation, as illustrated in **Figures S18–S22**.

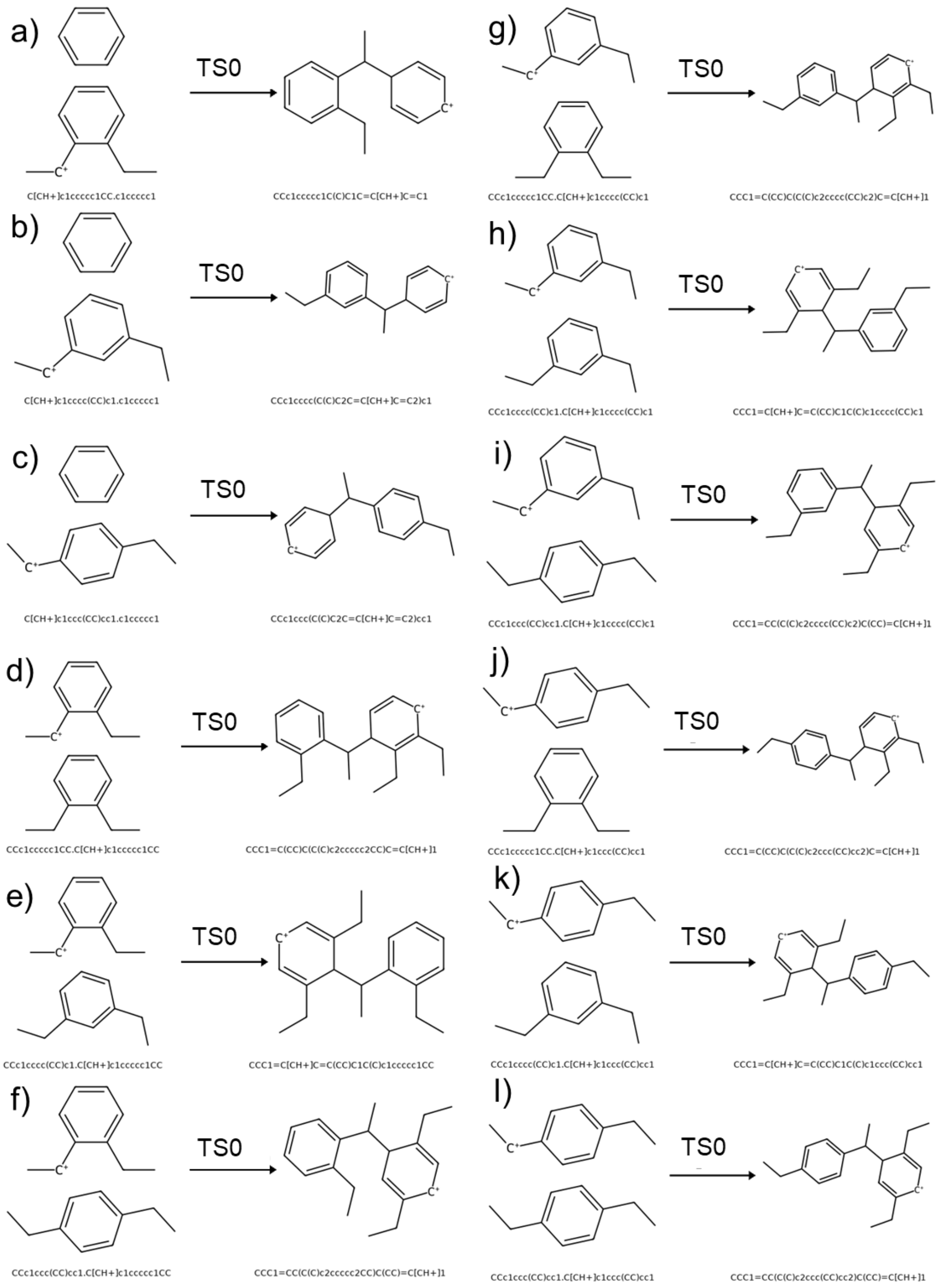

**Figure S18.** TS0 diaryl formation reactions are presented in panels (a–c) for ortho-, meta-, and para-$DEB^+$ with benzene; (d–f) and (g–i) for ortho- and meta-$DEB^+$ reactions, respectively; and (j–l) for para-$DEB^+$ reacting with ortho-, meta-, and para-DEB, respectively.

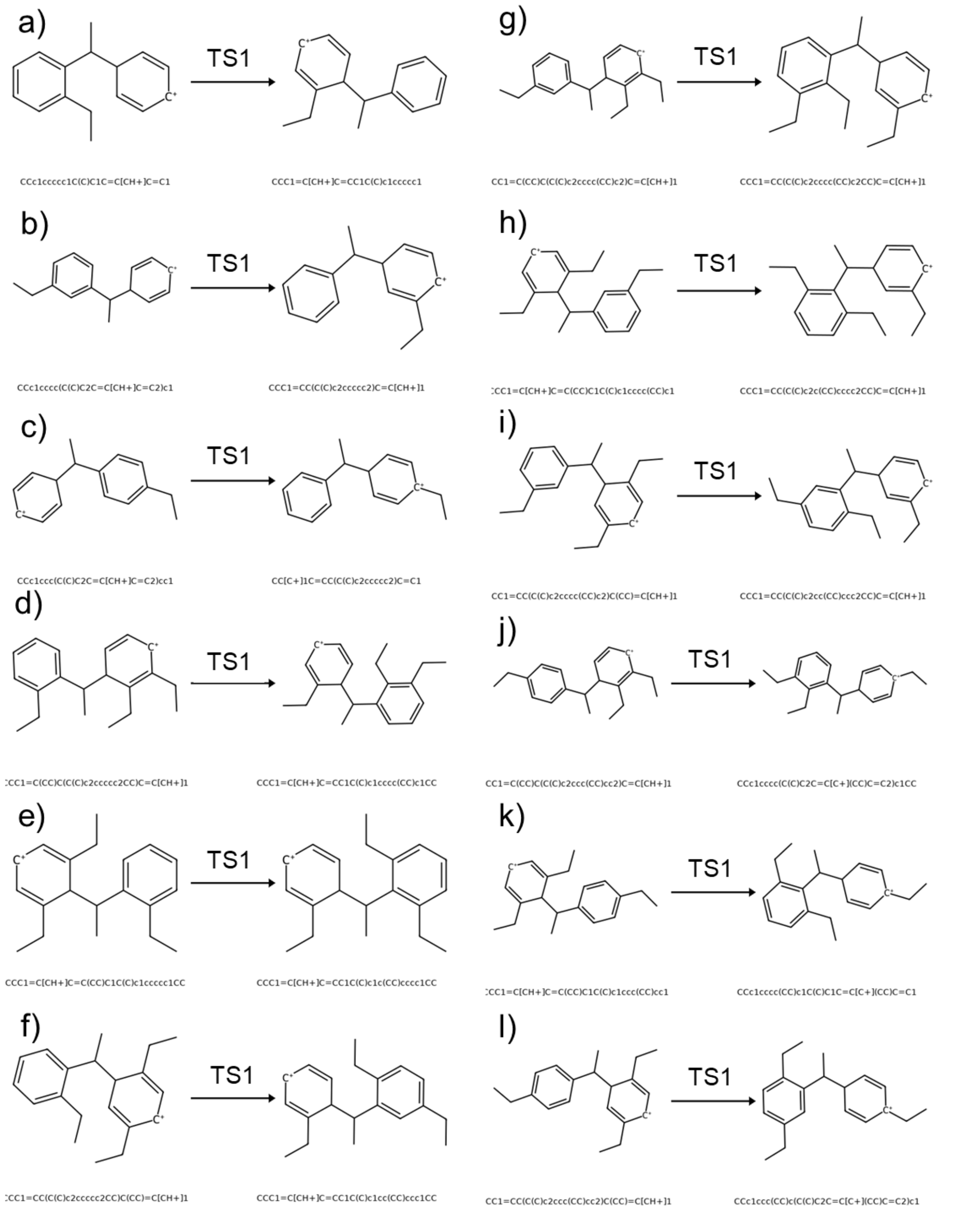

**Figure S19.** Direct path ring-to-ring hydrogen shift reactions via TS1 are presented in panels (a–c) for ortho-, meta-, and para-DEB$^+$ with benzene; (d–f) for ortho-DEB$^+$ with ortho-, meta-, and para-DEB; (g–i) for meta-DEB$^+$ with ortho-, meta-, and para-DEB; and (j–l) for para-DEB$^+$ with ortho-DEB, para-DEB, and para-DEB diaryl intermediates.

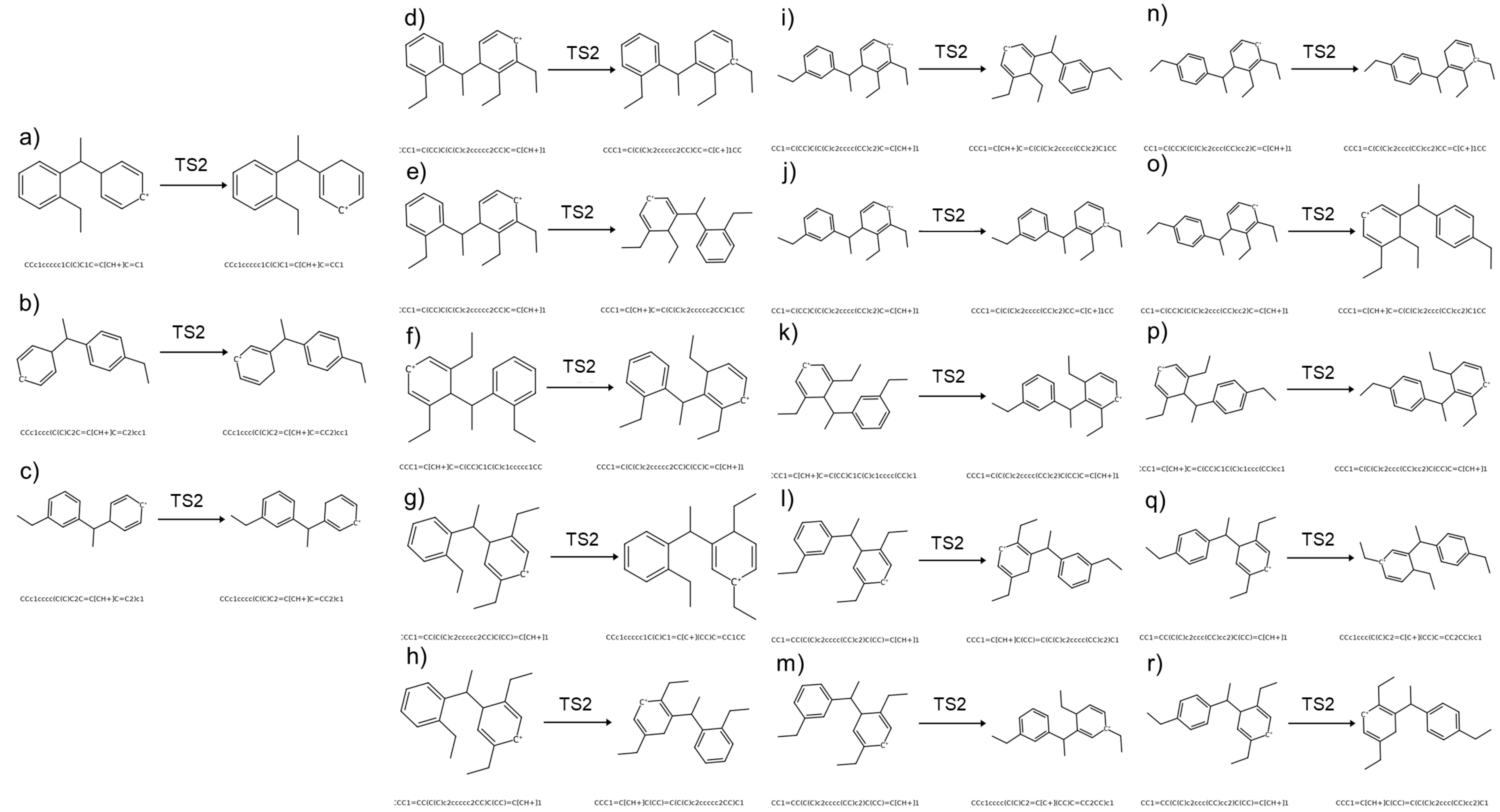

**Figure S20.** Multistep path intra-ring hydrogen shift reactions via TS2 are presented in panels (a–c) for ortho-, meta-, and para-DEB$^+$ with benzene; panels (d–e) for ortho-DEB$^+$ with ortho-DEB; panel (f) for ortho-DEB$^+$ with meta-DEB; panels (g–h) for ortho-DEB$^+$ with para-DEB; panels (i–j) for meta-DEB$^+$ with ortho-DEB; panel (k) for meta-DEB$^+$ with meta-DEB; panels (l–m) for meta-DEB$^+$ with para-DEB; panels (n–o) for para-DEB$^+$ with ortho-DEB; panel (p) for para-DEB$^+$ with para-DEB; and panels (q–r) for para-DEB$^+$ with para-DEB diaryl intermediates.

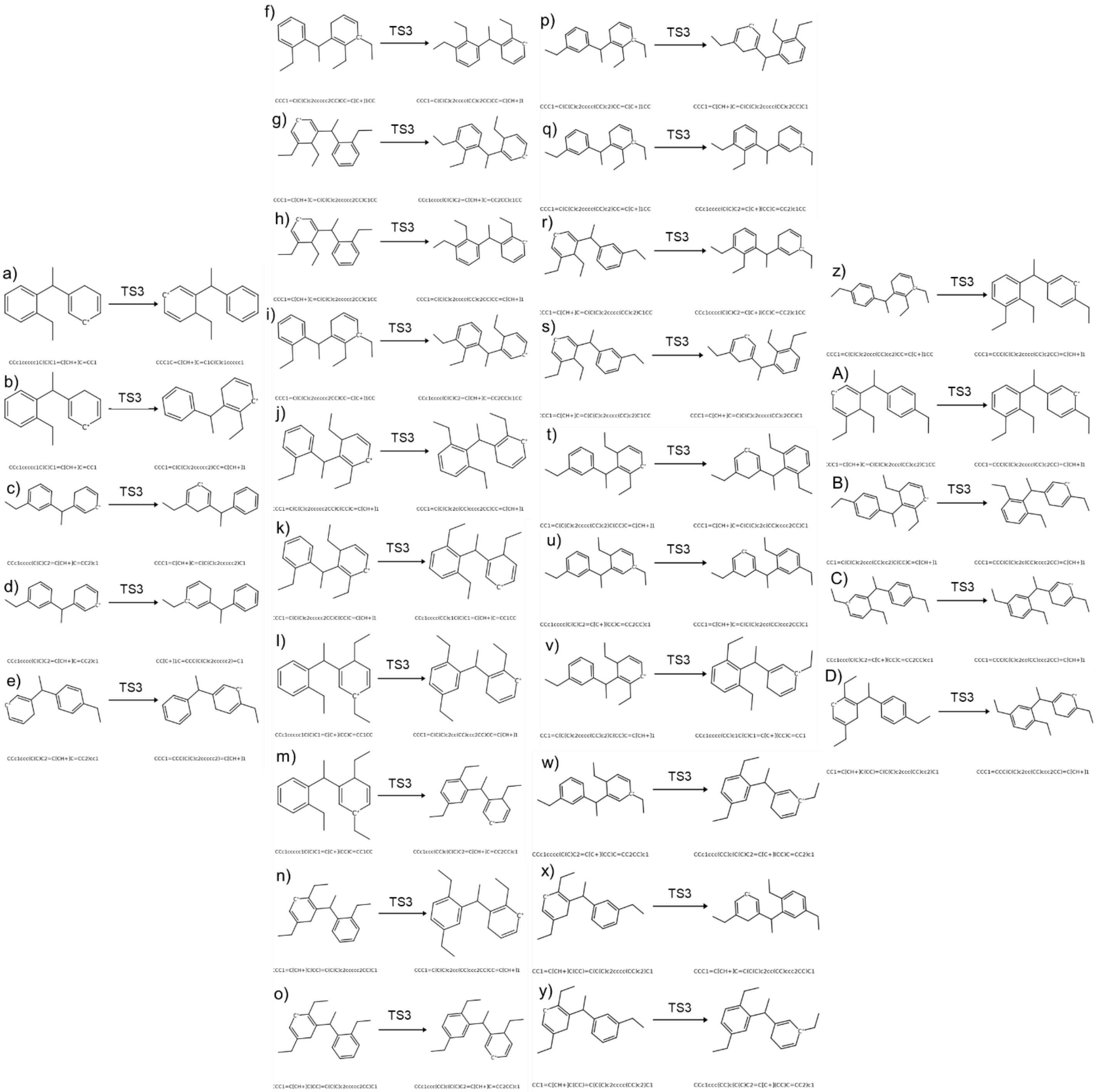

**Figure S21.** Multistep path ring-to-ring hydrogen shift reactions via TS3 are presented in panels (a–b) for ortho-DEB$^+$ with benzene, (c–d) for meta-DEB$^+$ with benzene, and (e) for para-DEB$^+$ with benzene; panels (f–i), (j–k), and (l–o) for ortho-DEB$^+$ reacting with ortho-, meta-, and para-DEB, respectively; panels (p–s), (t–v), and (w–y) for meta-DEB$^+$ reacting with ortho-, meta-, and para-DEB, respectively; and panels (z–A), (B), and (C–D) for para-DEB$^+$ reacting with ortho-DEB, para-DEB, and para-DEB diaryl intermediates, respectively.

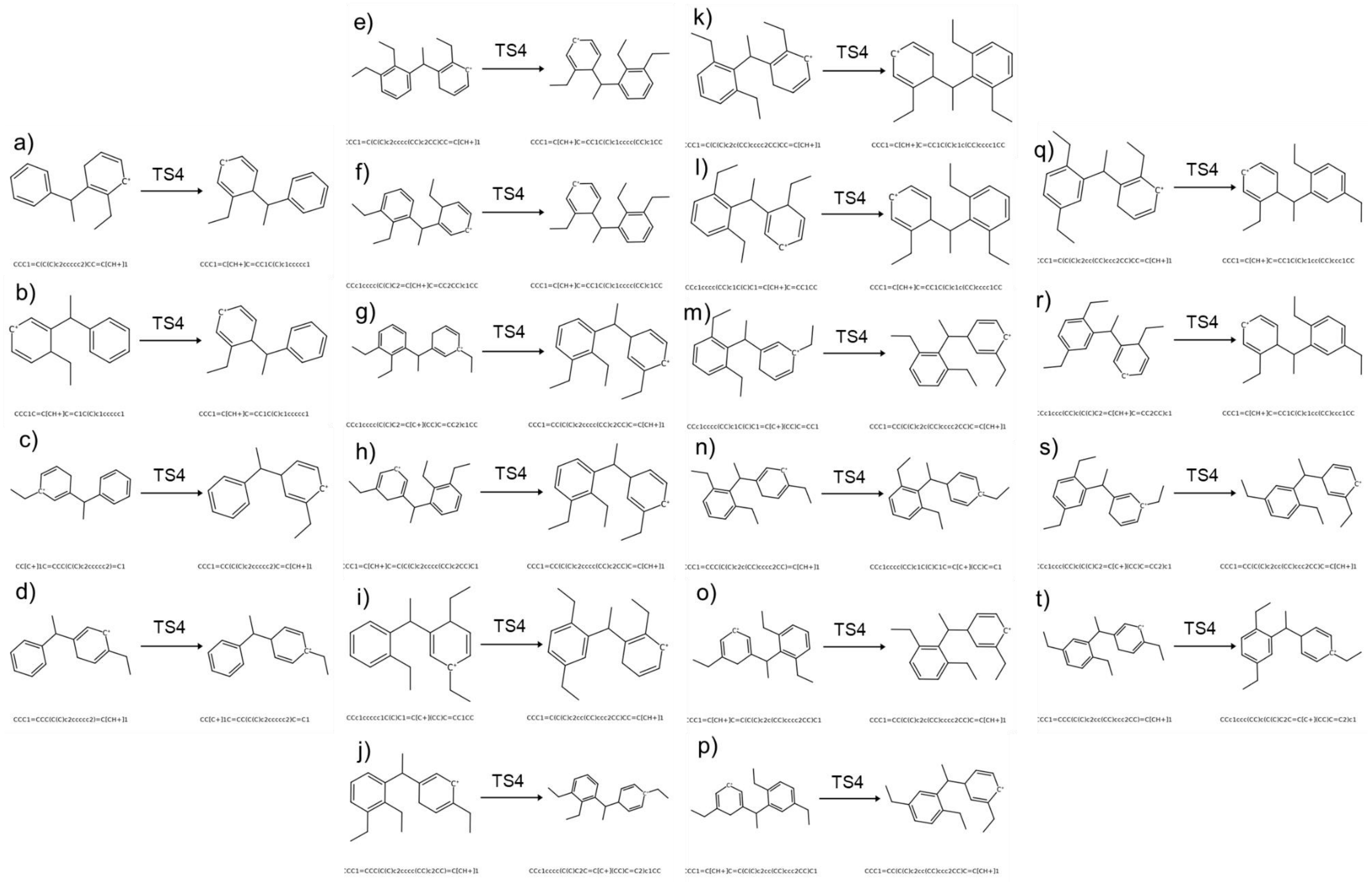

**Figure S22.** Multistep path intra-ring hydrogen shift reactions via TS4 are presented in panels (a–b) for ortho-DEB⁺ with benzene, panel (c) for meta-DEB⁺ with benzene, and panel (d) for para-DEB⁺ with benzene; panels (e–f), (g–h), and (i–j) for ortho-DEB⁺ reacting with ortho-DEB, meta-DEB, and para-DEB, respectively; panels (k–l), (m–n), and (o–p) for meta-DEB⁺ reacting with ortho-DEB, meta-DEB, and para-DEB, respectively; and panels (q–r), (s), and (t) for para-DEB⁺ reacting with ortho-DEB, para-DEB, and para-DEB diaryl intermediates, respectively.

## S2.2 DFT Gas Phase Study for Diethylbenzene Transalkylation

The gas-phase segment of the PoTS computational pipeline designed to identify TS gas-phase vibrational modes also enabled mechanistic reconstructions and provided valuable insights into the intrinsic kinetics of transalkylation and disproportionation reactions through the diaryl mechanism when zeolites are not involved. All diaryl cycle combinations in DEB transalkylation and disproportionation were systematically explored. In transalkylation, each ortho-, meta-, and para-$DEB^+$ reacts with benzene, while disproportionation involves each $DEB^+$ isomer interacting with each neutral DEB isomer, yielding nine distinct $DEB^+$–DEB pairs. These reactions begin via the TS0, for each of the 12 resulting $I1^+$ intermediates, see **Figure S24**. One direct pathway to $I4^+$ was identified, see **Figure S23 blue**, along with multiple multistep routes dictated by substituent arrangement, see **Figure S23** orange arrows and **Figure S25**, rendering a total of 42 unique energy paths, see **Figure S26a,b**. In the multistep paths, TS2 and TS4 involve hydrogen transfers between a secondary and a tertiary carbon or between two tertiary carbon atoms, whereas TS3 features proton transfer directly between opposite carbons or diagonally when the rings are tilted (see **Figure S25**).

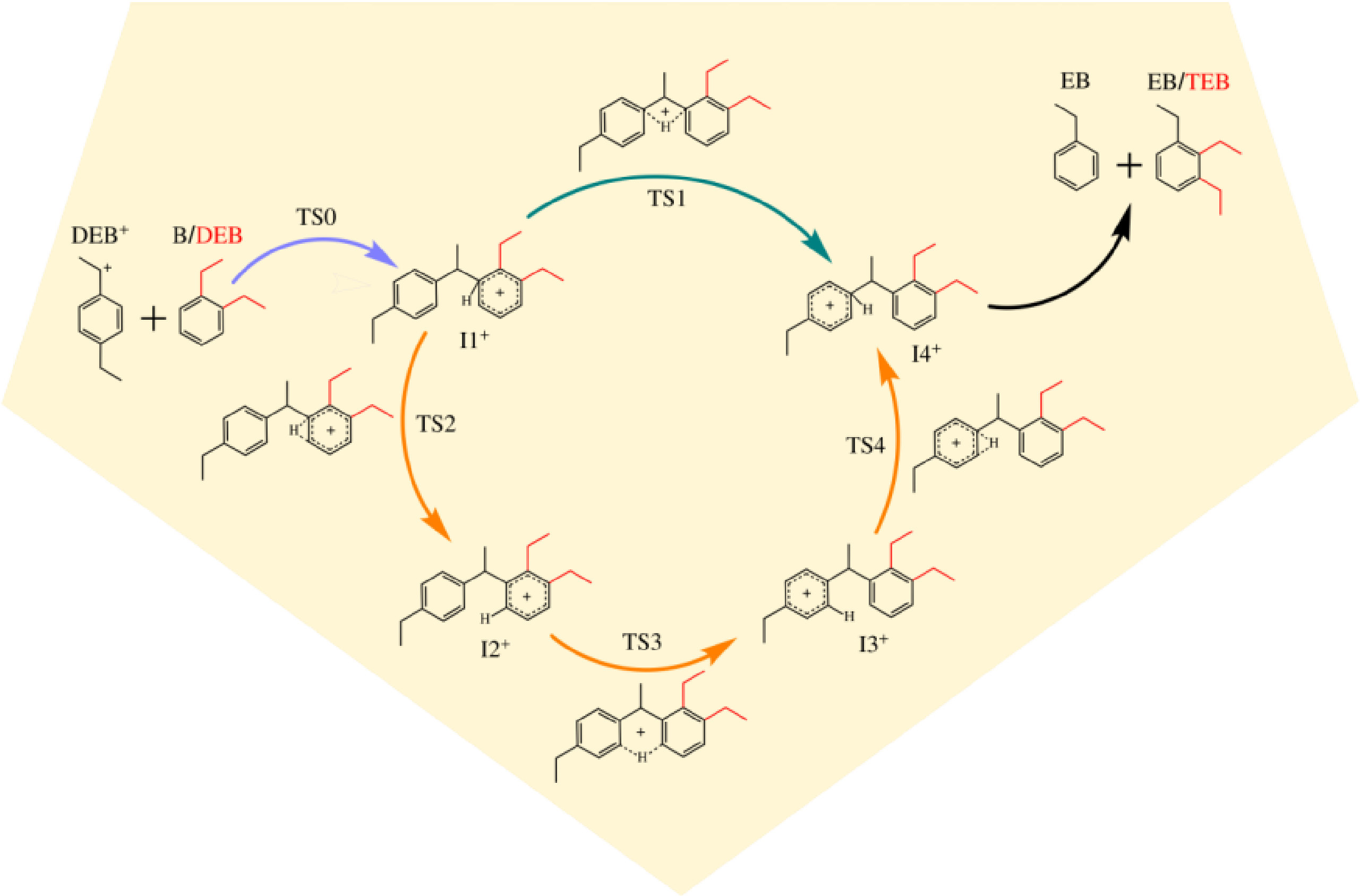

**Figure S23**. Diaryl-mediated pathway for the transalkylation of diethylbenzene with benzene and for the competing disproportionation yielding triethylbenzene, represented by the red bonds in all structures.

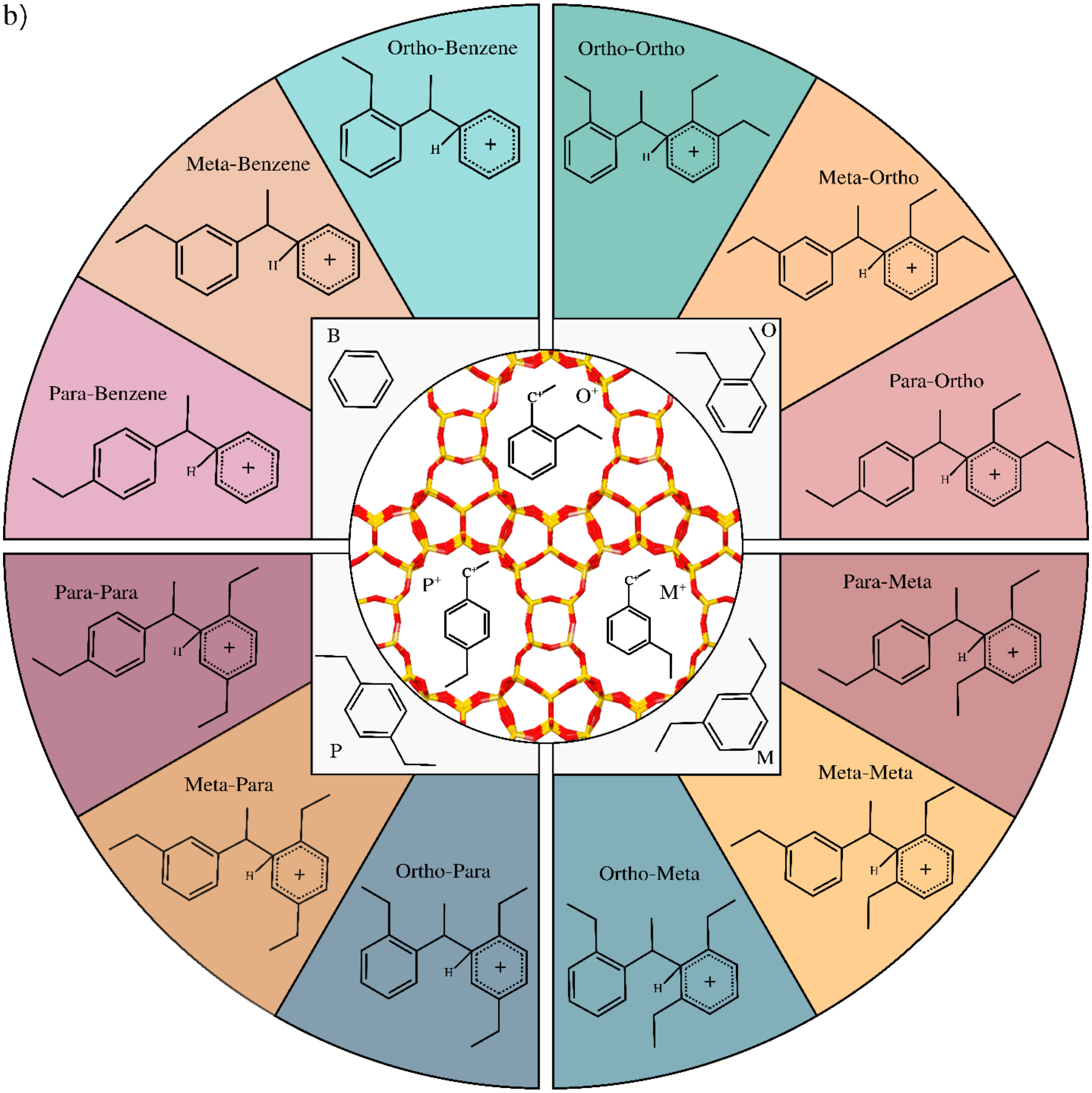

**Figure S24.** Schematic representation of the diaryl I1 isomeric pairs, color-coded to differentiate their positional relationships. Ortho-, meta-, and para-benzene reference isomers are shown in cyan, light orange, and pink, respectively. Ortho-ortho, meta-meta, and para-para isomers are represented in teal, bright orange, and burgundy, while mixed-position isomers, ortho-meta, ortho-para, meta-para, meta-ortho, para-ortho, and para-meta, are depicted in navy, dark blue, burnt orange, golden, magenta, and deep red. The central zeolite framework highlights the confined environment in which these intermediates interact, illustrating the spatial relationships between isomeric pairs within the pores.

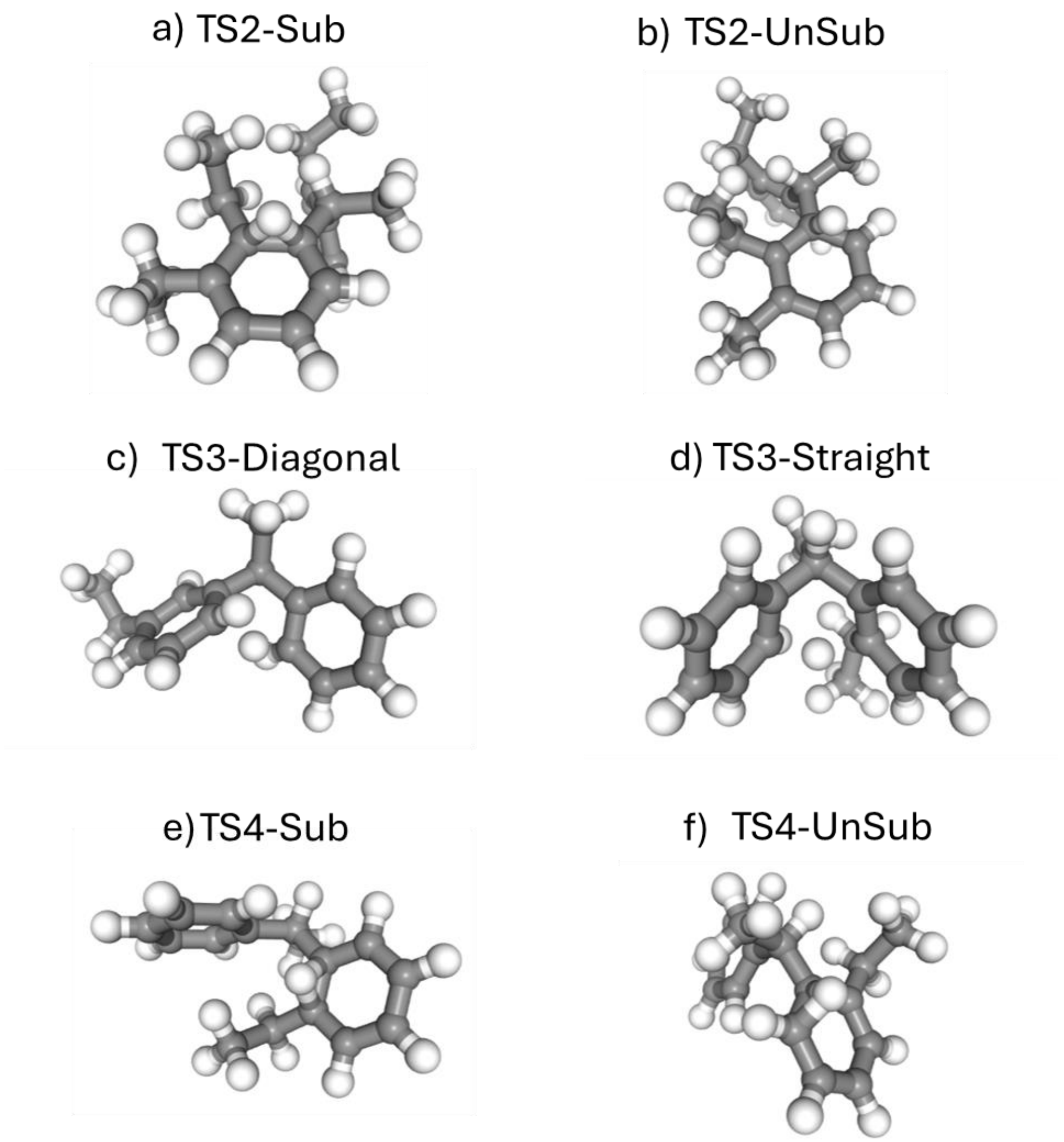

**Figure S25.** Diaryl Mechanism Multistep Path TS structure examples for a) TS2 between tertiary C atoms on the benzene ring b) TS2 between a tertiary C atom and a secondary C atom c) TS3 Diagonal d) TS3 Straight e) TS4 between tertiary C atoms f) TS4 between a tertiary C atom and a secondary C atom.

Although the zeolite study presented in the main manuscript shows a different scenario, in the gas phase the TS0 was identified, displaying low energy barriers that nevertheless represent the rate-determining step for each process, all barriers obtained are below 46kJ/mol. Along the multistep pathway, proton shifts between carbon atoms on the same aromatic ring via TS2 and TS4 occur rapidly. Consequently, the highest activation energy barrier in each pathway is attributed to the transfer of $H^+$ from one aromatic ring of the cationic diaryl intermediate to the other, proceeding through either TS1 or TS3, except for meta-benzene and meta-meta disproportionation, which will be discussed later in this section. Observing **Figure S26c**, which compares the energy barriers for each isomeric pair, it is evident that the barriers associated with direct path TS1 (purple) are consistently higher than those of the corresponding competing TS3 (green). This indicates that the multistep pathway, which involves a six-membered ring geometry at TS3, is kinetically favored over the direct pathway via TS1 because the more stable TS3 transition state reduces steric strain and electronic repulsion, thereby lowering the activation energy required for the proton transfer.

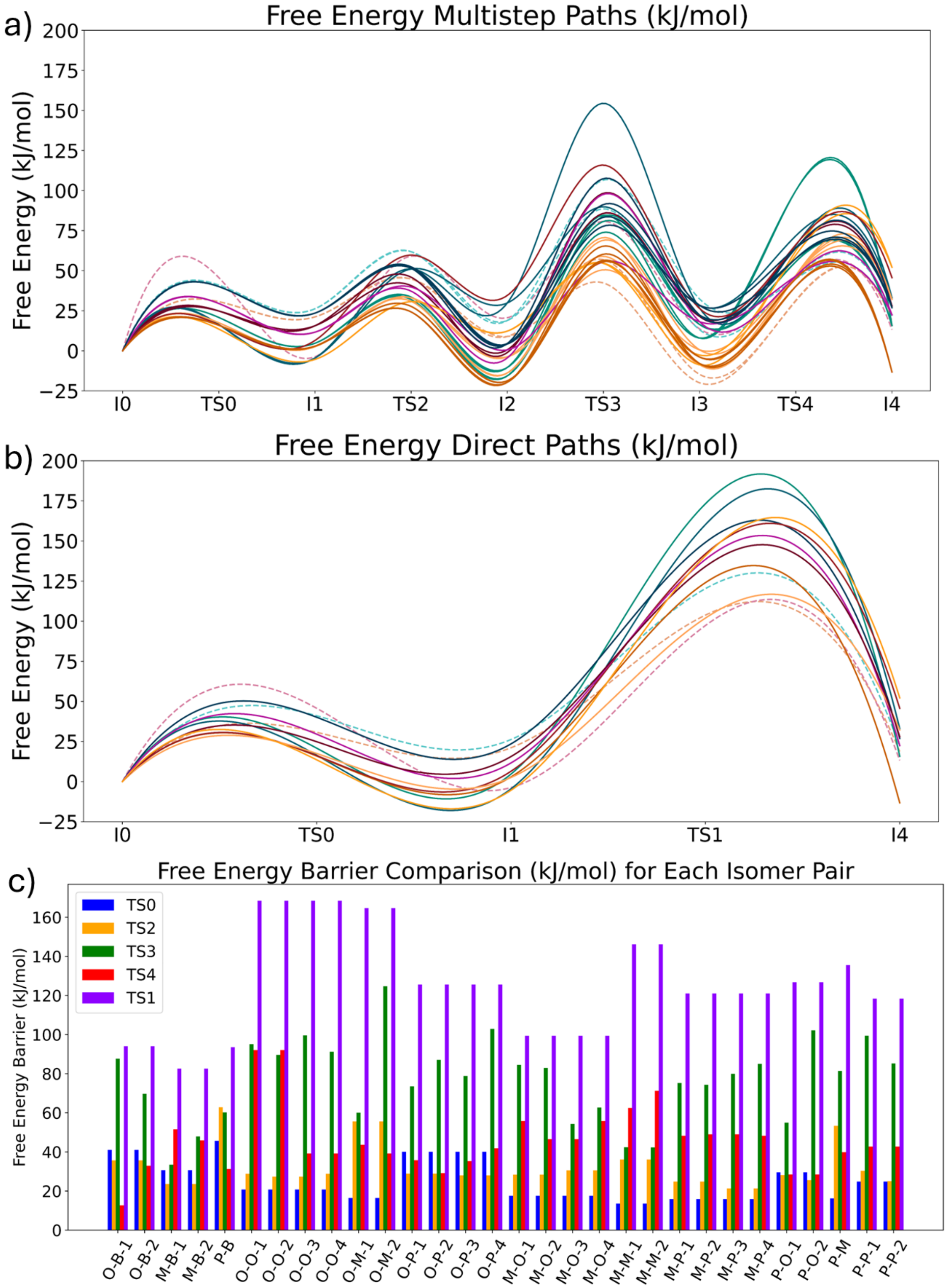

**Figure S26.** Gas-phase paths for ortho, meta, and para-DEB+ cation. a) Multistep paths b) Direct paths. Color palette used is described in **Figure 24** caption. Linestyles; Transalkylation dashed. Disproportionation solid.

Comparing transalkylation and disproportionation revealed that transalkylation exhibits lower activation barriers, particularly for TS1 and TS3, due to two main factors. First, transalkylation intermediates lack any substituents on the arene molecule being attacked by the $DEB^+$ cation. The absence of substituents on the benzene ring allows for more efficient ring folding and alignment during proton transfer at the transition state. In contrast, disproportionation intermediates must orient their ethyl chains to avoid steric interference with the proton transfer, which restricts optimal alignment and increases the transition state energy. As a result, transalkylation transition states are more stable than their disproportionation counterparts. This stability is reflected in the activation barriers, as illustrated by the case of ortho isomers: while the barrier for ortho-benzene TS1 is 94.05 kJ/mol, the corresponding values for disproportionation pathways are significantly higher, 168.47 kJ/mol for ortho-ortho, 164.71 kJ/mol for ortho-meta, and 125.69 kJ/mol for ortho-para.

Second, in the multistep pathways, the absence of substituents, which act as charge donors, on the benzene ring in all transalkylation intermediates results in less effective stabilization of the positive charge in I2, see **Figure S26**, and $\Delta G_{I3\text{-}I2}$ (kJ/mol) plotted in **Figure S28** for transalkylation and disproportionation species. This weaker stabilization shown by the majority of isomers considered in the study lowers the activation barrier for proton transfer through TS3. In contrast, charge delocalization is enhanced when the positive charge is transferred to a substituted benzene ring, as observed in transalkylation I3. This increased stabilization makes I3 more thermodynamically favorable than I2 for all transalkylation pathways except ortho-benzene-1, effectively making the reaction irreversible.

The stabilization of I3 also influences TS4 barriers, particularly for meta-isomers. For instance, in meta-benzene-1, the TS4 barrier is 51.54 kJ/mol, making it the rate-determining step, although it remains significantly lower than its TS1 barrier. In disproportionation, however, the stabilization effect of substituents occurs in both $I_2$ and $I_3$, leading to higher TS3 barriers and a less pronounced energetic bias between I3 and I4. This results in a more balanced energy landscape between the first and second stages of the multistep pathway, ultimately making disproportionation slower compared to transalkylation.

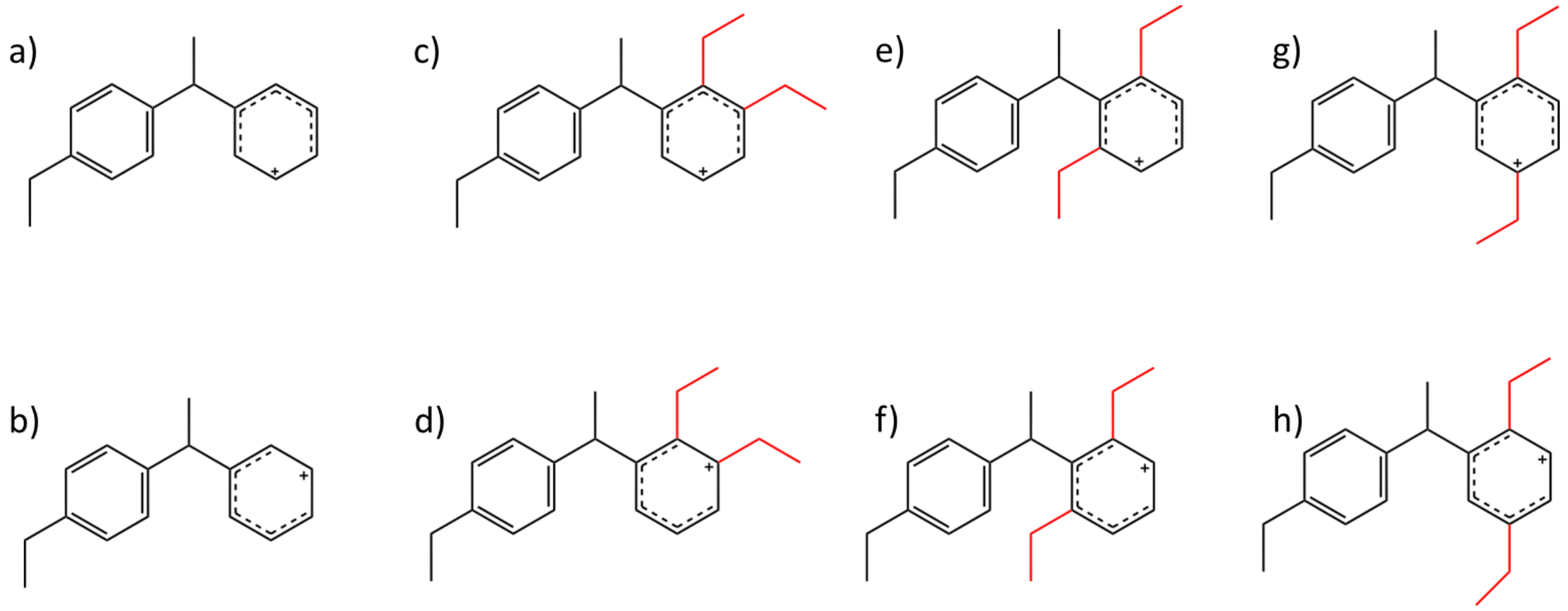

**Figure S27.** Depiction of I2 intermediates for para-DEB transalkylation a) & b), para-ortho disproportionation c) & d), para-meta disproportionation e) & f) para-para disproportionation g) & h).

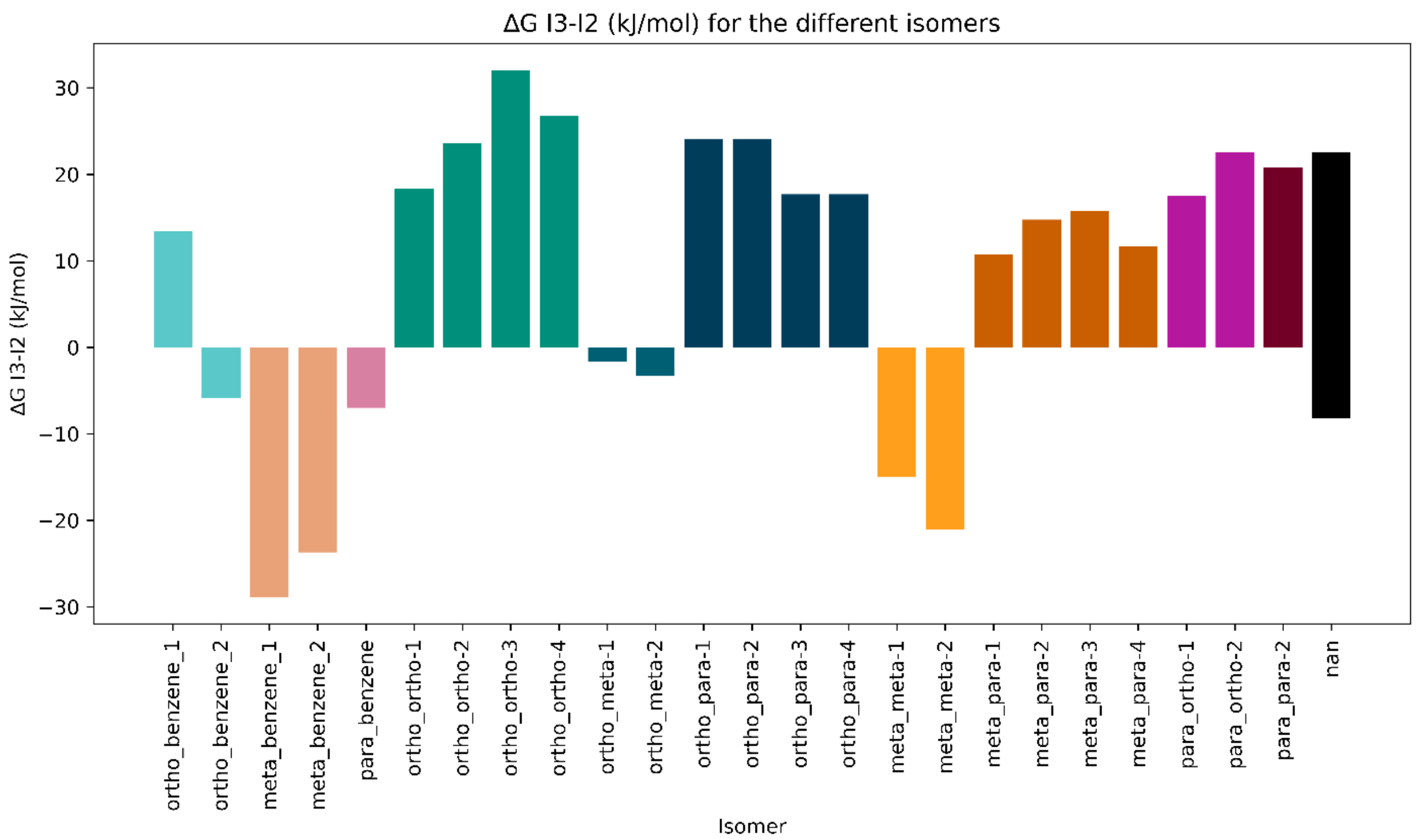

**Figure S28.** $\Delta G_{I3\text{-}I2}$ (kJ/mol) calculated for all different isomer combinations and multistep paths studied. Color palette corresponds to the one described in **Figure S24** caption.

Among the isomers, meta-benzene and meta-ortho disproportionation are the preferred routes in both the direct and multistep mechanisms, see **Figure S29.** In the transalkylation multistep pathway, the relatively poor stabilization of intermediate I2 and the enhanced stability of I3 lowers the barrier at TS3, while TS4 becomes rate-determining due to a high barrier associated with an overly stabilized I3. For the transalkylation direct mechanism, the meta isomer offers optimal branch positioning; in contrast, the para isomer overly stabilizes intermediates I1 and I4, thereby raising the TS1 barrier, and the ortho isomer suffers from steric hindrance due to its branch positioning. For disproportionation, the meta-ortho isomer exhibits a high activation barrier at TS1 (99.37 kJ/mol). However, in the multistep pathway meta-ortho-3, the reaction proceeds through the Diagonal TS3, where benzene ring tilting positions the ortho substituent away from the reactive site. This geometric adjustment facilitates proton transfer, resulting in a significantly lower barrier of 54.34 kJ/mol (see **Figure S29** for MEPs and **S30** for MEPs TS structures, **Figure S31** for Multistep and direct paths by isomer).

The comparison between DEB transalkylation and disproportionation highlights the crucial influence of branching degree and position on reaction kinetics. While transalkylation benefits from sterically unhindered intermediates and lower activation barriers, disproportionation is constrained by ethyl group positioning, leading to higher barriers and a less favorable energy landscape. These results indicate that even in the absence of any zeolite confinement effect, disproportionation is intrinsically more energetically demanding than transalkylation.

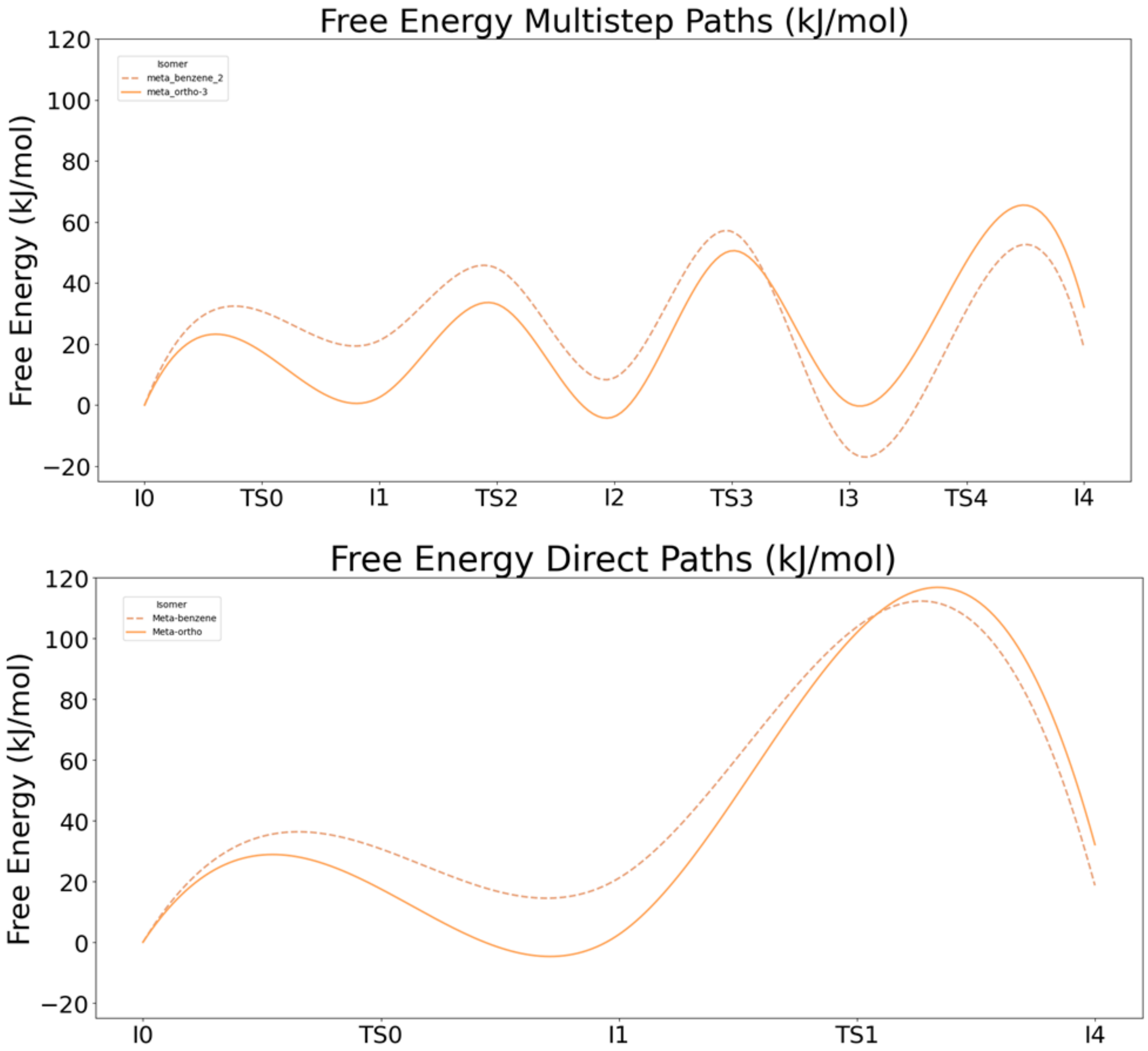

**Figure S29.** Gas-phase Minimum Energy Paths for a) Direct path b) Multistep Path, linestyles; Transalkylation dashed. Disproportionation solid. Color palette described in Figure S24 caption.

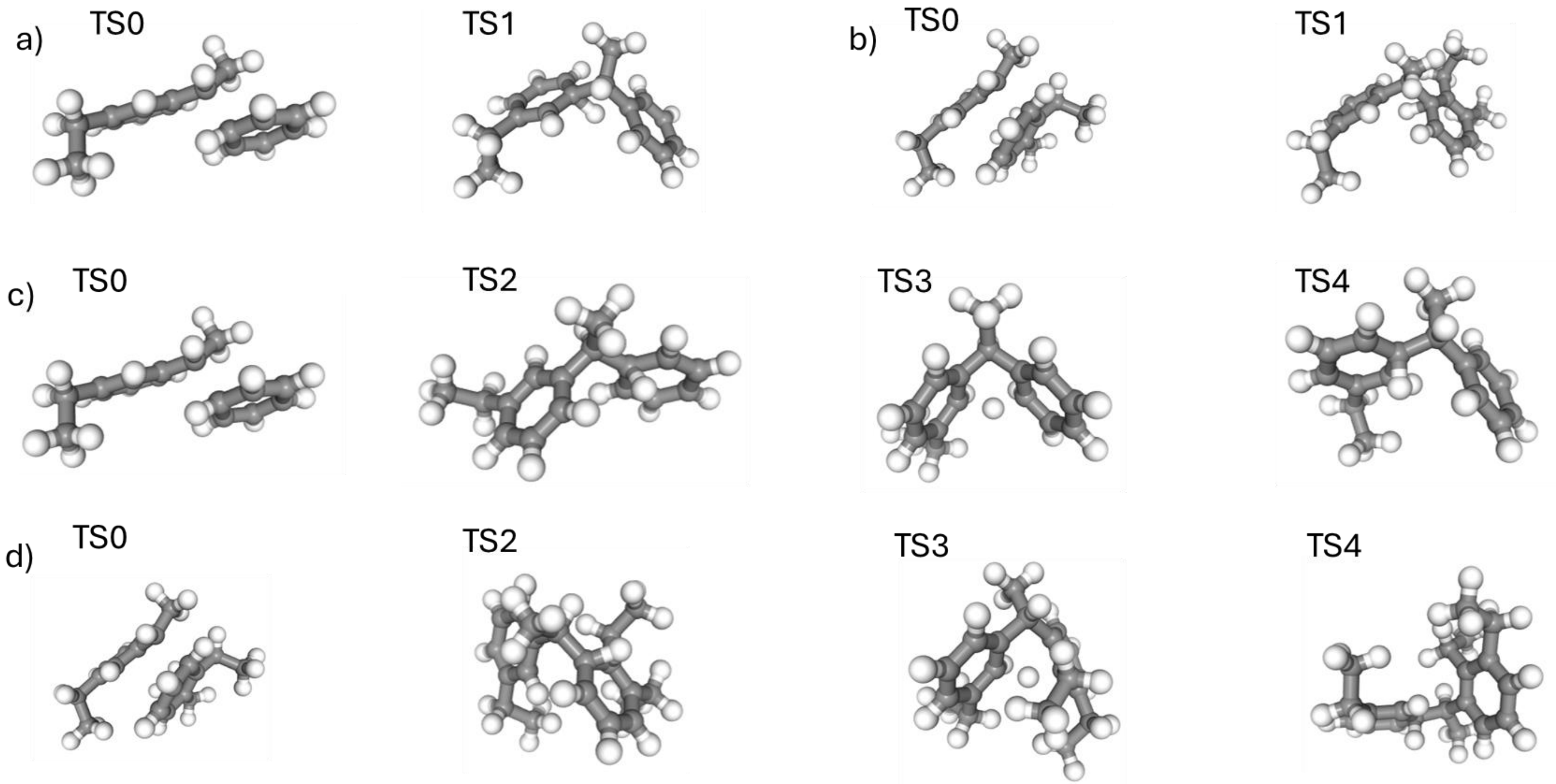

**Figure S30.** Gas-phase TS structure for diaryl mechanism MEP a) Direct Path for meta-benzene transalkylation b) Direct Path for meta-ortho disproportionation c) Multistep path for meta-benzene transalkylation d) Multistep path for meta-ortho disproportionation.

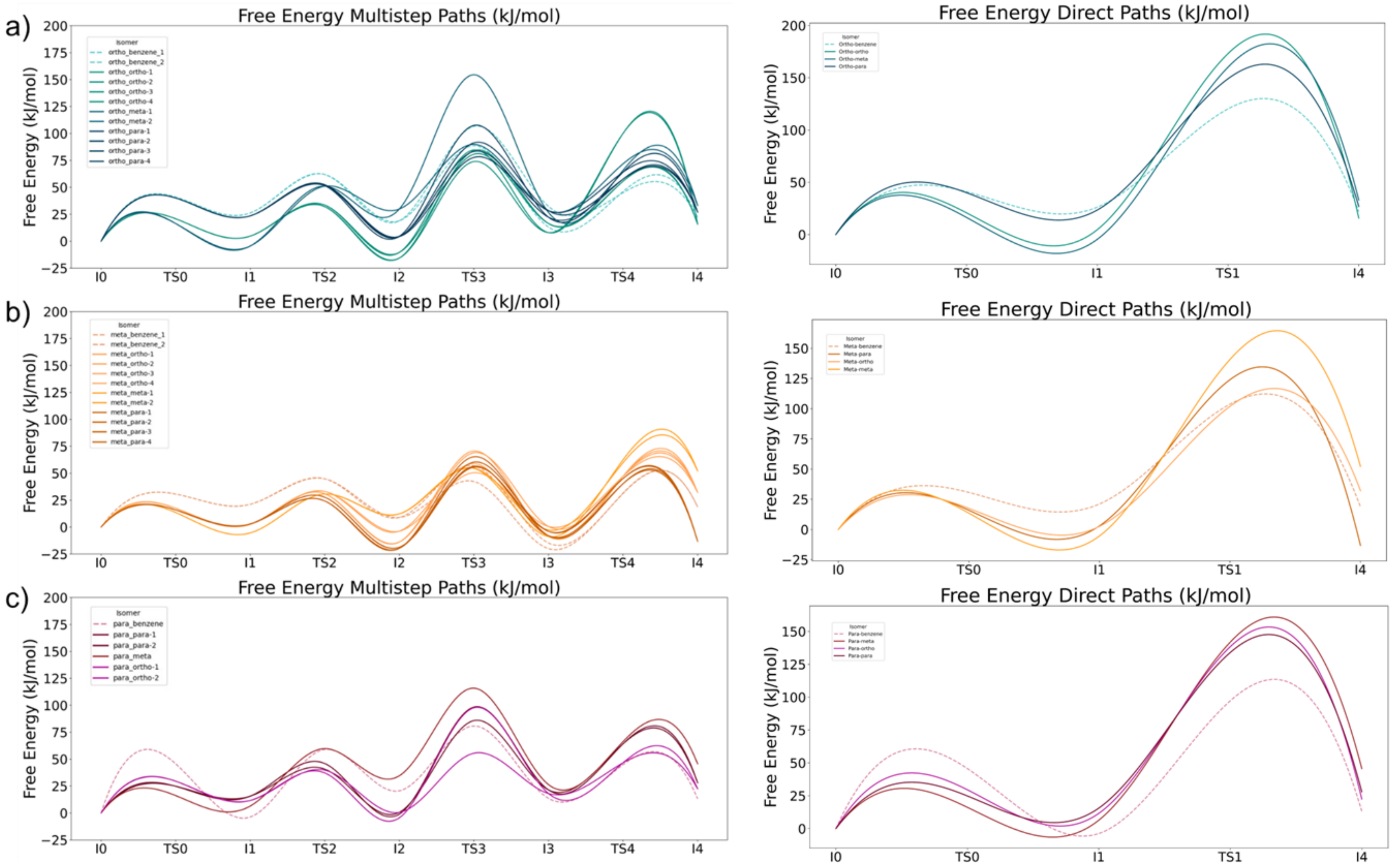

**Figure S31.** Gas-phase multistep and direct paths for a) ortho-DEB+ cation b) meta-DEB+ cation c) para-DEB+ cation. Blue shades denote $I1^+$ intermediates derived from ortho-$DEB^+$, orange from meta-$DEB^+$, and burgundy from para-$DEB^+$. Further details for the color palette are detailed in Figure S24 caption. Linestyles; Transalkylation dashed. Disproportionation solid.

## S2.3 DEB ZEOLITE STUDY; Reaction Mechanisms

This section presents the energy pathways identified for the different mechanisms discussed in the main paper, aiming to provide a clearer representation of these pathways given the large number of reaction profiles generated by the PoTS computational pipeline.

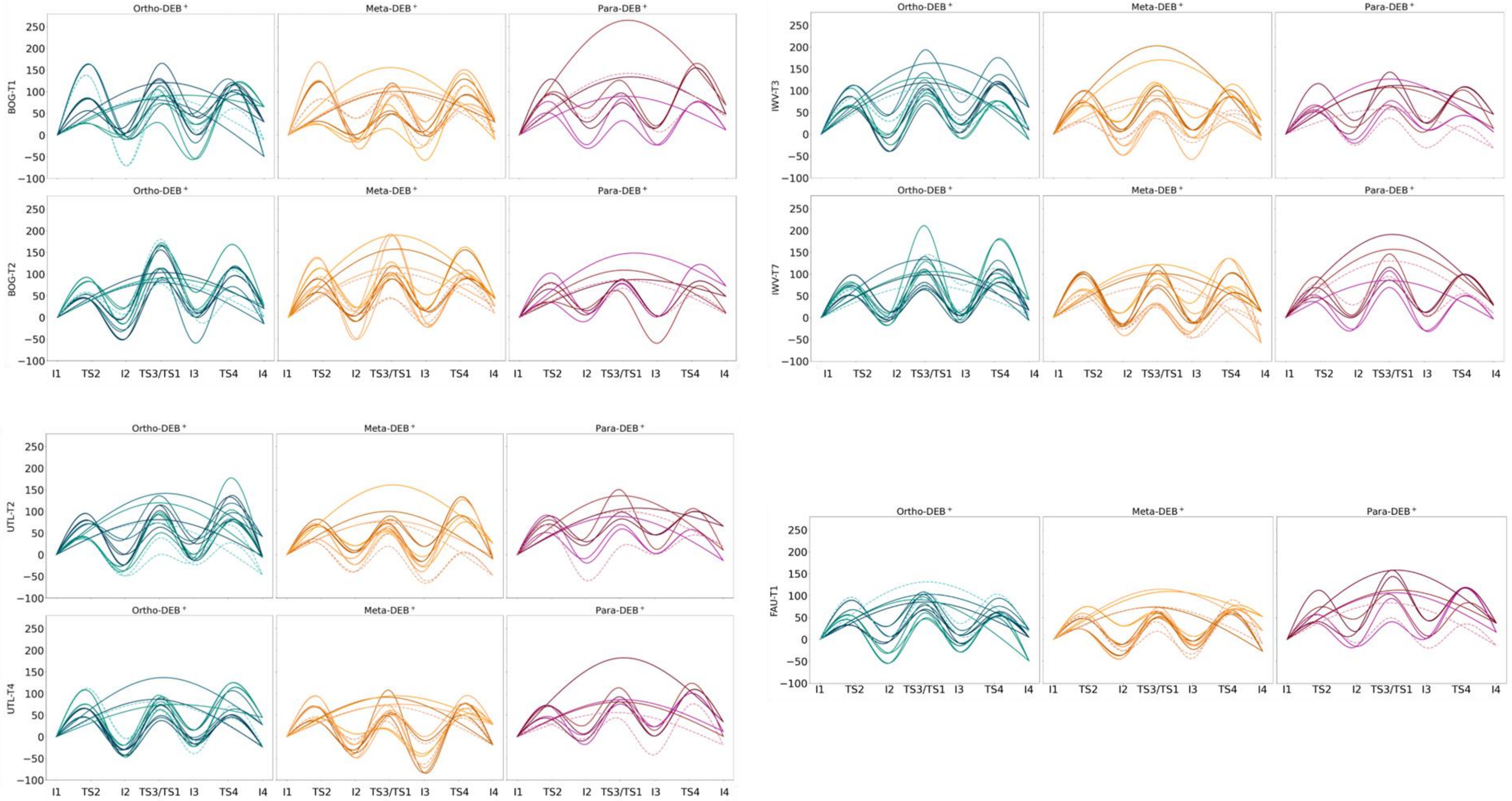

**Figure S32.** Zeolite multistep and direct paths for ortho, meta and para-DEB+ cation. Frameworks and Al T-sites indicated on the Y axis labels. Blue shades denote $I1^+$ intermediates derived from ortho-$DEB^+$, orange from meta-$DEB^+$, and burgundy from para-$DEB^+$. Further details for the color palette are detailed in Figure S24 caption. Linestyles; Transalkylation dashed. Disproportionation solid.

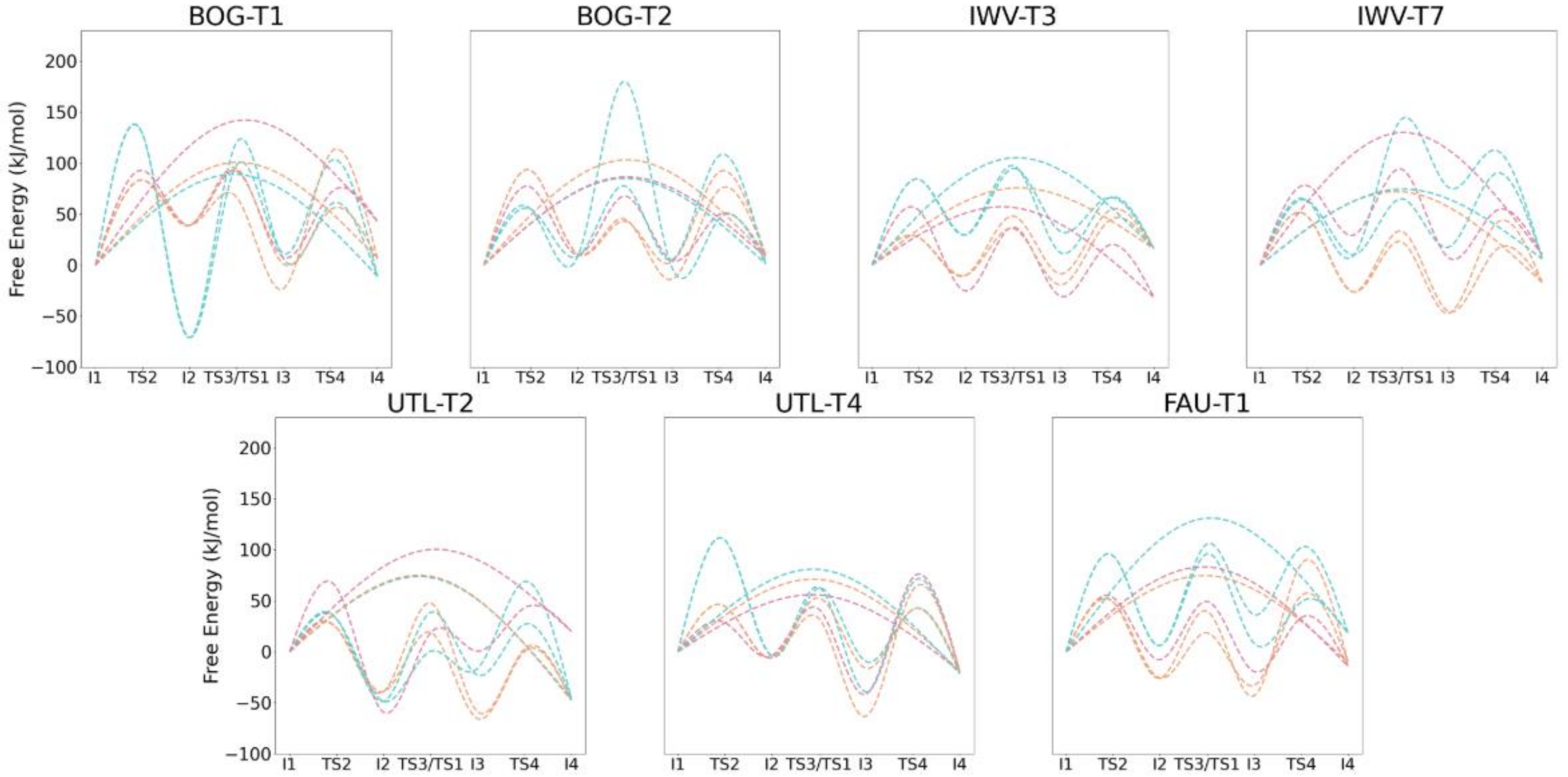

**Figure S33.** Transalkylation zeolite multistep and direct paths for ortho, meta and para-DEB+ cation. Frameworks and Al T-sites indicated on the plot titles. Blue shades denote **I1⁺** intermediates derived from ortho-DEB⁺, orange from meta-DEB⁺, and burgundy from para-DEB⁺. Further details for the color palette are detailed in **Figure S24** caption.

a

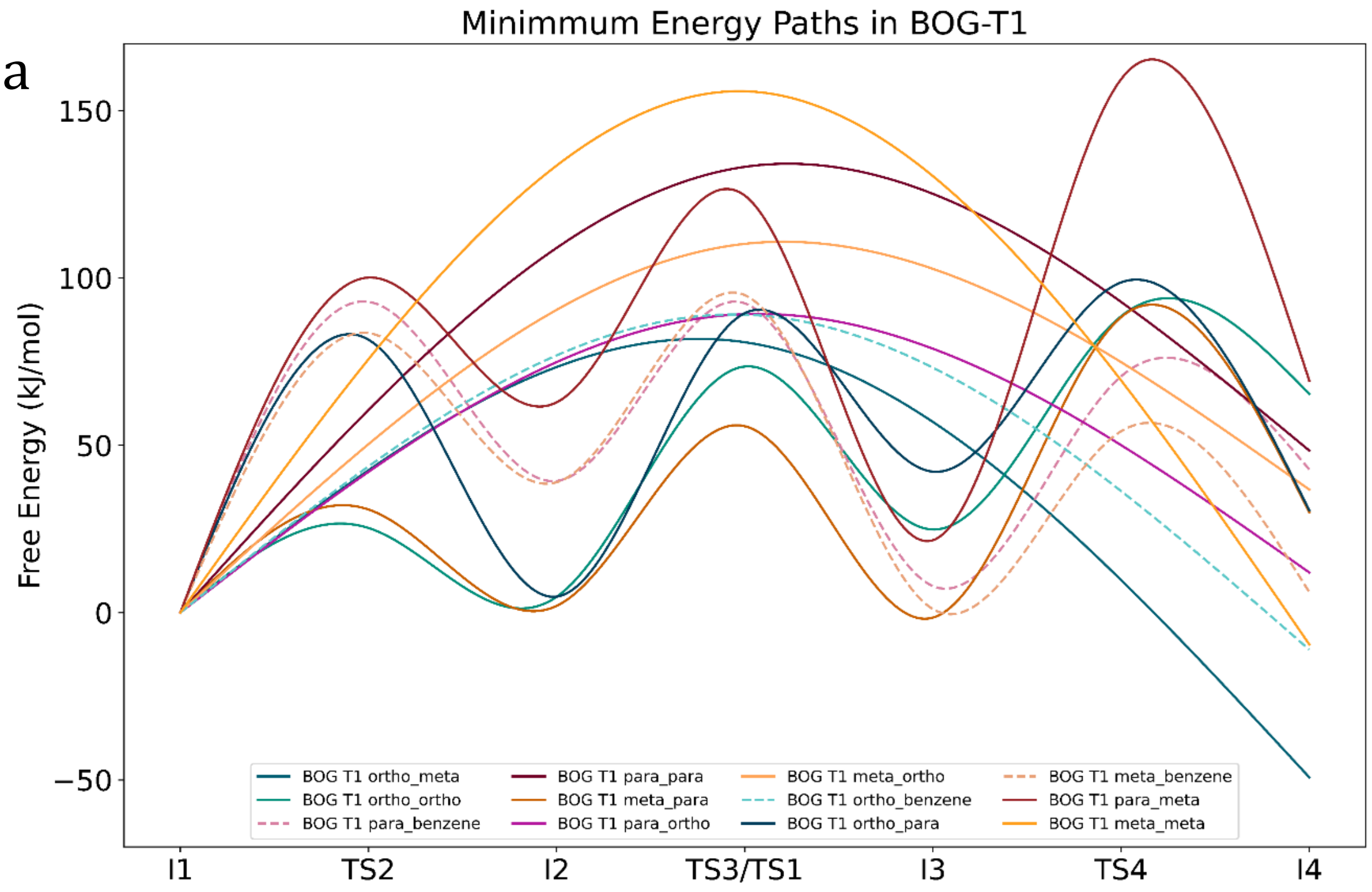

b

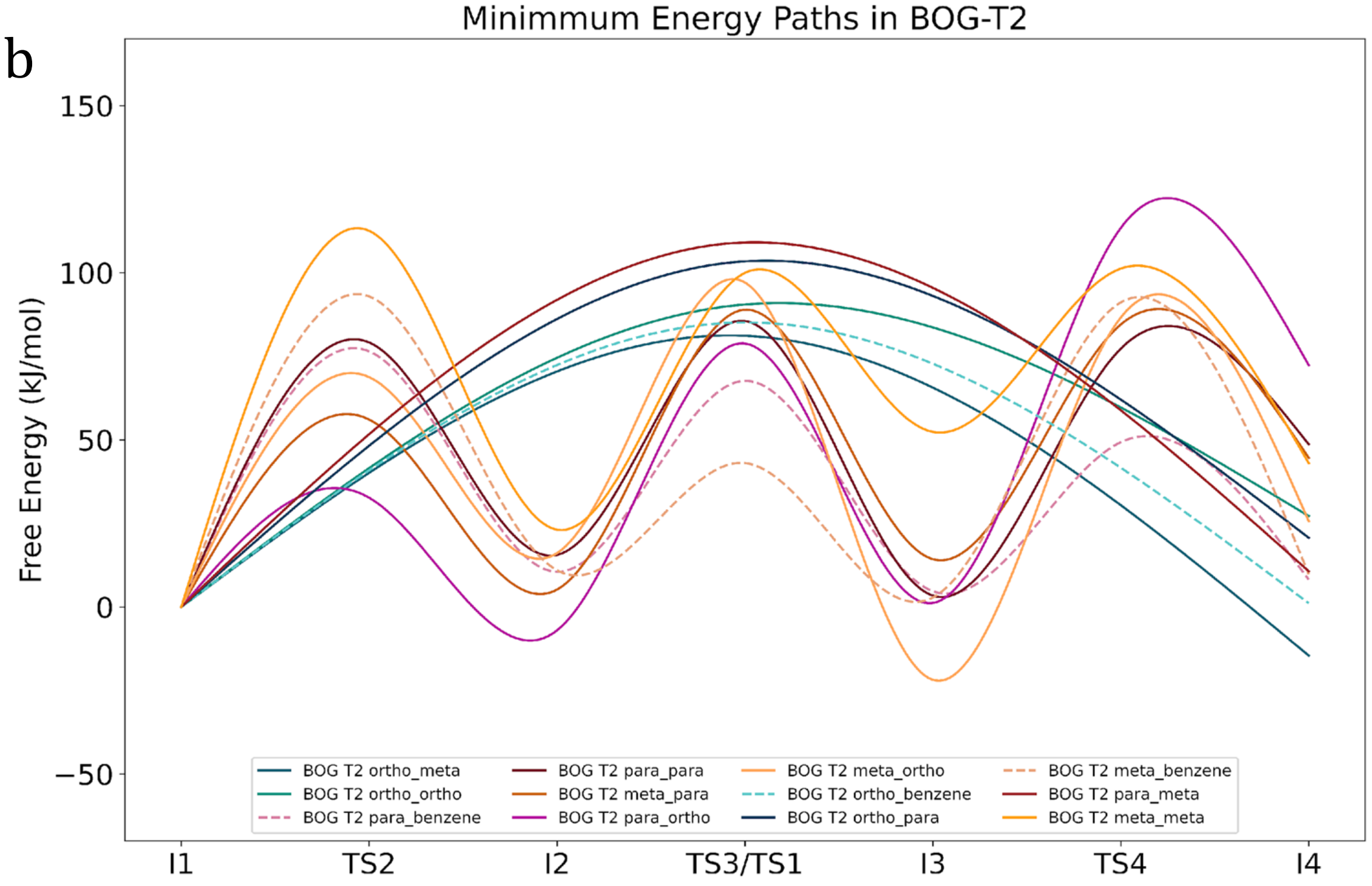

**Figure S34.** MEPs by isomer on a) BOG-T1 b) BOG-T2 models. Blue shades denote **I1⁺** intermediates derived from ortho-DEB⁺, orange from meta-DEB⁺, and burgundy from para-DEB⁺. Further details for the color palette are detailed in **Figure S24** caption.

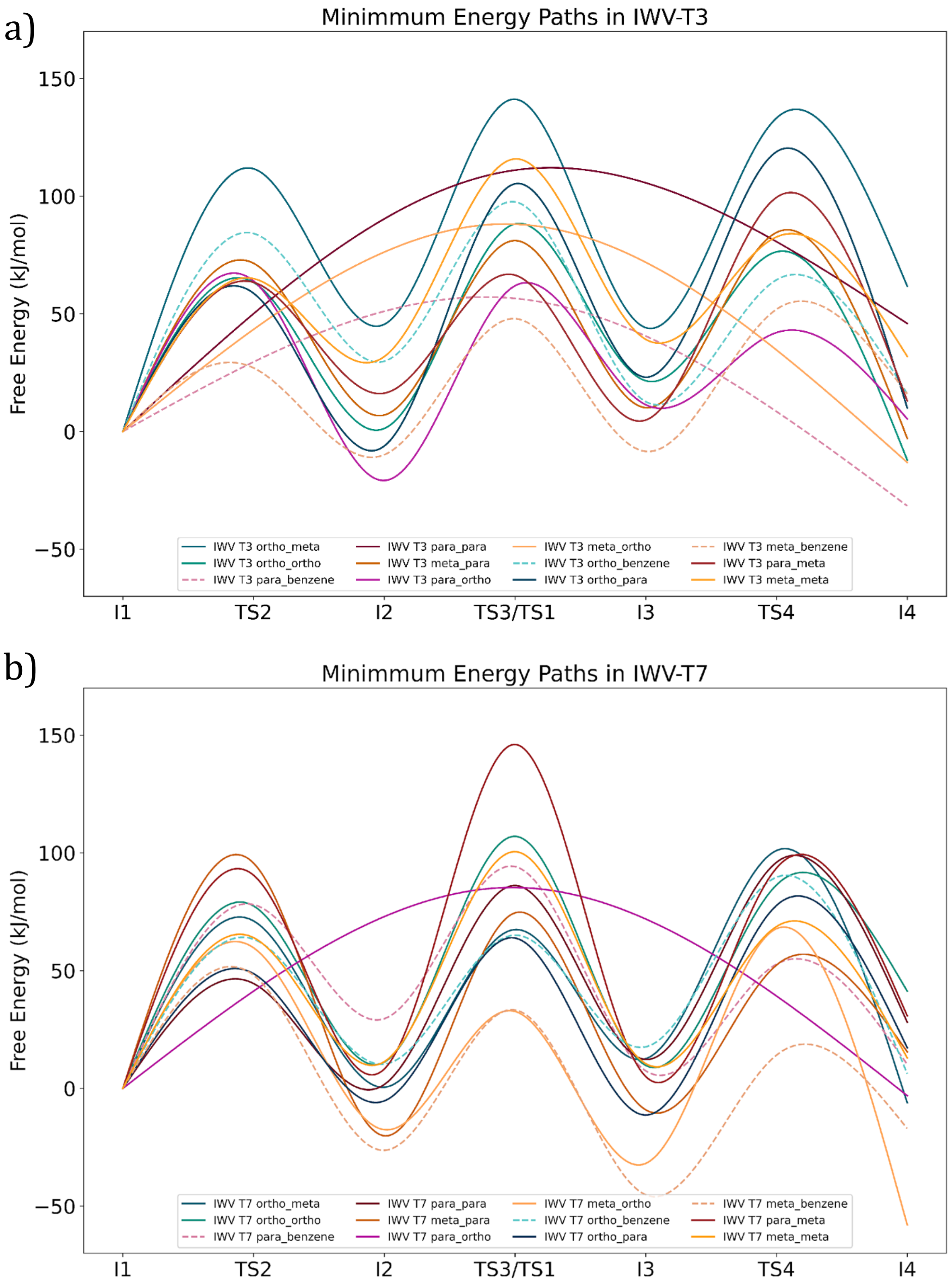

**Figure S35**. MEPs by isomer on a) IWV-T3 b) IWV-T7 models. Blue shades denote $I1^{+}$ intermediates derived from ortho-$DEB^{+}$, orange from meta-$DEB^{+}$, and burgundy from para-$DEB^{+}$. Further details for the color palette are detailed in **Figure S24** caption.

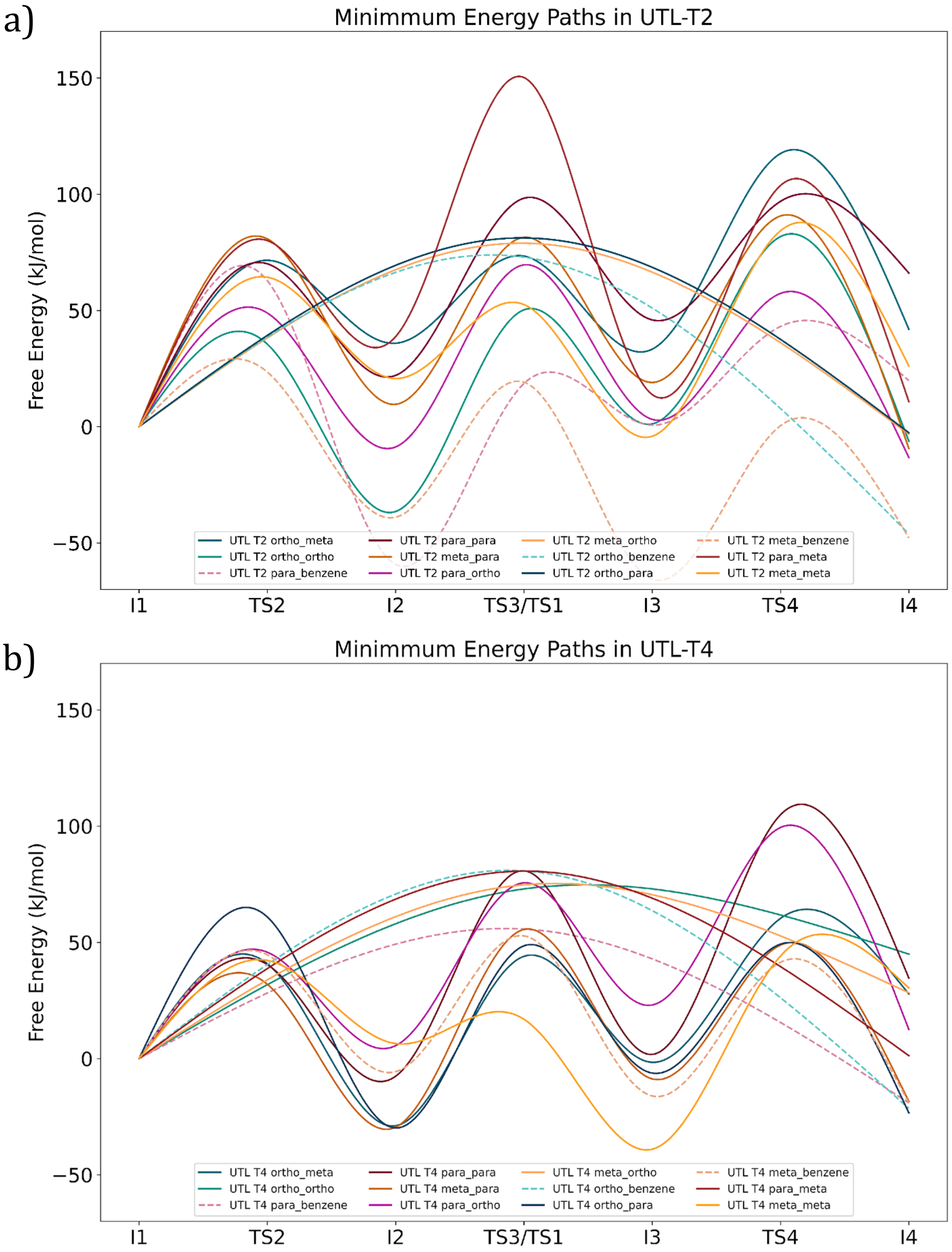

**Figure S36**. MEPs by isomer on a) UTL-T2 b) UTL-T4 models. Blue shades denote $I1^+$ intermediates derived from ortho-$DEB^+$, orange from meta-$DEB^+$, and burgundy from para-$DEB^+$. Further details for the color palette are detailed in **Figure S24** caption.

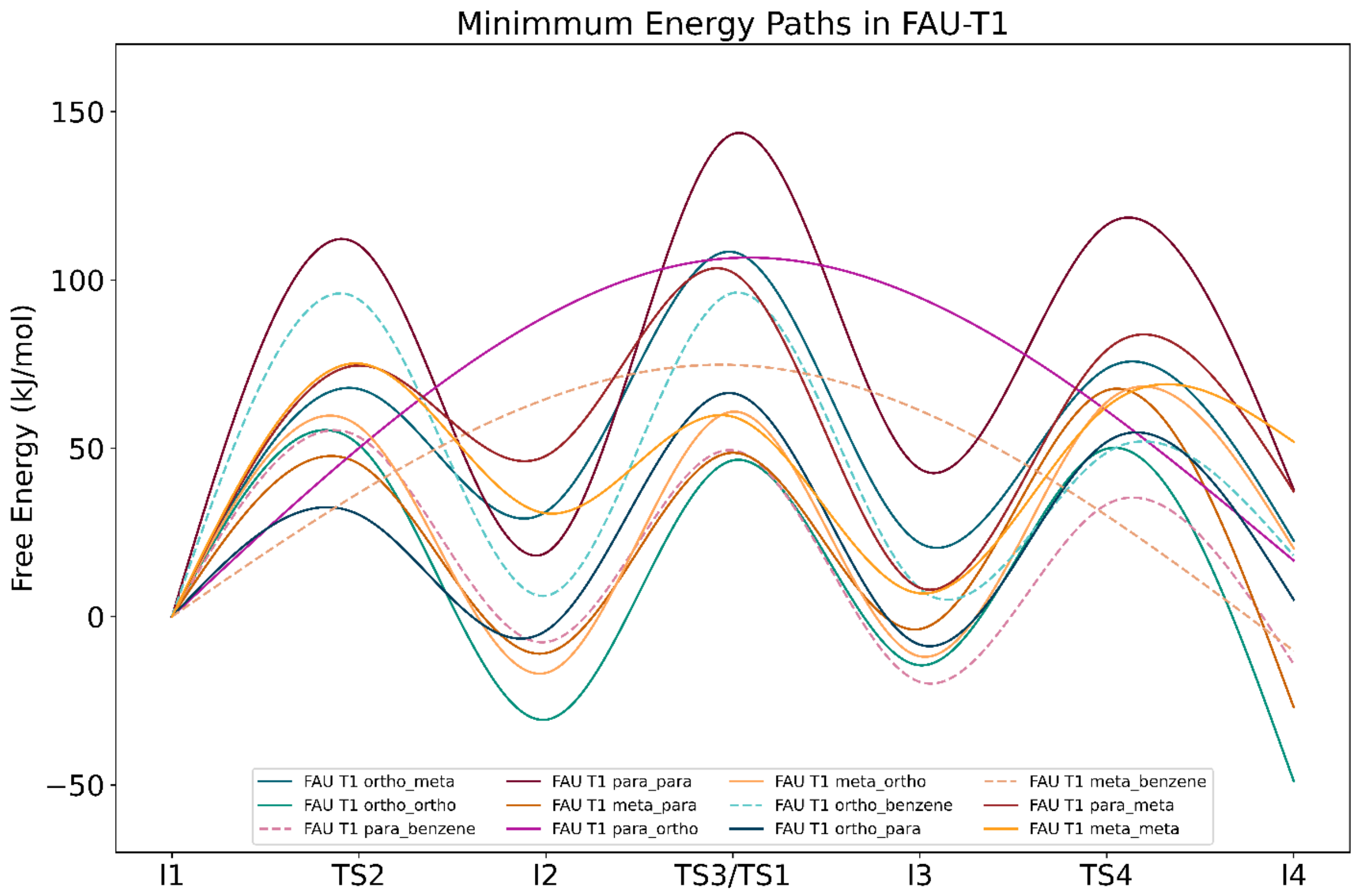

**Figure S37.** MEPS by isomer on FAU-T1 model. Blue shades denote I1⁺ intermediates derived from ortho-DEB⁺, orange from meta-DEB⁺, and burgundy from para-DEB⁺. Further details for the color palette are detailed in **Figure S24** caption.

## S2.4 ZEOLITE STUDY: Confinement

This section describes how zeolite confinement effects were quantified in this work. We established two metrics to translate confinement effects into numerical values. First, we calculated the minimum distances from each hydrogen atom in the diaryl disproportionation molecules docked within the zeolite channels to the nearest framework atom, either oxygen or silicon. To obtain the effective atomic distances, we subtracted the covalent radius of hydrogen, 0.37 Å and a fixed value of 1.35 Å for the corresponding framework atom, either Si or O, from each measured value, treating atoms as hard spheres with predefined radii. The 1.35 Å value for framework atoms independently of their nature was chosen to align with the same hard sphere approximation applied in the free pore area measurements, to be presented later, providing a unified criterion for evaluating confinement effects across different metrics. Each structure yielded a total of 35 minimum distances, which were averaged to obtain a single mean value per structure. Similarly, we determined the maximum distance from each hydrogen atom to the framework by identifying the framework atom in the opposite direction of the previously identified minimum distance vector **(Figure S38**). These maximum distances were also averaged into a mean value.

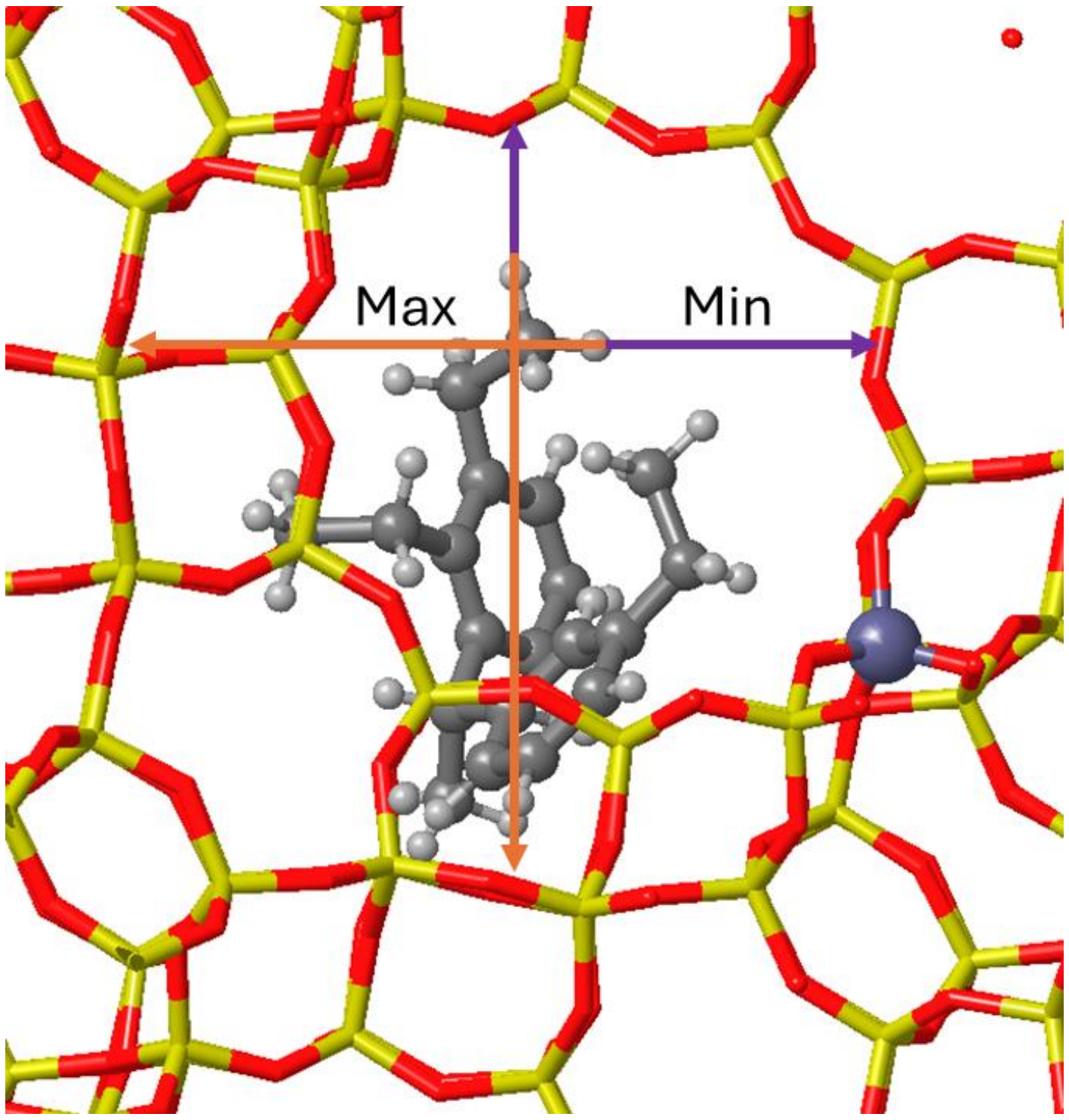

**Figure S38**. Schematic representation of the minimum (Min) and maximum (Max) distance calculations for a diaryl disproportionation intermediate confined within a zeolite pore. The minimum distance is measured from each hydrogen atom to the nearest framework atom (Si or O), while the maximum distance is determined in the opposite direction from the minimum distance vector. These values were used to quantify zeolite confinement effects, following a hard sphere approximation with predefined atomic radii.

For a clearer assessment of confinement effects, only the values for disproportionation intermediates from all isomeric pairs were plotted in **Figure S39**. These intermediates are the largest species within the zeolite pores and are therefore the most susceptible to overconfinement effects, making them the most suitable for comparing confinement across different frameworks.

The results of these confinement metrics reveal a clear trend across different zeolite frameworks. BOG, the narrowest system, exhibits the smallest mean minimum distances, with values clustering around 1.3–1.5 Å, indicating a tight interaction between the hydrogen atoms of the bulky diaryl disproportionation intermediates and the zeolite framework. IWV follows a similar trend but with slightly higher mean minimum distances (centered around 1.4–1.6 Å), reflecting the more open 12×12 MR topology compared to BOG's 12×10 MR structure. UTL, with its wider 14×12 MR channels, further increases these distances, reaching 1.6–1.8 Å, while FAU, the least confining framework, shows the highest mean minimum distances, with some values exceeding 2.0 Å.

A similar trend is observed in the mean maximum distances, which further illustrate how available space increases from BOG to FAU. BOG shows max distances between 3.2 and 3.9 Å, highlighting the tight confinement in its narrow channels. IWV presents max distances around 3.2–4.0 Å, still indicative of a constrained environment. UTL, with its larger channels, allows for max distances between 3.5 and 4.2 Å, while FAU, due to its spacious framework, enables significantly larger values, exceeding 5 Å.

Comparing Al placement in IWV shows that the mean maximum distance for disproportionation intermediates is 3.47 Å for T3, but only 3.26 Å for T7. Acid site T3 resides at the 12×12 MR intersection (8.54 Å in diameter, per the IZA database), whereas T7 is located in the narrower 12 MR channel (6.9×6.2 Å). This tighter environment for T7 leads to lower relative energies (see Stabilization section). Meanwhile, the mean minimum distances remain nearly identical for T3 (1.49 Å) and T7 (1.48 Å).

For disproportionation intermediates in UTL models, T2 and T4 exhibit similar mean minimum distances, 1.59 Å and 1.60 Å, respectively, which are slightly higher than IWV's, reflecting UTL's wider 14×12 MR channels. Mean maximum distances are also larger: 3.67 Å (T2) and 3.76 Å (T4). T2 sits at the 14×12 MR intersection (9.3 x 9.3 Å), whereas T4 occupies the 14 MR channel (9.5×7.1 Å), providing additional space for the disproportionation intermediates and thus better stabilization (see Stabilization section) which correlates with the higher yield of triethylbenzene observed experimentally compared to IWV catalyst.

It is important to note that these are mean values, meaning that individual maximum distances can be significantly higher, especially in FAU, where the spacious framework allows for much larger deviations beyond the reported averages. Data plotted in **Figure S39** is given in **Table S21**.

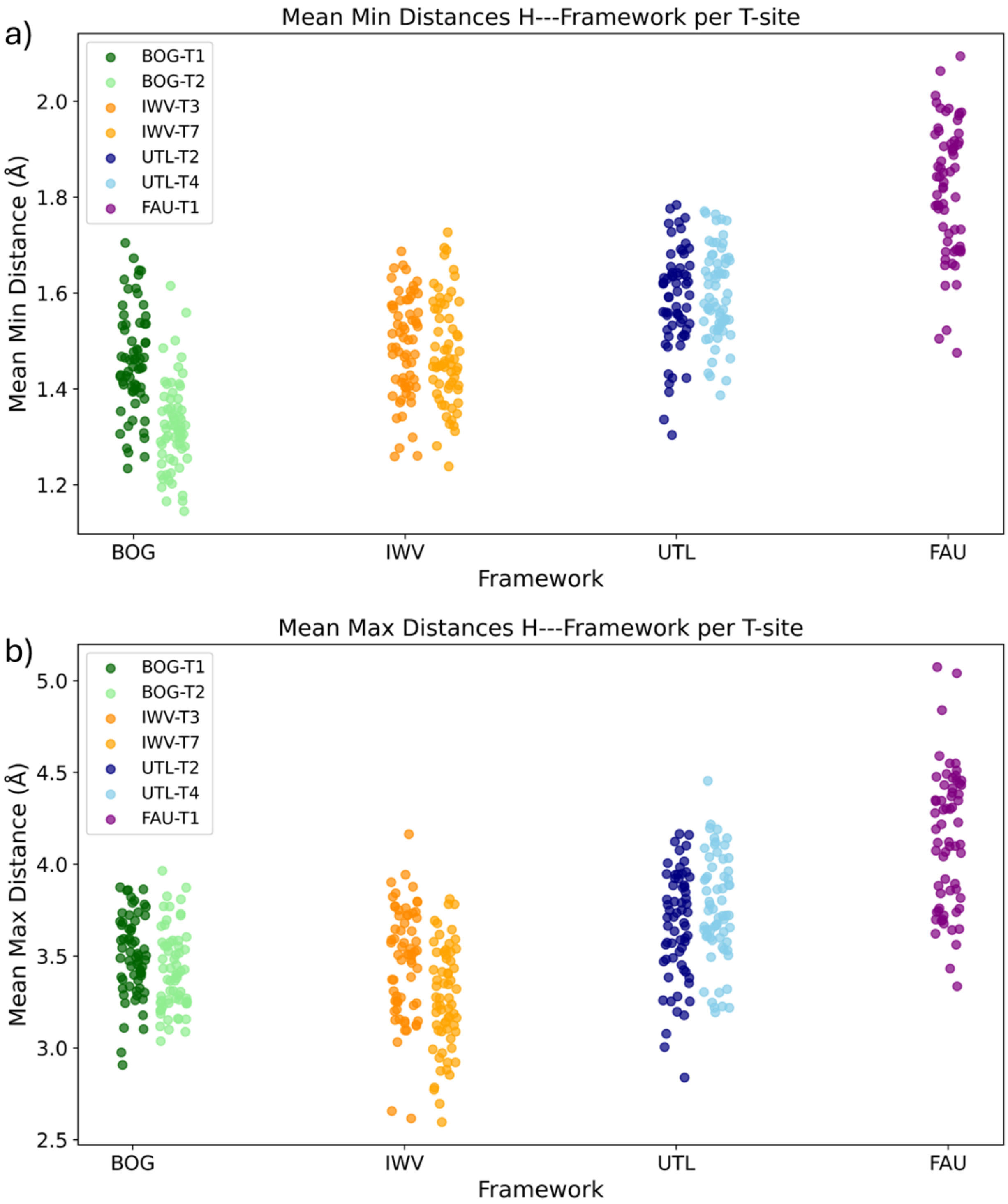

**Figure S39**. Mean distances calculated for each disproportionation diarylic intermediate inside the different zeolite model pores. a) minimum distances b) maximum distances. Color palette, BOG-T1 (darkgreen), BOG-T2 (light green), IWV-T3 (darkorange), IWV-T7 (light orange), UTL-T2 (darkblue), UTL-T4 (light blue), FAU (purple).

Second, we quantified the free pore area available when a diaryl molecule occupies a zeolite channel. To do so, we identified the three principal molecular axes (x, y, and z) based on the most distant atomic coordinates along each direction, see **Figure S40**. As both most distant atoms on each axis are always hydrogen atoms, twice the distance of the covalent H radius, 0.37 Å, was added to each distance. The longest axis was discarded, as it aligns with the channel direction and does not contribute to confinement constraints. The effective molecular cross-sectional area was then approximated using an elliptical model based on the middle and shortest axes.

The channel area was determined using the maximum sphere diameter that can fit within each pore, as reported in the IZA database: 8.05 Å for BOG, 8.54 Å for IWV, 9.3 Å for UTL, and 11.24 Å for FAU. We selected the sphere diameter from the IZA database instead of the reported channel dimensions to account for both channel constraints and intersection accessibility, providing a more representative measure of the effective pore space. According to IZA database definition these maximum sphere diameters were computed geometrically using Delaunay triangulation, assuming that framework T- and O-atoms are hard spheres with a diameter of 2.7 Å. The calculations were performed using the ideal $SiO_2$ framework coordinates in the highest possible symmetry, as provided in the Atlas of Zeolite Frameworks.[2]

The free pore area was then obtained by subtracting the molecular cross-sectional area from the channel area, see **Figure S40c**. A visual representation of these calculations is provided in **Figure S41**, and the corresponding data is summarized in **Table S20**.

The equations used for these calculations are as follows:

$$TS\ Area = \pi\ x \frac{Middle\ axis}{2}\ x\ \frac{Minimum\ Axis}{2}$$

$$Channel\ Area = \pi\ x\ \left(\frac{Max\ D\ Sphere\ IZA}{2}\right)^2$$

$$Free\ Pore\ Area\ =\ Channel\ Area - TS\ Area$$

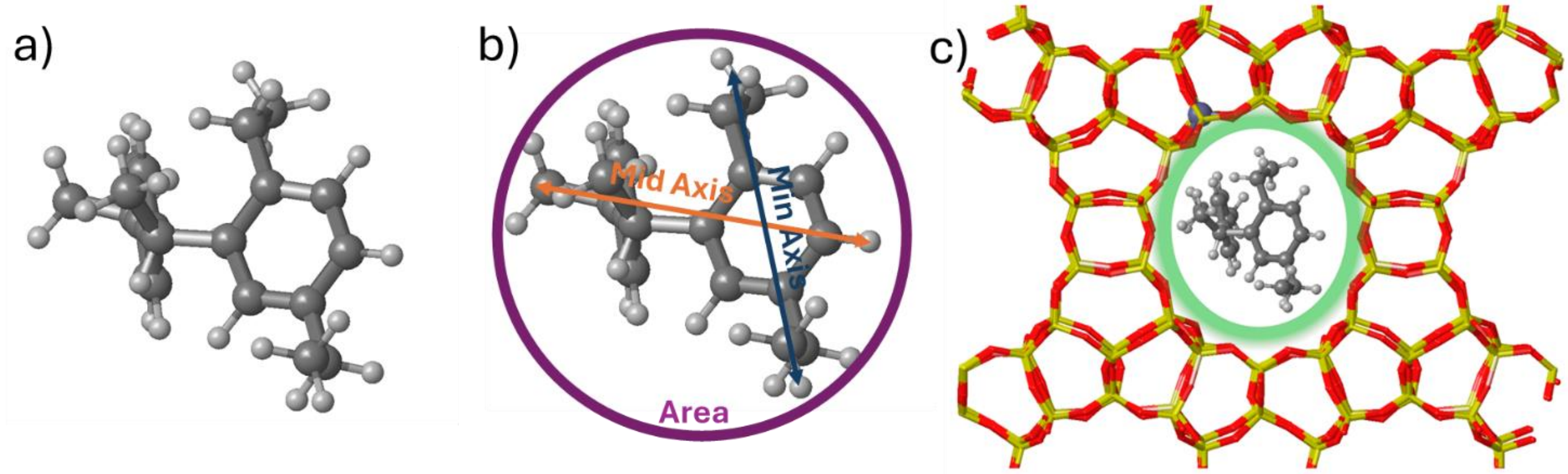

**Figure S40.** a) Transition state (TS) structure studied. b) Schematic representation of the TS structure, highlighting the minimum axis (blue), middle axis (orange), and molecule area (purple). c) Free pore area remaining in the zeolite channel after accommodating the molecule, shown in green. The longest axis is not included in the area calculation, as it is assumed to align with the channel direction and does not contribute to confinement constraints. The free pore area is calculated as: Channel Area = π x ((Max D Sphere IZA)/2)2, TS Area = π x (Middle axis/2) x (Minimum Axis)/2, Free Pore Area = Channel Area – TS Area.

Comparing the obtained free pore area values for transalkylation (**Figure S41a**) and disproportionation (**Figure S41b**), it is evident how channel size influences the confinement effects observed in the main paper's free energy barriers.

In the case of BOG (green), the narrowest system with a 12×10 MR topology, diaryl intermediates are restricted within a straight 12MR channel with narrow 10MR intersections. As a result, free pore area values approach zero or even negative values for disproportionation intermediates in 75 out of the 128 cases, due to the rough approximation used in molecular area calculations. This highlights the pronounced confinement of BOG, reflected in the highest free energy barriers observed in the PoTS pipeline, with an average free pore area of 7.7 Å$^2$ for transalkylation intermediates and -1.9 Å$^2$ for disproportionation intermediates.

For IWV (orange), a 12×12 MR bidimensional topology provides slightly more space and broader intersections. While disproportionation intermediates exhibit similar confinement behavior as in BOG, reaching free pore area values close to zero, only 17 negative values are observed among the 128 disproportionation intermediates measured, with an average free pore area of 9.1 Å$^2$. Transalkylation intermediates experience a more accommodating environment, with an average free pore area of 18.0 Å$^2$. This suggests that IWV offers optimal confinement for transalkylation and tight confinement for disproportionation intermediates.

UTL (blue), with its wider 14×12 MR channels and larger intersections, provides an average free pore area of 30.2 Å$^2$ for transalkylation intermediates. For disproportionation intermediates, free pore area decreases to an average value of 18.2 Å$^2$, and no negative values are observed even when molecule areas exceed 50 Å$^2$. This results in better confinement and lower barriers for disproportionation, particularly in the UTL-T4 model.

Finally, FAU (purple) is so spacious that it accommodates all intermediates without significant confinement effects, leading to lower transalkylation and disproportionation barriers, likely due to electron cloud van der Waals interactions.

Comparing the frameworks, an interesting trend emerges: the average free pore area for disproportionation intermediates in IWV is larger than the free pore area for transalkylation intermediates in BOG, despite the latter being smaller molecules. A similar pattern is observed when comparing UTL with IWV and FAU with UTL. This suggests that diaryl intermediates require over 20 Å$^2$ of free pore area to proceed with either the transalkylation or disproportionation mechanism. Notably, this threshold is only met for disproportionation in the UTL and FAU frameworks in a considerable amount of cases.

No clear differences were observed among ortho-, meta-, and para-DEB isomers in terms of free pore area. This can be attributed to the flexibility of diaryl intermediates, which ethyl branches can fold and benze rings are able to rotate within the channels to achieve favorable accommodations. As a result, confinement effects do not significantly vary between isomers, and no noticeable deviations or spikes in free pore area values are observed in the data.

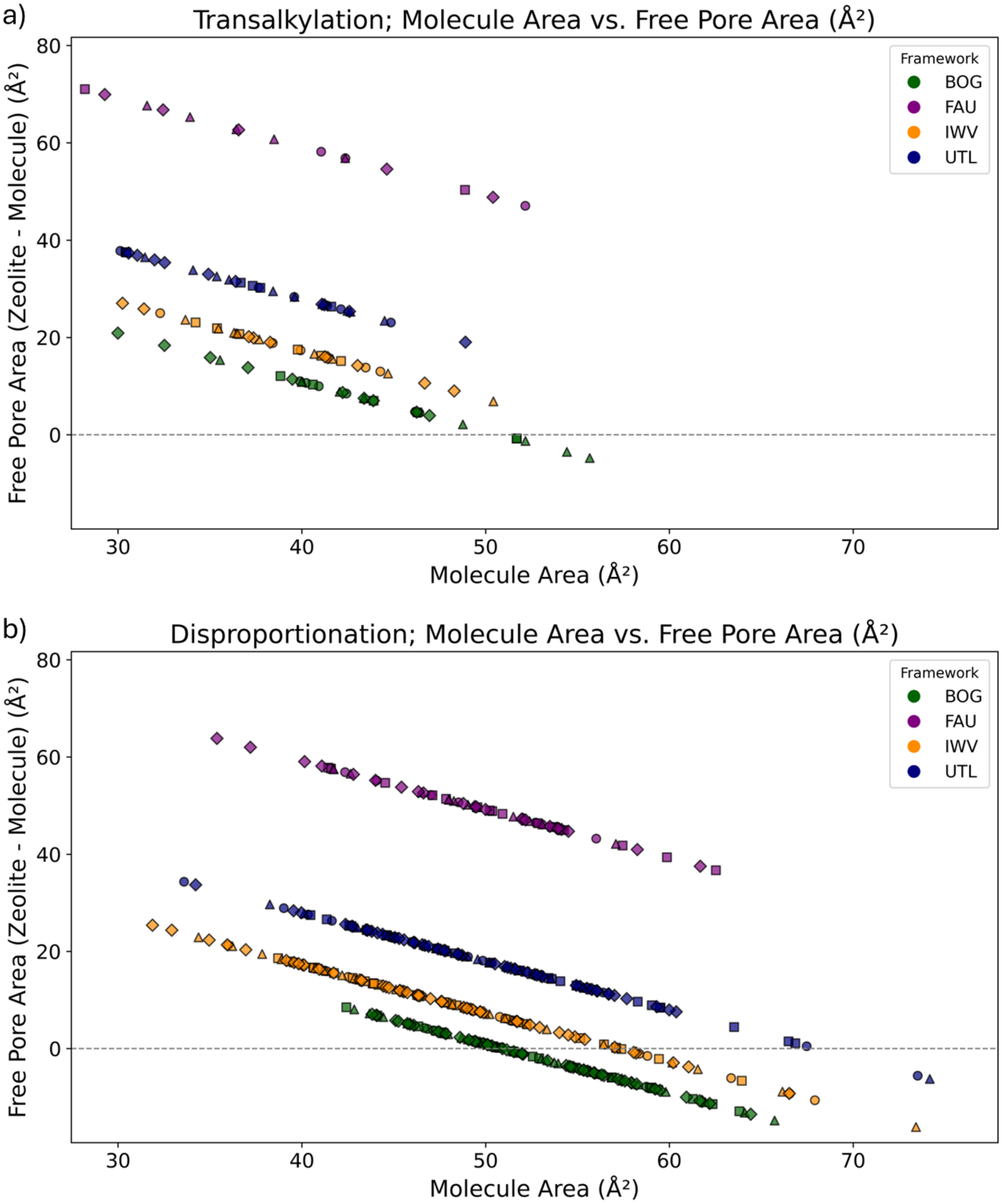

**Figure S41.** Free pore area on each zeolite channel, a) Transalkylation TS structures. b) Disproportionation TS structures. The free pore area is calculated as: Channel Area = π x ((Max D Sphere IZA)/2)$^2$, TS Area = π x (Middle axis/2) x (Minimum Axis)/2, Free Pore Area = Channel Area – TS Area. A visual representation of the method followed is provided in **Figure S40**, and the corresponding data is shown in **Table S20**. Max D Sphere values used according to IZA database website, IWV = 8.05 Å, IWV = 8.54 Å, UTL = 9.3 Å, FAU = 11.24 Å.

## S2.5 ZEOLITE STUDY: Al Placing Influences Stability

### Raw Free Energy Comparison

In this section, we compare stability differences among the BOG (T1 vs. T2), IWV (T3 vs. T7), and UTL (T4 vs. T2) models, each with different Al placements along the zeolite topologies, to clarify how Al site location influences reactivity. Specifically, in BOG (12×10 MR), T1 and T2 occupy distinct positions within its intersecting channels. In IWV, T3 resides at the more open 12×12 MR channel intersection (8.54 Å), whereas T7 is in the narrower 12 MR channel (6.9 × 6.2 Å). For UTL, T4 sits in the wide 14 MR channel (9.5 × 7.1 Å), while T2 is located in the wider 14×12 MR intersection (9.3 x 9.3 Å). **Figures S42a** and **S42b** show raw energy differences for the same intermediates and TS, with positive $G_{TA} - G_{TB}$ values indicating that site B is more stable and negative values indicating site A is more stable.

For BOG, no single site stands out, although T1 stabilizes TS3 disproportionation TS better than T2. However, ass discussed in the main paper, both sites provide relatively poor stabilization for diaryl intermediates and TS overall. In IWV, T3 outperforms T7 by consistently stabilizing both intermediates and TS, due to T3's location at the 12 × 12 MR channel intersection (8.54 Å), which offers more space than the narrower 12 MR channel (6.9 × 6.2 Å) used in T7. Finally, for UTL, T2 site (in the 14x12 MR intersections, 9.3 x 9.3 Å) stabilizes intermediates more effectively than T4 site (located in the straight 14 MR channel, 9.5 × 7.1 Å).

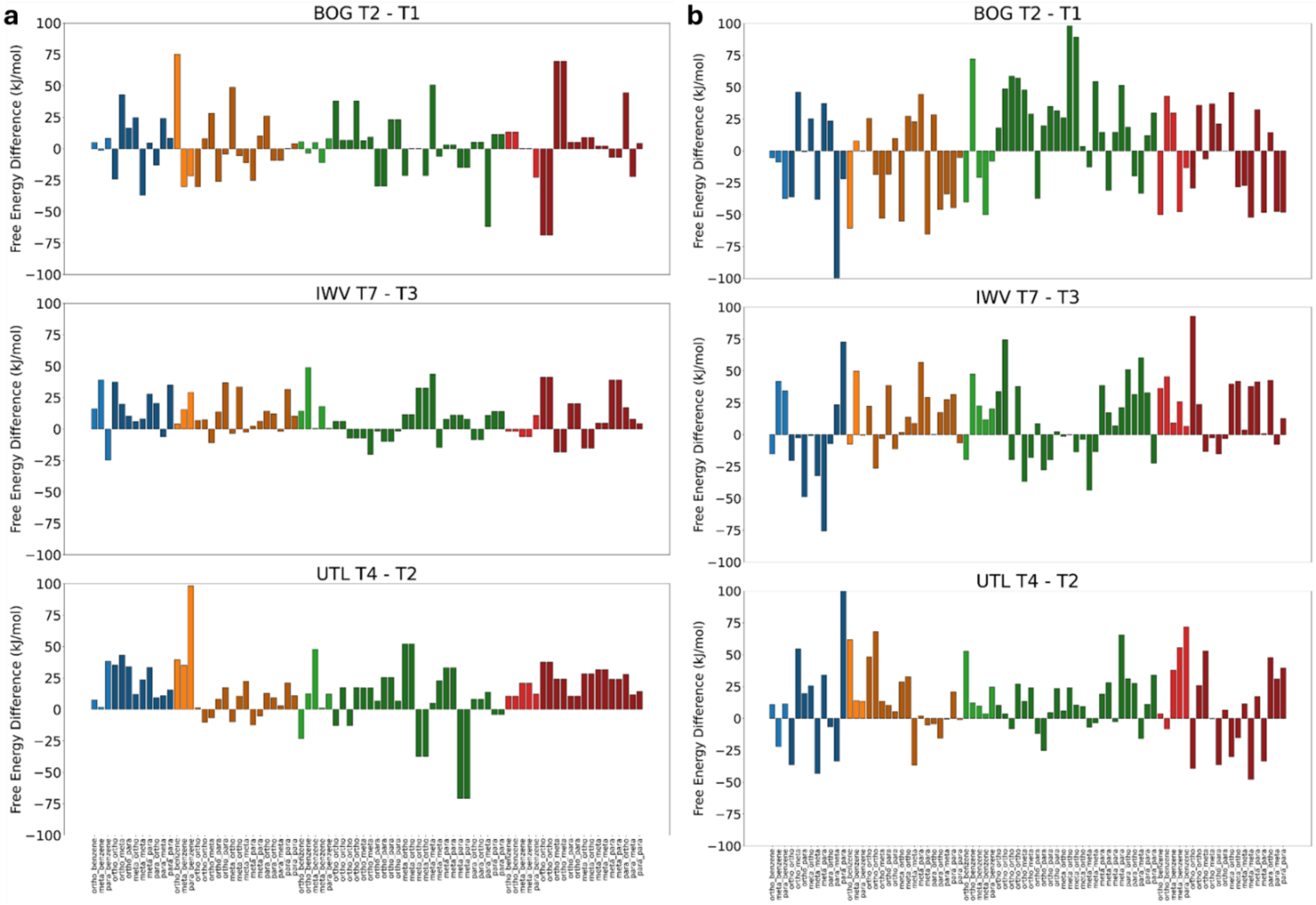

**Figure S42.** Relative free energy stabilities, $G_{TA} - G_{TB}$, for the different BOG, IWV and UTL models studied. a) Intermediates b) Transition States in kJ/mol. Blue bars account for I1 and TS1, orange bars account for I2 and TS, green bars account for I3 and TS3 and red bars account for I4 and TS4, in a and b plots respectively.

These stabilization differences also appear in the main paper discussion about barrier heights and rate-determining steps. In IWV, the T3 site supports two viable transalkylation routes, meta- and para-DEB$^+$, with barriers of 60 kJ/mol and 57 kJ/mol, respectively. By contrast, T7 only stabilizes the meta-DEB$^+$ path (60 kJ/mol). For UTL, the T4 site accommodates both meta- and para-DEB$^+$ transalkylation with barriers of 58 kJ/mol and 56 kJ/mol, as well as multiple disproportionation pathways in the 73–78 kJ/mol range. Meanwhile, T2 stabilizes only the meta-DEB$^+$ route (66 kJ/mol) and generally exhibits higher disproportionation barriers, most likely due to the better stabilization of intermediates observed in **Figure S42**.

**Scaled Pathways and Isomeric Study**

Next, we present transalkylation rescaled paths relative to the minimum I0 energy identified among the BOG (T1: meta-benzene), IWV (T7: ortho-benzene), and UTL (T2: para-benzene) models in **Figure S43**, along with only those paths having a negative free-energy difference between I4 and I1, therefore, those which are thermodynamically favorable, in **Figure S44**. BOG-T2 exhibits no thermodynamically favored pathway. Meanwhile, IWV-T3 stabilizes only the para-DEB$^+$ route, whereas IWV-T7 favors the meta-DEB$^+$ route. UTL-T2 supports both meta-DEB and ortho-DEB$^+$ paths, and UTL-T4, with the T site placed at the wider 14MR channel, stabilizes all three (ortho, meta, para) transalkylations. Details on barrier heights and rate-determining steps appear in the main paper.

Scaled disproportionation pathways are presented in **Figures S45** and **S46**, each referenced to the minimum I0 energy found among BOG (T1: meta–para), IWV (T7: meta–para), and UTL (T2: para–para). Although these plots do not reveal a decisive overall trend, disproportionation isomers reacting with neutral para-DEB emerge as the most stable among the twelve intermediates considered. Notably, the number of thermodynamically favored disproportionation pathways shown in **Figure S46** increases systematically with zeolite channel size. BOG models exhibit limited favorable pathways (one or two for BOG-T2 and BOG-T1, respectively), while IWV and FAU display slightly more. This number reaches a maximum for UTL-T2, reflecting that its 14×12 MR intersections provide optimal confinement to stabilize bulky intermediates and transition states, facilitating energetically favorable disproportionation pathways.

An isomer-specific analysis of the rescaled pathways further reveals clear selectivity trends across zeolite topologies. For transalkylation, IWV selectively stabilizes meta-DEB$^+$ and para-DEB$^+$ routes, with T3 favoring para and T7 favoring meta intermediates. UTL models accommodate all three isomers, but ortho pathways consistently exhibit higher energies, reflecting their less favorable accommodation even in wider channels. FAU, despite its spacious framework, only supports viable meta and para-DEB$^+$ pathways, with ortho again being disfavored. A similar trend is observed for disproportionation: meta and para intermediates consistently yield the most thermodynamically favorable pathways across IWV, UTL, and FAU, while ortho-derived routes are higher in energy and can be excluded from mechanistic relevance. These observations suggest that, regardless of the mechanism or framework, the ortho isomer faces persistent steric penalties, while meta and para intermediates benefit from more accessible and stable configurations within the pores.

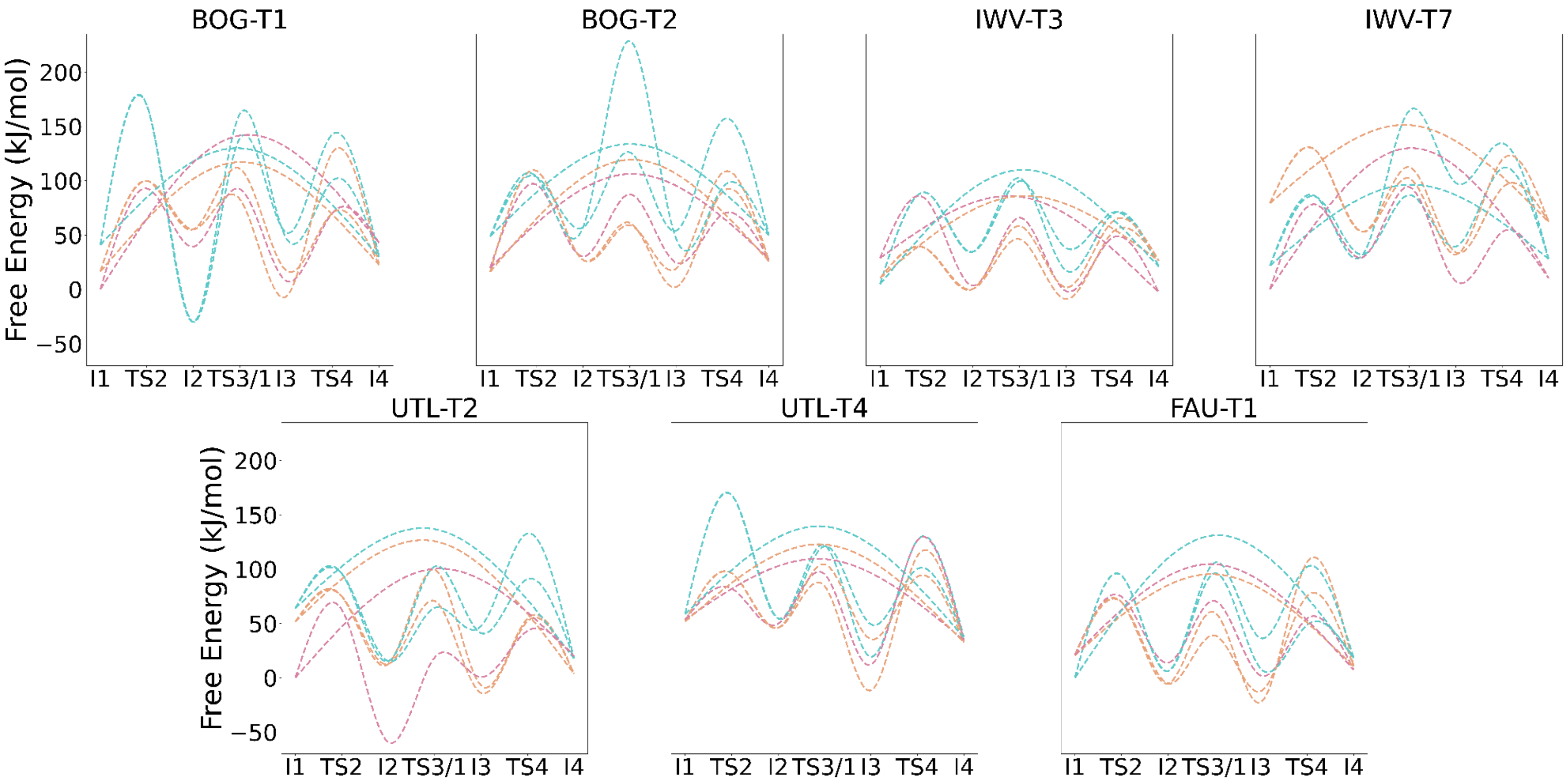

**Figure S43.** Transalkylation pathways scaled to the minimum I0 energy identified among BOG (T1), IWV (T7), and UTL (T2) models. Blue shades denote I1⁺ intermediates derived from ortho-DEB⁺, orange from meta-DEB⁺, and burgundy from para-DEB⁺. Further details for the color palette are detailed in **Figure S24** caption.

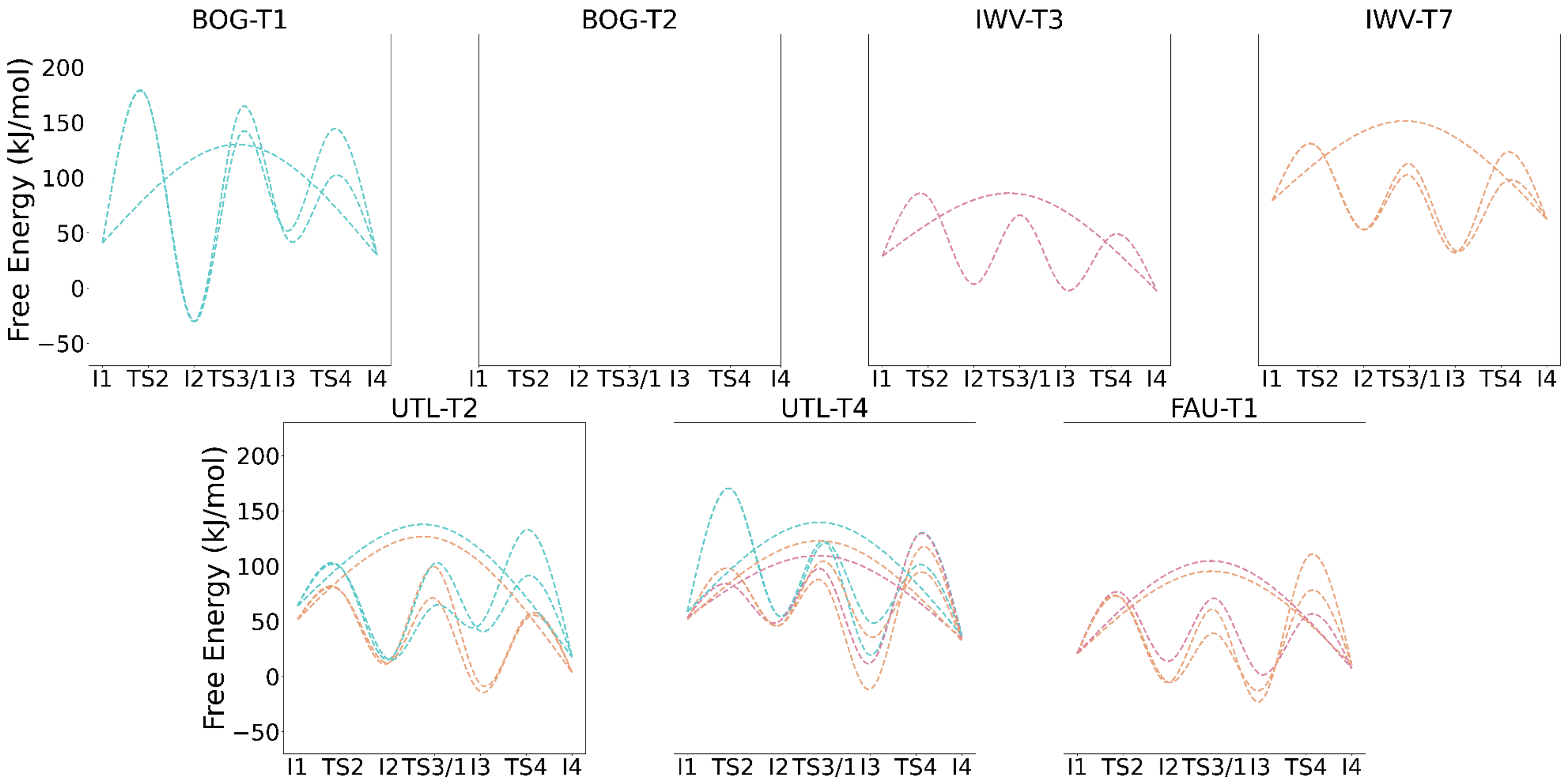

**Figure S44.** Thermodynamically favored transalkylation pathways, those with a negative free-energy difference between I4 and I1. Blue shades denote I1⁺ intermediates derived from ortho-DEB⁺, orange from meta-DEB⁺, and burgundy from para-DEB⁺. Further details for the color palette are detailed in **Figure S24** caption.

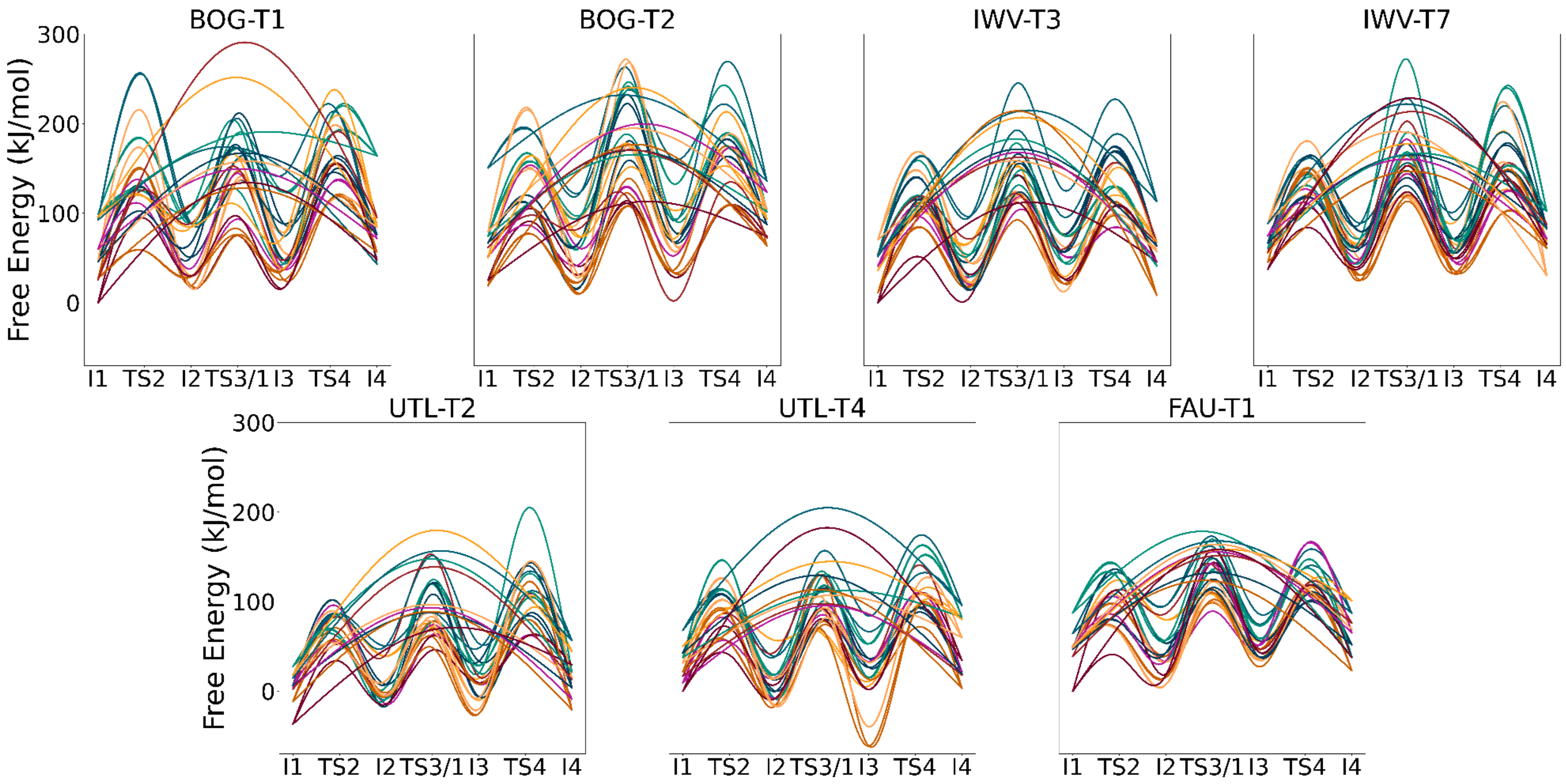

**Figure S45**. Disproportionation pathways scaled to the minimum I0 energy identified among BOG (T1), IWV (T7), and UTL (T2) models. Blue shades denote I1⁺ intermediates derived from ortho-DEB⁺, orange from meta-DEB⁺, and burgundy from para-DEB⁺. Further details for the color palette are detailed in **Figure S24** caption.

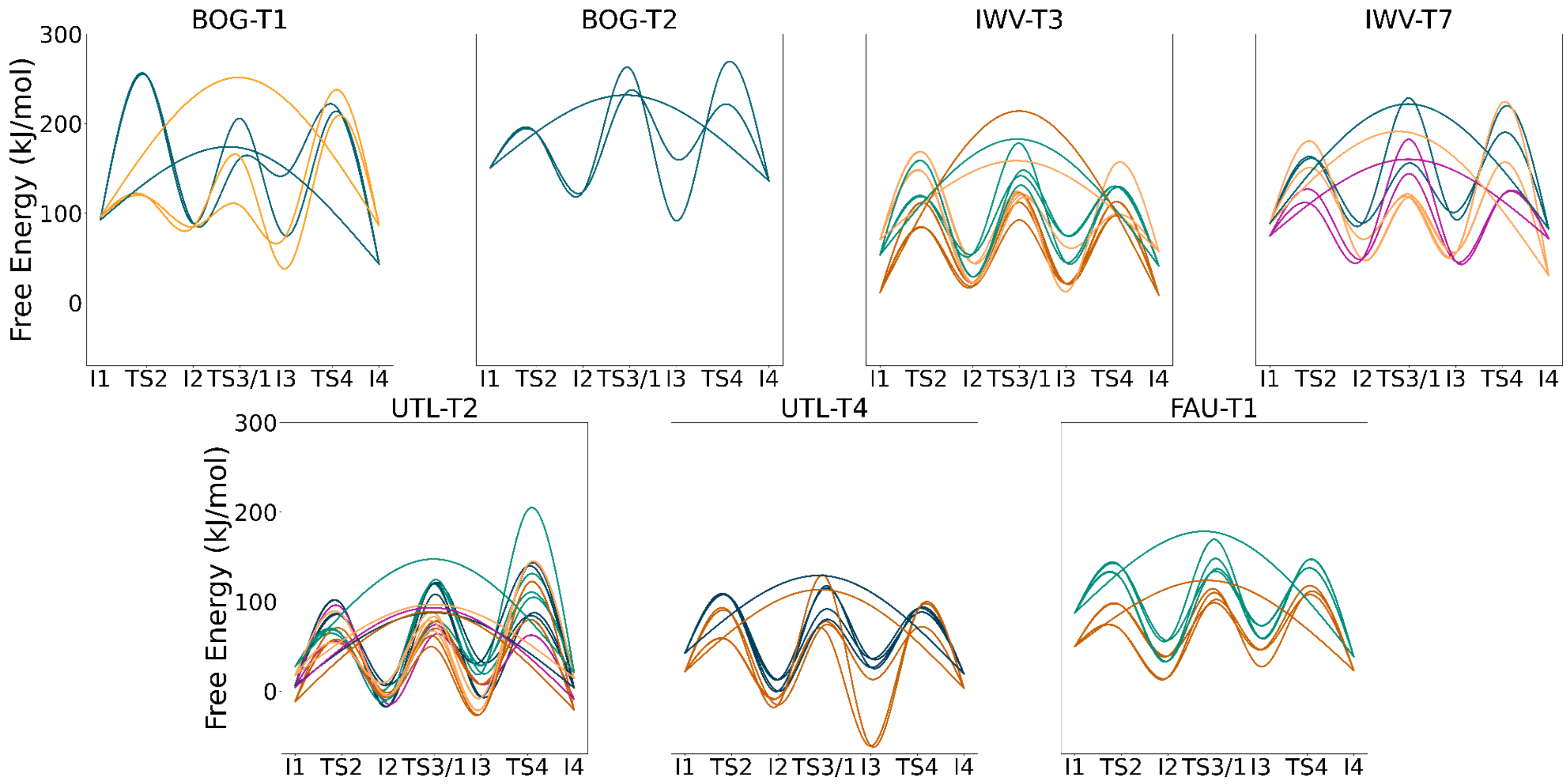

**Figure S46.** Thermodynamically favored disproportionation pathways, those with a negative free-energy difference between $I_4$ and $I_1$. Blue shades denote I1⁺ intermediates derived from ortho-DEB⁺, orange from meta-DEB⁺, and burgundy from para-DEB⁺. Further details for the color palette are detailed in **Figure S24** cap

# S4 Cyclohexene skeletal isomerization

## S4.1 Reactions and Transition States Studied

This section presents a detailed breakdown of all individual reaction steps investigated in this work for cyclohexene skeletal isomerization, as illustrated in **Figures S50–S52**.

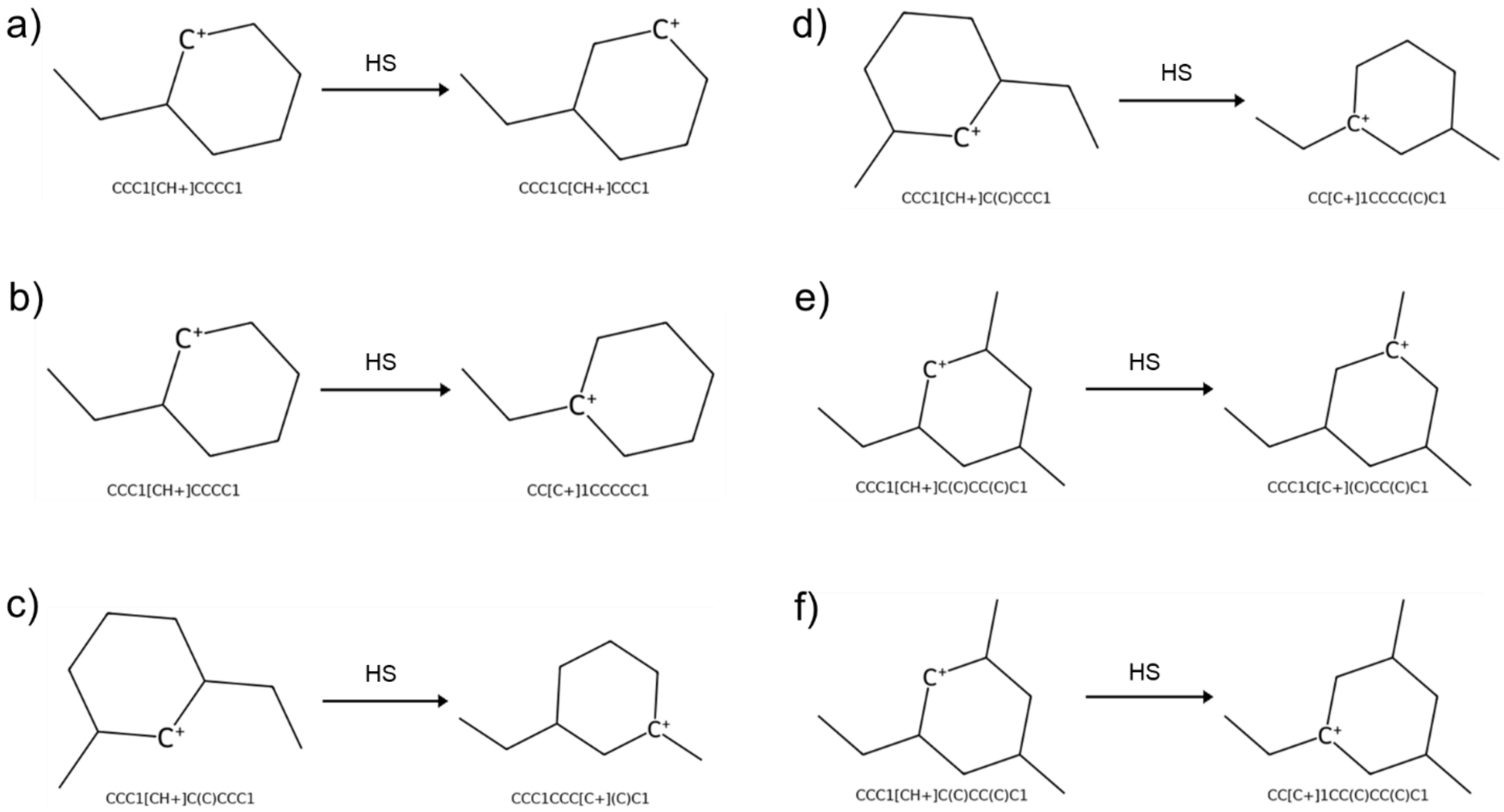

**Figure S47.** Hydrogen shift positional isomerizations studied for 1-ethyl-1-cyclohexyl are shown in panels (a–b); for 1-ethyl-2-cyclohexyl in panels (c–d); and for 1-ethyl-3-cyclohexyl in panels (e–f).

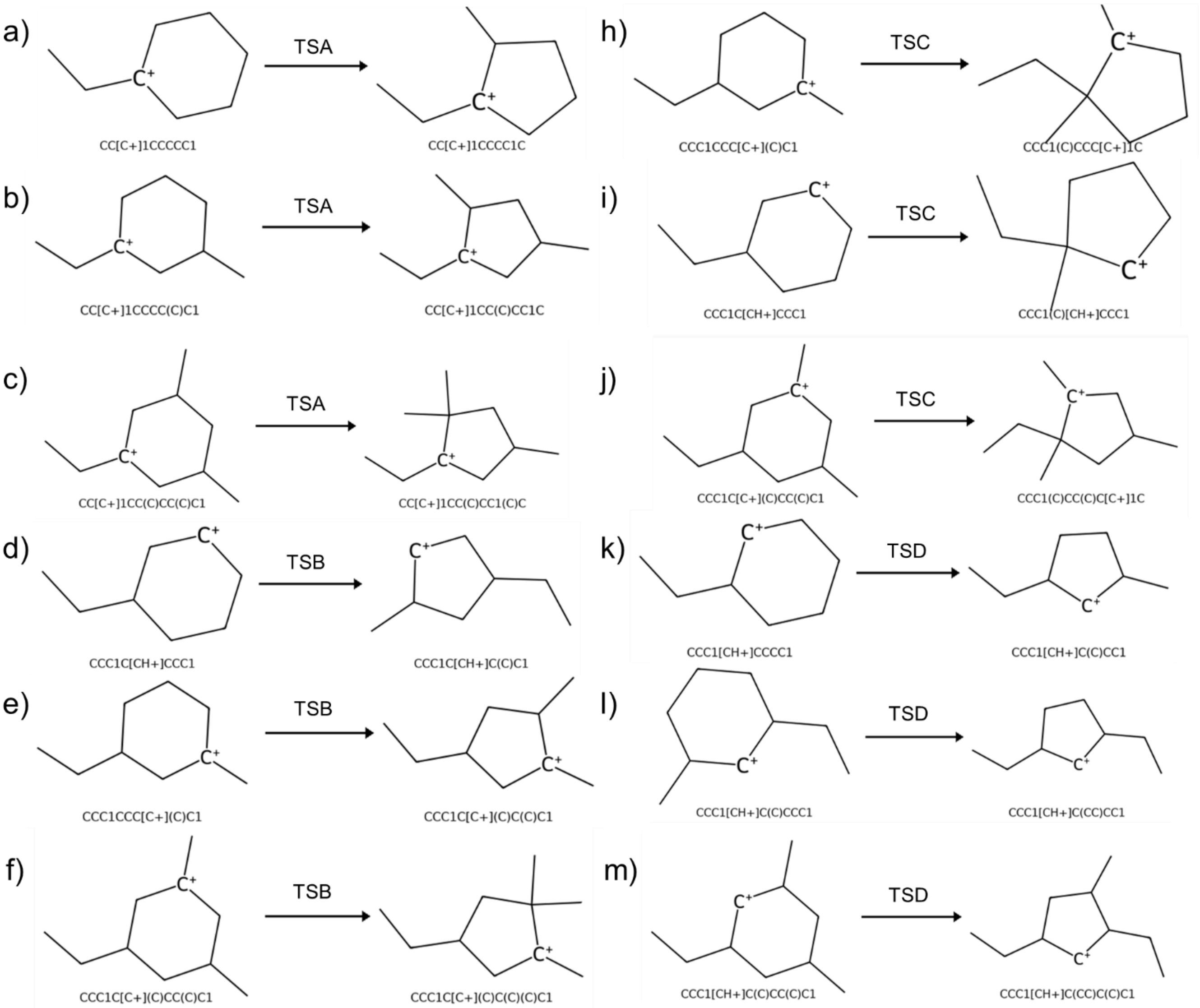

**Figure S48.** Ring contraction skeletal isomerizations studied for 1-ethyl-1-cyclohexyl, 1-ethyl-2-cyclohexyl, and 1-ethyl-3-cyclohexyl via TSA are shown in panels (a–c); via TSB in panels (d–f); via TSC in panels (h–j); and via TSD in panels (k–m).

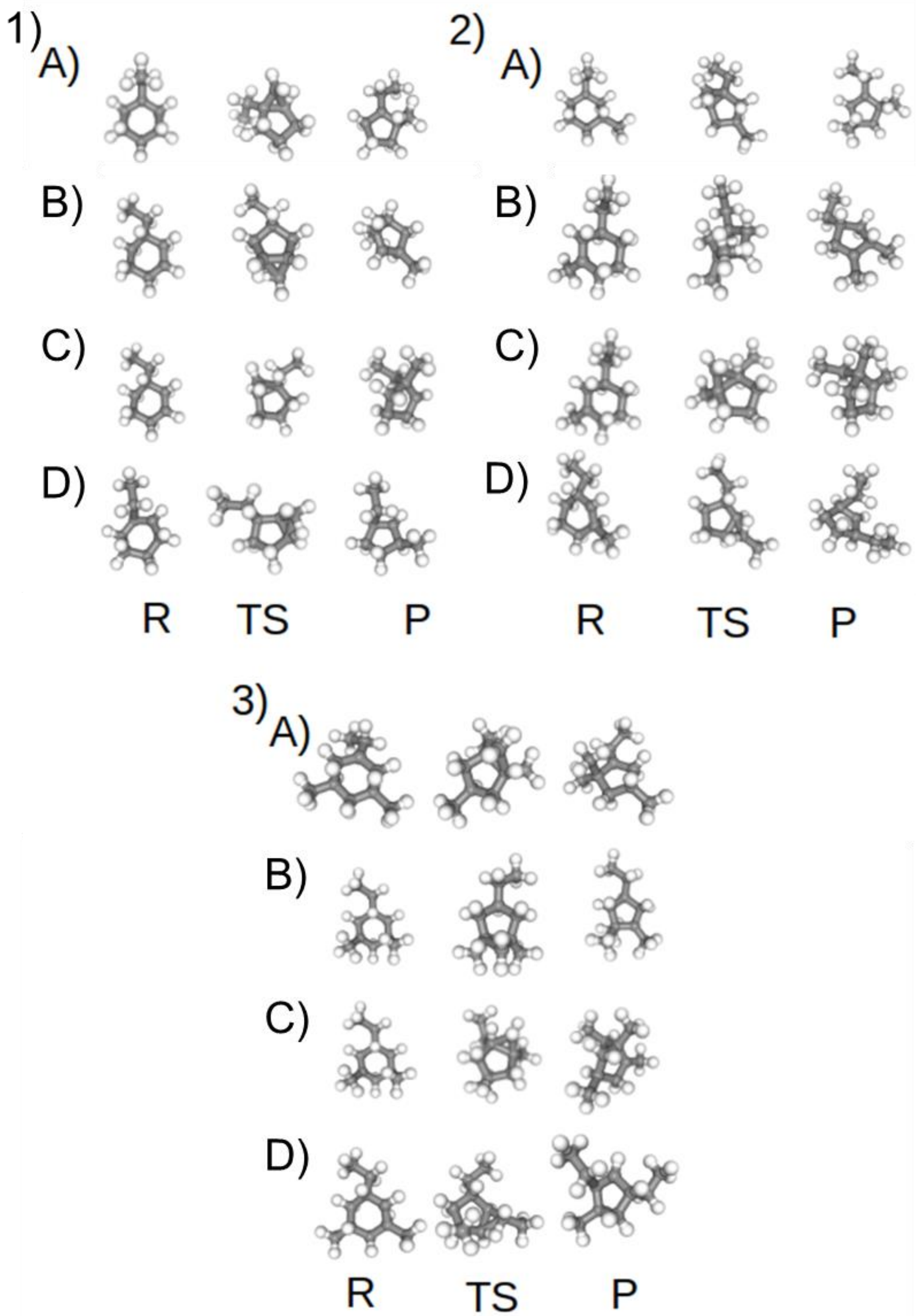

**Figure S49.** Ring contraction skeletal isomerizations TS structures obtained for 1) 1-ethyl-1-cyclohexyl, 2) 1-ethyl-2-cyclohexyl, and 3) 1-ethyl-3-cyclohexyl via TSA, TSB, TSC and TSD.

## S4.2 DFT Gas Phase Study for Cyclohexene Skeletal Isomerization

This study provides a comprehensive understanding of ring contraction mechanisms in alkyl-substituted cyclohexyl cations, revealing how charge localization, methyl substitution, and transition state stabilization govern reactivity. By focusing on a small but representative subset of the extensive ethylcyclohexene isomerization network, this work serves as a proof of concept to investigate key mechanistic features of ring contraction and carbocation stability. Given the vast number of possible isomerization pathways—including hydride and alkyl shifts alongside PCP-mediated rearrangements—we selectively examine tertiary carbenium ion intermediates, which are expected to be the most stable species, offering molecular-level insight into a fundamental aspect of this complex transformation.

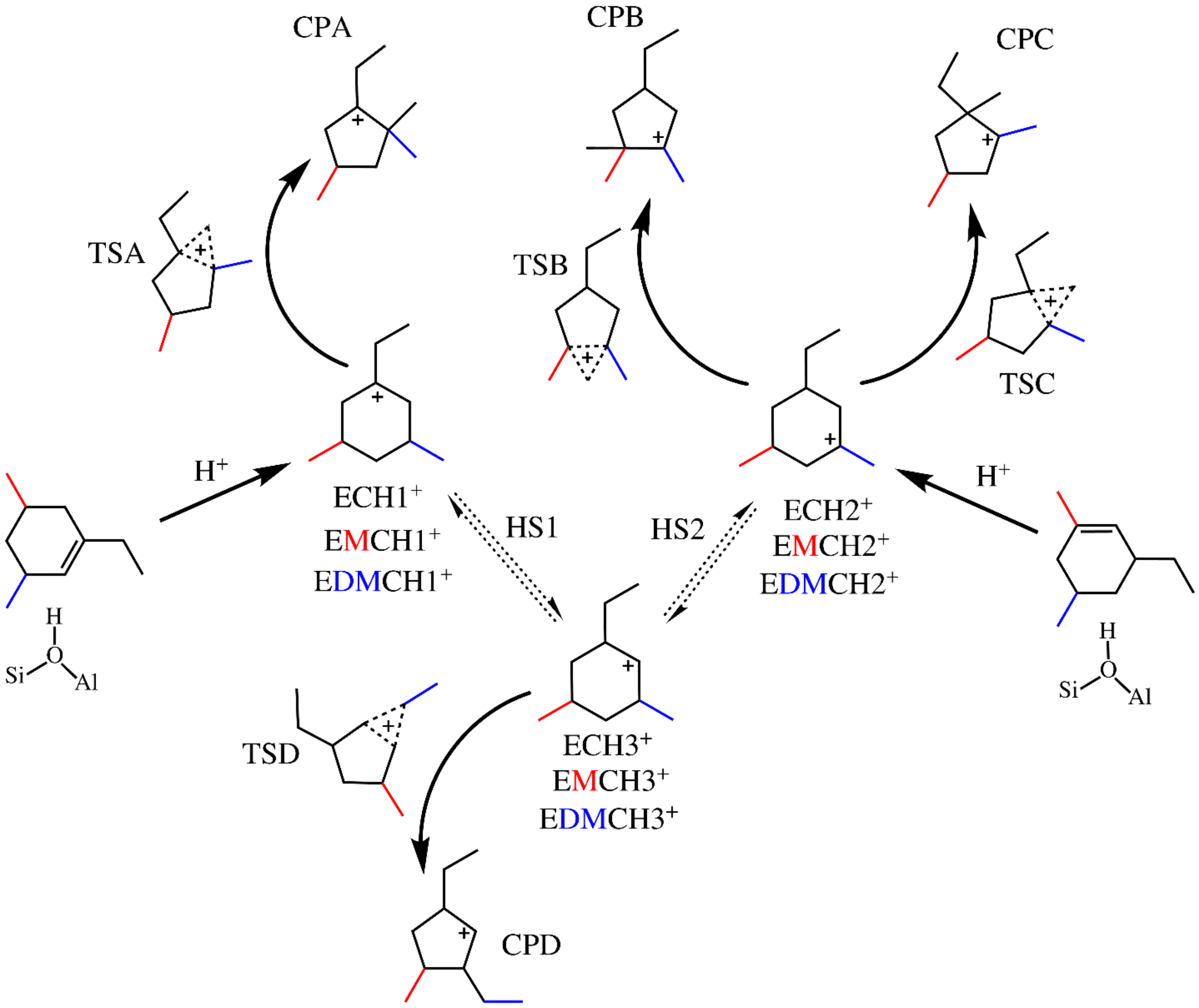

**Figure S50.** Ring contraction skeletal isomerizations, TSA, TSB, TSC and TSD, studied for 1-ethyl-1-cyclohexyl ($ECH^+$), 1-ethyl-3-methyl-cyclohexyl ($EMCH^+$) (blue sticks), and 1-ethyl-(3,5)-cyclohexyl ($EDMCH^+$) (blue and red sticks).

**1-Ethylcyclohexyl Isomers**

The isomerization of ethylcyclohexenes was investigated by examining three different ethylcyclohexyl cations, 1-ethyl-1-cyclohexyl ($ECH1^+$), 1-ethyl-2-cyclohexyl ($ECH3^+$), 1-ethyl-3-cyclohexyl ($ECH2^+$), focusing on the effects of charge placement. From these intermediates, a subset of all possible ring contraction proceeding via cyclopropane-like (PCP) TS was analyzed, see paths A, B, C and D in **Figure S51a**. Notable differences in activation barriers were observed, which correlate with the intrinsic stability of the initial cyclohexyl cations.

Charge localization plays a crucial role in stabilizing these intermediates. As the positive charge is positioned further from the ethyl-substituted carbon, the stability of the ethylcyclohexyl cations decreases. The 2-ethylcyclohexyl cation was found to be 38.49 kJ/mol less stable than the 1-ethylcyclohexyl counterpart, while the 3-ethylcyclohexyl cation exhibited an even greater destabilization of 64.40 kJ/mol. Interestingly, this trend in stability does not translate directly to the transition states, as all PCP-like TS structures remain within a narrower energy window of approximately 21 kJ/mol.

This stability trend significantly influences the ring contraction barriers. The least stable species, 1-ethyl-3-cyclohexyl, exhibits the lowest activation barriers, with TS-C and TS-D requiring only 34.17 kJ/mol and 17.34 kJ/mol, respectively, despite originating from a high-energy (+64.4 kJ/mol) conformation. Conversely, the highest activation barrier (87.44 kJ/mol) as observed for the most stable 1-ethylcyclohexyl undergoing contraction via TS-A, see Figure **S51a** and **Table S9**.

Notably, the same stability trend observed for the ethylcyclohexyl cations extends to their cyclopentyl contraction products. The 1-ethyl-2-methyl-1-cyclopentyl cation, formed via contraction path A, retains its tertiary carbocation nature, mirroring the stability of its reactant counterpart. This structural feature confers an increased stability of 40.23, 67.20, and 53.97 kJ/mol compared to the cyclopentyl cations formed via contraction paths B, C, and D, which instead generate secondary carbocations, leading to reduced stability. Finally, no preferential contraction pathways were identified based on the free energy differences between reactants and products. The computed $\Delta G_{P-R}$ values for contractions A, B, and C were 10.39, 12.14, and 13.19 kJ/mol, respectively, while contraction D resulted in $\Delta G_{P-R}$ = -0.07 kJ/mol, indicating a nearly thermoneutral transformation.

**1-Ethyl-3-methylcyclohexyl Isomers**

The ring contraction pathways A, B, C, and D were investigated for 1-ethyl-3-methylcyclohexene to assess the impact of methyl substitution on carbocation stability and TS barriers. The introduction of a methyl group at the 3-position significantly stabilizes the 1-ethyl-3-methyl-3-cyclohexyl ($EMCH3^+$) cation, as the positive charge is now accommodated at a tertiary carbon, increasing its intrinsic stability.

This increased stability directly correlates with higher activation barriers for the PCP-like transition states in paths C and D (**Figure S51b**). Specifically, the barrier for TS-C increases from 34.17 kJ/mol to 95.77 kJ/mol, and for TS-D, from 17.35 kJ/mol to 77.07 kJ/mol, marking an approximate 60 kJ/mol increase when transitioning from a secondary to a tertiary carbocation reactant.

In contrast, for 1-ethyl-3-methyl-1-cyclohexyl ($EMCH1^+$), the barrier for TS-A remains nearly unchanged, shifting only slightly from 87.44 to 87.94 kJ/mol. This suggests that methyl substitution at this position has minimal influence on the transition state energy, likely because it does not significantly alter charge distribution or steric effects in the PCP-like TS structure.

For TS-B, which undergoes contraction from a secondary to another secondary carbocation, both the reactant and product structures are approximately 58 kJ/mol less stable than their tertiary counterparts. This lower intrinsic stability results in a barrier of 37.96 kJ/mol, comparable to the barrier observed in the 1-ethylcyclohexyl system for secondary carbocations.

Among the ring contraction pathways involving tertiary-to-tertiary carbocation transitions (A, C, and D), TS-D exhibits the lowest activation barrier (77.07 kJ/mol). This can be attributed to its unique PCP-like transition state, where the cyclopropyl moiety is bonded to four different carbon atoms, enhancing charge delocalization and stabilizing the positive charge more effectively. This trend is further supported by the $G_{rel}$ comparison among the transition states, where TS-D is the most stable, followed by TS-A, TS-B, and TS-C, which are 13.8, 18.2, and 18.7 kJ/mol less stable, respectively.

**1-Ethyl-(3,5)-dimethylcyclohexyl Isomers**

The isomerization of 1-ethyl-3,5-dimethylcyclohexyl ($EDMCH^+$) introduces an additional methyl group compared to the 1-ethyl-3-methylcyclohexene system. However, since this methyl substitution does not directly influence charge placement, its effect on the activation barriers is minimal, unlike the pronounced changes observed when transitioning from 1-ethylcyclohexyl to 1-ethyl-3-methylcyclohexyl isomers (**Figure S51c**).

In this system, TS-B follows a pathway that proceeds through secondary-to-secondary carbocations, leading to a lower activation barrier of 45.54 kJ/mol. This aligns with the fact that both the reactant and product structures for this pathway are intrinsically less stable, with relative free energies of +61.85 kJ/mol for the reactant and +67.98 kJ/mol for the product.

In contrast, the contraction pathways that transition through tertiary carbocations (TS-A, TS-C, and TS-D) exhibit similar activation barriers of 84.69, 78.17, and 78.43 kJ/mol, respectively. This consistency arises from the shared structural nature of their PCP-like transition states, where the cyclopropyl moiety is bonded to four different carbon atoms. This bonding pattern facilitates charge delocalization and enhances carbocation stability, effectively equalizing the energy requirements for ring contraction.

Furthermore, this trend is reflected in the $G_{rel}$ comparison among the transition states, where TS-A, TS-C, and TS-D fall within a narrow 2 kJ/mol energy range, reinforcing their structural similarity. In contrast, TS-B, which lacks the tetra-carbon bonded PCP-like structure, is 21.8 kJ/mol less stable, highlighting the key role of transition state stabilization in governing barrier heights.

**Table S9.** Smiles for Reactant and Product molecules, ring contraction paths, relative free energy $G_{rel}$ kJ/mol for reactants (R), transitions states (TS) and products (P). Free energy barriers ($\Delta G_{TS-R}$), free energy difference between products and reactants ($\Delta G_{P-R}$) and free energy difference between TS structures ($\Delta G_{relTS}$), all values expressed in kJ/mol.

| 1-Ethyl-Cyclohexyl cation isomers ring contraction | | | | | | | | |
|---|---|---|---|---|---|---|---|---|
| Reactant | Product | Path | $G_{rel}$ R | $G_{rel}$ TS | $G_{rel}$ P | $\Delta G_{TS-R}$ | $\Delta G_{P-R}$ | $\Delta G_{relTS}$ |
| CC[C+]1CCCCC1 | CC[C+]1CCCC1C | A | 0.00 | 87.44 | 10.39 | 87.44 | 10.39 | 5.7 |
| CCC1C[CH+]CCC1 | CCC1C[CH+]C(C)C1 | B | 64.40 | 98.57 | 77.59 | 34.17 | 13.19 | 16.8 |
| CCC1C[CH+]CCC1 | CCC1(C)[CH+]CCC1 | C | 64.40 | 81.75 | 64.33 | 17.34 | -0.07 | 0.0 |
| CCC1[CH+]CCCC1 | CCC1[CH+]C(C)CC1 | D | 38.49 | 103.43 | 50.62 | 64.94 | 12.14 | 21.7 |
| 1-Ethyl-3-MethylCyclohexyl cation isomers ring contraction | | | | | | | | |
| Reactant | Product | Path | $G_{rel}$ R | $G_{rel}$ TS | $G_{rel}$ P | $\Delta G_{TS-R}$ | $\Delta G_{P-R}$ | $\Delta G_{relTS}$ |
| CC[C+]1CCCC(C)C1 | CC[C+]1CC(C)CC1C | A | 2.90 | 90.84 | 20.53 | 87.94 | 17.63 | 13.8 |
| CCC1CCC[C+](C)C1 | CCC1C[C+](C)C(C)C1 | B | 0.00 | 95.77 | 11.32 | 95.77 | 11.32 | 18.7 |
| CCC1CCC[C+](C)C1 | CCC1(C)CCC[C+]1C | C | 0.00 | 77.07 | 6.37 | 77.07 | 6.37 | 0.0 |
| CCC1[CH+]C(C)CCC1 | CCC1[CH+]C(CC)CC1 | D | 57.36 | 95.32 | 58.95 | 37.96 | 1.59 | 18.2 |
| 1-Ethyl-3,5-DimethylCyclohexyl cation isomers ring contraction | | | | | | | | |
| Reactant | Product | Path | $G_{rel}$ R | $G_{rel}$ TS | $G_{rel}$ P | $\Delta G_{TS-R}$ | $\Delta G_{P-R}$ | $\Delta G_{relTS}$ |
| CC[C+]1CC(C)CC(C)C1 | CC[C+]1CC(C)CC1(C)C | A | 0.00 | 84.69 | 10.34 | 84.69 | 10.34 | 2.1 |
| CCC1C[C+](C)CC(C)C1 | CCC1C[C+](C)C(C)(C)C1 | B | 4.47 | 82.64 | 14.20 | 78.17 | 9.74 | 0.0 |
| CCC1C[C+](C)CC(C)C1 | CCC1(C)CC(C)C[C+]1C | C | 4.47 | 83.80 | 10.72 | 78.43 | 6.25 | 1.2 |
| CCC1[CH+]C(C)CC(C)C1 | CCC1[CH+]C(CC)C(C)C1 | D | 61.85 | 107.40 | 67.98 | 45.54 | 6.13 | 24.8 |

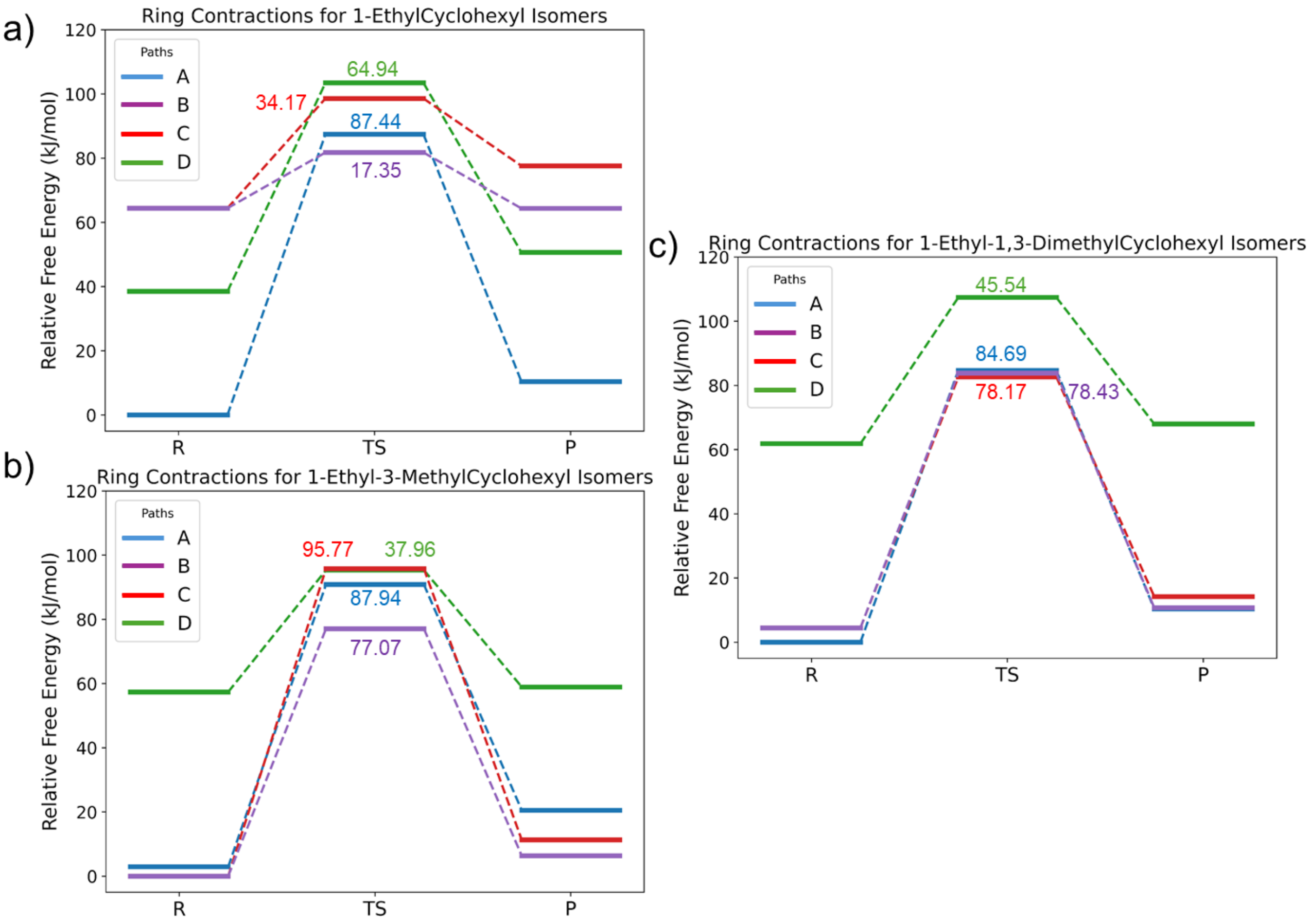

**Figure S51.** Free energy profiles for the different ring contraction paths, A (blue), B (purple), C (red), D (green) for a) 1-ethylcyclohexyl isomers, b) 1-ethyl-3-methylcyclohexyl isomers and c) 1-ethyl-(3,5)-dimethyl isomers.

## Gas-Phase Conclusions

The results highlight that reactant stability, rather than transition state energy, plays the primary role in determining ring contraction barriers through PCP-like TS. Less stable carbocations exhibit lower activation barriers but may have shorter lifetimes, limiting their likelihood of undergoing reaction before rearranging into more stable species through hydrogen shifts. The introduction of methyl substituents at different positions alters the stability of reactants but does not significantly affect the transition state energy landscape, as PCP-like TS structures remain within a relatively narrow energy window.

Results indicate that PCP-like TS are more stable when the two carbon atoms forming the new bond during ring contraction are each bonded to two other carbon atoms, one from the ring and one from an alkyl substituent, rather than when one of these positions is occupied by a hydrogen (C–H bond). The presence of additional carbon atoms around the TS moiety enhances charge delocalization, leading to greater stabilization of the positive charge and consequently lowering the activation barrier for ring contraction.

## S4.3 RC ZEOLITE STUDY; Secondary Carbocation TS structures Identified

As discussed in the main text on **Cyclohexene Skeletal Isomerization** results, secondary carbocations such as $ECH2^+$ and $ECH3^+$ exhibit intrinsic instability in zeolite environments, typically rearranging into more stable tertiary carbocations via rapid hydrogen or methyl shifts. Consequently, locating a well-defined minimum for these species requires imposing artificial bond constraints, which inevitably skew their computed energy barriers and forces these structures to be ruled out from the main paper discussion.

In this section, we present 9 out of the 36 different ring-contraction transition state structures obtained for these secondary carbocations without constrained conditions if **Figure S52**. Their inclusion illustrates the challenges of capturing short-lived carbocationic intermediates in rigid nanopores and underscores the need for careful validation of TS relevance when applying static DFT methods to heterogeneous catalytic systems.

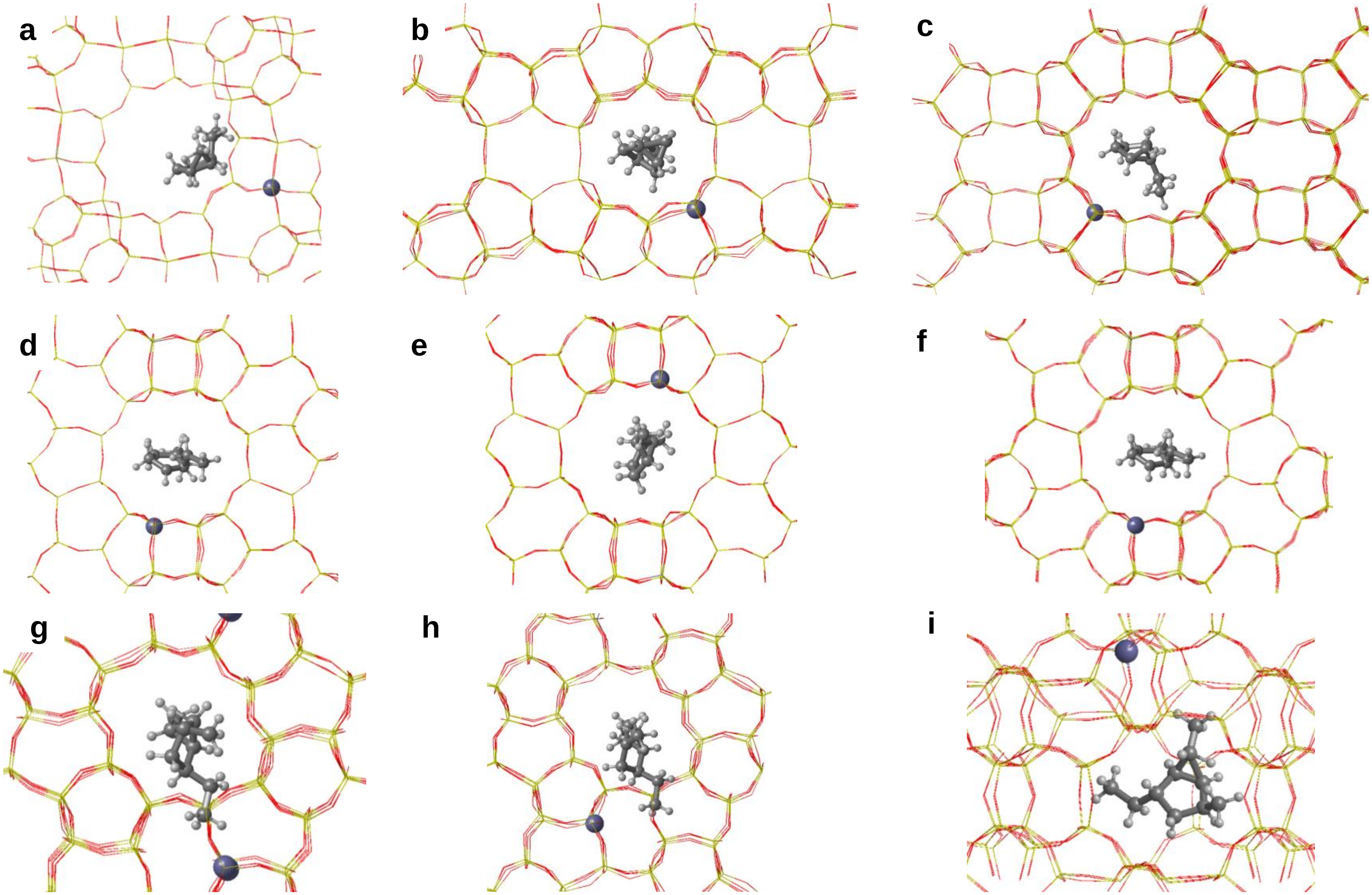

**Figure S52.** Ring-contraction transition state structures obtained for secondary carbocations without constrained conditions **a** FAU-T1 TSD $ECH^+$ **b** FER-T1 TSB $ECH^+$ **c** MOR-T2 TSC $ECH^+$ **d** BEA-T1a TSD $EMCH^+$ **e** BEA-T1b TSD $EMCH^+$ **f** BEA-T2 TSD $EMCH^+$ **g** MFI-T4 TSD $EDMCH^+$ **h** MFI-T10 TSD $EDMCH^+$ **i** MFI-T11 TSD $EDMCH^+$.

## S4.4 RC ZEOLITE STUDY; Reaction Mechanisms

This section presents the energy pathways identified for the different ring contraction mechanisms discussed in the main paper, aiming to provide a clearer representation of these pathways given the large number of reaction profiles generated by the PoTS computational pipeline. Free energy values plotted in kJ/mol in **Figures S53-55** are given in the main paper at **Table 1**.

In **Figure S53** it is possible to see how MFI narrow 10MR channel system returns the lowest barrier for TSA ring contraction.

In **Figures S54** and **S55** it is possible to see how TSB in blue color is generally the lowest barrier found for each zeolite model studied because of its lower steric hindrance as discussed in the main manuscript.

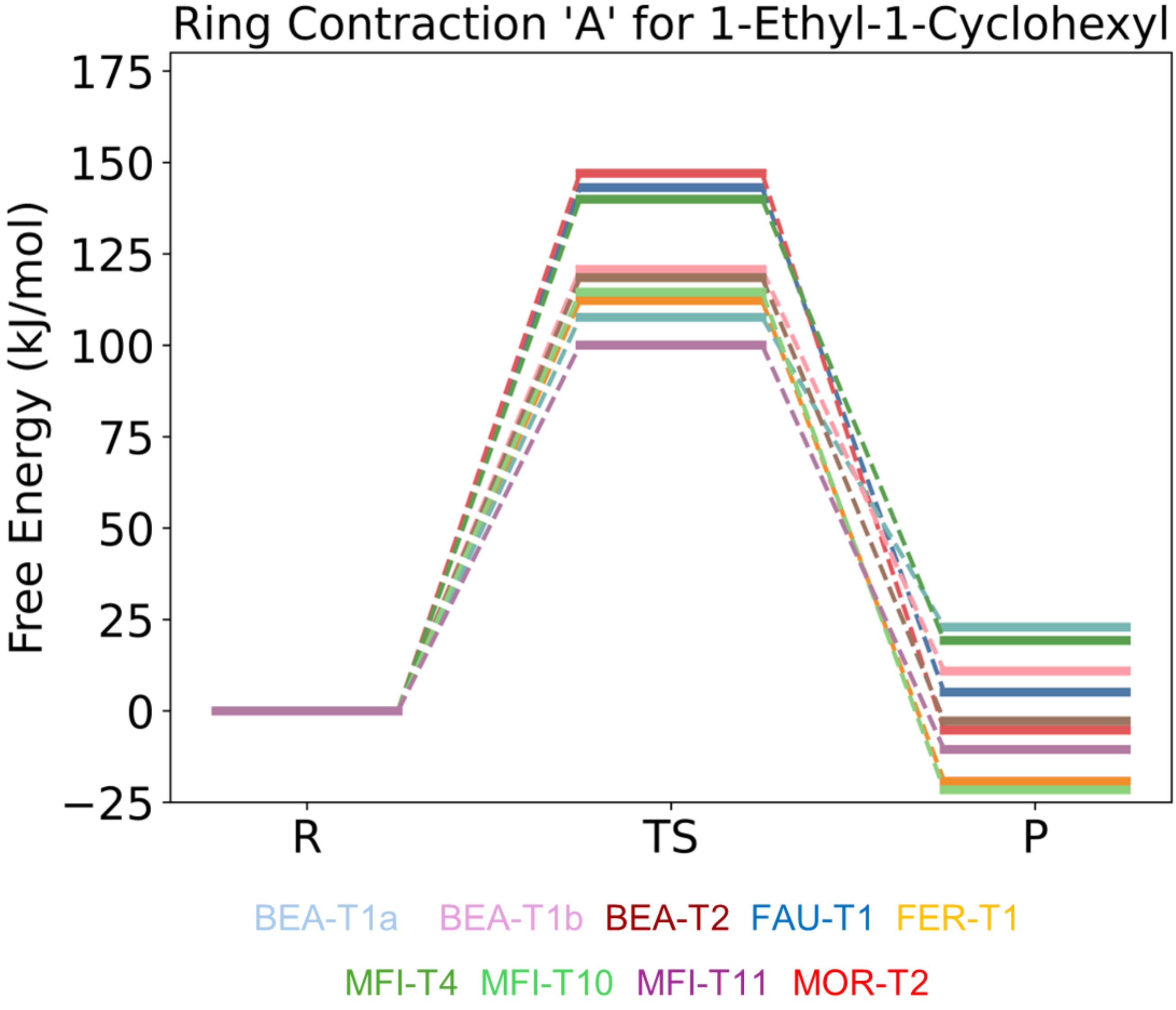

**Figure S53.** Free energy paths (kJ/mol) identified for 1-ethyl-cyclohexyl ($ECH^+$) through TSA. Energies obtained are shown in the main paper **Table 1**. The color scheme used to distinguish each (Framework, Aluminum) pair is: FAU T1 in blue, FER T2 in orange, MOR T1 in red, BEA T1a in teal, BEA T1b in pink, BEA T2 in brown, MFI T4 in green, MFI T10 in light green, and MFI T11 in purple.

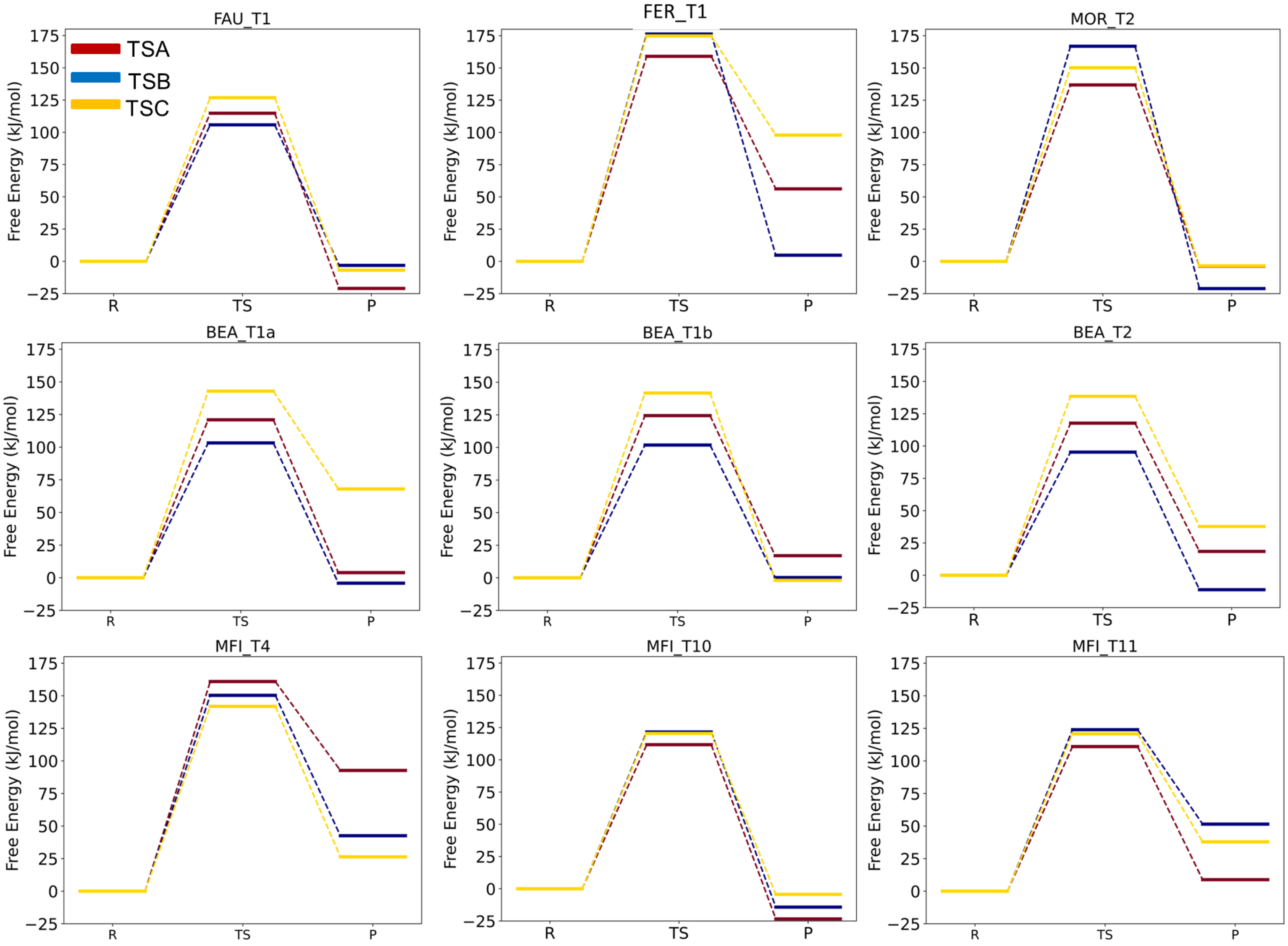

**Figure S54.** Free energy paths (kJ/mol) identified for 1-ethyl-3-methylcyclohexyl ($EMCH^+$) through TSA (red), TSB (blue) and TSC (gold). Energies obtained are shown in the main paper **Table 1**. Corresponding zeolite models indicated on the top of each subplot.

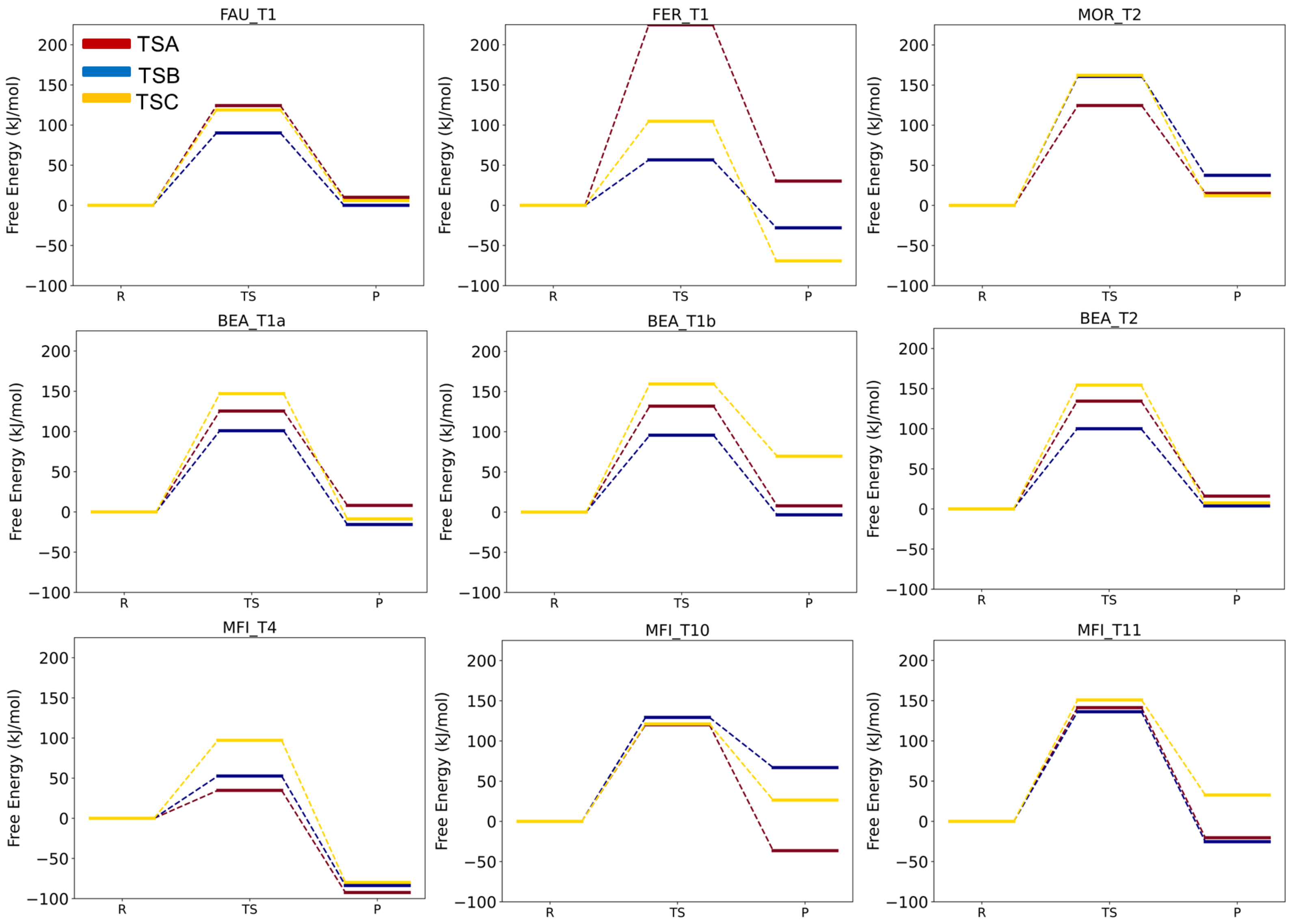

**Figure S55.** Free energy paths (kJ/mol) identified for 1-ethyl-(3,5)-dimethylcyclohexyl (EDMCH$^{+}$) through TSA (red), TSB (blue) and TSC (gold). Energies obtained are shown in the main paper **Table 1**. Corresponding zeolite models indicated on the top of each subplot.

## S4.5 RC ZEOLITE STUDY; Al Placing Influences Stability

## Scaled Reaction Mechanisms

In this section, we compare stability differences among the BEA (T1a, T1b, and T2) and MFI (T4, T10, and T11) zeolite models, each featuring different Al placements within their respective frameworks, to elucidate how Al site location and pore dimensions influence catalytic reactivity. Specifically, the BEA framework, consists of an irregular three-dimensional pore system with 12-membered ring (12MR) apertures ranging from 5.6–7.3 Å. Al placements at T1a and T2 were positioned within distinct straight 12MR channels, while the T1b site occupied a position at a 12MR intersection, allowing investigation into how spatial confinement affects stabilization.

For MFI, T4 was situated in the narrow sinusoidal 10MR channels (5.1 × 5.5 Å), T11 in the straight 10MR channels (5.3 × 5.6 Å), and T10 at the intersection of these channels, which provides a larger void space of approximately 9 Å. These placements were selected based on their intrinsic stabilities and the chemical relevance of the Al sites to assess their role in catalytic performance, see **S1**.

We present rescaled ring-contraction pathways relative to the lowest-energy reactant minima identified within each zeolite framework (BEA-T1a, BEA-T1b, BEA-T2; and MFI-T4, MFI-T10, MFI-T11), as summarized in **Table S10**. Minimum reactant structures among the different models are highlighted in green.

In the MFI framework, sites T10 and T11 exhibit significantly lower reaction barriers and more stable cyclopentyl products compared to T4, underscoring the enhanced stabilization offered at the straight channels and channel intersections over the narrow sinusoidal channels.

Within the BEA framework, T1a and T2 sites provide the lowest-energy minima for ring contraction reactions. For smaller cyclohexyl ring intermediates, such as ECH+ and EMCH+ isomers, energy differences between sites T1a and T2 rarely exceed 10 kJ/mol. However, for bulkier EDMCH+ isomers, these differences increase up to 29 kJ/mol, indicating the T2 site as particularly favorable for stabilizing larger intermediates. Despite this, the topological similarity among the BEA 12MR channels and the relatively minor energy differences compared to the more distinct case of MFI-T4 suggest that other T-sites should not be excluded from consideration. **Table S10** values are plotted in **Figure S56**. Detailed analyses of barrier heights, intermediate stabilities, and rate-determining steps for each specific pathway are discussed extensively in the main text.

**Table S10.** Relative Gibbs free energies (ΔG, kJ/mol) for ring contraction reactions of ethyl-substituted cyclohexyl cations in **a)** BEA and **b)** MFI zeolite models. Energies for reactant minima (R), transition states (TS), and products (P) are rescaled to the lowest-energy reactant minimum (set as 0.00 kJ/mol) within each group of zeolite frameworks (BEA-T1a, BEA-T1b, BEA-T2 in part a and MFI-T4, MFI-T10, MFI-T11 in part b). Entries marked 0.00 indicate the most stable reactant minima within each cationic species for the given set of zeolite models.

| **a)** | | **BEA-T1a** | | | **BEA-T1b** | | | **BEA-T2** | | |
|---|---|---|---|---|---|---|---|---|---|---|
| | | **R** | **TS** | **P** | **R** | **TS** | **P** | **R** | **TS** | **P** |
| **1-Ethyl-Cyclohexyl Cation Ring Contraction** | | | | | | | | | | |
| **$ECH1^+$** | A | 0.00 | 117.64 | 22.95 | 2.16 | 122.94 | 13.09 | 6.30 | 124.82 | 3.58 |
| **1-Ethyl-3-MethylCyclohexyl Cations Isomers Ring Contraction** | | | | | | | | | | |
| **$EMCH1^+$** | A | 5.53 | 126.49 | 9.42 | 1.30 | 125.80 | 18.28 | 0.00 | 117.70 | 18.45 |
| **$EMCH2^+$** | B | 10.56 | 113.78 | 6.40 | 9.91 | 111.75 | 10.12 | 10.27 | 105.57 | -0.88 |
| **$EMCH2^+$** | C | 10.56 | 153.43 | 78.54 | 9.91 | 151.66 | 7.96 | 10.27 | 148.72 | 48.05 |
| **1-Ethyl-3,5-DimethylCyclohexyl Cations Isomers Ring Contraction** | | | | | | | | | | |
| **$EDMCH1^+$** | A | 29.00 | 154.38 | 37.10 | 10.24 | 142.13 | 18.07 | 0.00 | 134.39 | 16.06 |
| **$EDMCH2^+$** | B | 27.73 | 128.66 | 12.12 | 29.21 | 124.89 | 25.79 | 18.97 | 118.95 | 22.78 |
| **$EDMCH2^+$** | C | 27.73 | 174.75 | 19.66 | 29.21 | 188.66 | 98.77 | 18.97 | 173.36 | 26.55 |

| **b)** | | **MFI-T4** | | | **MFI-T10** | | | **MFI-T11** | | |
|---|---|---|---|---|---|---|---|---|---|---|
| | | **R** | **TS** | **P** | **R** | **TS** | **P** | **R** | **TS** | **P** |
| **1-Ethyl-Cyclohexyl Cation Ring Contraction** | | | | | | | | | | |
| **$ECH1^+$** | A | 29.34 | 169.32 | 48.63 | 0.00 | 114.55 | -21.54 | 8.90 | 108.98 | -1.63 |
| **1-Ethyl-3-MethylCyclohexyl Cations Isomers Ring Contraction** | | | | | | | | | | |
| **$EMCH1^+$** | A | 34.09 | 194.98 | 126.76 | 20.00 | 131.72 | -3.36 | 38.36 | 149.34 | 47.19 |
| **$EMCH2^+$** | B | 12.70 | 163.03 | 55.28 | 10.11 | 131.69 | -4.10 | 0.00 | 123.96 | 51.49 |
| **$EMCH2^+$** | C | 12.70 | 154.59 | 39.07 | 10.11 | 130.57 | 5.75 | 0.00 | 120.79 | 37.88 |
| **1-Ethyl-3,5-DimethylCyclohexyl Cations Isomers Ring Contraction** | | | | | | | | | | |
| **$EDMCH1^+$** | A | 143.48 | 178.26 | 51.13 | 17.16 | 137.33 | -19.20 | 32.91 | 174.23 | 12.53 |
| **$EDMCH2^+$** | B | 215.80 | 268.46 | 132.10 | 0.00 | 129.34 | 66.93 | 6.29 | 142.48 | -19.01 |
| **$EDMCH2^+$** | C | 215.80 | 312.96 | 135.97 | 0.00 | 121.20 | 26.24 | 6.29 | 157.12 | 39.02 |

## Interaction Energies

The analysis of interaction energies (Eint), summarized in **Table S11** and calculated as Eint = E(zeo+molecule) – Emolecule_gas – Ezeolite, further supports the previous findings. For BEA models, interaction energies across the T1a, T1b, and T2 sites exhibit relatively small variations (e.g., $ECH^+$: T1a = -638.12 kJ/mol, T1b = -641.00 kJ/mol, T2 = -637.27 kJ/mol; $EMCH^+$: T1a = -690.07 kJ/mol, T1b = -680.91 kJ/mol, T2 = -684.85 kJ/mol), aligning with the small energy differences observed for these intermediates. However, slightly stronger interaction energies for $EDMCH^+$ intermediates at the T2 site ($EDMCH1^+$: -685.52 kJ/mol; $EDMCH2^+$: -663.84 kJ/mol) reinforce its enhanced capability for stabilizing bulkier species. For MFI models, notably larger differences are observed, with significantly weaker interactions at T4 ($EDMCH1^+$: -623.86 kJ/mol; $EDMCH2^+$: -564.27 kJ/mol) compared to T10 ($EDMCH1^+$: -738.57 kJ/mol; $EDMCH2^+$: -736.40 kJ/mol) and T11 ($EDMCH1^+$: -720.60 kJ/mol; $EDMCH2^+$: -750.95 kJ/mol), confirming that the intersection and straight channel locations (T10 and T11) provide superior stabilization.

Narrow-pore zeolites such as MFI and FER, along with the monodimensional MOR zeolite, consistently yield lower Eint values generally below 700 kJ/mol, indicating stronger interactions and thus better stabilization of cyclohexyl intermediates. Conversely, frameworks with wider channels or larger cavities, such as FAU and BEA, exhibit higher Eint values, reflecting weaker overall stabilization. Notably, however, the wider-channel frameworks offer lower energy barriers for bulkier intermediates, highlighting the importance of an optimal balance in confinement, neither too tight nor too loose.

**Table S11.** Interaction energies ($E_{int}$) calculated as $E_{int} = E_{(zeo+molecule)} – E_{molecule_gas} – E_{zeolite}$ for each model.

| | **FAU-T1** | **FER-T1** | **MOR-T2** | **BEA-T1a** | **BEA-T1b** | **BEA-T2** | **MFI-T4** | **MFI-T10** | **MFI-T11** |
|---|---|---|---|---|---|---|---|---|---|
| $ECH1^+$ | -658.21 | -739.91 | -742.79 | -638.12 | -641.00 | -637.27 | -705.80 | -726.98 | -726.67 |
| $EMCH1^+$ | -656.11 | -744.01 | -784.14 | -690.07 | -680.91 | -684.85 | -718.77 | -744.10 | -723.73 |
| $EMCH2^+$ | -658.71 | -702.87 | -769.68 | -678.75 | -679.31 | -674.00 | -691.29 | -728.86 | -751.84 |
| $EDMCH1^+$ | -666.42 | -747.60 | -764.67 | -678.88 | -661.20 | -685.52 | -623.86 | -738.57 | -720.60 |
| $EDMCH2^+$ | -657.52 | -636.00 | -780.57 | -663.20 | -675.50 | -663.84 | -564.27 | -736.40 | -750.95 |

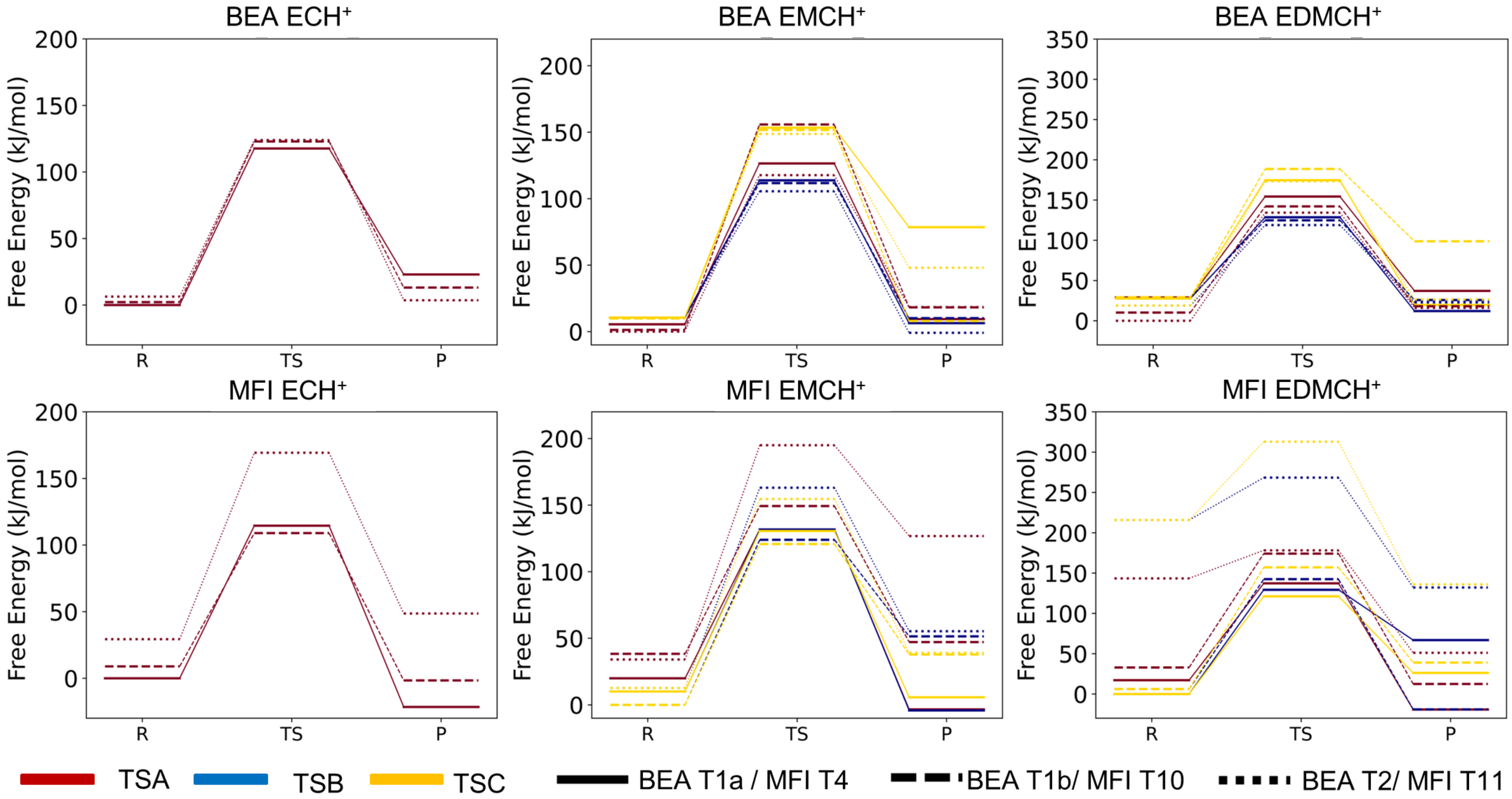

**Figure S56.** Ring-contraction pathways through TSA (red), TSB (blue), and TSC (gold), rescaled to the lowest reactant minima identified within the BEA (T1a for $ECH^+$, T2 for $EMCH^+$ and $EDMCH^+$) and MFI (T10 for $ECH^+$ and $EDMCH^+$, T11 for $EMCH^+$) frameworks. Solid lines correspond to BEA-T1a and MFI-T4 model barriers, dashed lines to BEA-T1b and MFI-T10 barriers, and dotted lines to BEA-T2 and MFI-T11 barriers.

## S4 Alkene Cracking

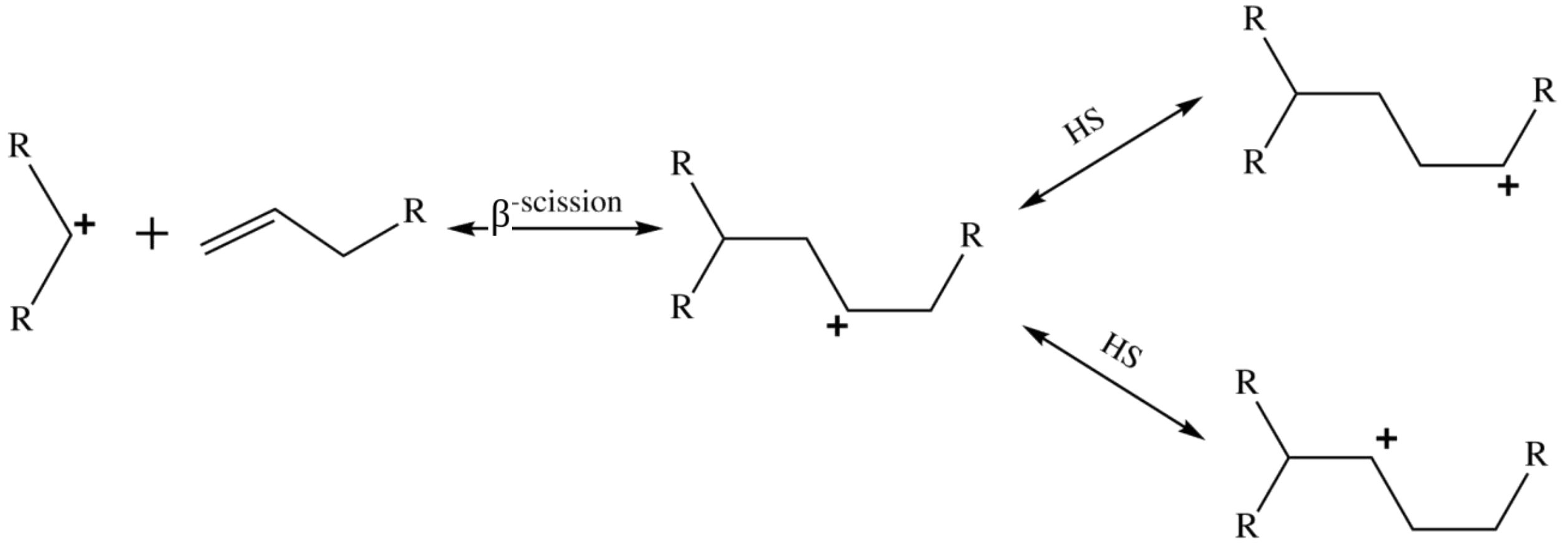

**Figure S57.** Schematic representation of the β-scission and hydrogen shift pathways investigated in this study, illustrating the hydrogen shift rearrangements originating from the secondary carbenium ion intermediates.

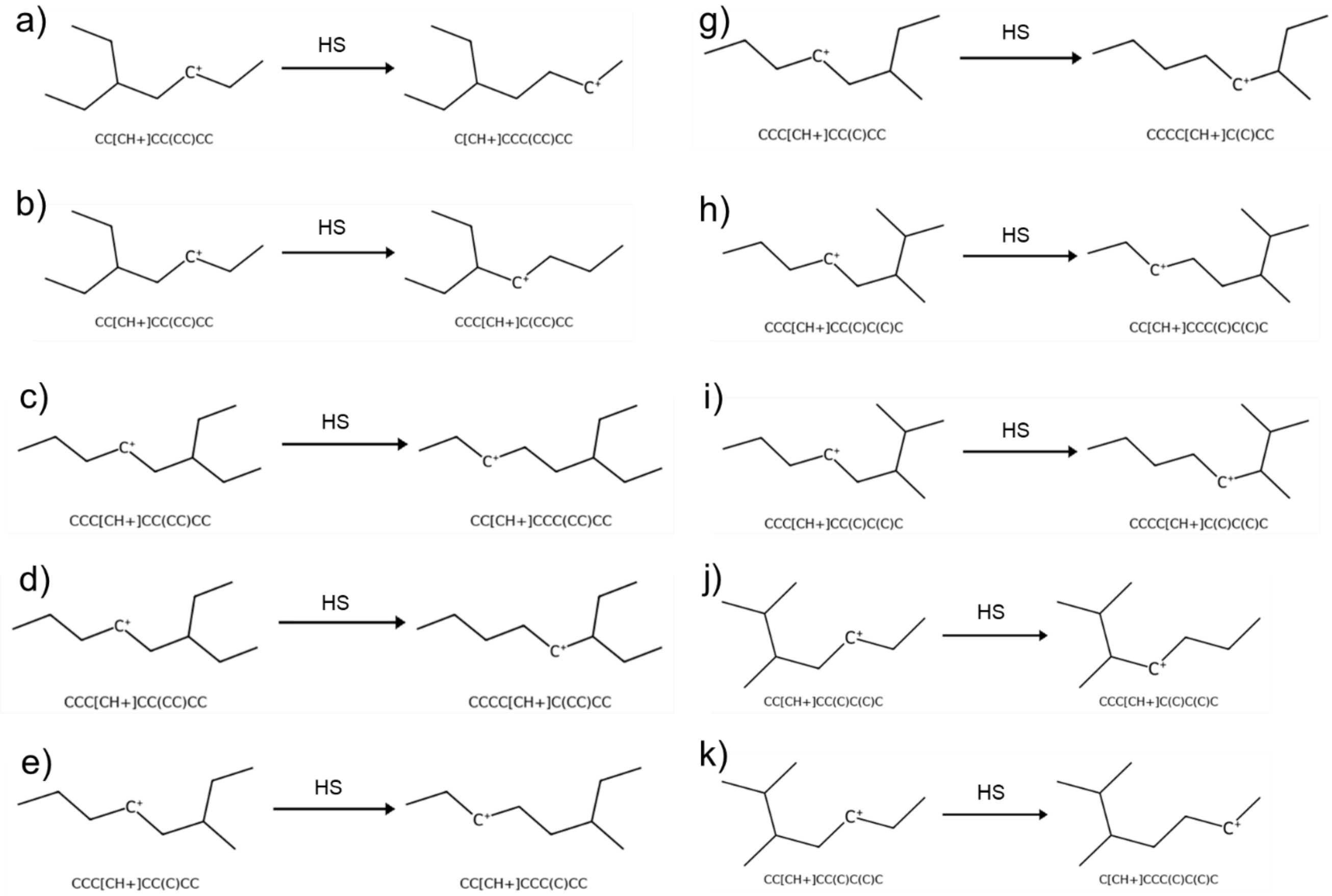

**Figure S58.** Schematic representation of hydrogen shift pathways among secondary carbocations. Each panel (a–h) depicts a hydrogen shit (HS) from an adjacent secondary carbon, resulting in the formation of a new secondary carbocation.

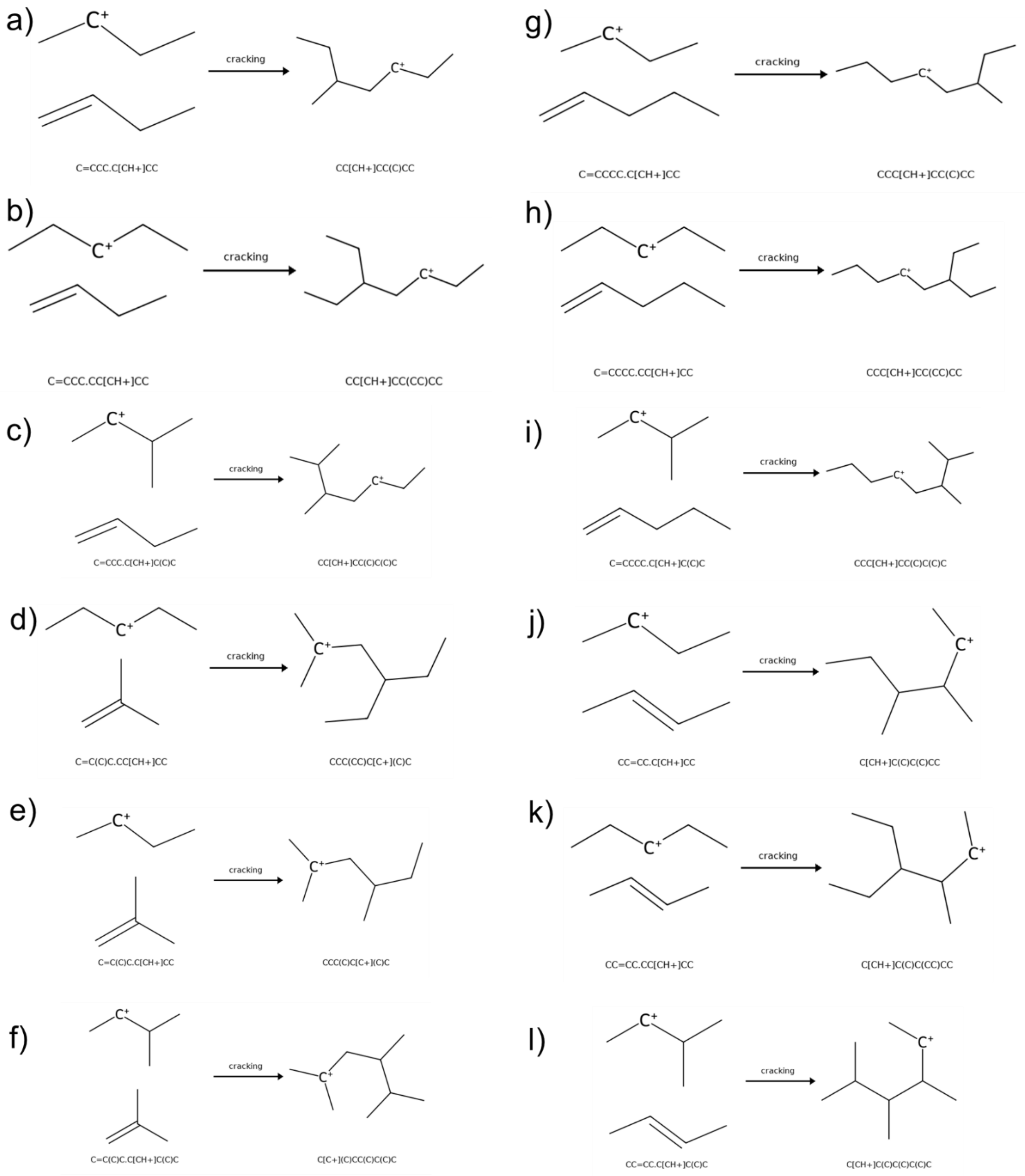

**Figure S59.** β-Scission reactions studied are presented in panels (a–c) for 1-butene, (d–f) for isobutene, (g–i) for 1-pentene, and (j–l) for 2-butene reacting with 2-butyl, 3-pentyl, and 3-methyl-2-butyl, respectively.

**Table S12**. Free energy paths (in kJ/mol) for the hydrogen shift (HS) transitions. Columns list, SMILES of the reactant (R) and product (P), zeolite framework (Frame), T-site, and the computed free energies (GR, GTS, GP).

| R smiles | P smiles | Frame | T-site | $G_R$ (kJ/mol) | $G_{TS}$ (kJ/mol) | $G_P$ (kJ/mol) |
|---|---|---|---|---|---|---|
| CCCC[CH+]C(CC)CC | CCC[CH+]CC(CC)CC | MFI | T11 | 0 | 85.10 | 9.93 |
| CC[CH+]CC(C)C(C)C | C[CH+]CCC(C)C(C)C | MFI | T11 | 0 | 87.89 | -23.10 |
| CC[CH+]CC(C)C(C)C | CCC[CH+]C(C)C(C)C | MFI | T4 | 0 | 136.91 | 22.1 |
| CC[CH+]CC(CC)CC | C[CH+]CCC(CC)CC | FAU | T1 | 0 | 49.74 | 19.31 |

## S5. Diethylbenzene Transalkylation Study Data Tables

This section presents the data used for the results shown in the diethylbenzene transalkylation study in the **Results** section of the main text and in **Section S4** of this document.

**Table S13.** Multistep Paths TS labeling for different isomers: TS2 (between substituted or unsubstituted C atoms), TS3 (diagonal or straight), and TS4 (between substituted or unsubstituted C atoms).

| Isomer | TS2 | TS3 | TS4 |
|---|---|---|---|
| Ortho-Benzene-1 | Unsub | Straight | Sub |
| Ortho-Benzene-2 | Unsub | Diagonal | Unsub |
| Meta-Benzene-1 | Unsub | Diagonal | Unsub |
| Meta-Benzene-2 | Unsub | Straight | Sub |
| Para-Benezene | Unsub | Diagonal | Unsub |
| Ortho-Ortho-1 | Sub | Straight | Unsub |
| Ortho-Ortho-2 | Unsub | Diagonal | Unsub |
| Ortho-Ortho-3 | Unsub | Straight | Sub |
| Ortho-Ortho-4 | Sub | Diagonal | Sub |
| Ortho-Meta-1 | Sub | Diagonal | Sub |
| Ortho-Meta-2 | Sub | Straight | Unsub |
| Ortho-Para-1 | Sub | Diagonal | Sub |
| Ortho-Para-2 | Sub | Straight | Unsub |
| Ortho-Para-3 | Unsub | Diagonal | Unsub |
| Ortho-Para-4 | Unsub | Straight | Sub |
| Meta-Ortho-1 | Unsub | Straight | Unsub |
| Meta-Ortho-2 | Unsub | Diagonal | Sub |
| Meta-Ortho-3 | Sub | Diagonal | Sub |
| Meta-Ortho-4 | Sub | Straight | Unsub |
| Meta-Meta-1 | Sub | Straight | Unsub |
| Meta-Meta-2 | Sub | Diagonal | Unsub |
| Meta-Para-1 | Unsub | Diagonal | Unsub |
| Meta-Para-2 | Unsub | Straight | Unsub |
| Meta-Para-3 | Sub | Diagonal | Unsub |
| Meta-Para-4 | Sub | Straight | Unsub |
| Para-Ortho-1 | Sub | Diagonal | Unsub |
| Para-Ortho-2 | Unsub | Straight | Unsub |
| Para-Meta | Sub | Straight | Unsub |
| Para-Para-1 | Unsub | Straight | Unsub |
| Para-Para-2 | Sub | Diagonal | Unsub |

**Table S14.** Gas-phase free energy barriers (kJ/mol) for Multistep TS2, TS3 and TS4, and Direct paths, TS1 for different isomers at transition states TS0, TS2, TS3, TS4, and TS1. TS0 is shared between both paths. Values depicted in **Figure S26**.

| Isomer | TS0 | TS2 | TS3 | TS4 | TS1 |
|---|---|---|---|---|---|
| Ortho-Benzene-1 | 41.03 | 35.65 | 87.71 | 12.71 | 94.05 |
| Ortho-Benzene-2 | 41.03 | 35.65 | 69.72 | 32.96 | 94.05 |
| Meta-Benzene-1 | 30.69 | 23.67 | 33.43 | 51.54 | 82.66 |
| Meta-Benzene-2 | 30.69 | 23.67 | 47.97 | 45.93 | 82.66 |
| Para-Benzene | 45.67 | 62.82 | 60.17 | 31.24 | 93.51 |
| Ortho-Ortho-1 | 20.81 | 28.81 | 95.12 | 92.07 | 168.47 |
| Ortho-Ortho-2 | 20.81 | 27.32 | 89.62 | 92.07 | 168.47 |
| Ortho-Ortho-3 | 20.81 | 27.32 | 99.68 | 39.22 | 168.47 |
| Ortho-Ortho-4 | 20.81 | 28.81 | 91.20 | 39.22 | 168.47 |
| Ortho-Meta-1 | 16.40 | 55.61 | 60.08 | 43.58 | 164.71 |
| Ortho-Meta-2 | 16.40 | 55.61 | 124.68 | 39.18 | 164.71 |
| Ortho-Para-1 | 40.09 | 28.89 | 73.58 | 35.76 | 125.59 |
| Ortho-Para-2 | 40.09 | 28.89 | 87.09 | 29.20 | 125.59 |
| Ortho-Para-3 | 40.09 | 28.06 | 78.84 | 35.33 | 125.59 |
| Ortho-Para-4 | 40.09 | 28.06 | 102.97 | 41.89 | 125.59 |
| Meta-Ortho-1 | 17.50 | 28.38 | 84.52 | 55.73 | 99.37 |
| Meta-Ortho-2 | 17.50 | 28.38 | 82.90 | 46.54 | 99.37 |
| Meta-Ortho-3 | 17.50 | 30.55 | 54.34 | 46.54 | 99.37 |
| Meta-Ortho-4 | 17.50 | 30.55 | 62.75 | 55.73 | 99.37 |
| Meta-Meta-1 | 13.54 | 36.12 | 42.43 | 62.46 | 146.19 |
| Meta-Meta-2 | 13.54 | 36.12 | 42.30 | 71.24 | 146.19 |
| Meta-Para-1 | 15.91 | 24.82 | 75.25 | 48.25 | 121.09 |
| Meta-Para-2 | 15.91 | 24.82 | 74.36 | 48.99 | 121.09 |
| Meta-Para-3 | 15.91 | 21.35 | 79.99 | 48.99 | 121.09 |
| Meta-Para-4 | 15.91 | 21.35 | 85.04 | 48.25 | 121.09 |
| Para-Ortho-1 | 29.54 | 28.16 | 54.99 | 28.52 | 126.83 |
| Para-Ortho-2 | 29.54 | 25.60 | 102.18 | 28.52 | 126.83 |
| Para-Meta | 16.19 | 53.31 | 81.39 | 39.89 | 135.50 |
| Para-Para-1 | 24.84 | 30.32 | 99.46 | 42.75 | 118.42 |
| Para-Para-2 | 24.84 | 25.08 | 85.23 | 42.75 | 118.42 |

**Table S15.** Gas-Phase Direct Path Energies in kJ/mol for different isomers at key intermediates (I0, I1, I4) and transition states (TS0, TS1). MEPs are highlighted in green color.

| Isomer | I0 | TS0 | I1 | TS1 | I4 |
|---|---|---|---|---|---|
| Ortho-benzene | 0.0 | 41.03 | 26.23 | 120.28 | 21.93 |
| Meta-benzene | 0.0 | 30.69 | 21.10 | 103.76 | 18.69 |
| Para-benzene | 0.0 | 45.67 | -3.72 | 97.28 | 13.22 |
| Ortho-ortho | 0.0 | 20.81 | 4.71 | 173.18 | 15.78 |
| Ortho-meta | 0.0 | 16.40 | -4.51 | 160.20 | 33.02 |
| Ortho-para | 0.0 | 40.09 | 23.52 | 149.11 | 26.91 |
| Meta-ortho | 0.0 | 17.50 | 2.55 | 101.92 | 32.14 |
| Meta-meta | 0.0 | 13.54 | -5.40 | 140.79 | 52.16 |
| Meta-para | 0.0 | 15.91 | 2.92 | 124.01 | -13.23 |
| Para-ortho | 0.0 | 29.54 | 11.64 | 138.47 | 22.44 |
| Para-meta | 0.0 | 16.19 | 6.45 | 141.95 | 45.66 |
| Para-para | 0.0 | 24.84 | 15.63 | 134.05 | 27.98 |

**Table S16.** Gas-Phase Multistep Path Energies in kJ/mol for different isomers at key intermediates (I0, I1, I2, I3, I4) and transition states (TS0, TS2, TS3, TS4). $\Delta G_{I3-I2}$ expressed in kJ/mol. MEPs are highlighted in green color.

| Isomer | I0 | TS0 | I1 | TS2 | I2 | TS3 | I3 | TS4 | I4 | $\Delta G_{I3-I2}$ |
|---|---|---|---|---|---|---|---|---|---|---|
| Ortho-Benzene-1 | 0.0 | 41.03 | 26.23 | 61.88 | 18.90 | 106.61 | 32.30 | 45.01 | 21.93 | 13.4 |
| Ortho-Benzene-2 | 0.0 | 41.03 | 26.23 | 61.88 | 18.90 | 88.62 | 13.09 | 46.05 | 21.93 | -5.81 |
| Meta-Benzene-1 | 0.0 | 30.69 | 21.10 | 44.84 | 8.92 | 42.35 | -20.00 | 31.54 | 18.69 | -28.92 |
| Meta-Benzene-2 | 0.0 | 30.69 | 21.10 | 44.84 | 8.92 | 56.89 | -14.77 | 31.16 | 18.69 | -23.69 |
| Para-Benzene | 0.0 | 45.67 | -3.72 | 59.10 | 20.61 | 80.78 | 13.64 | 44.88 | 13.22 | -6.97 |
| Ortho-Ortho-1 | 0.0 | 20.81 | 4.71 | 33.52 | -10.42 | 84.70 | 7.93 | 100.00 | 15.78 | 18.35 |
| Ortho-Ortho-2 | 0.0 | 20.81 | 4.71 | 32.03 | -15.67 | 73.95 | 7.93 | 100.00 | 15.78 | 23.6 |
| Ortho-Ortho-3 | 0.0 | 20.81 | 4.71 | 32.03 | -15.67 | 84.01 | 16.40 | 55.62 | 15.78 | 32.07 |
| Ortho-Ortho-4 | 0.0 | 20.81 | 4.71 | 33.52 | -10.42 | 80.78 | 16.40 | 55.62 | 15.78 | 26.82 |
| Ortho-Meta-1 | 0.0 | 16.40 | -4.51 | 51.10 | 29.84 | 89.92 | 28.20 | 71.78 | 33.02 | -1.64 |
| Ortho-Meta-2 | 0.0 | 16.40 | -4.51 | 51.10 | 29.84 | 154.52 | 26.59 | 65.77 | 33.02 | -3.25 |
| Ortho-Para-1 | 0.0 | 40.09 | 23.52 | 52.41 | 4.23 | 77.81 | 28.36 | 64.12 | 26.91 | 24.13 |
| Ortho-Para-2 | 0.0 | 40.09 | 23.52 | 52.41 | 4.23 | 91.32 | 28.36 | 57.56 | 26.91 | 24.13 |
| Ortho-Para-3 | 0.0 | 40.09 | 23.52 | 51.58 | 4.52 | 83.36 | 22.23 | 57.56 | 26.91 | 17.71 |
| Ortho-Para-4 | 0.0 | 40.09 | 23.52 | 51.58 | 4.52 | 107.49 | 22.23 | 64.12 | 26.91 | 17.71 |
| Meta-Ortho-1 | 0.0 | 17.50 | 2.55 | 30.93 | -13.84 | 70.68 | -8.64 | 47.09 | 32.14 | 5.2 |
| Meta-Ortho-2 | 0.0 | 17.50 | 2.55 | 30.93 | -13.84 | 69.06 | 0.56 | 47.10 | 32.14 | 14.4 |
| Meta-Ortho-3 | 0.0 | 17.50 | 2.55 | 33.10 | -3.77 | 50.57 | 0.56 | 47.10 | 32.14 | 4.33 |
| Meta-Ortho-4 | 0.0 | 17.50 | 2.55 | 33.10 | -3.77 | 58.98 | -8.64 | 47.09 | 32.14 | -4.87 |
| Meta-Meta-1 | 0.0 | 13.54 | -5.40 | 30.72 | 12.28 | 54.71 | -2.70 | 59.76 | 52.16 | -14.98 |
| Meta-Meta-2 | 0.0 | 13.54 | -5.40 | 30.72 | 12.28 | 54.58 | -8.80 | 62.44 | 52.16 | -21.08 |
| Meta-Para-1 | 0.0 | 15.91 | 2.92 | 27.74 | -18.72 | 56.53 | -7.99 | 40.26 | -13.23 | 10.73 |
| Meta-Para-2 | 0.0 | 15.91 | 2.92 | 27.74 | -18.72 | 55.64 | -3.94 | 45.05 | -13.23 | 14.78 |
| Meta-Para-3 | 0.0 | 15.91 | 2.92 | 24.27 | -19.67 | 60.32 | -3.94 | 45.05 | -13.23 | 15.73 |
| Meta-Para-4 | 0.0 | 15.91 | 2.92 | 24.27 | -19.67 | 65.37 | -7.99 | 40.26 | -13.23 | 11.68 |
| Para-Ortho-1 | 0.0 | 29.54 | 11.64 | 39.80 | 0.55 | 55.54 | 18.10 | 46.62 | 22.44 | 17.55 |
| Para-Ortho-2 | 0.0 | 29.54 | 11.64 | 37.24 | -4.46 | 97.72 | 18.10 | 46.62 | 22.44 | 22.56 |
| Para-Meta | 0.0 | 16.19 | 6.45 | 59.76 | 34.56 | 115.95 | 26.36 | 66.25 | 45.66 | -8.2 |
| Para-Para-1 | 0.0 | 24.84 | 15.63 | 45.95 | -1.09 | 98.37 | 21.44 | 64.19 | 27.98 | 22.53 |
| Para-Para-2 | 0.0 | 24.84 | 15.63 | 40.71 | 0.61 | 85.84 | 21.44 | 64.19 | 27.98 | 20.83 |

**Table S17.** Zeolite Multistep Path relative free energies in kJ/mol. Global MEPs starred at Figure 3 in the main paper are highlighted in green color.

| Framework | Isomer | T-Site | I1 | TS2 | I2 | TS3 | I3 | TS4 | I4 | Track TS2 | Track TS3 | Track TS4 |
|---|---|---|---|---|---|---|---|---|---|---|---|---|
| BOG | ortho_benzene | T1 | 0 | 128,55 | -70,97 | 97,78 | 11,47 | 101,99 | -11,06 | Unsub | Diagonal | Unsub |
| BOG | ortho_benzene | T1 | 0 | 128,55 | -70,97 | 120,36 | 6,3 | 59,3 | -11,06 | Unsub | Straight | Sub |
| BOG | meta_benzene | T1 | 0 | 83,51 | 38,95 | 66,79 | -23,01 | 110,04 | 6,15 | Unsub | Diagonal | Unsub |
| BOG | meta_benzene | T1 | 0 | 83,51 | 38,95 | 94,88 | 1,04 | 54,19 | 6,15 | Unsub | Straight | Sub |
| BOG | para_benzene | T1 | 0 | 92,86 | 39,34 | 92,43 | 7,92 | 70,59 | 42,8 | Unsub | Straight | Unsub |
| BOG | ortho_ortho | T1 | 0 | 83,03 | -10,08 | 90,71 | 24,83 | 88,55 | 65,37 | Sub | Diagonal | Sub |
| BOG | ortho_ortho | T1 | 0 | 83,03 | -10,08 | 27,41 | -55,1 | 106,39 | 65,37 | Sub | Straight | Unsub |
| BOG | ortho_ortho | T1 | 0 | 25,29 | 4,72 | 73,53 | 24,83 | 88,55 | 65,37 | Unsub | Straight | Sub |
| BOG | ortho_ortho | T1 | 0 | 25,29 | 4,72 | 102,41 | -55,1 | 106,39 | 65,37 | Unsub | Diagonal | Unsub |
| BOG | ortho_meta | T1 | 0 | 160,85 | -3,67 | 113,46 | -18,1 | 119,71 | -49,22 | Sub | Straight | Unsub |
| BOG | ortho_meta | T1 | 0 | 160,85 | -3,67 | 68,76 | 51,1 | 129,27 | -49,22 | Sub | Diagonal | Sub |
| BOG | ortho_para | T1 | 0 | 81,56 | 4,77 | 165,26 | 42,11 | 112,93 | 30,58 | Unsub | Straight | Sub |
| BOG | ortho_para | T1 | 0 | 81,56 | 4,77 | 89,49 | 42,11 | 98,67 | 30,58 | Unsub | Diagonal | Unsub |
| BOG | ortho_para | T1 | 0 | 55 | 20,84 | 126,54 | 1,43 | 112,93 | 30,58 | Sub | Diagonal | Sub |
| BOG | ortho_para | T1 | 0 | 55 | 20,84 | 129,99 | 1,43 | 98,67 | 30,58 | Sub | Straight | Unsub |
| BOG | meta_ortho | T1 | 0 | 121,97 | -16,86 | 111,33 | 31,23 | 148,47 | 36,77 | Unsub | Diagonal | Sub |
| BOG | meta_ortho | T1 | 0 | 121,97 | -16,86 | 113,45 | 9,15 | 65,56 | 36,77 | Unsub | Straight | Unsub |
| BOG | meta_ortho | T1 | 0 | 163,32 | -30,76 | 87,67 | 31,23 | 148,47 | 36,77 | Sub | Diagonal | Sub |
| BOG | meta_ortho | T1 | 0 | 163,32 | -30,76 | 91,28 | 9,15 | 65,56 | 36,77 | Sub | Straight | Unsub |
| BOG | meta_meta | T1 | 0 | 22,63 | -11,04 | 13,18 | -20,54 | 140,17 | -9,49 | Sub | Diagonal | Unsub |
| BOG | meta_meta | T1 | 0 | 22,63 | -11,04 | 67,55 | -57,66 | 107,24 | -9,49 | Sub | Straight | Sub |
| BOG | meta_para | T1 | 0 | 120,26 | -7,8 | 48,86 | 10,18 | 126,81 | 30 | Sub | Straight | Unsub |
| BOG | meta_para | T1 | 0 | 120,26 | -7,8 | 119,8 | -1,5 | 88,28 | 30 | Sub | Diagonal | Sub |
| BOG | meta_para | T1 | 0 | 30,81 | 2,01 | 55,71 | -1,5 | 88,28 | 30 | Unsub | Straight | Sub |
| BOG | meta_para | T1 | 0 | 30,81 | 2,01 | 48,46 | 10,18 | 126,81 | 30 | Unsub | Diagonal | Unsub |
| BOG | para_ortho | T1 | 0 | 47,37 | -30,73 | 33,19 | -22,39 | 72,26 | 11,99 | Sub | Diagonal | Unsub |
| BOG | para_ortho | T1 | 0 | 73,12 | -21,71 | 84,4 | -22,39 | 72,26 | 11,99 | Unsub | Straight | Unsub |
| BOG | para_meta | T1 | 0 | 100,12 | 62,82 | 124,63 | 21,64 | 159,39 | 69,34 | Sub | Straight | Unsub |
| BOG | para_para | T1 | 0 | 128,2 | 16,52 | 74,86 | 16,89 | 151,16 | 48,43 | Unsub | Straight | Unsub |
| BOG | para_para | T1 | 0 | 94,43 | 29,95 | 96,79 | 16,89 | 151,16 | 48,43 | Sub | Straight | Unsub |
| BOG | ortho_benzene | T2 | 0 | 54,78 | 8,54 | 77,56 | 5,33 | 107,02 | 1,14 | Unsub | Straight | Sub |
| BOG | ortho_benzene | T2 | 0 | 54,78 | 8,54 | 180,05 | -1,69 | 44,98 | 1,14 | Unsub | Diagonal | Unsub |
| BOG | meta_benzene | T2 | 0 | 92,92 | 10,98 | 43,05 | 2,84 | 91,13 | 9,8 | Unsub | Straight | Sub |
| BOG | meta_benzene | T2 | 0 | 92,92 | 10,98 | 45,2 | -13,95 | 73,55 | 9,8 | Unsub | Diagonal | Unsub |
| BOG | para_benzene | T2 | 0 | 76,48 | 10,52 | 67,65 | 4,71 | 49,3 | 8,21 | Unsub | Straight | Unsub |
| BOG | ortho_ortho | T2 | 0 | 81,87 | 23,4 | 167,02 | 58 | 167,08 | 27,21 | Sub | Diagonal | Sub |
| BOG | ortho_ortho | T2 | 0 | 81,87 | 23,4 | 113,74 | 18,68 | 110,43 | 27,21 | Sub | Straight | Unsub |
| BOG | ortho_ortho | T2 | 0 | 87,41 | -11,53 | 171,71 | 18,68 | 110,43 | 27,21 | Unsub | Diagonal | Unsub |
| BOG | ortho_ortho | T2 | 0 | 87,41 | -11,53 | 163,31 | 58 | 167,08 | 27,21 | Unsub | Straight | Sub |
| BOG | ortho_meta | T2 | 0 | 38,63 | -26,58 | 86,54 | 9,43 | 70,39 | -14,53 | Sub | Diagonal | Sub |
| BOG | ortho_meta | T2 | 0 | 38,63 | -26,58 | 111,37 | -59,1 | 111,51 | -14,53 | Sub | Straight | Unsub |
| BOG | ortho_para | T2 | 0 | 36,93 | -48,11 | 91,84 | 0,96 | 111,74 | 20,7 | Unsub | Diagonal | Unsub |
| BOG | ortho_para | T2 | 0 | 36,93 | -48,11 | 113,34 | 0,96 | 92,16 | 20,7 | Unsub | Straight | Sub |
| BOG | ortho_para | T2 | 0 | 49,46 | 3,76 | 165,78 | 13,11 | 111,74 | 20,7 | Sub | Straight | Unsub |
| BOG | ortho_para | T2 | 0 | 49,46 | 3,76 | 155,59 | 13,11 | 92,16 | 20,7 | Sub | Diagonal | Sub |
| BOG | meta_ortho | T2 | 0 | 68,76 | 16,12 | 192,35 | -12,53 | 100,44 | 25,72 | Unsub | Straight | Unsub |
| BOG | meta_ortho | T2 | 0 | 68,76 | 16,12 | 97,06 | -21,79 | 86,61 | 25,72 | Unsub | Diagonal | Sub |
| BOG | meta_ortho | T2 | 0 | 129,06 | -49,59 | 188,4 | -12,53 | 100,44 | 25,72 | Sub | Straight | Unsub |
| BOG | meta_ortho | T2 | 0 | 129,06 | -49,59 | 96,94 | -21,79 | 86,61 | 25,72 | Sub | Diagonal | Sub |
| BOG | meta_meta | T2 | 0 | 112,54 | 23,11 | 127,64 | 20,95 | 158,42 | 43,02 | Sub | Diagonal | Unsub |
| BOG | meta_meta | T2 | 0 | 112,54 | 23,11 | 99,98 | 52,38 | 101,24 | 43,02 | Sub | Straight | Sub |
| BOG | meta_para | T2 | 0 | 85,54 | -8,53 | 119,27 | 14,3 | 84,52 | 44,63 | Sub | Diagonal | Unsub |
| BOG | meta_para | T2 | 0 | 85,54 | -8,53 | 88,43 | 17,3 | 151,98 | 44,63 | Sub | Straight | Sub |
| BOG | meta_para | T2 | 0 | 56,07 | 5,21 | 103,32 | 17,3 | 151,98 | 44,63 | Unsub | Diagonal | Sub |
| BOG | meta_para | T2 | 0 | 56,07 | 5,21 | 88,94 | 14,3 | 84,52 | 44,63 | Unsub | Straight | Unsub |
| BOG | para_ortho | T2 | 0 | 32,68 | -6,99 | 78,82 | 1,2 | 113,4 | 72,35 | Sub | Straight | Unsub |
| BOG | para_ortho | T2 | 0 | 100,93 | 9,19 | 77,54 | 1,2 | 113,4 | 72,35 | Unsub | Cross | Unsub |
| BOG | para_meta | T2 | 0 | 35,91 | 22,25 | 55,56 | -59,65 | 64,21 | 10,71 | Sub | Straight | Unsub |
| BOG | para_para | T2 | 0 | 79,13 | 15,59 | 85,58 | 3,39 | 77,28 | 48,67 | Unsub | Straight | Unsub |
| BOG | para_para | T2 | 0 | 63,94 | 6,83 | 88,5 | 3,39 | 77,28 | 48,67 | Sub | Straight | Unsub |
| IWV | ortho_benzene | T3 | 0 | 84,15 | 29,85 | 97,58 | 12,68 | 64,45 | 16,09 | Unsub | Diagonal | Unsub |
| IWV | ortho_benzene | T3 | 0 | 84,15 | 29,85 | 94,75 | 33,43 | 65,81 | 16,09 | Unsub | Straight | Sub |
| IWV | meta_benzene | T3 | 0 | 27,03 | -9,87 | 47,97 | -8,43 | 51,94 | 15,99 | Unsub | Straight | Unsub |
| IWV | meta_benzene | T3 | 0 | 27,03 | -9,87 | 35,99 | -19,3 | 41,98 | 15,99 | Unsub | Diagonal | Sub |
| IWV | para_benzene | T3 | 0 | 53,68 | -25,28 | 37,19 | -30,32 | 18,62 | -31,62 | Unsub | Straight | Unsub |
| IWV | ortho_ortho | T3 | 0 | 63,35 | 1,29 | 88,23 | 21,56 | 76,35 | -12,13 | Sub | Diagonal | Sub |
| IWV | ortho_ortho | T3 | 0 | 63,35 | 1,29 | 124,71 | -8,27 | 73,45 | -12,13 | Sub | Straight | Unsub |
| IWV | ortho_ortho | T3 | 0 | 101,34 | -24,39 | 93,05 | 21,56 | 76,35 | -12,13 | Unsub | Straight | Sub |
| IWV | ortho_ortho | T3 | 0 | 101,34 | -24,39 | 77,6 | -8,27 | 73,45 | -12,13 | Unsub | Diagonal | Unsub |
| IWV | ortho_meta | T3 | 0 | 111,65 | 45,55 | 141,2 | 44,11 | 133,22 | 61,75 | Sub | Diagonal | Unsub |
| IWV | ortho_meta | T3 | 0 | 111,65 | 45,55 | 193,56 | 73,97 | 173,46 | 61,75 | Sub | Straight | Sub |
| IWV | ortho_para | T3 | 0 | 58,75 | -6,16 | 112,6 | 23,08 | 113,26 | 9,98 | Unsub | Diagonal | Sub |
| IWV | ortho_para | T3 | 0 | 58,75 | -6,16 | 105,26 | 23,08 | 118,89 | 9,98 | Unsub | Straight | Unsub |
| IWV | ortho_para | T3 | 0 | 80,78 | -38,92 | 101,72 | 3,45 | 113,26 | 9,98 | Sub | Straight | Sub |
| IWV | ortho_para | T3 | 0 | 80,78 | -38,92 | 69,58 | 3,45 | 118,89 | 9,98 | Sub | Diagonal | Unsub |
| IWV | meta_ortho | T3 | 0 | 94,49 | -26,08 | 49,67 | -7,86 | 27,34 | -13,14 | Unsub | Diagonal | Sub |
| IWV | meta_ortho | T3 | 0 | 94,49 | -26,08 | 53,35 | -57,96 | 80,6 | -13,14 | Unsub | Straight | Unsub |
| IWV | meta_ortho | T3 | 0 | 71,59 | -47,96 | 45,46 | -7,86 | 27,34 | -13,14 | Sub | Diagonal | Sub |
| IWV | meta_ortho | T3 | 0 | 71,59 | -47,96 | 51,62 | -57,96 | 80,6 | -13,14 | Sub | Straight | Unsub |
| IWV | meta_meta | T3 | 0 | 64,96 | 31,84 | 117,95 | -6,98 | 108,89 | 31,89 | Sub | Diagonal | Unsub |
| IWV | meta_meta | T3 | 0 | 64,96 | 31,84 | 115,84 | 39,14 | 82,68 | 31,89 | Sub | Straight | Sub |
| IWV | meta_para | T3 | 0 | 98,77 | 11,35 | 110,86 | 10,11 | 84,59 | -2,93 | Sub | Straight | Unsub |
| IWV | meta_para | T3 | 0 | 98,77 | 11,35 | 100,71 | 10,21 | 99,91 | -2,93 | Sub | Diagonal | Sub |
| IWV | meta_para | T3 | 0 | 71,49 | 7,01 | 111,39 | 10,21 | 99,91 | -2,93 | Unsub | Straight | Sub |
| IWV | meta_para | T3 | 0 | 71,49 | 7,01 | 81,13 | 10,11 | 84,59 | -2,93 | Unsub | Diagonal | Unsub |
| IWV | para_ortho | T3 | 0 | 63,94 | -20,82 | 61,95 | 11,37 | 42,05 | 5,36 | Sub | Straight | Unsub |
| IWV | para_ortho | T3 | 0 | 52,52 | -11,97 | 76,41 | 11,37 | 42,05 | 5,36 | Unsub | Diagonal | Unsub |
| IWV | para_meta | T3 | 0 | 63,4 | 16,29 | 66,46 | 4,88 | 99,48 | 13 | Sub | Straight | Unsub |
| IWV | para_para | T3 | 0 | 48,88 | 7,81 | 142,56 | 26,79 | 104,57 | 45,89 | Unsub | Straight | Unsub |
| IWV | para_para | T3 | 0 | 115,48 | 31,52 | 111,67 | 26,79 | 104,57 | 45,89 | Sub | Straight | Unsub |
| IWV | ortho_benzene | T7 | 0 | 63,35 | 10,29 | 142,79 | 76,37 | 112,71 | 6,22 | Unsub | Straight | Sub |
| IWV | ortho_benzene | T7 | 0 | 63,35 | 10,29 | 65,05 | 17,63 | 89,66 | 6,22 | Unsub | Diagonal | Unsub |
| IWV | meta_benzene | T7 | 0 | 48,04 | -26,31 | 33,31 | -45,2 | 14.12 | -17,03 | Unsub | Straight | Sub |
| IWV | meta_benzene | T7 | 0 | 48,04 | -26,31 | 22,93 | -47,09 | 40,06 | -17,03 | Unsub | Diagonal | Unsub |
| IWV | para_benzene | T7 | 0 | 78,04 | 29,6 | 94,19 | 7,44 | 52,75 | 10,19 | Unsub | Straight | Unsub |
| IWV | ortho_ortho | T7 | 0 | 77,46 | 11,15 | 107,03 | 9,75 | 86,37 | 41,28 | Sub | Diagonal | Sub |
| IWV | ortho_ortho | T7 | 0 | 77,46 | 11,15 | 210,56 | -4,53 | 172,62 | 41,28 | Sub | Straight | Unsub |
| IWV | ortho_ortho | T7 | 0 | 65,23 | -15,34 | 128,83 | 9,75 | 86,37 | 41,28 | Unsub | Diagonal | Unsub |
| IWV | ortho_ortho | T7 | 0 | 65,23 | -15,34 | 103,37 | -4,53 | 172,62 | 41,28 | Unsub | Straight | Sub |
| IWV | ortho_meta | T7 | 0 | 71,11 | 0,5 | 140,02 | 4,27 | 128,66 | -6,09 | Sub | Straight | Sub |
| IWV | ortho_meta | T7 | 0 | 71,11 | 0,5 | 67,43 | 12,6 | 101 | -6,09 | Sub | Diagonal | Unsub |
| IWV | ortho_para | T7 | 0 | 96,86 | 12,89 | 72,86 | 4,76 | 77,68 | 17,26 | Unsub | Straight | Unsub |
| IWV | ortho_para | T7 | 0 | 96,86 | 12,89 | 80,76 | 4,76 | 106,61 | 17,26 | Unsub | Diagonal | Sub |
| IWV | ortho_para | T7 | 0 | 48,72 | -5,32 | 110,82 | -11,33 | 106,61 | 17,26 | Sub | Straight | Unsub |
| IWV | ortho_para | T7 | 0 | 48,72 | -5,32 | 63,86 | -11,33 | 77,68 | 17,26 | Sub | Diagonal | Sub |
| IWV | meta_ortho | T7 | 0 | 59,7 | -17,54 | 32,7 | -31,86 | 67,83 | -57,92 | Unsub | Diagonal | Unsub |
| IWV | meta_ortho | T7 | 0 | 59,7 | -17,54 | 29,21 | -33,57 | 134,86 | -57,92 | Unsub | Straight | Sub |
| IWV | meta_ortho | T7 | 0 | 87,36 | -40,45 | 29,84 | -31,86 | 67,83 | -57,92 | Sub | Diagonal | Unsub |
| IWV | meta_ortho | T7 | 0 | 87,36 | -40,45 | 28,8 | -33,57 | 134,86 | -57,92 | Sub | Straight | Sub |
| IWV | meta_meta | T7 | 0 | 64,01 | 11,35 | 100,52 | 10,49 | 68,72 | 12,94 | Sub | Straight | Unsub |
| IWV | meta_meta | T7 | 0 | 64,01 | 11,35 | 88,33 | 33,66 | 135,14 | 12,94 | Sub | Diagonal | Sub |
| IWV | meta_para | T7 | 0 | 101,05 | -14,83 | 107,49 | -9,12 | 52,36 | 15,54 | Sub | Straight | Unsub |
| IWV | meta_para | T7 | 0 | 101,05 | -14,83 | 67,74 | -12,78 | 97,68 | 15,54 | Sub | Diagonal | Sub |
| IWV | meta_para | T7 | 0 | 95,65 | -20,14 | 119,88 | -12,78 | 97,68 | 15,54 | Unsub | Straight | Sub |
| IWV | meta_para | T7 | 0 | 95,65 | -20,14 | 74,55 | -9,12 | 52,36 | 15,54 | Unsub | Diagonal | Unsub |
| IWV | para_ortho | T7 | 0 | 47,38 | -25,44 | 69,24 | -28,68 | 44,88 | -3,09 | Sub | Diagonal | Unsub |
| IWV | para_ortho | T7 | 0 | 30,53 | -25,25 | 107,59 | -28,68 | 44,88 | -3,09 | Unsub | Straight | Unsub |
| IWV | para_meta | T7 | 0 | 90,52 | 8,22 | 146,03 | 5,01 | 93,56 | 30,82 | Sub | Straight | Unsub |
| IWV | para_para | T7 | 0 | 44,53 | 1,66 | 86,16 | 12,27 | 95,99 | 28,1 | Unsub | Straight | Unsub |
| IWV | para_para | T7 | 0 | 66,49 | 4,61 | 115,54 | 12,27 | 95,99 | 28,1 | Sub | Straight | Unsub |
| UTL | ortho_benzene | T2 | 0 | 32,19 | -48,54 | 0,53 | -16,28 | 68,93 | -46,11 | Unsub | Diagonal | Unsub |
| UTL | ortho_benzene | T2 | 0 | 32,19 | -48,54 | 38,58 | -22,75 | 27,02 | -46,11 | Unsub | Straight | Sub |
| UTL | meta_benzene | T2 | 0 | 23,49 | -38,92 | 47,83 | -58,65 | 3,02 | -47,89 | Unsub | Diagonal | Unsub |
| UTL | meta_benzene | T2 | 0 | 23,49 | -38,92 | 19,15 | -65,53 | 0,46 | -47,89 | Unsub | Straight | Sub |
| UTL | para_benzene | T2 | 0 | 62,79 | -59,22 | 18,60 | 0,67 | 43,15 | 19,81 | Unsub | Straight | Unsub |
| UTL | ortho_ortho | T2 | 0 | 31,48 | -23,01 | 93,09 | 1,22 | 81,81 | -6,31 | Sub | Diagonal | Sub |
| UTL | ortho_ortho | T2 | 0 | 31,48 | -23,01 | 92,09 | -7,43 | 74,55 | -6,31 | Sub | Straight | Unsub |
| UTL | ortho_ortho | T2 | 0 | 35,51 | -36,46 | 96,81 | -7,43 | 74,55 | -6,31 | Unsub | Diagonal | Sub |
| UTL | ortho_ortho | T2 | 0 | 35,51 | -36,46 | 50,28 | 1,22 | 81,81 | -6,31 | Unsub | Straight | Unsub |
| UTL | ortho_meta | T2 | 0 | 71,58 | 35,9 | 136,38 | 38,39 | 94,53 | 41,9 | Sub | Straight | Unsub |
| UTL | ortho_meta | T2 | 0 | 71,58 | 35,9 | 73,41 | 33,36 | 117,53 | 41,9 | Sub | Diagonal | Sub |
| UTL | ortho_para | T2 | 0 | 77,72 | 1,05 | 114,63 | 27,12 | 77,45 | -2,6 | Unsub | Diagonal | Sub |
| UTL | ortho_para | T2 | 0 | 77,72 | 1,05 | 63,47 | 27,12 | 133,19 | -2,6 | Unsub | Straight | Unsub |
| UTL | ortho_para | T2 | 0 | 90,42 | -23,15 | 101,75 | -12,67 | 77,45 | -2,6 | Sub | Straight | Sub |
| UTL | ortho_para | T2 | 0 | 90,42 | -23,15 | 114,05 | -12,67 | 133,19 | -2,6 | Sub | Diagonal | Unsub |
| UTL | meta_ortho | T2 | 0 | 67,8 | -22,69 | 49,21 | -37,9 | 122,9 | -3,18 | Unsub | Diagonal | Sub |
| UTL | meta_ortho | T2 | 0 | 67,8 | -22,69 | 65,22 | -25,21 | 123,42 | -3,18 | Unsub | Straight | Unsub |
| UTL | meta_ortho | T2 | 0 | 33,94 | -6,23 | 59,11 | -37,9 | 122,9 | -3,18 | Sub | Diagonal | Sub |
| UTL | meta_ortho | T2 | 0 | 33,94 | -6,23 | 69,98 | -25,21 | 123,42 | -3,18 | Sub | Straight | Unsub |
| UTL | meta_meta | T2 | 0 | 114,31 | 20,71 | 54,61 | -28,27 | 69,09 | 26,13 | Sub | Straight | Unsub |
| UTL | meta_meta | T2 | 0 | 114,31 | 20,71 | 52,54 | -4,1 | 84,32 | 26,13 | Sub | Diagonal | Sub |
| UTL | meta_para | T2 | 0 | 80,79 | 9,64 | 72,1 | -13,83 | 131,29 | -9,39 | Sub | Diagonal | Unsub |
| UTL | meta_para | T2 | 0 | 80,79 | 9,64 | 81,47 | 19,04 | 90,68 | -9,39 | Sub | Straight | Sub |
| UTL | meta_para | T2 | 0 | 67,53 | 5,22 | 59,89 | -13,83 | 131,29 | -9,39 | Unsub | Straight | Sub |
| UTL | meta_para | T2 | 0 | 67,53 | 5,22 | 89,6 | 19,04 | 90,68 | -9,39 | Unsub | Diagonal | Unsub |
| UTL | para_ortho | T2 | 0 | 48,91 | -8,7 | 69,62 | 3,15 | 57,36 | -13,27 | Sub | Straight | Unsub |
| UTL | para_ortho | T2 | 0 | 88,53 | -18,89 | 58,92 | 3,15 | 57,36 | -13,27 | Unsub | Diagonal | Unsub |
| UTL | para_meta | T2 | 0 | 80,05 | 38,76 | 150,2 | 13,8 | 103,79 | 10,74 | Sub | Straight | Unsub |
| UTL | para_para | T2 | 0 | 89,03 | 30,07 | 82,48 | 45,93 | 96,96 | 66,08 | Unsub | Straight | Unsub |
| UTL | para_para | T2 | 0 | 70,07 | 22,07 | 98,29 | 45,93 | 96,96 | 66,08 | Sub | Straight | Unsub |
| UTL | ortho_benzene | T4 | 0 | 109,61 | -3,87 | 63,35 | -39,18 | 68,12 | -21,43 | Unsub | Diagonal | Unsub |
| UTL | ortho_benzene | T4 | 0 | 109,61 | -3,87 | 62,49 | -9,31 | 41,6 | -21,43 | Unsub | Straight | Sub |
| UTL | meta_benzene | T4 | 0 | 44,66 | -5,59 | 32,85 | -63,16 | 60,53 | -19,04 | Unsub | Diagonal | Unsub |
| UTL | meta_benzene | T4 | 0 | 44,66 | -5,59 | 52,82 | -16,05 | 41,46 | -19,04 | Unsub | Straight | Sub |
| UTL | para_benzene | T4 | 0 | 28,71 | -3,78 | 41,86 | -41,54 | 72,8 | -19,03 | Unsub | Straight | Unsub |
| UTL | ortho_ortho | T4 | 0 | 101,99 | -46,73 | 90,93 | 15,1 | 120,82 | 44,96 | Sub | Straight | Sub |
| UTL | ortho_ortho | T4 | 0 | 101,99 | -46,73 | 62,46 | -22,57 | 105,55 | 44,96 | Sub | Diagonal | Unsub |
| UTL | ortho_ortho | T4 | 0 | 71,36 | -19,1 | 96,14 | -22,57 | 105,55 | 44,96 | Unsub | Straight | Sub |
| UTL | ortho_ortho | T4 | 0 | 71,36 | -19,1 | 80,09 | 15,1 | 120,82 | 44,96 | Unsub | Cross | Unsub |
| UTL | ortho_meta | T4 | 0 | 40,82 | -28,85 | 88,32 | 15,88 | 103,54 | 27,8 | Sub | Straight | Unsub |
| UTL | ortho_meta | T4 | 0 | 40,82 | -28,85 | 44,01 | -1,6 | 60,33 | 27,8 | Sub | Diagonal | Sub |
| UTL | ortho_para | T4 | 0 | 60,92 | -29,83 | 48,51 | -6,24 | 49,26 | -23,33 | Unsub | Straight | Sub |
| UTL | ortho_para | T4 | 0 | 60,92 | -29,83 | 71,87 | -6,24 | 44,52 | -23,33 | Unsub | Diagonal | Unsub |
| UTL | ortho_para | T4 | 0 | 59,73 | -42,67 | 36,73 | -16,28 | 44,52 | -23,33 | Sub | Straight | Sub |
| UTL | ortho_para | T4 | 0 | 59,73 | -42,67 | 74,4 | -16,28 | 49,26 | -23,33 | Sub | Diagonal | Unsub |
| UTL | meta_ortho | T4 | 0 | 88,13 | -48,59 | 58,96 | -71,47 | 81,72 | 28,32 | Unsub | Straight | Sub |
| UTL | meta_ortho | T4 | 0 | 88,13 | -48,59 | 48,05 | 0,18 | 72,91 | 28,32 | Unsub | Diagonal | Unsub |
| UTL | meta_ortho | T4 | 0 | 66,6 | -17,63 | 72,49 | -71,47 | 81,72 | 28,32 | Sub | Straight | Sub |
| UTL | meta_ortho | T4 | 0 | 66,6 | -17,63 | 60,89 | 0,18 | 72,91 | 28,32 | Sub | Diagonal | Unsub |
| UTL | meta_meta | T4 | 0 | 42,22 | 6,47 | 16,93 | -38,97 | 44,16 | 30,4 | Sub | Straight | Unsub |
| UTL | meta_meta | T4 | 0 | 42,22 | 6,47 | 14,03 | -44,31 | 56,89 | 30,4 | Sub | Diagonal | Sub |
| UTL | meta_para | T4 | 0 | 31,97 | -29,47 | 46,67 | -83,21 | 67,61 | -18,4 | Sub | Straight | Unsub |
| UTL | meta_para | T4 | 0 | 31,97 | -29,47 | 55,56 | -8,72 | 48,83 | -18,4 | Sub | Straight | Sub |
| UTL | meta_para | T4 | 0 | 63,7 | -37,29 | 106,98 | -83,21 | 67,61 | -18,4 | Unsub | Straight | Sub |
| UTL | meta_para | T4 | 0 | 63,7 | -37,29 | 51,66 | -8,72 | 48,83 | -18,4 | Unsub | Diagonal | Unsub |
| UTL | para_ortho | T4 | 0 | 45,76 | 5,58 | 75,63 | 23,07 | 99,44 | 12,55 | Sub | Straight | Unsub |
| UTL | para_ortho | T4 | 0 | 69,12 | -17,57 | 85,53 | 23,07 | 99,44 | 12,55 | Unsub | Diagonal | Unsub |
| UTL | para_meta | T4 | 0 | 69,49 | 27,42 | 112,78 | 14,98 | 122,15 | 1,27 | Sub | Straight | Unsub |
| UTL | para_para | T4 | 0 | 40,43 | -7,57 | 80,82 | 1,83 | 104,91 | 34,65 | Unsub | Straight | Unsub |
| UTL | para_para | T4 | 0 | 70,84 | 7,53 | 92,23 | 1,83 | 104,91 | 34,65 | Sub | Straight | Unsub |
| FAU | ortho_benzene | T1 | 0 | 94,18 | 6,18 | 105,77 | 36,5 | 102,27 | 18,14 | Unsub | Straight | Sub |
| FAU | ortho_benzene | T1 | 0 | 94,18 | 6,18 | 96,11 | 8,33 | 48,7 | 18,14 | Unsub | Diagonal | Unsub |
| FAU | meta_benzene | T1 | 0 | 49,02 | -25,79 | 18,47 | -32,92 | 54,79 | -10,13 | Unsub | Diagonal | Sub |
| FAU | meta_benzene | T1 | 0 | 49,02 | -25,79 | 39,81 | -43 | 86,05 | -10,13 | Unsub | Straight | Unsub |
| FAU | para_benzene | T1 | 0 | 53,36 | -7,57 | 49,39 | -19,39 | 33,33 | -14,1 | Unsub | Straight | Unsub |
| FAU | ortho_ortho | T1 | 0 | 37,63 | -53,91 | 60,56 | -28,36 | 58,44 | -48,78 | Sub | Diagonal | Sub |
| FAU | ortho_ortho | T1 | 0 | 37,63 | -53,91 | 48,44 | -14,45 | 49,86 | -48,78 | Sub | Straight | Unsub |
| FAU | ortho_ortho | T1 | 0 | 51,02 | -30,63 | 82,02 | -28,36 | 58,44 | -48,78 | Unsub | Straight | Unsub |
| FAU | ortho_ortho | T1 | 0 | 51,02 | -30,63 | 46,41 | -14,45 | 49,86 | -48,78 | Unsub | Diagonal | Sub |
| FAU | ortho_meta | T1 | 0 | 67,6 | 31,13 | 108,31 | 21,85 | 73,67 | 22,57 | Sub | Straight | Sub |
| FAU | ortho_meta | T1 | 0 | 67,6 | 31,13 | 96,94 | 10,18 | 91,24 | 22,57 | Sub | Diagonal | Unsub |
| FAU | ortho_para | T1 | 0 | 88,07 | 6,61 | 78,06 | 9,7 | 59,87 | 5,05 | Unsub | Straight | Unsub |
| FAU | ortho_para | T1 | 0 | 88,07 | 6,61 | 68,77 | 9,7 | 51,88 | 5,05 | Unsub | Diagonal | Sub |
| FAU | ortho_para | T1 | 0 | 30,28 | -4,23 | 66,34 | -8,33 | 51,88 | 5,05 | Sub | Straight | Unsub |
| FAU | ortho_para | T1 | 0 | 30,28 | -4,23 | 102,25 | -8,33 | 59,87 | 5,05 | Sub | Diagonal | Sub |
| FAU | meta_ortho | T1 | 0 | 56,51 | -16,74 | 60,87 | -11,8 | 63,68 | 20,28 | Unsub | Straight | Unsub |
| FAU | meta_ortho | T1 | 0 | 56,51 | -16,74 | 51,14 | -14,43 | 65,08 | 20,28 | Unsub | Diagonal | Sub |
| FAU | meta_ortho | T1 | 0 | 16,31 | -41,91 | 73,03 | -14,43 | 65,08 | 20,28 | Sub | Diagonal | Unsub |
| FAU | meta_ortho | T1 | 0 | 16,31 | -41,91 | 63,49 | -11,8 | 63,68 | 20,28 | Sub | Straight | Sub |
| FAU | meta_meta | T1 | 0 | 75,18 | 30,68 | 59,38 | 6,92 | 62,31 | 51,95 | Sub | Straight | Unsub |
| FAU | meta_meta | T1 | 0 | 75,18 | 30,68 | 60,24 | -12,2 | 69,22 | 51,95 | Sub | Diagonal | Sub |
| FAU | meta_para | T1 | 0 | 45,42 | -10,85 | 64,23 | -22,76 | 59,69 | -26,87 | Sub | Straight | Unsub |
| FAU | meta_para | T1 | 0 | 45,42 | -10,85 | 48,67 | -3,71 | 67,21 | -26,87 | Sub | Diagonal | Sub |
| FAU | meta_para | T1 | 0 | 17,81 | -34,7 | 60,1 | -22,76 | 59,69 | -26,87 | Unsub | Diagonal | Sub |
| FAU | meta_para | T1 | 0 | 17,81 | -34,7 | 51,92 | -3,71 | 57,21 | -26,87 | Unsub | Straight | Unsub |
| FAU | para_ortho | T1 | 0 | 54,46 | -15,91 | 40,41 | 2,18 | 114,77 | 16,66 | Sub | Diagonal | Unsub |
| FAU | para_ortho | T1 | 0 | 33,12 | -14,96 | 93,49 | 2,18 | 114,77 | 16,66 | Unsub | Straight | Unsub |
| FAU | para_meta | T1 | 0 | 74,6 | 47,72 | 102,2 | 8,56 | 79,14 | 37,18 | Sub | Straight | Unsub |
| FAU | para_para | T1 | 0 | 110,37 | 18,89 | 143,34 | 43,89 | 116,33 | 37,76 | Unsub | Straight | Unsub |
| FAU | para_para | T1 | 0 | 38,8 | 17,73 | 158,05 | 43,89 | 116,33 | 37,76 | Sub | Straight | Unsub |

**Table S18.** Zeolite Multstep Path free energy barriers in kJ/mol. Global MEPs starred at **Figure 5** in the main paper are highlighted in green color.

| Framework | Isomers | Al T-Site | TS2 | TS3 | TS4 | Track TS2 | Track TS3 | Track TS4 |
|---|---|---|---|---|---|---|---|---|
| BOG | ortho_benzene | T1 | 128,55 | 168,75 | 90,52 | Unsub | Diagonal | Unsub |
| BOG | ortho_benzene | T1 | 128,55 | 191,33 | 53 | Unsub | Straight | Sub |
| BOG | meta_benzene | T1 | 83,51 | 27,84 | 133,06 | Unsub | Diagonal | Unsub |
| BOG | meta_benzene | T1 | 83,51 | 55,93 | 53,15 | Unsub | Straight | Sub |
| BOG | para_benzene | T1 | 92,86 | 53,09 | 62,67 | Unsub | Straight | Unsub |
| BOG | ortho_ortho | T1 | 83,03 | 100,79 | 63,72 | Sub | Diagonal | Sub |
| BOG | ortho_ortho | T1 | 83,03 | 37,5 | 161,49 | Sub | Straight | Unsub |
| BOG | ortho_ortho | T1 | 25,29 | 68,81 | 63,72 | Unsub | Straight | Sub |
| BOG | ortho_ortho | T1 | 25,29 | 97,69 | 161,49 | Unsub | Diagonal | Unsub |
| BOG | ortho_meta | T1 | 160,85 | 117,14 | 137,81 | Sub | Straight | Unsub |
| BOG | ortho_meta | T1 | 160,85 | 72,43 | 78,17 | Sub | Diagonal | Sub |
| BOG | ortho_para | T1 | 81,56 | 160,49 | 70,82 | Unsub | Straight | Sub |
| BOG | ortho_para | T1 | 81,56 | 84,72 | 56,56 | Unsub | Diagonal | Unsub |
| BOG | ortho_para | T1 | 55 | 105,7 | 111,5 | Sub | Diagonal | Sub |
| BOG | ortho_para | T1 | 55 | 109,15 | 97,23 | Sub | Straight | Unsub |
| BOG | meta_ortho | T1 | 121,97 | 128,19 | 117,24 | Unsub | Diagonal | Sub |
| BOG | meta_ortho | T1 | 121,97 | 130,31 | 56,41 | Unsub | Straight | Unsub |
| BOG | meta_ortho | T1 | 163,32 | 118,42 | 117,24 | Sub | Diagonal | Sub |
| BOG | meta_ortho | T1 | 163,32 | 122,04 | 56,41 | Sub | Straight | Unsub |
| BOG | meta_meta | T1 | 22,63 | 24,22 | 160,71 | Sub | Diagonal | Unsub |
| BOG | meta_meta | T1 | 22,63 | 78,59 | 164,9 | Sub | Straight | Sub |
| BOG | meta_para | T1 | 120,26 | 56,66 | 116,63 | Sub | Straight | Unsub |
| BOG | meta_para | T1 | 120,26 | 127,61 | 89,77 | Sub | Diagonal | Sub |
| BOG | meta_para | T1 | 30,81 | 53,7 | 89,77 | Unsub | Straight | Sub |
| BOG | meta_para | T1 | 30,81 | 46,45 | 116,63 | Unsub | Diagonal | Unsub |
| BOG | para_ortho | T1 | 47,37 | 63,93 | 94,66 | Sub | Diagonal | Unsub |
| BOG | para_ortho | T1 | 73,12 | 106,1 | 94,66 | Unsub | Straight | Unsub |
| BOG | para_meta | T1 | 100,12 | 61,81 | 137,75 | Sub | Straight | Unsub |
| BOG | para_para | T1 | 128,2 | 58,33 | 134,27 | Unsub | Straight | Unsub |
| BOG | para_para | T1 | 94,43 | 66,84 | 134,27 | Sub | Straight | Unsub |
| BOG | ortho_benzene | T2 | 54,78 | 69,03 | 101,69 | Unsub | Straight | Sub |
| BOG | ortho_benzene | T2 | 54,78 | 171,51 | 46,67 | Unsub | Diagonal | Unsub |
| BOG | meta_benzene | T2 | 92,92 | 32,07 | 88,3 | Unsub | Straight | Sub |
| BOG | meta_benzene | T2 | 92,92 | 34,23 | 87,5 | Unsub | Diagonal | Unsub |
| BOG | para_benzene | T2 | 76,58 | 57,13 | 44,58 | Unsub | Straight | Unsub |
| BOG | ortho_ortho | T2 | 81,87 | 143,62 | 109,08 | Sub | Diagonal | Sub |
| BOG | ortho_ortho | T2 | 81,87 | 90,34 | 91,75 | Sub | Straight | Unsub |
| BOG | ortho_ortho | T2 | 87,41 | 183,24 | 91,75 | Unsub | Diagonal | Unsub |
| BOG | ortho_ortho | T2 | 87,41 | 174,84 | 109,08 | Unsub | Straight | Sub |
| BOG | ortho_meta | T2 | 38,63 | 113,12 | 60,96 | Sub | Diagonal | Sub |
| BOG | ortho_meta | T2 | 38,63 | 137,96 | 170,61 | Sub | Straight | Unsub |
| BOG | ortho_para | T2 | 36,93 | 139,95 | 110,77 | Unsub | Diagonal | Unsub |
| BOG | ortho_para | T2 | 36,93 | 161,45 | 91,2 | Unsub | Straight | Sub |
| BOG | ortho_para | T2 | 49,46 | 162,02 | 98,62 | Sub | Straight | Unsub |
| BOG | ortho_para | T2 | 49,46 | 151,83 | 79,05 | Sub | Diagonal | Sub |
| BOG | meta_ortho | T2 | 68,76 | 176,23 | 112,97 | Unsub | Straight | Unsub |
| BOG | meta_ortho | T2 | 68,76 | 80,93 | 108,4 | Unsub | Diagonal | Sub |
| BOG | meta_ortho | T2 | 129,06 | 238 | 112,97 | Sub | Straight | Unsub |
| BOG | meta_ortho | T2 | 129,06 | 146,53 | 108,4 | Sub | Diagonal | Sub |
| BOG | meta_meta | T2 | 112,54 | 104,53 | 137,47 | Sub | Diagonal | Unsub |
| BOG | meta_meta | T2 | 112,54 | 76,86 | 48,86 | Sub | Straight | Sub |
| BOG | meta_para | T2 | 85,54 | 127,8 | 70,22 | Sub | Diagonal | Unsub |
| BOG | meta_para | T2 | 85,54 | 96,97 | 134,68 | Sub | Straight | Sub |
| BOG | meta_para | T2 | 56,07 | 98,11 | 134,68 | Unsub | Diagonal | Sub |
| BOG | meta_para | T2 | 56,07 | 83,72 | 70,22 | Unsub | Straight | Unsub |
| BOG | para_ortho | T2 | 32,68 | 85,81 | 112,2 | Sub | Straight | Unsub |
| BOG | para_ortho | T2 | 100,93 | 68,35 | 112,2 | Unsub | Cross | Unsub |
| BOG | para_meta | T2 | 35,91 | 33,31 | 123,86 | Sub | Straight | Unsub |
| BOG | para_para | T2 | 79,13 | 69,99 | 73,89 | Unsub | Straight | Unsub |
| BOG | para_para | T2 | 63,94 | 81,66 | 73,89 | Sub | Straight | Unsub |
| IWV | ortho_benzene | T3 | 84,15 | 67,73 | 51,77 | Unsub | Diagonal | Unsub |
| IWV | ortho_benzene | T3 | 84,15 | 64,9 | 32,38 | Unsub | Straight | Sub |
| IWV | meta_benzene | T3 | 27,03 | 57,84 | 60,38 | Unsub | Straight | Unsub |
| IWV | meta_benzene | T3 | 27,03 | 45,86 | 61,28 | Unsub | Diagonal | Sub |
| IWV | para_benzene | T3 | 53,68 | 62,48 | 48,94 | Unsub | Straight | Unsub |
| IWV | ortho_ortho | T3 | 63,35 | 86,94 | 54,79 | Sub | Diagonal | Sub |
| IWV | ortho_ortho | T3 | 63,35 | 123,43 | 81,72 | Sub | Straight | Unsub |
| IWV | ortho_ortho | T3 | 101,34 | 117,44 | 54,79 | Unsub | Straight | Sub |
| IWV | ortho_ortho | T3 | 101,34 | 102 | 81,72 | Unsub | Diagonal | Unsub |
| IWV | ortho_meta | T3 | 111,65 | 95,65 | 89,11 | Sub | Diagonal | Unsub |
| IWV | ortho_meta | T3 | 111,65 | 148,02 | 99,49 | Sub | Straight | Sub |
| IWV | ortho_para | T3 | 58,75 | 118,76 | 90,18 | Unsub | Diagonal | Sub |
| IWV | ortho_para | T3 | 58,75 | 111,41 | 95,81 | Unsub | Straight | Unsub |
| IWV | ortho_para | T3 | 80,78 | 140,64 | 109,81 | Sub | Straight | Sub |
| IWV | ortho_para | T3 | 80,78 | 108,5 | 115,44 | Sub | Diagonal | Unsub |
| IWV | meta_ortho | T3 | 94,49 | 75,75 | 35,2 | Unsub | Diagonal | Sub |
| IWV | meta_ortho | T3 | 94,49 | 79,42 | 138,55 | Unsub | Straight | Unsub |
| IWV | meta_ortho | T3 | 71,59 | 93,42 | 35,2 | Sub | Diagonal | Sub |
| IWV | meta_ortho | T3 | 71,59 | 99,57 | 138,55 | Sub | Straight | Unsub |
| IWV | meta_meta | T3 | 64,96 | 86,11 | 115,87 | Sub | Diagonal | Unsub |
| IWV | meta_meta | T3 | 64,96 | 84,01 | 43,54 | Sub | Straight | Sub |
| IWV | meta_para | T3 | 98,77 | 99,51 | 74,48 | Sub | Straight | Unsub |
| IWV | meta_para | T3 | 98,77 | 89,36 | 89,7 | Sub | Diagonal | Sub |
| IWV | meta_para | T3 | 71,49 | 104,38 | 89,7 | Unsub | Straight | Sub |
| IWV | meta_para | T3 | 71,49 | 74,13 | 74,48 | Unsub | Diagonal | Unsub |
| IWV | para_ortho | T3 | 63,94 | 82,78 | 30,68 | Sub | Straight | Unsub |
| IWV | para_ortho | T3 | 52,52 | 88,38 | 30,68 | Unsub | Diagonal | Unsub |
| IWV | para_meta | T3 | 63,4 | 50,17 | 94,6 | Sub | Straight | Unsub |
| IWV | para_para | T3 | 48,88 | 134,76 | 77,78 | Unsub | Straight | Unsub |
| IWV | para_para | T3 | 115,48 | 80,14 | 77,78 | Sub | Straight | Unsub |
| IWV | ortho_benzene | T7 | 63,35 | 132,5 | 36,34 | Unsub | Straight | Sub |
| IWV | ortho_benzene | T7 | 63,35 | 54,77 | 72,04 | Unsub | Diagonal | Unsub |
| IWV | meta_benzene | T7 | 48,04 | 59,62 | 59,32 | Unsub | Straight | Sub |
| IWV | meta_benzene | T7 | 48,04 | 49,24 | 87,15 | Unsub | Diagonal | Unsub |
| IWV | para_benzene | T7 | 78,04 | 64,59 | 45,31 | Unsub | Straight | Unsub |
| IWV | ortho_ortho | T7 | 77,46 | 95,88 | 76,62 | Sub | Diagonal | Sub |
| IWV | ortho_ortho | T7 | 77,46 | 199,41 | 177,15 | Sub | Straight | Unsub |
| IWV | ortho_ortho | T7 | 65,23 | 144,17 | 76,62 | Unsub | Diagonal | Unsub |
| IWV | ortho_ortho | T7 | 65,23 | 118,71 | 177,15 | Unsub | Straight | Sub |
| IWV | ortho_meta | T7 | 71,11 | 139,53 | 124,39 | Sub | Straight | Sub |
| IWV | ortho_meta | T7 | 71,11 | 66,93 | 88,4 | Sub | Diagonal | Unsub |
| IWV | ortho_para | T7 | 96,86 | 59,97 | 72,92 | Unsub | Straight | Unsub |
| IWV | ortho_para | T7 | 96,86 | 67,87 | 101,86 | Unsub | Diagonal | Sub |
| IWV | ortho_para | T7 | 48,72 | 116,13 | 117,95 | Sub | Straight | Unsub |
| IWV | ortho_para | T7 | 48,72 | 69,17 | 89,01 | Sub | Diagonal | Sub |
| IWV | meta_ortho | T7 | 59,7 | 50,24 | 99,69 | Unsub | Diagonal | Unsub |
| IWV | meta_ortho | T7 | 59,7 | 46,75 | 168,43 | Unsub | Straight | Sub |
| IWV | meta_ortho | T7 | 87,36 | 70,3 | 99,69 | Sub | Diagonal | Unsub |
| IWV | meta_ortho | T7 | 87,36 | 69,25 | 168,43 | Sub | Straight | Sub |
| IWV | meta_meta | T7 | 64,01 | 89,17 | 58,23 | Sub | Straight | Unsub |
| IWV | meta_meta | T7 | 64,01 | 76,98 | 101,49 | Sub | Diagonal | Sub |
| IWV | meta_para | T7 | 101,05 | 122,32 | 61,48 | Sub | Straight | Unsub |
| IWV | meta_para | T7 | 101,05 | 82,57 | 110,46 | Sub | Diagonal | Sub |
| IWV | meta_para | T7 | 95,65 | 140,02 | 110,46 | Unsub | Straight | Sub |
| IWV | meta_para | T7 | 95,65 | 94,69 | 61,48 | Unsub | Diagonal | Unsub |
| IWV | para_ortho | T7 | 47,38 | 94,68 | 73,56 | Sub | Diagonal | Unsub |
| IWV | para_ortho | T7 | 30,53 | 132,83 | 73,56 | Unsub | Straight | Unsub |
| IWV | para_meta | T7 | 90,52 | 137,82 | 88,55 | Sub | Straight | Unsub |
| IWV | para_para | T7 | 44,53 | 84,49 | 83,72 | Unsub | Straight | Unsub |
| IWV | para_para | T7 | 66,49 | 110,93 | 83,72 | Sub | Straight | Unsub |
| UTL | ortho_benzene | T2 | 32,19 | 49,07 | 85,21 | Unsub | Diagonal | Unsub |
| UTL | ortho_benzene | T2 | 32,19 | 87,13 | 49,77 | Unsub | Straight | Sub |
| UTL | meta_benzene | T2 | 23,49 | 86,76 | 61,67 | Unsub | Diagonal | Unsub |
| UTL | meta_benzene | T2 | 23,49 | 58,08 | 65,99 | Unsub | Straight | Sub |
| UTL | para_benzene | T2 | 62,79 | 77,82 | 42,48 | Unsub | Straight | Unsub |
| UTL | ortho_ortho | T2 | 31,48 | 116,1 | 83.03 | Sub | Diagonal | Sub |
| UTL | ortho_ortho | T2 | 31,48 | 115,1 | 81,98 | Sub | Straight | Unsub |
| UTL | ortho_ortho | T2 | 35,51 | 133,27 | 81,98 | Unsub | Diagonal | Sub |
| UTL | ortho_ortho | T2 | 35,51 | 86,73 | 83.03 | Unsub | Straight | Unsub |
| UTL | ortho_meta | T2 | 71,58 | 100,48 | 56,14 | Sub | Straight | Unsub |
| UTL | ortho_meta | T2 | 71,58 | 37,51 | 84,17 | Sub | Diagonal | Sub |
| UTL | ortho_para | T2 | 77,72 | 113,58 | 50,33 | Unsub | Diagonal | Sub |
| UTL | ortho_para | T2 | 77,72 | 62,43 | 106,06 | Unsub | Straight | Unsub |
| UTL | ortho_para | T2 | 90,42 | 124,89 | 90,12 | Sub | Straight | Sub |
| UTL | ortho_para | T2 | 90,42 | 137,19 | 145,85 | Sub | Diagonal | Unsub |
| UTL | meta_ortho | T2 | 67,8 | 71,91 | 160,8 | Unsub | Diagonal | Sub |
| UTL | meta_ortho | T2 | 67,8 | 87,91 | 148,63 | Unsub | Straight | Unsub |
| UTL | meta_ortho | T2 | 33,94 | 65,34 | 160,8 | Sub | Diagonal | Sub |
| UTL | meta_ortho | T2 | 33,94 | 76,21 | 148,63 | Sub | Straight | Unsub |
| UTL | meta_meta | T2 | 64,31 | 33,89 | 97,36 | Sub | Straight | Unsub |
| UTL | meta_meta | T2 | 64,31 | 31,83 | 88,42 | Sub | Diagonal | Sub |
| UTL | meta_para | T2 | 80,79 | 62,46 | 145,12 | Sub | Diagonal | Unsub |
| UTL | meta_para | T2 | 80,79 | 71,83 | 71,63 | Sub | Straight | Sub |
| UTL | meta_para | T2 | 67,53 | 54,67 | 145,12 | Unsub | Straight | Sub |
| UTL | meta_para | T2 | 67,53 | 84,39 | 71,63 | Unsub | Diagonal | Unsub |
| UTL | para_ortho | T2 | 48,91 | 78,31 | 54,21 | Sub | Straight | Unsub |
| UTL | para_ortho | T2 | 88,53 | 77,81 | 54,21 | Unsub | Diagonal | Unsub |
| UTL | para_meta | T2 | 80,05 | 111,44 | 89,98 | Sub | Straight | Unsub |
| UTL | para_para | T2 | 89,03 | 52,41 | 51,03 | Unsub | Straight | Unsub |
| UTL | para_para | T2 | 70,07 | 76,22 | 51,03 | Sub | Straight | Unsub |
| UTL | ortho_benzene | T4 | 109,61 | 67,22 | 107,3 | Unsub | Diagonal | Unsub |
| UTL | ortho_benzene | T4 | 109,61 | 66,36 | 50,92 | Unsub | Straight | Sub |
| UTL | meta_benzene | T4 | 44,66 | 38,43 | 123,69 | Unsub | Diagonal | Unsub |
| UTL | meta_benzene | T4 | 44,66 | 58,4 | 57,5 | Unsub | Straight | Sub |
| UTL | para_benzene | T4 | 28,71 | 45,65 | 114,34 | Unsub | Straight | Unsub |
| UTL | ortho_ortho | T4 | 101,99 | 137,66 | 105,73 | Sub | Straight | Sub |
| UTL | ortho_ortho | T4 | 101,99 | 109,19 | 128,12 | Sub | Diagonal | Unsub |
| UTL | ortho_ortho | T4 | 71,36 | 115,24 | 128,12 | Unsub | Straight | Sub |
| UTL | ortho_ortho | T4 | 71,36 | 99,19 | 105,73 | Unsub | Cross | Unsub |
| UTL | ortho_meta | T4 | 40,82 | 117,17 | 87,66 | Sub | Straight | Unsub |
| UTL | ortho_meta | T4 | 40,82 | 72,86 | 61,93 | Sub | Diagonal | Sub |
| UTL | ortho_para | T4 | 60,92 | 78,34 | 55,5 | Unsub | Straight | Sub |
| UTL | ortho_para | T4 | 60,92 | 101,71 | 50,76 | Unsub | Diagonal | Unsub |
| UTL | ortho_para | T4 | 59,73 | 79,4 | 60,8 | Sub | Straight | Sub |
| UTL | ortho_para | T4 | 59,73 | 117,07 | 65,55 | Sub | Diagonal | Unsub |
| UTL | meta_ortho | T4 | 88,13 | 107,55 | 153,19 | Unsub | Straight | Sub |
| UTL | meta_ortho | T4 | 88,13 | 96,63 | 72,73 | Unsub | Diagonal | Unsub |
| UTL | meta_ortho | T4 | 66,6 | 90,13 | 153,19 | Sub | Straight | Sub |
| UTL | meta_ortho | T4 | 66,6 | 78,53 | 72,73 | Sub | Diagonal | Unsub |
| UTL | meta_meta | T4 | 42,22 | 10,46 | 83,13 | Sub | Straight | Unsub |
| UTL | meta_meta | T4 | 42,22 | 7,56 | 101,2 | Sub | Diagonal | Sub |
| UTL | meta_para | T4 | 31,97 | 76,13 | 150,82 | Sub | Straight | Unsub |
| UTL | meta_para | T4 | 31,97 | 85,03 | 57,55 | Sub | Straight | Sub |
| UTL | meta_para | T4 | 63,7 | 144,26 | 150,82 | Unsub | Straight | Sub |
| UTL | meta_para | T4 | 63,7 | 88,95 | 57,55 | Unsub | Diagonal | Unsub |
| UTL | para_ortho | T4 | 45,76 | 70,05 | 76,37 | Sub | Straight | Unsub |
| UTL | para_ortho | T4 | 69,12 | 103,09 | 76,37 | Unsub | Diagonal | Unsub |
| UTL | para_meta | T4 | 69,49 | 85,36 | 107,17 | Sub | Straight | Unsub |
| UTL | para_para | T4 | 40,43 | 88,39 | 103,08 | Unsub | Straight | Unsub |
| UTL | para_para | T4 | 70,84 | 84,7 | 103,08 | Sub | Straight | Unsub |
| FAU | ortho_benzene | T1 | 94,18 | 99,59 | 65,78 | Unsub | Straight | Sub |
| FAU | ortho_benzene | T1 | 94,18 | 89,93 | 40,37 | Unsub | Diagonal | Unsub |
| FAU | meta_benzene | T1 | 49,02 | 44,26 | 87,72 | Unsub | Diagonal | Sub |
| FAU | meta_benzene | T1 | 49,02 | 65,6 | 129,05 | Unsub | Straight | Unsub |
| FAU | para_benzene | T1 | 53,36 | 56,97 | 52,72 | Unsub | Straight | Unsub |
| FAU | ortho_ortho | T1 | 37,63 | 114,48 | 86,8 | Sub | Diagonal | Sub |
| FAU | ortho_ortho | T1 | 37,63 | 102,35 | 64,31 | Sub | Straight | Unsub |
| FAU | ortho_ortho | T1 | 51,02 | 112,65 | 86,8 | Unsub | Straight | Unsub |
| FAU | ortho_ortho | T1 | 51,02 | 77,05 | 64,31 | Unsub | Diagonal | Sub |
| FAU | ortho_meta | T1 | 67,6 | 77,18 | 51,83 | Sub | Straight | Sub |
| FAU | ortho_meta | T1 | 67,6 | 65,81 | 81,06 | Sub | Diagonal | Unsub |
| FAU | ortho_para | T1 | 88,07 | 71,45 | 50,17 | Unsub | Straight | Unsub |
| FAU | ortho_para | T1 | 88,07 | 62,17 | 42,18 | Unsub | Diagonal | Sub |
| FAU | ortho_para | T1 | 30,28 | 70,57 | 60,22 | Sub | Straight | Unsub |
| FAU | ortho_para | T1 | 30,28 | 106,48 | 68,2 | Sub | Diagonal | Sub |
| FAU | meta_ortho | T1 | 56,51 | 77,61 | 75,48 | Unsub | Straight | Unsub |
| FAU | meta_ortho | T1 | 56,51 | 67,87 | 79,51 | Unsub | Diagonal | Sub |
| FAU | meta_ortho | T1 | 16,31 | 114,94 | 79,51 | Sub | Diagonal | Unsub |
| FAU | meta_ortho | T1 | 16,31 | 105,39 | 75,48 | Sub | Straight | Sub |
| FAU | meta_meta | T1 | 75,18 | 28,69 | 55,39 | Sub | Straight | Unsub |
| FAU | meta_meta | T1 | 75,18 | 29,56 | 81,42 | Sub | Diagonal | Sub |
| FAU | meta_para | T1 | 45,42 | 75,09 | 82,45 | Sub | Straight | Unsub |
| FAU | meta_para | T1 | 45,42 | 59,52 | 70,92 | Sub | Diagonal | Sub |
| FAU | meta_para | T1 | 17,81 | 94,8 | 82,45 | Unsub | Diagonal | Sub |
| FAU | meta_para | T1 | 17,81 | 86,61 | 60,92 | Unsub | Straight | Unsub |
| FAU | para_ortho | T1 | 54,46 | 56,32 | 112,59 | Sub | Diagonal | Unsub |
| FAU | para_ortho | T1 | 33,12 | 108,45 | 112,59 | Unsub | Straight | Unsub |
| FAU | para_meta | T1 | 74,6 | 54,48 | 70,58 | Sub | Straight | Unsub |
| FAU | para_para | T1 | 110,37 | 124,45 | 72,44 | Unsub | Straight | Unsub |
| FAU | para_para | T1 | 38,8 | 140,32 | 72,44 | Sub | Straight | Unsub |

**Table S19.** Zeolite Direct Path relative free energies in kJ/mol. Free energy barriers reported on the main manuscript correspond to the TS1 energy value. Global MEPs starred at **Figure 5** in the main paper are highlighted in green color.

| Framework | Isomer | T-Site | I1 | TS1 | I4 |
|---|---|---|---|---|---|
| BOG | ortho_benzene | T1 | 0 | 89,03 | -11,06 |
| BOG | meta_benzene | T1 | 0 | 100,8 | 6,15 |
| BOG | para_benzene | T1 | 0 | 141,63 | 42,8 |
| BOG | ortho_ortho | T1 | 0 | 88,7 | 65,37 |
| BOG | ortho_meta | T1 | 0 | 80,79 | -49,22 |
| BOG | ortho_para | T1 | 0 | 120,75 | 30,58 |
| BOG | meta_ortho | T1 | 0 | 110,19 | 36,77 |
| BOG | meta_meta | T1 | 0 | 155,78 | -9,49 |
| BOG | meta_para | T1 | 0 | 100,49 | 30 |
| BOG | para_ortho | T1 | 0 | 89,18 | 11,99 |
| BOG | para_meta | T1 | 0 | 263,99 | 69,34 |
| BOG | para_para | T1 | 0 | 133,2 | 48,43 |
| BOG | ortho_benzene | T2 | 0 | 85,1 | 1,14 |
| BOG | meta_benzene | T2 | 0 | 103,24 | 9,8 |
| BOG | para_benzene | T2 | 0 | 86,46 | 8,21 |
| BOG | ortho_ortho | T2 | 0 | 90,49 | 27,21 |
| BOG | ortho_meta | T2 | 0 | 81,14 | -14,53 |
| BOG | ortho_para | T2 | 0 | 103,4 | 20,7 |
| BOG | meta_ortho | T2 | 0 | 115,4 | 25,72 |
| BOG | meta_meta | T2 | 0 | 189,22 | 43,02 |
| BOG | meta_para | T2 | 0 | 157 | 44,63 |
| BOG | para_ortho | T2 | 0 | 146,44 | 72,35 |
| BOG | para_meta | T2 | 0 | 109,05 | 10,71 |
| BOG | para_para | T2 | 0 | 86,08 | 48,67 |
| IWV | ortho_benzene | T3 | 0 | 105,24 | 16,09 |
| IWV | meta_benzene | T3 | 0 | 75,65 | 15,99 |
| IWV | para_benzene | T3 | 0 | 56,56 | -31,62 |
| IWV | ortho_ortho | T3 | 0 | 129,38 | -12,13 |
| IWV | ortho_meta | T3 | 0 | 162,03 | 61,75 |
| IWV | ortho_para | T3 | 0 | 118,06 | 9,98 |
| IWV | meta_ortho | T3 | 0 | 88,1 | -13,14 |
| IWV | meta_meta | T3 | 0 | 170,41 | 31,89 |
| IWV | meta_para | T3 | 0 | 202,77 | -2,93 |
| IWV | para_ortho | T3 | 0 | 126,69 | 5,36 |
| IWV | para_meta | T3 | 0 | 107,25 | 13 |
| IWV | para_para | T3 | 0 | 111,07 | 45,89 |
| IWV | ortho_benzene | T7 | 0 | 74,59 | 6,22 |
| IWV | meta_benzene | T7 | 0 | 71,97 | -17,03 |
| IWV | para_benzene | T7 | 0 | 130,03 | 10,19 |
| IWV | ortho_ortho | T7 | 0 | 105 | 41,28 |
| IWV | ortho_meta | T7 | 0 | 133,34 | -6,09 |
| IWV | ortho_para | T7 | 0 | 98,25 | 17,26 |
| IWV | meta_ortho | T7 | 0 | 101,71 | -57,92 |
| IWV | meta_meta | T7 | 0 | 121,91 | 12,94 |
| IWV | meta_para | T7 | 0 | 101,1 | 15,54 |
| IWV | para_ortho | T7 | 0 | 85,29 | -3,09 |
| IWV | para_meta | T7 | 0 | 156,46 | 30,82 |
| IWV | para_para | T7 | 0 | 190,67 | 28,1 |
| UTL | ortho_benzene | T2 | 0 | 72,93 | -46,11 |
| UTL | meta_benzene | T2 | 0 | 74,09 | -47,89 |
| UTL | para_benzene | T2 | 0 | 100,35 | 19,81 |
| UTL | ortho_ortho | T2 | 0 | 119,9 | -6,31 |
| UTL | ortho_meta | T2 | 0 | 141,22 | 41,9 |
| UTL | ortho_para | T2 | 0 | 81,21 | -2,6 |
| UTL | meta_ortho | T2 | 0 | 78,93 | -3,18 |
| UTL | meta_meta | T2 | 0 | 161,03 | 26,13 |
| UTL | meta_para | T2 | 0 | 100,11 | -9,39 |
| UTL | para_ortho | T2 | 0 | 88,79 | -13,27 |
| UTL | para_meta | T2 | 0 | 136,33 | 10,74 |
| UTL | para_para | T2 | 0 | 105,11 | 66,08 |
| UTL | ortho_benzene | T4 | 0 | 80,76 | -21,43 |
| UTL | meta_benzene | T4 | 0 | 71,02 | -19,04 |
| UTL | para_benzene | T4 | 0 | 55,73 | -19,03 |
| UTL | ortho_ortho | T4 | 0 | 72,98 | 44,96 |
| UTL | ortho_meta | T4 | 0 | 136,45 | 27,8 |
| UTL | ortho_para | T4 | 0 | 86,15 | -23,33 |
| UTL | meta_ortho | T4 | 0 | 74,76 | 28,32 |
| UTL | meta_meta | T4 | 0 | 94,22 | 30,4 |
| UTL | meta_para | T4 | 0 | 91,32 | -18,4 |
| UTL | para_ortho | T4 | 0 | 85,65 | 12,55 |
| UTL | para_meta | T4 | 0 | 80,66 | 1,27 |
| UTL | para_para | T4 | 0 | 182,19 | 34,65 |
| FAU | ortho_benzene | T1 | 0 | 131,09 | 18,14 |
| FAU | meta_benzene | T1 | 0 | 74,75 | -10,13 |
| FAU | para_benzene | T1 | 0 | 83,17 | -14,1 |
| FAU | ortho_ortho | T1 | 0 | 90,1 | -48,78 |
| FAU | ortho_meta | T1 | 0 | 102,86 | 22,57 |
| FAU | ortho_para | T1 | 0 | 85,36 | 5,05 |
| FAU | meta_ortho | T1 | 0 | 114,61 | 20,28 |
| FAU | meta_meta | T1 | 0 | 107,97 | 51,95 |
| FAU | meta_para | T1 | 0 | 73,17 | -26,87 |
| FAU | para_ortho | T1 | 0 | 106,5 | 16,66 |
| FAU | para_meta | T1 | 0 | 111,94 | 37,18 |
| FAU | para_para | T1 | 0 | 157,62 | 37,76 |

**Table S20.** Free pore area (Å$^2$) available in each zeolite channel after accommodating each TS structure. The free pore area is calculated as: Channel Area = π x ((Max D Sphere IZA)/2)$^2$, TS Area = π x (Middle axis/2) x (Minimum Axis)/2 , Free Pore Area = Channel Area – TS Area. A visual representation of these calculations is provided in **Figure S40**, and the corresponding data are illustrated in **Figure S41**. Max D Sphere values used according to IZA database website, IWV = 8.05 Å, IWV = 8.54 Å, UTL = 9.3 Å, FAU = 11.24 Å.

| Isomer | TS | BOG-T1 | BOG-T2 | IWV-T3 | IWV-T7 | UTL-T2 | UTL-T4 | FAU-T1 |
|---|---|---|---|---|---|---|---|---|
| ortho_benzene | TS1 | 32,3 | 32 | 30,7 | 35,2 | 36,5 | 34 | 33,1 |
| ortho_benzene | TS2 | 32,7 | 35,6 | 31,9 | 29,1 | 29,2 | 33,6 | 40,1 |
| ortho_benzene | TS3 | 31,6 | 38,4 | 39,6 | 23,4 | 24 | 27,6 | 22,6 |
| ortho_benzene | TS3 | 23 | 29,5 | 29,5 | 29,7 | 33,2 | 28,9 | 36,3 |
| ortho_benzene | TS4 | 32,2 | 35,1 | 36,2 | 33,6 | 34,3 | 34,5 | 28,9 |
| ortho_benzene | TS4 | 32,1 | 28,1 | 28,8 | 30 | 30,7 | 26,8 | 26,7 |
| meta_benzene | TS1 | 32,9 | 37,9 | 32 | 35,9 | 31,7 | 33,3 | 43,1 |
| meta_benzene | TS2 | 37,8 | 31,1 | 26,9 | 33 | 29,5 | 23,6 | 21,7 |
| meta_benzene | TS3 | 27,6 | 25,4 | 34,9 | 33,4 | 23,6 | 34,4 | 25,4 |
| meta_benzene | TS3 | 34,1 | 35,2 | 33,2 | 30,5 | 25,4 | 40,1 | 41,5 |
| meta_benzene | TS4 | 45,2 | 34 | 28,7 | 26,4 | 31,7 | 33 | 30,8 |
| meta_benzene | TS4 | 35,2 | 40 | 41,5 | 28,9 | 28 | 36,2 | 34,3 |
| para_benzene | TS1 | 37,7 | 34,3 | 33,3 | 25,3 | 30 | 23,3 | 34 |
| para_benzene | TS2 | 42,6 | 42,7 | 27,6 | 33,9 | 23,6 | 30 | 18,8 |
| para_benzene | TS3 | 37,6 | 35,6 | 38 | 24,5 | 33 | 25 | 29 |
| para_benzene | TS4 | 43,1 | 46,3 | 32,6 | 28 | 24,5 | 28,6 | 24,6 |
| ortho_ortho | TS1 | 37,2 | 42,5 | 42,1 | 33,2 | 35,5 | 36,5 | 46,6 |
| ortho_ortho | TS2 | 46,1 | 51,4 | 49,7 | 34,6 | 46,1 | 53,5 | 42 |
| ortho_ortho | TS2 | 39,1 | 45,5 | 43,2 | 39,7 | 41,3 | 38,8 | 39,2 |
| ortho_ortho | TS3 | 45,1 | 41,4 | 36,3 | 28,5 | 37,6 | 38,8 | 34,6 |
| ortho_ortho | TS3 | 35,8 | 41,3 | 50,4 | 32,1 | 36,1 | 47,2 | 40,1 |
| ortho_ortho | TS3 | 41,9 | 37,6 | 36,8 | 37,8 | 37,1 | 49,6 | 37,9 |
| ortho_ortho | TS3 | 54,3 | 45,3 | 33,7 | 31,4 | 36,7 | 42,5 | 29,6 |
| ortho_ortho | TS4 | 41,6 | 52,1 | 42,5 | 40,6 | 39,1 | 38,7 | 43,3 |
| ortho_ortho | TS4 | 37,5 | 40,2 | 41,1 | 33,3 | 36,8 | 35,3 | 39,5 |
| ortho_meta | TS1 | 38,2 | 49,1 | 40 | 34,2 | 31,1 | 38,8 | 38,5 |
| ortho_meta | TS2 | 40,4 | 35,8 | 37 | 32,6 | 36,3 | 35,3 | 33,4 |
| ortho_meta | TS3 | 43 | 45,7 | 37,8 | 24,9 | 43,2 | 42,1 | 44,7 |
| ortho_meta | TS3 | 45,5 | 40,6 | 40,5 | 40,1 | 46,1 | 37,4 | 42,9 |
| ortho_meta | TS4 | 40,7 | 44,5 | 34,8 | 32,1 | 42,7 | 41,3 | 33,7 |
| ortho_meta | TS4 | 34,7 | 39,3 | 44,2 | 30,2 | 38,2 | 43,6 | 33,7 |
| ortho_para | TS1 | 45,4 | 46,8 | 48,7 | 42,8 | 41 | 47 | 43,6 |
| ortho_para | TS2 | 46 | 42,7 | 40,1 | 33,6 | 33,3 | 34,4 | 41,5 |
| ortho_para | TS2 | 40,6 | 41,8 | 35,2 | 39,5 | 39,5 | 44,2 | 33,5 |
| ortho_para | TS3 | 46 | 42,9 | 37,3 | 37,1 | 39,9 | 36,8 | 32,1 |
| ortho_para | TS3 | 41 | 49,8 | 38,4 | 36,7 | 42 | 35,8 | 35,7 |
| ortho_para | TS3 | 46,2 | 45,4 | 40,8 | 29,3 | 46,4 | 50,6 | 44,8 |
| ortho_para | TS3 | 42,9 | 37,4 | 44,8 | 40,4 | 37,6 | 35,2 | 36,9 |
| ortho_para | TS4 | 35,3 | 48 | 62,6 | 33,1 | 36,6 | 40,2 | 40,5 |
| ortho_para | TS4 | 37,3 | 50 | 43,2 | 41,3 | 45,5 | 39,3 | 40,1 |
| meta_ortho | TS1 | 38,6 | 45,7 | 57,5 | 41,8 | 40,2 | 37,5 | 44,4 |
| meta_ortho | TS2 | 41,8 | 37,8 | 42,6 | 40,2 | 44,8 | 43,1 | 47,9 |
| meta_ortho | TS2 | 53,8 | 43,5 | 35,6 | 37,8 | 36,6 | 43,8 | 43,7 |
| meta_ortho | TS3 | 42,5 | 51,8 | 34,9 | 36,1 | 32,1 | 34,2 | 48,6 |
| meta_ortho | TS3 | 47 | 48,6 | 42,6 | 25,8 | 42,3 | 35,3 | 42,8 |
| meta_ortho | TS3 | 35,8 | 52,2 | 56,3 | 35 | 26,9 | 35,3 | 28 |
| meta_ortho | TS3 | 36,7 | 36,7 | 47,5 | 32,1 | 37,9 | 44 | 43,1 |
| meta_ortho | TS4 | 47,3 | 48,7 | 55,9 | 51,7 | 34,7 | 32,2 | 44 |
| meta_ortho | TS4 | 41,8 | 44,2 | 42,1 | 34,9 | 34,7 | 36,7 | 42,5 |
| meta_meta | TS1 | 42,8 | 42,9 | 33,3 | 42,8 | 41,4 | 42,7 | 43,6 |
| meta_meta | TS2 | 42,5 | 34,3 | 32,9 | 35,6 | 36,2 | 34,6 | 40,7 |
| meta_meta | TS3 | 47,9 | 35,6 | 36,3 | 31,6 | 43,5 | 34,6 | 51,8 |
| meta_meta | TS3 | 44,1 | 49,6 | 32 | 34,8 | 38,1 | 38,3 | 37,8 |
| meta_meta | TS4 | 46,3 | 49,4 | 31,1 | 39,5 | 38,2 | 39,9 | 47,6 |
| meta_meta | TS4 | 36,1 | 41,1 | 34,7 | 34 | 38,1 | 46 | 39,3 |
| meta_para | TS1 | 46,4 | 40,6 | 53,2 | 49,1 | 36,2 | 62,5 | 43,9 |
| meta_para | TS2 | 40,9 | 43,8 | 47,6 | 37,7 | 45,5 | 49,8 | 45 |
| meta_para | TS2 | 39 | 41 | 47,8 | 32,6 | 46,4 | 43,5 | 44,4 |
| meta_para | TS3 | 39,9 | 39,1 | 45,5 | 38,9 | 45,8 | 41,6 | 41,1 |
| meta_para | TS3 | 38,8 | 46 | 45,2 | 37,1 | 39,8 | 43,6 | 44,7 |
| meta_para | TS3 | 47,7 | 41,2 | 42,7 | 48,6 | 43,8 | 35,3 | 44,9 |
| meta_para | TS3 | 46,6 | 48,1 | 50,9 | 43,3 | 39,2 | 50,2 | 44,2 |
| meta_para | TS4 | 44,5 | 54 | 32,8 | 27,1 | 36,5 | 40 | 39,2 |
| meta_para | TS4 | 46,1 | 46,6 | 47,6 | 40,2 | 36,4 | 34,7 | 38 |
| para_ortho | TS1 | 46,2 | 38,9 | 56,3 | 28,5 | 43,6 | 26,4 | 35,8 |
| para_ortho | TS2 | 49,7 | 38,6 | 45,8 | 35,9 | 44,4 | 49,3 | 52,6 |
| para_ortho | TS2 | 47,3 | 49,2 | 31,7 | 35,6 | 39,1 | 42,8 | 41,3 |
| para_ortho | TS3 | 46,5 | 37,2 | 40,9 | 43,7 | 42,6 | 31,7 | 40,6 |
| para_ortho | TS3 | 48,4 | 51,8 | 48,5 | 27,6 | 45,6 | 39,8 | 45,2 |
| para_ortho | TS4 | 45,4 | 55,5 | 50,3 | 47 | 63,2 | 42,1 | 40,7 |
| para_meta | TS1 | 38,1 | 42,6 | 42,3 | 39 | 33,6 | 57,1 | 39,8 |
| para_meta | TS2 | 42,4 | 41,2 | 47 | 31 | 39 | 56,2 | 36,2 |
| para_meta | TS3 | 40,6 | 46,9 | 39 | 33 | 46,6 | 46,6 | 44,7 |
| para_meta | TS4 | 41,5 | 43,8 | 28,7 | 42,6 | 44,3 | 41,3 | 34,5 |
| para_para | TS1 | 47,7 | 41,1 | 40,9 | 37,7 | 36,4 | 32,4 | 34,2 |
| para_para | TS2 | 47,4 | 52,3 | 53,8 | 39,2 | 56,5 | 32,5 | 38,4 |
| para_para | TS2 | 49,2 | 41,4 | 31,4 | 39,1 | 48,7 | 49,7 | 50 |
| para_para | TS3 | 48,4 | 42,1 | 40,2 | 41 | 47,2 | 48,1 | 40,6 |
| para_para | TS3 | 51 | 40,3 | 37,8 | 46 | 47,4 | 38,9 | 33 |
| para_para | TS4 | 39,9 | 39,1 | 36,7 | 40 | 30,5 | 40,7 | 44,9 |

**Table S21.** Framework, isomer combination, T-site, TS label, mean minium distance and mean maximum distance in Amstrongs (Å).

| Framework | Isomer | T-Site | TS | mean_min | mean_max |
|---|---|---|---|---|---|
| BOG | ortho_benzene | T1 | TS1 | 1.5 | 3.5 |
| BOG | ortho_benzene | T1 | TS2 | 1.6 | 3.4 |
| BOG | ortho_benzene | T1 | TS4 | 1.8 | 4.0 |
| BOG | ortho_benzene | T1 | TS3 | 1.4 | 3.5 |
| BOG | ortho_benzene | T1 | TS3 | 1.4 | 3.7 |
| BOG | ortho_benzene | T1 | TS4 | 1.5 | 3.6 |
| BOG | meta_benzene | T1 | TS3 | 1.5 | 3.7 |
| BOG | meta_benzene | T1 | TS1 | 1.5 | 3.6 |
| BOG | meta_benzene | T1 | TS4 | 1.4 | 3.6 |
| BOG | meta_benzene | T1 | TS3 | 1.4 | 3.2 |
| BOG | meta_benzene | T1 | TS2 | 1.5 | 3.7 |
| BOG | meta_benzene | T1 | TS4 | 1.7 | 3.6 |
| BOG | para_benzene | T1 | TS1 | 1.2 | 3.4 |
| BOG | para_benzene | T1 | TS4 | 1.5 | 3.5 |
| BOG | para_benzene | T1 | TS2 | 1.6 | 4.1 |
| BOG | para_benzene | T1 | TS3 | 1.3 | 3.5 |
| BOG | ortho_ortho | T1 | TS2 | 1.5 | 3.6 |
| BOG | ortho_ortho | T1 | TS2 | 1.3 | 3.1 |
| BOG | ortho_ortho | T1 | TS1 | 1.4 | 3.5 |
| BOG | ortho_ortho | T1 | TS3 | 1.3 | 3.0 |
| BOG | ortho_ortho | T1 | TS4 | 1.5 | 3.4 |
| BOG | ortho_ortho | T1 | TS4 | 1.5 | 3.6 |
| BOG | ortho_ortho | T1 | TS3 | 1.4 | 3.4 |
| BOG | ortho_ortho | T1 | TS3 | 1.5 | 3.5 |
| BOG | ortho_ortho | T1 | TS3 | 1.4 | 3.7 |
| BOG | ortho_meta | T1 | TS1 | 1.5 | 3.6 |
| BOG | ortho_meta | T1 | TS4 | 1.6 | 3.6 |
| BOG | ortho_meta | T1 | TS2 | 1.3 | 3.3 |
| BOG | ortho_meta | T1 | TS3 | 1.4 | 3.5 |
| BOG | ortho_meta | T1 | TS3 | 1.4 | 3.7 |
| BOG | ortho_meta | T1 | TS4 | 1.5 | 3.5 |
| BOG | ortho_para | T1 | TS2 | 1.4 | 3.3 |
| BOG | ortho_para | T1 | TS3 | 1.4 | 3.5 |
| BOG | ortho_para | T1 | TS1 | 1.4 | 3.4 |
| BOG | ortho_para | T1 | TS3 | 1.4 | 3.3 |
| BOG | ortho_para | T1 | TS2 | 1.6 | 3.8 |
| BOG | ortho_para | T1 | TS4 | 1.5 | 3.5 |
| BOG | ortho_para | T1 | TS4 | 1.5 | 3.6 |
| BOG | ortho_para | T1 | TS3 | 1.6 | 3.4 |
| BOG | ortho_para | T1 | TS3 | 1.4 | 3.2 |
| BOG | meta_ortho | T1 | TS3 | 1.5 | 3.7 |
| BOG | meta_ortho | T1 | TS2 | 1.2 | 3.2 |
| BOG | meta_ortho | T1 | TS1 | 1.4 | 3.4 |
| BOG | meta_ortho | T1 | TS2 | 1.4 | 3.3 |
| BOG | meta_ortho | T1 | TS3 | 1.5 | 3.4 |
| BOG | meta_ortho | T1 | TS3 | 1.3 | 3.5 |
| BOG | meta_ortho | T1 | TS3 | 1.5 | 3.4 |
| BOG | meta_ortho | T1 | TS4 | 1.4 | 3.4 |
| BOG | meta_ortho | T1 | TS4 | 1.5 | 3.3 |
| BOG | meta_meta | T1 | TS1 | 1.4 | 3.7 |
| BOG | meta_meta | T1 | TS3 | 1.5 | 3.3 |
| BOG | meta_meta | T1 | TS4 | 1.5 | 3.5 |
| BOG | meta_meta | T1 | TS4 | 1.4 | 3.3 |
| BOG | meta_meta | T1 | TS2 | 1.5 | 3.5 |
| BOG | meta_meta | T1 | TS3 | 1.6 | 3.6 |
| BOG | meta_para | T1 | TS2 | 1.6 | 3.1 |
| BOG | meta_para | T1 | TS3 | 1.5 | 3.5 |
| BOG | meta_para | T1 | TS1 | 1.3 | 3.7 |
| BOG | meta_para | T1 | TS2 | 1.4 | 3.5 |
| BOG | meta_para | T1 | TS4 | 1.6 | 3.8 |
| BOG | meta_para | T1 | TS3 | 1.4 | 3.4 |
| BOG | meta_para | T1 | TS4 | 1.5 | 3.7 |
| BOG | meta_para | T1 | TS3 | 1.7 | 3.9 |
| BOG | meta_para | T1 | TS3 | 1.6 | 3.8 |
| BOG | para_ortho | T1 | TS3 | 1.7 | 3.9 |
| BOG | para_ortho | T1 | TS2 | 1.6 | 3.5 |
| BOG | para_ortho | T1 | TS2 | 1.3 | 3.3 |
| BOG | para_ortho | T1 | TS1 | 1.4 | 3.9 |
| BOG | para_ortho | T1 | TS3 | 1.4 | 3.7 |
| BOG | para_ortho | T1 | TS4 | 1.6 | 3.9 |
| BOG | para_meta | T1 | TS1 | 1.3 | 2.9 |
| BOG | para_meta | T1 | TS4 | 1.3 | 3.4 |
| BOG | para_meta | T1 | TS2 | 1.6 | 3.8 |
| BOG | para_meta | T1 | TS3 | 1.5 | 3.6 |
| BOG | para_para | T1 | TS1 | 1.5 | 3.7 |
| BOG | para_para | T1 | TS4 | 1.3 | 3.3 |
| BOG | para_para | T1 | TS2 | 1.4 | 3.6 |
| BOG | para_para | T1 | TS2 | 1.4 | 3.8 |
| BOG | para_para | T1 | TS3 | 1.4 | 3.4 |
| BOG | para_para | T1 | TS3 | 1.6 | 3.7 |
| BOG | ortho_benzene | T2 | TS3 | 1.4 | 3.8 |
| BOG | ortho_benzene | T2 | TS1 | 1.3 | 3.6 |
| BOG | ortho_benzene | T2 | TS4 | 1.4 | 3.5 |
| BOG | ortho_benzene | T2 | TS2 | 1.3 | 3.7 |
| BOG | ortho_benzene | T2 | TS3 | 1.2 | 3.0 |
| BOG | ortho_benzene | T2 | TS4 | 1.2 | 3.5 |
| BOG | meta_benzene | T2 | TS2 | 1.5 | 3.6 |
| BOG | meta_benzene | T2 | TS3 | 1.2 | 3.9 |
| BOG | meta_benzene | T2 | TS3 | 1.2 | 3.1 |
| BOG | meta_benzene | T2 | TS4 | 1.4 | 3.5 |
| BOG | meta_benzene | T2 | TS1 | 1.2 | 3.0 |
| BOG | meta_benzene | T2 | TS4 | 1.4 | 3.8 |
| BOG | para_benzene | T2 | TS3 | 1.5 | 3.9 |
| BOG | para_benzene | T2 | TS1 | 1.4 | 3.7 |
| BOG | para_benzene | T2 | TS2 | 1.5 | 3.8 |
| BOG | para_benzene | T2 | TS4 | 1.4 | 3.6 |
| BOG | ortho_ortho | T2 | TS2 | 1.3 | 3.5 |
| BOG | ortho_ortho | T2 | TS2 | 1.2 | 3.2 |
| BOG | ortho_ortho | T2 | TS3 | 1.2 | 3.2 |
| BOG | ortho_ortho | T2 | TS1 | 1.4 | 3.4 |
| BOG | ortho_ortho | T2 | TS4 | 1.3 | 3.3 |
| BOG | ortho_ortho | T2 | TS3 | 1.2 | 3.2 |
| BOG | ortho_ortho | T2 | TS3 | 1.6 | 3.6 |
| BOG | ortho_ortho | T2 | TS4 | 1.4 | 3.5 |
| BOG | ortho_ortho | T2 | TS3 | 1.3 | 3.6 |
| BOG | ortho_meta | T2 | TS2 | 1.3 | 3.4 |
| BOG | ortho_meta | T2 | TS4 | 1.3 | 3.3 |
| BOG | ortho_meta | T2 | TS1 | 1.2 | 3.1 |
| BOG | ortho_meta | T2 | TS3 | 1.2 | 3.2 |
| BOG | ortho_meta | T2 | TS3 | 1.3 | 3.4 |
| BOG | ortho_meta | T2 | TS4 | 1.1 | 3.2 |
| BOG | ortho_para | T2 | TS2 | 1.4 | 3.7 |
| BOG | ortho_para | T2 | TS1 | 1.2 | 3.3 |
| BOG | ortho_para | T2 | TS3 | 1.4 | 3.2 |
| BOG | ortho_para | T2 | TS4 | 1.3 | 3.4 |
| BOG | ortho_para | T2 | TS3 | 1.5 | 3.6 |
| BOG | ortho_para | T2 | TS3 | 1.3 | 3.3 |
| BOG | ortho_para | T2 | TS4 | 1.3 | 3.5 |
| BOG | ortho_para | T2 | TS3 | 1.4 | 3.3 |
| BOG | ortho_para | T2 | TS2 | 1.5 | 3.7 |
| BOG | meta_ortho | T2 | TS2 | 1.4 | 3.4 |
| BOG | meta_ortho | T2 | TS3 | 1.2 | 3.2 |
| BOG | meta_ortho | T2 | TS4 | 1.3 | 3.2 |
| BOG | meta_ortho | T2 | TS3 | 1.3 | 3.3 |
| BOG | meta_ortho | T2 | TS3 | 1.2 | 3.1 |
| BOG | meta_ortho | T2 | TS4 | 1.3 | 3.2 |
| BOG | meta_ortho | T2 | TS1 | 1.3 | 3.4 |
| BOG | meta_ortho | T2 | TS2 | 1.2 | 3.2 |
| BOG | meta_ortho | T2 | TS3 | 1.3 | 3.3 |
| BOG | meta_meta | T2 | TS3 | 1.3 | 3.4 |
| BOG | meta_meta | T2 | TS1 | 1.3 | 3.1 |
| BOG | meta_meta | T2 | TS4 | 1.2 | 3.3 |
| BOG | meta_meta | T2 | TS2 | 1.3 | 3.8 |
| BOG | meta_meta | T2 | TS3 | 1.3 | 3.5 |
| BOG | meta_meta | T2 | TS4 | 1.2 | 3.0 |
| BOG | meta_para | T2 | TS2 | 1.3 | 3.7 |
| BOG | meta_para | T2 | TS2 | 1.3 | 3.6 |
| BOG | meta_para | T2 | TS3 | 1.4 | 3.6 |
| BOG | meta_para | T2 | TS3 | 1.3 | 3.3 |
| BOG | meta_para | T2 | TS1 | 1.3 | 3.3 |
| BOG | meta_para | T2 | TS4 | 1.3 | 3.2 |
| BOG | meta_para | T2 | TS3 | 1.3 | 3.8 |
| BOG | meta_para | T2 | TS4 | 1.3 | 3.7 |
| BOG | meta_para | T2 | TS3 | 1.2 | 3.4 |
| BOG | para_ortho | T2 | TS4 | 1.4 | 3.5 |
| BOG | para_ortho | T2 | TS1 | 1.4 | 3.6 |
| BOG | para_ortho | T2 | TS2 | 1.3 | 3.2 |
| BOG | para_ortho | T2 | TS3 | 1.3 | 3.6 |
| BOG | para_ortho | T2 | TS3 | 1.3 | 3.3 |
| BOG | para_ortho | T2 | TS2 | 1.4 | 3.4 |
| BOG | para_meta | T2 | TS3 | 1.3 | 3.6 |
| BOG | para_meta | T2 | TS4 | 1.5 | 3.5 |
| BOG | para_meta | T2 | TS1 | 1.4 | 3.2 |
| BOG | para_meta | T2 | TS2 | 1.4 | 3.8 |
| BOG | para_para | T2 | TS2 | 1.4 | 3.8 |
| BOG | para_para | T2 | TS2 | 1.4 | 3.4 |
| BOG | para_para | T2 | TS3 | 1.4 | 3.7 |
| BOG | para_para | T2 | TS1 | 1.6 | 4.0 |
| BOG | para_para | T2 | TS3 | 1.3 | 3.5 |
| BOG | para_para | T2 | TS4 | 1.4 | 3.9 |
| IWV | ortho_benzene | T3 | TS1 | 1.6 | 3.9 |
| IWV | ortho_benzene | T3 | TS3 | 1.7 | 3.5 |
| IWV | ortho_benzene | T3 | TS3 | 1.4 | 3.8 |
| IWV | ortho_benzene | T3 | TS2 | 1.5 | 3.1 |
| IWV | ortho_benzene | T3 | TS4 | 1.5 | 3.3 |
| IWV | ortho_benzene | T3 | TS4 | 1.4 | 3.2 |
| IWV | meta_benzene | T3 | TS3 | 1.4 | 3.6 |
| IWV | meta_benzene | T3 | TS1 | 1.6 | 3.3 |
| IWV | meta_benzene | T3 | TS4 | 1.6 | 3.4 |
| IWV | meta_benzene | T3 | TS3 | 1.6 | 3.8 |
| IWV | meta_benzene | T3 | TS2 | 1.3 | 3.1 |
| IWV | meta_benzene | T3 | TS4 | 1.7 | 3.9 |
| IWV | para_benzene | T3 | TS4 | 1.5 | 3.8 |
| IWV | para_benzene | T3 | TS3 | 1.4 | 3.8 |
| IWV | para_benzene | T3 | TS1 | 1.7 | 3.9 |
| IWV | para_benzene | T3 | TS2 | 1.6 | 4.0 |
| IWV | ortho_ortho | T3 | TS2 | 1.6 | 3.8 |
| IWV | ortho_ortho | T3 | TS4 | 1.3 | 3.3 |
| IWV | ortho_ortho | T3 | TS4 | 1.5 | 3.7 |
| IWV | ortho_ortho | T3 | TS1 | 1.5 | 3.8 |
| IWV | ortho_ortho | T3 | TS3 | 1.5 | 3.2 |
| IWV | ortho_ortho | T3 | TS2 | 1.4 | 3.5 |
| IWV | ortho_ortho | T3 | TS3 | 1.6 | 3.8 |
| IWV | ortho_ortho | T3 | TS3 | 1.4 | 3.1 |
| IWV | ortho_ortho | T3 | TS3 | 1.6 | 3.4 |
| IWV | ortho_meta | T3 | TS4 | 1.4 | 3.6 |
| IWV | ortho_meta | T3 | TS1 | 1.5 | 3.5 |
| IWV | ortho_meta | T3 | TS3 | 1.4 | 3.1 |
| IWV | ortho_meta | T3 | TS3 | 1.5 | 3.2 |
| IWV | ortho_meta | T3 | TS2 | 1.5 | 3.1 |
| IWV | ortho_meta | T3 | TS4 | 1.6 | 3.3 |
| IWV | ortho_para | T3 | TS3 | 1.5 | 3.2 |
| IWV | ortho_para | T3 | TS3 | 1.5 | 3.6 |
| IWV | ortho_para | T3 | TS3 | 1.4 | 3.2 |
| IWV | ortho_para | T3 | TS2 | 1.5 | 3.6 |
| IWV | ortho_para | T3 | TS1 | 1.5 | 3.6 |
| IWV | ortho_para | T3 | TS4 | 1.4 | 3.6 |
| IWV | ortho_para | T3 | TS2 | 1.5 | 3.8 |
| IWV | ortho_para | T3 | TS3 | 1.4 | 3.3 |
| IWV | ortho_para | T3 | TS4 | 1.6 | 4.2 |
| IWV | meta_ortho | T3 | TS3 | 1.6 | 3.8 |
| IWV | meta_ortho | T3 | TS1 | 1.6 | 3.5 |
| IWV | meta_ortho | T3 | TS4 | 1.6 | 3.9 |
| IWV | meta_ortho | T3 | TS2 | 1.4 | 3.4 |
| IWV | meta_ortho | T3 | TS3 | 1.5 | 3.5 |
| IWV | meta_ortho | T3 | TS3 | 1.5 | 3.7 |
| IWV | meta_ortho | T3 | TS2 | 1.4 | 3.1 |
| IWV | meta_ortho | T3 | TS3 | 1.5 | 3.7 |
| IWV | meta_ortho | T3 | TS4 | 1.4 | 3.5 |
| IWV | meta_meta | T3 | TS4 | 1.6 | 3.1 |
| IWV | meta_meta | T3 | TS2 | 1.6 | 3.2 |
| IWV | meta_meta | T3 | TS1 | 1.3 | 3.4 |
| IWV | meta_meta | T3 | TS3 | 1.3 | 2.7 |
| IWV | meta_meta | T3 | TS4 | 1.4 | 3.3 |
| IWV | meta_meta | T3 | TS3 | 1.5 | 3.3 |
| IWV | meta_para | T3 | TS2 | 1.3 | 3.0 |
| IWV | meta_para | T3 | TS4 | 1.6 | 3.6 |
| IWV | meta_para | T3 | TS4 | 1.4 | 3.2 |
| IWV | meta_para | T3 | TS3 | 1.4 | 3.1 |
| IWV | meta_para | T3 | TS1 | 1.4 | 3.7 |
| IWV | meta_para | T3 | TS2 | 1.4 | 3.5 |
| IWV | meta_para | T3 | TS3 | 1.3 | 3.9 |
| IWV | meta_para | T3 | TS3 | 1.5 | 3.6 |
| IWV | meta_para | T3 | TS3 | 1.7 | 3.7 |
| IWV | para_ortho | T3 | TS2 | 1.7 | 3.8 |
| IWV | para_ortho | T3 | TS1 | 1.5 | 3.4 |
| IWV | para_ortho | T3 | TS2 | 1.6 | 3.5 |
| IWV | para_ortho | T3 | TS3 | 1.6 | 3.8 |
| IWV | para_ortho | T3 | TS4 | 1.5 | 3.5 |
| IWV | para_ortho | T3 | TS3 | 1.7 | 3.6 |
| IWV | para_meta | T3 | TS4 | 1.6 | 3.1 |
| IWV | para_meta | T3 | TS2 | 1.5 | 3.7 |
| IWV | para_meta | T3 | TS3 | 1.5 | 3.8 |
| IWV | para_meta | T3 | TS1 | 1.6 | 3.5 |
| IWV | para_para | T3 | TS2 | 1.6 | 3.6 |
| IWV | para_para | T3 | TS1 | 1.6 | 3.9 |
| IWV | para_para | T3 | TS4 | 1.6 | 3.7 |
| IWV | para_para | T3 | TS2 | 1.4 | 3.3 |
| IWV | para_para | T3 | TS3 | 1.5 | 3.5 |
| IWV | para_para | T3 | TS3 | 1.3 | 2.6 |
| IWV | ortho_benzene | T7 | TS1 | 1.6 | 3.3 |
| IWV | ortho_benzene | T7 | TS4 | 1.1 | 3.0 |
| IWV | ortho_benzene | T7 | TS2 | 1.5 | 3.8 |
| IWV | ortho_benzene | T7 | TS3 | 1.4 | 3.0 |
| IWV | ortho_benzene | T7 | TS3 | 1.5 | 3.6 |
| IWV | ortho_benzene | T7 | TS4 | 1.3 | 3.2 |
| IWV | meta_benzene | T7 | TS1 | 1.5 | 3.5 |
| IWV | meta_benzene | T7 | TS4 | 1.4 | 3.5 |
| IWV | meta_benzene | T7 | TS4 | 1.4 | 3.1 |
| IWV | meta_benzene | T7 | TS3 | 1.7 | 3.3 |
| IWV | meta_benzene | T7 | TS2 | 1.8 | 3.5 |
| IWV | meta_benzene | T7 | TS3 | 1.7 | 3.2 |
| IWV | para_benzene | T7 | TS3 | 1.4 | 3.4 |
| IWV | para_benzene | T7 | TS2 | 1.6 | 3.1 |
| IWV | para_benzene | T7 | TS4 | 1.7 | 3.0 |
| IWV | para_benzene | T7 | TS1 | 1.5 | 3.2 |
| IWV | ortho_ortho | T7 | TS3 | 1.6 | 3.1 |
| IWV | ortho_ortho | T7 | TS3 | 1.5 | 3.2 |
| IWV | ortho_ortho | T7 | TS1 | 1.4 | 3.2 |
| IWV | ortho_ortho | T7 | TS4 | 1.5 | 3.1 |
| IWV | ortho_ortho | T7 | TS2 | 1.6 | 3.5 |
| IWV | ortho_ortho | T7 | TS2 | 1.4 | 3.5 |
| IWV | ortho_ortho | T7 | TS3 | 1.4 | 3.2 |
| IWV | ortho_ortho | T7 | TS3 | 1.5 | 3.3 |
| IWV | ortho_ortho | T7 | TS4 | 1.5 | 3.6 |
| IWV | ortho_meta | T7 | TS2 | 1.4 | 3.1 |
| IWV | ortho_meta | T7 | TS4 | 1.4 | 3.2 |
| IWV | ortho_meta | T7 | TS4 | 1.5 | 2.8 |
| IWV | ortho_meta | T7 | TS3 | 1.4 | 3.3 |
| IWV | ortho_meta | T7 | TS3 | 1.3 | 3.1 |
| IWV | ortho_meta | T7 | TS1 | 1.3 | 2.9 |
| IWV | ortho_para | T7 | TS4 | 1.4 | 2.9 |
| IWV | ortho_para | T7 | TS1 | 1.5 | 3.4 |
| IWV | ortho_para | T7 | TS3 | 1.4 | 3.1 |
| IWV | ortho_para | T7 | TS2 | 1.4 | 3.2 |
| IWV | ortho_para | T7 | TS2 | 1.7 | 3.8 |
| IWV | ortho_para | T7 | TS3 | 1.4 | 3.2 |
| IWV | ortho_para | T7 | TS3 | 1.5 | 2.9 |
| IWV | ortho_para | T7 | TS4 | 1.4 | 3.5 |
| IWV | ortho_para | T7 | TS3 | 1.6 | 3.0 |
| IWV | meta_ortho | T7 | TS3 | 1.5 | 3.6 |
| IWV | meta_ortho | T7 | TS4 | 1.3 | 3.4 |
| IWV | meta_ortho | T7 | TS1 | 1.4 | 3.5 |
| IWV | meta_ortho | T7 | TS3 | 1.7 | 3.5 |
| IWV | meta_ortho | T7 | TS4 | 1.4 | 2.9 |
| IWV | meta_ortho | T7 | TS2 | 1.5 | 3.5 |
| IWV | meta_ortho | T7 | TS3 | 1.5 | 3.7 |
| IWV | meta_ortho | T7 | TS2 | 1.3 | 2.9 |
| IWV | meta_ortho | T7 | TS3 | 1.5 | 3.3 |
| IWV | meta_meta | T7 | TS3 | 1.6 | 3.3 |
| IWV | meta_meta | T7 | TS3 | 1.4 | 3.1 |
| IWV | meta_meta | T7 | TS2 | 1.6 | 3.2 |
| IWV | meta_meta | T7 | TS4 | 1.2 | 2.9 |
| IWV | meta_meta | T7 | TS1 | 1.4 | 3.1 |
| IWV | meta_meta | T7 | TS4 | 1.5 | 3.2 |
| IWV | meta_para | T7 | TS3 | 1.6 | 3.4 |
| IWV | meta_para | T7 | TS3 | 1.4 | 3.3 |
| IWV | meta_para | T7 | TS3 | 1.4 | 3.1 |
| IWV | meta_para | T7 | TS4 | 1.6 | 3.2 |
| IWV | meta_para | T7 | TS1 | 1.5 | 3.4 |
| IWV | meta_para | T7 | TS4 | 1.5 | 3.6 |
| IWV | meta_para | T7 | TS3 | 1.6 | 3.5 |
| IWV | meta_para | T7 | TS2 | 1.4 | 2.8 |
| IWV | meta_para | T7 | TS2 | 1.5 | 3.6 |
| IWV | para_ortho | T7 | TS2 | 1.6 | 3.2 |
| IWV | para_ortho | T7 | TS1 | 1.6 | 3.4 |
| IWV | para_ortho | T7 | TS4 | 1.7 | 3.8 |
| IWV | para_ortho | T7 | TS3 | 1.6 | 3.7 |
| IWV | para_ortho | T7 | TS3 | 1.3 | 3.0 |
| IWV | para_ortho | T7 | TS2 | 1.6 | 3.4 |
| IWV | para_meta | T7 | TS1 | 1.3 | 2.6 |
| IWV | para_meta | T7 | TS4 | 1.6 | 3.8 |
| IWV | para_meta | T7 | TS2 | 1.4 | 3.4 |
| IWV | para_meta | T7 | TS3 | 1.4 | 3.4 |
| IWV | para_para | T7 | TS1 | 1.4 | 3.1 |
| IWV | para_para | T7 | TS3 | 1.3 | 2.7 |
| IWV | para_para | T7 | TS3 | 1.4 | 3.0 |
| IWV | para_para | T7 | TS4 | 1.7 | 3.5 |
| IWV | para_para | T7 | TS2 | 1.5 | 3.3 |
| IWV | para_para | T7 | TS2 | 1.5 | 3.6 |
| UTL | ortho_benzene | T2 | TS3 | 1.4 | 3.7 |
| UTL | ortho_benzene | T2 | TS1 | 1.4 | 3.8 |
| UTL | ortho_benzene | T2 | TS4 | 1.6 | 3.4 |
| UTL | ortho_benzene | T2 | TS4 | 1.8 | 3.9 |
| UTL | ortho_benzene | T2 | TS3 | 1.6 | 3.9 |
| UTL | ortho_benzene | T2 | TS2 | 1.7 | 3.9 |
| UTL | meta_benzene | T2 | TS2 | 1.3 | 3.4 |
| UTL | meta_benzene | T2 | TS3 | 1.3 | 3.3 |
| UTL | meta_benzene | T2 | TS4 | 1.6 | 3.9 |
| UTL | meta_benzene | T2 | TS1 | 1.8 | 4.0 |
| UTL | meta_benzene | T2 | TS3 | 1.9 | 4.1 |
| UTL | meta_benzene | T2 | TS4 | 1.7 | 4.1 |
| UTL | para_benzene | T2 | TS4 | 1.5 | 4.3 |
| UTL | para_benzene | T2 | TS3 | 1.5 | 3.6 |
| UTL | para_benzene | T2 | TS1 | 1.4 | 3.5 |
| UTL | para_benzene | T2 | TS2 | 1.4 | 3.8 |
| UTL | ortho_ortho | T2 | TS3 | 1.5 | 3.5 |
| UTL | ortho_ortho | T2 | TS2 | 1.6 | 4.0 |
| UTL | ortho_ortho | T2 | TS3 | 1.6 | 3.9 |
| UTL | ortho_ortho | T2 | TS1 | 1.5 | 3.3 |
| UTL | ortho_ortho | T2 | TS2 | 1.5 | 3.7 |
| UTL | ortho_ortho | T2 | TS4 | 1.3 | 2.8 |
| UTL | ortho_ortho | T2 | TS4 | 1.6 | 3.5 |
| UTL | ortho_ortho | T2 | TS3 | 1.7 | 3.8 |
| UTL | ortho_ortho | T2 | TS3 | 1.7 | 3.9 |
| UTL | ortho_meta | T2 | TS1 | 1.6 | 3.5 |
| UTL | ortho_meta | T2 | TS4 | 1.5 | 3.8 |
| UTL | ortho_meta | T2 | TS3 | 1.3 | 3.3 |
| UTL | ortho_meta | T2 | TS3 | 1.7 | 3.9 |
| UTL | ortho_meta | T2 | TS4 | 1.6 | 3.9 |
| UTL | ortho_meta | T2 | TS2 | 1.7 | 3.6 |
| UTL | ortho_para | T2 | TS4 | 1.5 | 3.6 |
| UTL | ortho_para | T2 | TS3 | 1.7 | 3.7 |
| UTL | ortho_para | T2 | TS1 | 1.7 | 4.2 |
| UTL | ortho_para | T2 | TS3 | 1.6 | 3.5 |
| UTL | ortho_para | T2 | TS2 | 1.7 | 4.1 |
| UTL | ortho_para | T2 | TS2 | 1.6 | 3.7 |
| UTL | ortho_para | T2 | TS4 | 1.8 | 3.9 |
| UTL | ortho_para | T2 | TS3 | 1.6 | 3.0 |
| UTL | ortho_para | T2 | TS3 | 1.8 | 3.7 |
| UTL | meta_ortho | T2 | TS3 | 1.6 | 4.1 |
| UTL | meta_ortho | T2 | TS3 | 1.5 | 3.8 |
| UTL | meta_ortho | T2 | TS4 | 1.4 | 3.3 |
| UTL | meta_ortho | T2 | TS3 | 1.6 | 3.6 |
| UTL | meta_ortho | T2 | TS4 | 1.4 | 3.3 |
| UTL | meta_ortho | T2 | TS2 | 1.6 | 3.6 |
| UTL | meta_ortho | T2 | TS3 | 1.7 | 3.5 |
| UTL | meta_ortho | T2 | TS1 | 1.6 | 3.8 |
| UTL | meta_ortho | T2 | TS2 | 1.6 | 3.4 |
| UTL | meta_meta | T2 | TS4 | 1.7 | 3.7 |
| UTL | meta_meta | T2 | TS1 | 1.6 | 3.7 |
| UTL | meta_meta | T2 | TS2 | 1.5 | 3.6 |
| UTL | meta_meta | T2 | TS3 | 1.8 | 3.8 |
| UTL | meta_meta | T2 | TS3 | 1.7 | 3.9 |
| UTL | meta_meta | T2 | TS4 | 1.5 | 3.1 |
| UTL | meta_para | T2 | TS3 | 1.6 | 3.5 |
| UTL | meta_para | T2 | TS4 | 1.6 | 3.8 |
| UTL | meta_para | T2 | TS3 | 1.7 | 3.6 |
| UTL | meta_para | T2 | TS1 | 1.6 | 3.4 |
| UTL | meta_para | T2 | TS4 | 1.7 | 3.6 |
| UTL | meta_para | T2 | TS2 | 1.6 | 3.9 |
| UTL | meta_para | T2 | TS2 | 1.6 | 3.8 |
| UTL | meta_para | T2 | TS3 | 1.6 | 3.6 |
| UTL | meta_para | T2 | TS3 | 1.6 | 3.7 |
| UTL | para_ortho | T2 | TS1 | 1.6 | 3.9 |
| UTL | para_ortho | T2 | TS3 | 1.6 | 4.1 |
| UTL | para_ortho | T2 | TS2 | 1.5 | 3.7 |
| UTL | para_ortho | T2 | TS3 | 1.5 | 3.9 |
| UTL | para_ortho | T2 | TS4 | 1.6 | 3.9 |
| UTL | para_ortho | T2 | TS2 | 1.4 | 3.2 |
| UTL | para_meta | T2 | TS2 | 1.6 | 3.8 |
| UTL | para_meta | T2 | TS1 | 1.5 | 4.0 |
| UTL | para_meta | T2 | TS3 | 1.4 | 3.2 |
| UTL | para_meta | T2 | TS4 | 1.7 | 4.0 |
| UTL | para_para | T2 | TS3 | 1.6 | 3.4 |
| UTL | para_para | T2 | TS3 | 1.5 | 4.0 |
| UTL | para_para | T2 | TS1 | 1.6 | 4.2 |
| UTL | para_para | T2 | TS4 | 1.6 | 3.9 |
| UTL | para_para | T2 | TS2 | 1.5 | 3.4 |
| UTL | para_para | T2 | TS2 | 1.4 | 3.4 |
| UTL | ortho_benzene | T4 | TS3 | 1.4 | 3.3 |
| UTL | ortho_benzene | T4 | TS4 | 1.6 | 3.7 |
| UTL | ortho_benzene | T4 | TS2 | 1.8 | 4.1 |
| UTL | ortho_benzene | T4 | TS1 | 1.7 | 4.3 |
| UTL | ortho_benzene | T4 | TS4 | 1.6 | 3.5 |
| UTL | ortho_benzene | T4 | TS3 | 1.6 | 3.8 |
| UTL | meta_benzene | T4 | TS1 | 1.5 | 2.9 |
| UTL | meta_benzene | T4 | TS4 | 1.8 | 4.0 |
| UTL | meta_benzene | T4 | TS3 | 1.9 | 4.2 |
| UTL | meta_benzene | T4 | TS2 | 1.3 | 3.5 |
| UTL | meta_benzene | T4 | TS4 | 1.5 | 3.3 |
| UTL | meta_benzene | T4 | TS3 | 1.6 | 3.9 |
| UTL | para_benzene | T4 | TS2 | 1.8 | 3.4 |
| UTL | para_benzene | T4 | TS3 | 1.6 | 3.7 |
| UTL | para_benzene | T4 | TS1 | 1.5 | 3.8 |
| UTL | para_benzene | T4 | TS4 | 1.2 | 3.1 |
| UTL | ortho_ortho | T4 | TS3 | 1.6 | 4.2 |
| UTL | ortho_ortho | T4 | TS1 | 1.7 | 3.5 |
| UTL | ortho_ortho | T4 | TS4 | 1.4 | 3.9 |
| UTL | ortho_ortho | T4 | TS3 | 1.6 | 3.5 |
| UTL | ortho_ortho | T4 | TS3 | 1.7 | 4.0 |
| UTL | ortho_ortho | T4 | TS2 | 1.5 | 3.6 |
| UTL | ortho_ortho | T4 | TS4 | 1.5 | 3.9 |
| UTL | ortho_ortho | T4 | TS2 | 1.6 | 3.9 |
| UTL | ortho_ortho | T4 | TS3 | 1.6 | 3.7 |
| UTL | ortho_meta | T4 | TS3 | 1.7 | 4.1 |
| UTL | ortho_meta | T4 | TS4 | 1.5 | 4.1 |
| UTL | ortho_meta | T4 | TS4 | 1.6 | 3.6 |
| UTL | ortho_meta | T4 | TS3 | 1.5 | 3.2 |
| UTL | ortho_meta | T4 | TS1 | 1.5 | 3.2 |
| UTL | ortho_meta | T4 | TS2 | 1.5 | 3.5 |
| UTL | ortho_para | T4 | TS3 | 1.6 | 3.6 |
| UTL | ortho_para | T4 | TS2 | 1.7 | 4.1 |
| UTL | ortho_para | T4 | TS1 | 1.7 | 4.1 |
| UTL | ortho_para | T4 | TS4 | 1.7 | 3.6 |
| UTL | ortho_para | T4 | TS3 | 1.7 | 3.7 |
| UTL | ortho_para | T4 | TS4 | 1.8 | 4.0 |
| UTL | ortho_para | T4 | TS3 | 1.8 | 4.1 |
| UTL | ortho_para | T4 | TS3 | 1.7 | 3.9 |
| UTL | ortho_para | T4 | TS2 | 1.5 | 3.3 |
| UTL | meta_ortho | T4 | TS3 | 1.8 | 4.5 |
| UTL | meta_ortho | T4 | TS1 | 1.7 | 4.1 |
| UTL | meta_ortho | T4 | TS3 | 1.6 | 3.3 |
| UTL | meta_ortho | T4 | TS3 | 1.8 | 3.6 |
| UTL | meta_ortho | T4 | TS4 | 1.5 | 3.8 |
| UTL | meta_ortho | T4 | TS2 | 1.6 | 3.7 |
| UTL | meta_ortho | T4 | TS3 | 1.6 | 3.6 |
| UTL | meta_ortho | T4 | TS4 | 1.6 | 3.9 |
| UTL | meta_ortho | T4 | TS2 | 1.6 | 3.8 |
| UTL | meta_meta | T4 | TS3 | 1.6 | 3.9 |
| UTL | meta_meta | T4 | TS4 | 1.4 | 3.2 |
| UTL | meta_meta | T4 | TS3 | 1.6 | 3.7 |
| UTL | meta_meta | T4 | TS4 | 1.6 | 4.0 |
| UTL | meta_meta | T4 | TS2 | 1.6 | 3.9 |
| UTL | meta_meta | T4 | TS1 | 1.7 | 3.9 |
| UTL | meta_para | T4 | TS3 | 1.6 | 3.5 |
| UTL | meta_para | T4 | TS1 | 1.6 | 3.8 |
| UTL | meta_para | T4 | TS4 | 1.8 | 4.1 |
| UTL | meta_para | T4 | TS4 | 1.7 | 4.0 |
| UTL | meta_para | T4 | TS3 | 1.5 | 3.7 |
| UTL | meta_para | T4 | TS2 | 1.5 | 3.7 |
| UTL | meta_para | T4 | TS3 | 1.6 | 3.6 |
| UTL | meta_para | T4 | TS2 | 1.5 | 3.7 |
| UTL | meta_para | T4 | TS3 | 1.7 | 3.6 |
| UTL | para_ortho | T4 | TS1 | 1.8 | 4.2 |
| UTL | para_ortho | T4 | TS3 | 1.5 | 3.8 |
| UTL | para_ortho | T4 | TS4 | 1.7 | 3.7 |
| UTL | para_ortho | T4 | TS3 | 1.5 | 3.7 |
| UTL | para_ortho | T4 | TS2 | 1.6 | 3.6 |
| UTL | para_ortho | T4 | TS2 | 1.7 | 3.7 |
| UTL | para_meta | T4 | TS1 | 1.4 | 3.9 |
| UTL | para_meta | T4 | TS4 | 1.5 | 3.2 |
| UTL | para_meta | T4 | TS2 | 1.5 | 3.5 |
| UTL | para_meta | T4 | TS3 | 1.6 | 3.9 |
| UTL | para_para | T4 | TS4 | 1.6 | 4.0 |
| UTL | para_para | T4 | TS1 | 1.5 | 3.2 |
| UTL | para_para | T4 | TS3 | 1.6 | 3.6 |
| UTL | para_para | T4 | TS3 | 1.5 | 3.7 |
| UTL | para_para | T4 | TS2 | 1.5 | 4.2 |
| UTL | para_para | T4 | TS2 | 1.4 | 3.3 |
| FAU | ortho_benzene | T1 | TS4 | 1.5 | 3.5 |
| FAU | ortho_benzene | T1 | TS3 | 1.8 | 4.0 |
| FAU | ortho_benzene | T1 | TS3 | 1.5 | 3.8 |
| FAU | ortho_benzene | T1 | TS2 | 2.3 | 5.3 |
| FAU | ortho_benzene | T1 | TS1 | 2.1 | 5.2 |
| FAU | ortho_benzene | T1 | TS4 | 1.7 | 3.7 |
| FAU | meta_benzene | T1 | TS2 | 1.7 | 4.3 |
| FAU | meta_benzene | T1 | TS3 | 1.9 | 3.9 |
| FAU | meta_benzene | T1 | TS4 | 1.9 | 4.2 |
| FAU | meta_benzene | T1 | TS4 | 1.3 | 3.3 |
| FAU | meta_benzene | T1 | TS1 | 1.7 | 4.0 |
| FAU | meta_benzene | T1 | TS3 | 1.7 | 4.2 |
| FAU | para_benzene | T1 | TS2 | 1.9 | 4.5 |
| FAU | para_benzene | T1 | TS3 | 2.0 | 4.7 |
| FAU | para_benzene | T1 | TS4 | 1.7 | 3.6 |
| FAU | para_benzene | T1 | TS1 | 1.8 | 4.1 |
| FAU | ortho_ortho | T1 | TS3 | 1.8 | 3.8 |
| FAU | ortho_ortho | T1 | TS1 | 1.7 | 3.8 |
| FAU | ortho_ortho | T1 | TS2 | 1.8 | 4.1 |
| FAU | ortho_ortho | T1 | TS4 | 2.0 | 4.5 |
| FAU | ortho_ortho | T1 | TS4 | 2.0 | 4.1 |
| FAU | ortho_ortho | T1 | TS2 | 1.8 | 4.2 |
| FAU | ortho_ortho | T1 | TS3 | 2.1 | 5.0 |
| FAU | ortho_ortho | T1 | TS3 | 1.8 | 4.1 |
| FAU | ortho_ortho | T1 | TS3 | 1.8 | 4.3 |
| FAU | ortho_meta | T1 | TS2 | 1.7 | 4.3 |
| FAU | ortho_meta | T1 | TS3 | 1.9 | 4.0 |
| FAU | ortho_meta | T1 | TS4 | 1.9 | 4.3 |
| FAU | ortho_meta | T1 | TS3 | 1.5 | 3.4 |
| FAU | ortho_meta | T1 | TS4 | 1.9 | 4.3 |
| FAU | ortho_meta | T1 | TS1 | 1.7 | 3.8 |
| FAU | ortho_para | T1 | TS2 | 1.7 | 4.4 |
| FAU | ortho_para | T1 | TS2 | 1.9 | 4.3 |
| FAU | ortho_para | T1 | TS1 | 2.0 | 4.3 |
| FAU | ortho_para | T1 | TS4 | 1.8 | 4.8 |
| FAU | ortho_para | T1 | TS4 | 1.8 | 4.4 |
| FAU | ortho_para | T1 | TS3 | 1.9 | 4.1 |
| FAU | ortho_para | T1 | TS3 | 1.5 | 3.7 |
| FAU | ortho_para | T1 | TS3 | 1.9 | 4.2 |
| FAU | ortho_para | T1 | TS3 | 1.9 | 3.7 |
| FAU | meta_ortho | T1 | TS4 | 1.9 | 4.4 |
| FAU | meta_ortho | T1 | TS2 | 1.5 | 3.6 |
| FAU | meta_ortho | T1 | TS2 | 1.8 | 4.4 |
| FAU | meta_ortho | T1 | TS4 | 1.9 | 4.6 |
| FAU | meta_ortho | T1 | TS3 | 1.9 | 3.9 |
| FAU | meta_ortho | T1 | TS3 | 1.9 | 4.5 |
| FAU | meta_ortho | T1 | TS1 | 1.7 | 3.6 |
| FAU | meta_ortho | T1 | TS3 | 1.7 | 3.9 |
| FAU | meta_ortho | T1 | TS3 | 2.0 | 4.3 |
| FAU | meta_meta | T1 | TS4 | 1.7 | 3.7 |
| FAU | meta_meta | T1 | TS3 | 2.0 | 4.5 |
| FAU | meta_meta | T1 | TS1 | 1.8 | 4.1 |
| FAU | meta_meta | T1 | TS2 | 1.6 | 3.7 |
| FAU | meta_meta | T1 | TS4 | 1.9 | 3.9 |
| FAU | meta_meta | T1 | TS3 | 1.9 | 3.7 |
| FAU | meta_para | T1 | TS2 | 1.7 | 3.7 |
| FAU | meta_para | T1 | TS3 | 2.0 | 4.5 |
| FAU | meta_para | T1 | TS2 | 1.8 | 3.8 |
| FAU | meta_para | T1 | TS1 | 2.0 | 4.5 |
| FAU | meta_para | T1 | TS3 | 1.9 | 4.4 |
| FAU | meta_para | T1 | TS3 | 1.9 | 4.5 |
| FAU | meta_para | T1 | TS4 | 2.0 | 4.4 |
| FAU | meta_para | T1 | TS3 | 1.8 | 4.4 |
| FAU | meta_para | T1 | TS4 | 1.9 | 4.5 |
| FAU | para_ortho | T1 | TS2 | 1.7 | 4.1 |
| FAU | para_ortho | T1 | TS3 | 1.7 | 3.3 |
| FAU | para_ortho | T1 | TS4 | 1.7 | 3.7 |
| FAU | para_ortho | T1 | TS3 | 1.7 | 3.6 |
| FAU | para_ortho | T1 | TS1 | 2.0 | 4.5 |
| FAU | para_ortho | T1 | TS2 | 1.7 | 4.1 |
| FAU | para_meta | T1 | TS2 | 1.8 | 4.1 |
| FAU | para_meta | T1 | TS3 | 1.8 | 4.2 |
| FAU | para_meta | T1 | TS1 | 2.1 | 5.1 |
| FAU | para_meta | T1 | TS4 | 1.9 | 4.3 |
| FAU | para_para | T1 | TS2 | 1.7 | 3.9 |
| FAU | para_para | T1 | TS1 | 1.9 | 4.4 |
| FAU | para_para | T1 | TS3 | 1.7 | 3.7 |
| FAU | para_para | T1 | TS2 | 1.9 | 4.5 |
| FAU | para_para | T1 | TS4 | 1.6 | 3.9 |
| FAU | para_para | T1 | TS3 | 1.7 | 3.6 |

**Table S22**. Zeolite rescaled Multstep Path free energy barriers in kJ/mol. Transalkylation paths have been rescaled relative to the minimum I0 energy identified among the BOG (T1: meta-benzene), IWV (T7: ortho-benzene), and UTL (T2: para-benzene). Scaled disproportionation pathways have been referenced to the minimum I0 energy found among BOG (T1: meta–para), IWV (T7: meta–para), and UTL (T2: para–para).

| Framework | T-Site | Isomer | I0 | TS2 | I2 | TS3 | I3 | TS4 | I4 |
|---|---|---|---|---|---|---|---|---|---|
| BOG | T1 | ortho_benzene | 40.93 | 169.48 | -30.04 | 138.71 | 52.40 | 142.92 | 29.87 |
| BOG | T1 | ortho_benzene | 40.93 | 169.48 | -30.04 | 161.29 | 47.23 | 100.23 | 29.87 |
| BOG | T1 | ortho_benzene | 40.93 | 169.48 | -30.04 | 138.71 | 52.40 | 142.92 | 29.87 |
| BOG | T1 | ortho_benzene | 40.93 | 169.48 | -30.04 | 161.29 | 47.23 | 100.23 | 29.87 |
| BOG | T1 | ortho_benzene | 40.93 | 169.48 | -30.04 | 138.71 | 52.40 | 142.92 | 29.87 |
| BOG | T1 | ortho_benzene | 40.93 | 169.48 | -30.04 | 161.29 | 47.23 | 100.23 | 29.87 |
| BOG | T1 | ortho_benzene | 40.93 | 169.48 | -30.04 | 138.71 | 52.40 | 142.92 | 29.87 |
| BOG | T1 | ortho_benzene | 40.93 | 169.48 | -30.04 | 161.29 | 47.23 | 100.23 | 29.87 |
| BOG | T1 | meta_benzene | 16.36 | 99.87 | 55.31 | 83.15 | -6.65 | 126.40 | 22.51 |
| BOG | T1 | meta_benzene | 16.36 | 99.87 | 55.31 | 111.24 | 17.40 | 70.55 | 22.51 |
| BOG | T1 | meta_benzene | 16.36 | 99.87 | 55.31 | 83.15 | -6.65 | 126.40 | 22.51 |
| BOG | T1 | meta_benzene | 16.36 | 99.87 | 55.31 | 111.24 | 17.40 | 70.55 | 22.51 |
| BOG | T1 | meta_benzene | 16.36 | 99.87 | 55.31 | 83.15 | -6.65 | 126.40 | 22.51 |
| BOG | T1 | meta_benzene | 16.36 | 99.87 | 55.31 | 111.24 | 17.40 | 70.55 | 22.51 |
| BOG | T1 | meta_benzene | 16.36 | 99.87 | 55.31 | 83.15 | -6.65 | 126.40 | 22.51 |
| BOG | T1 | meta_benzene | 16.36 | 99.87 | 55.31 | 111.24 | 17.40 | 70.55 | 22.51 |
| BOG | T1 | para_benzene | 0.00 | 92.86 | 39.34 | 92.43 | 7.92 | 70.59 | 42.80 |
| BOG | T1 | para_benzene | 0.00 | 92.86 | 39.34 | 92.43 | 7.92 | 70.59 | 42.80 |
| BOG | T1 | para_benzene | 0.00 | 92.86 | 39.34 | 92.43 | 7.92 | 70.59 | 42.80 |
| BOG | T1 | para_benzene | 0.00 | 92.86 | 39.34 | 92.43 | 7.92 | 70.59 | 42.80 |
| BOG | T1 | ortho_ortho | 98.75 | 181.78 | 88.67 | 189.46 | 123.58 | 187.30 | 164.12 |
| BOG | T1 | ortho_ortho | 98.75 | 181.78 | 88.67 | 126.16 | 43.65 | 205.14 | 164.12 |
| BOG | T1 | ortho_ortho | 98.75 | 124.04 | 103.47 | 172.28 | 123.58 | 187.30 | 164.12 |
| BOG | T1 | ortho_ortho | 98.75 | 124.04 | 103.47 | 201.16 | 43.65 | 205.14 | 164.12 |
| BOG | T1 | ortho_ortho | 98.75 | 181.78 | 88.67 | 189.46 | 123.58 | 187.30 | 164.12 |
| BOG | T1 | ortho_ortho | 98.75 | 181.78 | 88.67 | 126.16 | 43.65 | 205.14 | 164.12 |
| BOG | T1 | ortho_ortho | 98.75 | 124.04 | 103.47 | 172.28 | 123.58 | 187.30 | 164.12 |
| BOG | T1 | ortho_ortho | 98.75 | 124.04 | 103.47 | 201.16 | 43.65 | 205.14 | 164.12 |
| BOG | T1 | ortho_ortho | 98.75 | 181.78 | 88.67 | 189.46 | 123.58 | 187.30 | 164.12 |
| BOG | T1 | ortho_ortho | 98.75 | 181.78 | 88.67 | 126.16 | 43.65 | 205.14 | 164.12 |
| BOG | T1 | ortho_ortho | 98.75 | 124.04 | 103.47 | 172.28 | 123.58 | 187.30 | 164.12 |
| BOG | T1 | ortho_ortho | 98.75 | 124.04 | 103.47 | 201.16 | 43.65 | 205.14 | 164.12 |
| BOG | T1 | ortho_ortho | 98.75 | 181.78 | 88.67 | 189.46 | 123.58 | 187.30 | 164.12 |
| BOG | T1 | ortho_ortho | 98.75 | 181.78 | 88.67 | 126.16 | 43.65 | 205.14 | 164.12 |
| BOG | T1 | ortho_ortho | 98.75 | 124.04 | 103.47 | 172.28 | 123.58 | 187.30 | 164.12 |
| BOG | T1 | ortho_ortho | 98.75 | 124.04 | 103.47 | 201.16 | 43.65 | 205.14 | 164.12 |
| BOG | T1 | ortho_meta | 92.40 | 253.25 | 88.73 | 205.86 | 74.30 | 212.11 | 43.18 |
| BOG | T1 | ortho_meta | 92.40 | 253.25 | 88.73 | 161.16 | 143.50 | 221.67 | 43.18 |
| BOG | T1 | ortho_meta | 92.40 | 253.25 | 88.73 | 205.86 | 74.30 | 212.11 | 43.18 |
| BOG | T1 | ortho_meta | 92.40 | 253.25 | 88.73 | 161.16 | 143.50 | 221.67 | 43.18 |
| BOG | T1 | ortho_meta | 92.40 | 253.25 | 88.73 | 205.86 | 74.30 | 212.11 | 43.18 |
| BOG | T1 | ortho_meta | 92.40 | 253.25 | 88.73 | 161.16 | 143.50 | 221.67 | 43.18 |
| BOG | T1 | ortho_para | 46.07 | 127.63 | 50.84 | 211.33 | 88.18 | 159.00 | 76.65 |
| BOG | T1 | ortho_para | 46.07 | 127.63 | 50.84 | 135.56 | 88.18 | 144.74 | 76.65 |
| BOG | T1 | ortho_para | 46.07 | 101.07 | 66.91 | 172.61 | 47.50 | 159.00 | 76.65 |
| BOG | T1 | ortho_para | 46.07 | 101.07 | 66.91 | 176.06 | 47.50 | 144.74 | 76.65 |
| BOG | T1 | ortho_para | 46.07 | 127.63 | 50.84 | 211.33 | 88.18 | 159.00 | 76.65 |
| BOG | T1 | ortho_para | 46.07 | 127.63 | 50.84 | 135.56 | 88.18 | 144.74 | 76.65 |
| BOG | T1 | ortho_para | 46.07 | 101.07 | 66.91 | 172.61 | 47.50 | 159.00 | 76.65 |
| BOG | T1 | ortho_para | 46.07 | 101.07 | 66.91 | 176.06 | 47.50 | 144.74 | 76.65 |
| BOG | T1 | ortho_para | 46.07 | 127.63 | 50.84 | 211.33 | 88.18 | 159.00 | 76.65 |
| BOG | T1 | ortho_para | 46.07 | 127.63 | 50.84 | 135.56 | 88.18 | 144.74 | 76.65 |
| BOG | T1 | ortho_para | 46.07 | 101.07 | 66.91 | 172.61 | 47.50 | 159.00 | 76.65 |
| BOG | T1 | ortho_para | 46.07 | 101.07 | 66.91 | 176.06 | 47.50 | 144.74 | 76.65 |
| BOG | T1 | meta_ortho | 47.41 | 169.38 | 30.55 | 158.74 | 78.64 | 195.88 | 84.18 |
| BOG | T1 | meta_ortho | 47.41 | 169.38 | 30.55 | 160.86 | 56.56 | 112.97 | 84.18 |
| BOG | T1 | meta_ortho | 47.41 | 210.73 | 16.65 | 135.08 | 78.64 | 195.88 | 84.18 |
| BOG | T1 | meta_ortho | 47.41 | 210.73 | 16.65 | 138.69 | 56.56 | 112.97 | 84.18 |
| BOG | T1 | meta_ortho | 47.41 | 169.38 | 30.55 | 158.74 | 78.64 | 195.88 | 84.18 |
| BOG | T1 | meta_ortho | 47.41 | 169.38 | 30.55 | 160.86 | 56.56 | 112.97 | 84.18 |
| BOG | T1 | meta_ortho | 47.41 | 210.73 | 16.65 | 135.08 | 78.64 | 195.88 | 84.18 |
| BOG | T1 | meta_ortho | 47.41 | 210.73 | 16.65 | 138.69 | 56.56 | 112.97 | 84.18 |
| BOG | T1 | meta_ortho | 47.41 | 169.38 | 30.55 | 158.74 | 78.64 | 195.88 | 84.18 |
| BOG | T1 | meta_ortho | 47.41 | 169.38 | 30.55 | 160.86 | 56.56 | 112.97 | 84.18 |
| BOG | T1 | meta_ortho | 47.41 | 210.73 | 16.65 | 135.08 | 78.64 | 195.88 | 84.18 |
| BOG | T1 | meta_ortho | 47.41 | 210.73 | 16.65 | 138.69 | 56.56 | 112.97 | 84.18 |
| BOG | T1 | meta_meta | 95.75 | 118.38 | 84.71 | 108.93 | 75.21 | 235.92 | 86.26 |
| BOG | T1 | meta_meta | 95.75 | 118.38 | 84.71 | 163.30 | 38.09 | 202.99 | 86.26 |
| BOG | T1 | meta_meta | 95.75 | 118.38 | 84.71 | 108.93 | 75.21 | 235.92 | 86.26 |
| BOG | T1 | meta_meta | 95.75 | 118.38 | 84.71 | 163.30 | 38.09 | 202.99 | 86.26 |
| BOG | T1 | meta_meta | 95.75 | 118.38 | 84.71 | 108.93 | 75.21 | 235.92 | 86.26 |
| BOG | T1 | meta_meta | 95.75 | 118.38 | 84.71 | 163.30 | 38.09 | 202.99 | 86.26 |
| BOG | T1 | meta_meta | 95.75 | 118.38 | 84.71 | 108.93 | 75.21 | 235.92 | 86.26 |
| BOG | T1 | meta_meta | 95.75 | 118.38 | 84.71 | 163.30 | 38.09 | 202.99 | 86.26 |
| BOG | T1 | meta_para | 27.06 | 147.32 | 19.26 | 75.92 | 37.24 | 153.87 | 57.06 |
| BOG | T1 | meta_para | 27.06 | 147.32 | 19.26 | 146.86 | 25.56 | 115.34 | 57.06 |
| BOG | T1 | meta_para | 27.06 | 57.87 | 29.07 | 82.77 | 25.56 | 115.34 | 57.06 |
| BOG | T1 | meta_para | 27.06 | 57.87 | 29.07 | 75.52 | 37.24 | 153.87 | 57.06 |
| BOG | T1 | meta_para | 27.06 | 147.32 | 19.26 | 75.92 | 37.24 | 153.87 | 57.06 |
| BOG | T1 | meta_para | 27.06 | 147.32 | 19.26 | 146.86 | 25.56 | 115.34 | 57.06 |
| BOG | T1 | meta_para | 27.06 | 57.87 | 29.07 | 82.77 | 25.56 | 115.34 | 57.06 |
| BOG | T1 | meta_para | 27.06 | 57.87 | 29.07 | 75.52 | 37.24 | 153.87 | 57.06 |
| BOG | T1 | meta_para | 27.06 | 147.32 | 19.26 | 75.92 | 37.24 | 153.87 | 57.06 |
| BOG | T1 | meta_para | 27.06 | 147.32 | 19.26 | 146.86 | 25.56 | 115.34 | 57.06 |
| BOG | T1 | meta_para | 27.06 | 57.87 | 29.07 | 82.77 | 25.56 | 115.34 | 57.06 |
| BOG | T1 | meta_para | 27.06 | 57.87 | 29.07 | 75.52 | 37.24 | 153.87 | 57.06 |
| BOG | T1 | para_ortho | 60.12 | 107.49 | 29.39 | 93.31 | 37.73 | 132.38 | 72.11 |
| BOG | T1 | para_ortho | 60.12 | 133.24 | 38.41 | 144.52 | 37.73 | 132.38 | 72.11 |
| BOG | T1 | para_ortho | 60.12 | 107.49 | 29.39 | 93.31 | 37.73 | 132.38 | 72.11 |
| BOG | T1 | para_ortho | 60.12 | 133.24 | 38.41 | 144.52 | 37.73 | 132.38 | 72.11 |
| BOG | T1 | para_ortho | 60.12 | 107.49 | 29.39 | 93.31 | 37.73 | 132.38 | 72.11 |
| BOG | T1 | para_ortho | 60.12 | 133.24 | 38.41 | 144.52 | 37.73 | 132.38 | 72.11 |
| BOG | T1 | para_meta | 25.91 | 126.03 | 88.73 | 150.54 | 47.55 | 185.30 | 95.25 |
| BOG | T1 | para_meta | 25.91 | 126.03 | 88.73 | 150.54 | 47.55 | 185.30 | 95.25 |
| BOG | T1 | para_meta | 25.91 | 126.03 | 88.73 | 150.54 | 47.55 | 185.30 | 95.25 |
| BOG | T1 | para_para | 0.00 | 128.20 | 16.52 | 74.86 | 16.89 | 151.16 | 48.43 |
| BOG | T1 | para_para | 0.00 | 94.43 | 29.95 | 96.79 | 16.89 | 151.16 | 48.43 |
| BOG | T1 | para_para | 0.00 | 128.20 | 16.52 | 74.86 | 16.89 | 151.16 | 48.43 |
| BOG | T1 | para_para | 0.00 | 94.43 | 29.95 | 96.79 | 16.89 | 151.16 | 48.43 |
| BOG | T1 | para_para | 0.00 | 128.20 | 16.52 | 74.86 | 16.89 | 151.16 | 48.43 |
| BOG | T1 | para_para | 0.00 | 94.43 | 29.95 | 96.79 | 16.89 | 151.16 | 48.43 |
| BOG | T1 | para_para | 0.00 | 128.20 | 16.52 | 74.86 | 16.89 | 151.16 | 48.43 |
| BOG | T1 | para_para | 0.00 | 94.43 | 29.95 | 96.79 | 16.89 | 151.16 | 48.43 |
| BOG | T2 | ortho_benzene | 48.67 | 103.45 | 57.21 | 126.23 | 54.00 | 155.69 | 49.81 |
| BOG | T2 | ortho_benzene | 48.67 | 103.45 | 57.21 | 228.72 | 46.98 | 93.65 | 49.81 |
| BOG | T2 | ortho_benzene | 48.67 | 103.45 | 57.21 | 126.23 | 54.00 | 155.69 | 49.81 |
| BOG | T2 | ortho_benzene | 48.67 | 103.45 | 57.21 | 228.72 | 46.98 | 93.65 | 49.81 |
| BOG | T2 | ortho_benzene | 48.67 | 103.45 | 57.21 | 126.23 | 54.00 | 155.69 | 49.81 |
| BOG | T2 | ortho_benzene | 48.67 | 103.45 | 57.21 | 228.72 | 46.98 | 93.65 | 49.81 |
| BOG | T2 | ortho_benzene | 48.67 | 103.45 | 57.21 | 126.23 | 54.00 | 155.69 | 49.81 |
| BOG | T2 | ortho_benzene | 48.67 | 103.45 | 57.21 | 228.72 | 46.98 | 93.65 | 49.81 |
| BOG | T2 | meta_benzene | 16.12 | 109.04 | 27.10 | 59.17 | 18.96 | 107.25 | 25.92 |
| BOG | T2 | meta_benzene | 16.12 | 109.04 | 27.10 | 61.32 | 2.17 | 89.67 | 25.92 |
| BOG | T2 | meta_benzene | 16.12 | 109.04 | 27.10 | 59.17 | 18.96 | 107.25 | 25.92 |
| BOG | T2 | meta_benzene | 16.12 | 109.04 | 27.10 | 61.32 | 2.17 | 89.67 | 25.92 |
| BOG | T2 | meta_benzene | 16.12 | 109.04 | 27.10 | 59.17 | 18.96 | 107.25 | 25.92 |
| BOG | T2 | meta_benzene | 16.12 | 109.04 | 27.10 | 61.32 | 2.17 | 89.67 | 25.92 |
| BOG | T2 | meta_benzene | 16.12 | 109.04 | 27.10 | 59.17 | 18.96 | 107.25 | 25.92 |
| BOG | T2 | meta_benzene | 16.12 | 109.04 | 27.10 | 61.32 | 2.17 | 89.67 | 25.92 |
| BOG | T2 | para_benzene | 19.82 | 96.30 | 30.34 | 87.47 | 24.53 | 69.12 | 28.03 |
| BOG | T2 | para_benzene | 19.82 | 96.30 | 30.34 | 87.47 | 24.53 | 69.12 | 28.03 |
| BOG | T2 | para_benzene | 19.82 | 96.30 | 30.34 | 87.47 | 24.53 | 69.12 | 28.03 |
| BOG | T2 | para_benzene | 19.82 | 96.30 | 30.34 | 87.47 | 24.53 | 69.12 | 28.03 |
| BOG | T2 | ortho_ortho | 74.55 | 156.42 | 97.95 | 241.57 | 132.55 | 241.63 | 101.76 |
| BOG | T2 | ortho_ortho | 74.55 | 156.42 | 97.95 | 188.29 | 93.23 | 184.98 | 101.76 |
| BOG | T2 | ortho_ortho | 74.55 | 161.96 | 63.02 | 246.26 | 93.23 | 184.98 | 101.76 |
| BOG | T2 | ortho_ortho | 74.55 | 161.96 | 63.02 | 237.86 | 132.55 | 241.63 | 101.76 |
| BOG | T2 | ortho_ortho | 74.55 | 156.42 | 97.95 | 241.57 | 132.55 | 241.63 | 101.76 |
| BOG | T2 | ortho_ortho | 74.55 | 156.42 | 97.95 | 188.29 | 93.23 | 184.98 | 101.76 |
| BOG | T2 | ortho_ortho | 74.55 | 161.96 | 63.02 | 246.26 | 93.23 | 184.98 | 101.76 |
| BOG | T2 | ortho_ortho | 74.55 | 161.96 | 63.02 | 237.86 | 132.55 | 241.63 | 101.76 |
| BOG | T2 | ortho_ortho | 74.55 | 156.42 | 97.95 | 241.57 | 132.55 | 241.63 | 101.76 |
| BOG | T2 | ortho_ortho | 74.55 | 156.42 | 97.95 | 188.29 | 93.23 | 184.98 | 101.76 |
| BOG | T2 | ortho_ortho | 74.55 | 161.96 | 63.02 | 246.26 | 93.23 | 184.98 | 101.76 |
| BOG | T2 | ortho_ortho | 74.55 | 161.96 | 63.02 | 237.86 | 132.55 | 241.63 | 101.76 |
| BOG | T2 | ortho_ortho | 74.55 | 156.42 | 97.95 | 241.57 | 132.55 | 241.63 | 101.76 |
| BOG | T2 | ortho_ortho | 74.55 | 156.42 | 97.95 | 188.29 | 93.23 | 184.98 | 101.76 |
| BOG | T2 | ortho_ortho | 74.55 | 161.96 | 63.02 | 246.26 | 93.23 | 184.98 | 101.76 |
| BOG | T2 | ortho_ortho | 74.55 | 161.96 | 63.02 | 237.86 | 132.55 | 241.63 | 101.76 |
| BOG | T2 | ortho_meta | 150.65 | 189.28 | 124.07 | 237.19 | 160.08 | 221.04 | 136.12 |
| BOG | T2 | ortho_meta | 150.65 | 189.28 | 124.07 | 262.02 | 91.55 | 262.16 | 136.12 |
| BOG | T2 | ortho_meta | 150.65 | 189.28 | 124.07 | 237.19 | 160.08 | 221.04 | 136.12 |
| BOG | T2 | ortho_meta | 150.65 | 189.28 | 124.07 | 262.02 | 91.55 | 262.16 | 136.12 |
| BOG | T2 | ortho_meta | 150.65 | 189.28 | 124.07 | 237.19 | 160.08 | 221.04 | 136.12 |
| BOG | T2 | ortho_meta | 150.65 | 189.28 | 124.07 | 262.02 | 91.55 | 262.16 | 136.12 |
| BOG | T2 | ortho_para | 66.74 | 103.67 | 18.63 | 158.58 | 67.70 | 178.48 | 87.44 |
| BOG | T2 | ortho_para | 66.74 | 103.67 | 18.63 | 180.08 | 67.70 | 158.90 | 87.44 |
| BOG | T2 | ortho_para | 66.74 | 116.20 | 70.50 | 232.52 | 79.85 | 178.48 | 87.44 |
| BOG | T2 | ortho_para | 66.74 | 116.20 | 70.50 | 222.33 | 79.85 | 158.90 | 87.44 |
| BOG | T2 | ortho_para | 66.74 | 103.67 | 18.63 | 158.58 | 67.70 | 178.48 | 87.44 |
| BOG | T2 | ortho_para | 66.74 | 103.67 | 18.63 | 180.08 | 67.70 | 158.90 | 87.44 |
| BOG | T2 | ortho_para | 66.74 | 116.20 | 70.50 | 232.52 | 79.85 | 178.48 | 87.44 |
| BOG | T2 | ortho_para | 66.74 | 116.20 | 70.50 | 222.33 | 79.85 | 158.90 | 87.44 |
| BOG | T2 | ortho_para | 66.74 | 103.67 | 18.63 | 158.58 | 67.70 | 178.48 | 87.44 |
| BOG | T2 | ortho_para | 66.74 | 103.67 | 18.63 | 180.08 | 67.70 | 158.90 | 87.44 |
| BOG | T2 | ortho_para | 66.74 | 116.20 | 70.50 | 232.52 | 79.85 | 178.48 | 87.44 |
| BOG | T2 | ortho_para | 66.74 | 116.20 | 70.50 | 222.33 | 79.85 | 158.90 | 87.44 |
| BOG | T2 | meta_ortho | 79.14 | 147.90 | 95.26 | 271.49 | 66.61 | 179.58 | 104.86 |
| BOG | T2 | meta_ortho | 79.14 | 147.90 | 95.26 | 176.20 | 57.35 | 165.75 | 104.86 |
| BOG | T2 | meta_ortho | 79.14 | 208.20 | 29.55 | 267.54 | 66.61 | 179.58 | 104.86 |
| BOG | T2 | meta_ortho | 79.14 | 208.20 | 29.55 | 176.08 | 57.35 | 165.75 | 104.86 |
| BOG | T2 | meta_ortho | 79.14 | 147.90 | 95.26 | 271.49 | 66.61 | 179.58 | 104.86 |
| BOG | T2 | meta_ortho | 79.14 | 147.90 | 95.26 | 176.20 | 57.35 | 165.75 | 104.86 |
| BOG | T2 | meta_ortho | 79.14 | 208.20 | 29.55 | 267.54 | 66.61 | 179.58 | 104.86 |
| BOG | T2 | meta_ortho | 79.14 | 208.20 | 29.55 | 176.08 | 57.35 | 165.75 | 104.86 |
| BOG | T2 | meta_ortho | 79.14 | 147.90 | 95.26 | 271.49 | 66.61 | 179.58 | 104.86 |
| BOG | T2 | meta_ortho | 79.14 | 147.90 | 95.26 | 176.20 | 57.35 | 165.75 | 104.86 |
| BOG | T2 | meta_ortho | 79.14 | 208.20 | 29.55 | 267.54 | 66.61 | 179.58 | 104.86 |
| BOG | T2 | meta_ortho | 79.14 | 208.20 | 29.55 | 176.08 | 57.35 | 165.75 | 104.86 |
| BOG | T2 | meta_meta | 50.64 | 163.18 | 73.75 | 178.28 | 71.59 | 209.06 | 93.66 |
| BOG | T2 | meta_meta | 50.64 | 163.18 | 73.75 | 150.62 | 103.02 | 151.88 | 93.66 |
| BOG | T2 | meta_meta | 50.64 | 163.18 | 73.75 | 178.28 | 71.59 | 209.06 | 93.66 |
| BOG | T2 | meta_meta | 50.64 | 163.18 | 73.75 | 150.62 | 103.02 | 151.88 | 93.66 |
| BOG | T2 | meta_meta | 50.64 | 163.18 | 73.75 | 178.28 | 71.59 | 209.06 | 93.66 |
| BOG | T2 | meta_meta | 50.64 | 163.18 | 73.75 | 150.62 | 103.02 | 151.88 | 93.66 |
| BOG | T2 | meta_meta | 50.64 | 163.18 | 73.75 | 178.28 | 71.59 | 209.06 | 93.66 |
| BOG | T2 | meta_meta | 50.64 | 163.18 | 73.75 | 150.62 | 103.02 | 151.88 | 93.66 |
| BOG | T2 | meta_para | 19.18 | 104.72 | 10.65 | 138.45 | 33.48 | 103.70 | 63.81 |
| BOG | T2 | meta_para | 19.18 | 104.72 | 10.65 | 107.61 | 36.48 | 171.16 | 63.81 |
| BOG | T2 | meta_para | 19.18 | 75.25 | 24.39 | 122.50 | 36.48 | 171.16 | 63.81 |
| BOG | T2 | meta_para | 19.18 | 75.25 | 24.39 | 108.12 | 33.48 | 103.70 | 63.81 |
| BOG | T2 | meta_para | 19.18 | 104.72 | 10.65 | 138.45 | 33.48 | 103.70 | 63.81 |
| BOG | T2 | meta_para | 19.18 | 104.72 | 10.65 | 107.61 | 36.48 | 171.16 | 63.81 |
| BOG | T2 | meta_para | 19.18 | 75.25 | 24.39 | 122.50 | 36.48 | 171.16 | 63.81 |
| BOG | T2 | meta_para | 19.18 | 75.25 | 24.39 | 108.12 | 33.48 | 103.70 | 63.81 |
| BOG | T2 | meta_para | 19.18 | 104.72 | 10.65 | 138.45 | 33.48 | 103.70 | 63.81 |
| BOG | T2 | meta_para | 19.18 | 104.72 | 10.65 | 107.61 | 36.48 | 171.16 | 63.81 |
| BOG | T2 | meta_para | 19.18 | 75.25 | 24.39 | 122.50 | 36.48 | 171.16 | 63.81 |
| BOG | T2 | meta_para | 19.18 | 75.25 | 24.39 | 108.12 | 33.48 | 103.70 | 63.81 |
| BOG | T2 | para_ortho | 51.20 | 83.88 | 44.21 | 130.02 | 52.40 | 164.60 | 123.55 |
| BOG | T2 | para_ortho | 51.20 | 152.13 | 60.39 | 128.74 | 52.40 | 164.60 | 123.55 |
| BOG | T2 | para_ortho | 51.20 | 83.88 | 44.21 | 130.02 | 52.40 | 164.60 | 123.55 |
| BOG | T2 | para_ortho | 51.20 | 152.13 | 60.39 | 128.74 | 52.40 | 164.60 | 123.55 |
| BOG | T2 | para_ortho | 51.20 | 83.88 | 44.21 | 130.02 | 52.40 | 164.60 | 123.55 |
| BOG | T2 | para_ortho | 51.20 | 152.13 | 60.39 | 128.74 | 52.40 | 164.60 | 123.55 |
| BOG | T2 | para_meta | 61.72 | 97.63 | 83.97 | 117.28 | 2.07 | 125.93 | 72.43 |
| BOG | T2 | para_meta | 61.72 | 97.63 | 83.97 | 117.28 | 2.07 | 125.93 | 72.43 |
| BOG | T2 | para_meta | 61.72 | 97.63 | 83.97 | 117.28 | 2.07 | 125.93 | 72.43 |
| BOG | T2 | para_para | 25.31 | 104.44 | 40.90 | 110.89 | 28.70 | 102.59 | 73.98 |
| BOG | T2 | para_para | 25.31 | 89.25 | 32.14 | 113.81 | 28.70 | 102.59 | 73.98 |
| BOG | T2 | para_para | 25.31 | 104.44 | 40.90 | 110.89 | 28.70 | 102.59 | 73.98 |
| BOG | T2 | para_para | 25.31 | 89.25 | 32.14 | 113.81 | 28.70 | 102.59 | 73.98 |
| BOG | T2 | para_para | 25.31 | 104.44 | 40.90 | 110.89 | 28.70 | 102.59 | 73.98 |
| BOG | T2 | para_para | 25.31 | 89.25 | 32.14 | 113.81 | 28.70 | 102.59 | 73.98 |
| BOG | T2 | para_para | 25.31 | 104.44 | 40.90 | 110.89 | 28.70 | 102.59 | 73.98 |
| BOG | T2 | para_para | 25.31 | 89.25 | 32.14 | 113.81 | 28.70 | 102.59 | 73.98 |
| IWV | T3 | ortho_benzene | 4.80 | 88.95 | 34.65 | 102.38 | 17.48 | 69.25 | 20.89 |
| IWV | T3 | ortho_benzene | 4.80 | 88.95 | 34.65 | 99.55 | 38.23 | 70.61 | 20.89 |
| IWV | T3 | ortho_benzene | 4.80 | 88.95 | 34.65 | 102.38 | 17.48 | 69.25 | 20.89 |
| IWV | T3 | ortho_benzene | 4.80 | 88.95 | 34.65 | 99.55 | 38.23 | 70.61 | 20.89 |
| IWV | T3 | ortho_benzene | 4.80 | 88.95 | 34.65 | 102.38 | 17.48 | 69.25 | 20.89 |
| IWV | T3 | ortho_benzene | 4.80 | 88.95 | 34.65 | 99.55 | 38.23 | 70.61 | 20.89 |
| IWV | T3 | ortho_benzene | 4.80 | 88.95 | 34.65 | 102.38 | 17.48 | 69.25 | 20.89 |
| IWV | T3 | ortho_benzene | 4.80 | 88.95 | 34.65 | 99.55 | 38.23 | 70.61 | 20.89 |
| IWV | T3 | meta_benzene | 10.32 | 37.35 | 0.45 | 58.29 | 1.89 | 62.26 | 26.31 |
| IWV | T3 | meta_benzene | 10.32 | 37.35 | 0.45 | 46.31 | -8.98 | 52.30 | 26.31 |
| IWV | T3 | meta_benzene | 10.32 | 37.35 | 0.45 | 58.29 | 1.89 | 62.26 | 26.31 |
| IWV | T3 | meta_benzene | 10.32 | 37.35 | 0.45 | 46.31 | -8.98 | 52.30 | 26.31 |
| IWV | T3 | meta_benzene | 10.32 | 37.35 | 0.45 | 58.29 | 1.89 | 62.26 | 26.31 |
| IWV | T3 | meta_benzene | 10.32 | 37.35 | 0.45 | 46.31 | -8.98 | 52.30 | 26.31 |
| IWV | T3 | meta_benzene | 10.32 | 37.35 | 0.45 | 58.29 | 1.89 | 62.26 | 26.31 |
| IWV | T3 | meta_benzene | 10.32 | 37.35 | 0.45 | 46.31 | -8.98 | 52.30 | 26.31 |
| IWV | T3 | para_benzene | 28.80 | 82.48 | 3.52 | 65.99 | -1.52 | 47.42 | -2.82 |
| IWV | T3 | para_benzene | 28.80 | 82.48 | 3.52 | 65.99 | -1.52 | 47.42 | -2.82 |
| IWV | T3 | para_benzene | 28.80 | 82.48 | 3.52 | 65.99 | -1.52 | 47.42 | -2.82 |
| IWV | T3 | para_benzene | 28.80 | 82.48 | 3.52 | 65.99 | -1.52 | 47.42 | -2.82 |
| IWV | T3 | ortho_ortho | 53.36 | 116.71 | 54.65 | 141.59 | 74.92 | 129.71 | 41.23 |
| IWV | T3 | ortho_ortho | 53.36 | 116.71 | 54.65 | 178.07 | 45.09 | 126.81 | 41.23 |
| IWV | T3 | ortho_ortho | 53.36 | 154.70 | 28.97 | 146.41 | 74.92 | 129.71 | 41.23 |
| IWV | T3 | ortho_ortho | 53.36 | 154.70 | 28.97 | 130.96 | 45.09 | 126.81 | 41.23 |
| IWV | T3 | ortho_ortho | 53.36 | 116.71 | 54.65 | 141.59 | 74.92 | 129.71 | 41.23 |
| IWV | T3 | ortho_ortho | 53.36 | 116.71 | 54.65 | 178.07 | 45.09 | 126.81 | 41.23 |
| IWV | T3 | ortho_ortho | 53.36 | 154.70 | 28.97 | 146.41 | 74.92 | 129.71 | 41.23 |
| IWV | T3 | ortho_ortho | 53.36 | 154.70 | 28.97 | 130.96 | 45.09 | 126.81 | 41.23 |
| IWV | T3 | ortho_ortho | 53.36 | 116.71 | 54.65 | 141.59 | 74.92 | 129.71 | 41.23 |
| IWV | T3 | ortho_ortho | 53.36 | 116.71 | 54.65 | 178.07 | 45.09 | 126.81 | 41.23 |
| IWV | T3 | ortho_ortho | 53.36 | 154.70 | 28.97 | 146.41 | 74.92 | 129.71 | 41.23 |
| IWV | T3 | ortho_ortho | 53.36 | 154.70 | 28.97 | 130.96 | 45.09 | 126.81 | 41.23 |
| IWV | T3 | ortho_ortho | 53.36 | 116.71 | 54.65 | 141.59 | 74.92 | 129.71 | 41.23 |
| IWV | T3 | ortho_ortho | 53.36 | 116.71 | 54.65 | 178.07 | 45.09 | 126.81 | 41.23 |
| IWV | T3 | ortho_ortho | 53.36 | 154.70 | 28.97 | 146.41 | 74.92 | 129.71 | 41.23 |
| IWV | T3 | ortho_ortho | 53.36 | 154.70 | 28.97 | 130.96 | 45.09 | 126.81 | 41.23 |
| IWV | T3 | ortho_meta | 51.58 | 163.23 | 97.13 | 192.78 | 95.69 | 184.80 | 113.33 |
| IWV | T3 | ortho_meta | 51.58 | 163.23 | 97.13 | 245.14 | 125.55 | 225.04 | 113.33 |
| IWV | T3 | ortho_meta | 51.58 | 163.23 | 97.13 | 192.78 | 95.69 | 184.80 | 113.33 |
| IWV | T3 | ortho_meta | 51.58 | 163.23 | 97.13 | 245.14 | 125.55 | 225.04 | 113.33 |
| IWV | T3 | ortho_meta | 51.58 | 163.23 | 97.13 | 192.78 | 95.69 | 184.80 | 113.33 |
| IWV | T3 | ortho_meta | 51.58 | 163.23 | 97.13 | 245.14 | 125.55 | 225.04 | 113.33 |
| IWV | T3 | ortho_para | 53.40 | 112.15 | 47.24 | 166.00 | 76.48 | 166.66 | 63.38 |
| IWV | T3 | ortho_para | 53.40 | 112.15 | 47.24 | 158.66 | 76.48 | 172.29 | 63.38 |
| IWV | T3 | ortho_para | 53.40 | 134.18 | 14.48 | 155.12 | 56.85 | 166.66 | 63.38 |
| IWV | T3 | ortho_para | 53.40 | 134.18 | 14.48 | 122.98 | 56.85 | 172.29 | 63.38 |
| IWV | T3 | ortho_para | 53.40 | 112.15 | 47.24 | 166.00 | 76.48 | 166.66 | 63.38 |
| IWV | T3 | ortho_para | 53.40 | 112.15 | 47.24 | 158.66 | 76.48 | 172.29 | 63.38 |
| IWV | T3 | ortho_para | 53.40 | 134.18 | 14.48 | 155.12 | 56.85 | 166.66 | 63.38 |
| IWV | T3 | ortho_para | 53.40 | 134.18 | 14.48 | 122.98 | 56.85 | 172.29 | 63.38 |
| IWV | T3 | ortho_para | 53.40 | 112.15 | 47.24 | 166.00 | 76.48 | 166.66 | 63.38 |
| IWV | T3 | ortho_para | 53.40 | 112.15 | 47.24 | 158.66 | 76.48 | 172.29 | 63.38 |
| IWV | T3 | ortho_para | 53.40 | 134.18 | 14.48 | 155.12 | 56.85 | 166.66 | 63.38 |
| IWV | T3 | ortho_para | 53.40 | 134.18 | 14.48 | 122.98 | 56.85 | 172.29 | 63.38 |
| IWV | T3 | meta_ortho | 70.37 | 164.86 | 44.29 | 120.04 | 62.51 | 97.71 | 57.23 |
| IWV | T3 | meta_ortho | 70.37 | 164.86 | 44.29 | 123.72 | 12.41 | 150.97 | 57.23 |
| IWV | T3 | meta_ortho | 70.37 | 141.96 | 22.41 | 115.83 | 62.51 | 97.71 | 57.23 |
| IWV | T3 | meta_ortho | 70.37 | 141.96 | 22.41 | 121.99 | 12.41 | 150.97 | 57.23 |
| IWV | T3 | meta_ortho | 70.37 | 164.86 | 44.29 | 120.04 | 62.51 | 97.71 | 57.23 |
| IWV | T3 | meta_ortho | 70.37 | 164.86 | 44.29 | 123.72 | 12.41 | 150.97 | 57.23 |
| IWV | T3 | meta_ortho | 70.37 | 141.96 | 22.41 | 115.83 | 62.51 | 97.71 | 57.23 |
| IWV | T3 | meta_ortho | 70.37 | 141.96 | 22.41 | 121.99 | 12.41 | 150.97 | 57.23 |
| IWV | T3 | meta_ortho | 70.37 | 164.86 | 44.29 | 120.04 | 62.51 | 97.71 | 57.23 |
| IWV | T3 | meta_ortho | 70.37 | 164.86 | 44.29 | 123.72 | 12.41 | 150.97 | 57.23 |
| IWV | T3 | meta_ortho | 70.37 | 141.96 | 22.41 | 115.83 | 62.51 | 97.71 | 57.23 |
| IWV | T3 | meta_ortho | 70.37 | 141.96 | 22.41 | 121.99 | 12.41 | 150.97 | 57.23 |
| IWV | T3 | meta_meta | 36.23 | 101.19 | 68.07 | 154.18 | 29.25 | 145.12 | 68.12 |
| IWV | T3 | meta_meta | 36.23 | 101.19 | 68.07 | 152.07 | 75.37 | 118.91 | 68.12 |
| IWV | T3 | meta_meta | 36.23 | 101.19 | 68.07 | 154.18 | 29.25 | 145.12 | 68.12 |
| IWV | T3 | meta_meta | 36.23 | 101.19 | 68.07 | 152.07 | 75.37 | 118.91 | 68.12 |
| IWV | T3 | meta_meta | 36.23 | 101.19 | 68.07 | 154.18 | 29.25 | 145.12 | 68.12 |
| IWV | T3 | meta_meta | 36.23 | 101.19 | 68.07 | 152.07 | 75.37 | 118.91 | 68.12 |
| IWV | T3 | meta_meta | 36.23 | 101.19 | 68.07 | 154.18 | 29.25 | 145.12 | 68.12 |
| IWV | T3 | meta_meta | 36.23 | 101.19 | 68.07 | 152.07 | 75.37 | 118.91 | 68.12 |
| IWV | T3 | meta_para | 11.33 | 110.10 | 22.68 | 122.19 | 21.44 | 95.92 | 8.40 |
| IWV | T3 | meta_para | 11.33 | 110.10 | 22.68 | 112.04 | 21.54 | 111.24 | 8.40 |
| IWV | T3 | meta_para | 11.33 | 82.82 | 18.34 | 122.72 | 21.54 | 111.24 | 8.40 |
| IWV | T3 | meta_para | 11.33 | 82.82 | 18.34 | 92.46 | 21.44 | 95.92 | 8.40 |
| IWV | T3 | meta_para | 11.33 | 110.10 | 22.68 | 122.19 | 21.44 | 95.92 | 8.40 |
| IWV | T3 | meta_para | 11.33 | 110.10 | 22.68 | 112.04 | 21.54 | 111.24 | 8.40 |
| IWV | T3 | meta_para | 11.33 | 82.82 | 18.34 | 122.72 | 21.54 | 111.24 | 8.40 |
| IWV | T3 | meta_para | 11.33 | 82.82 | 18.34 | 92.46 | 21.44 | 95.92 | 8.40 |
| IWV | T3 | meta_para | 11.33 | 110.10 | 22.68 | 122.19 | 21.44 | 95.92 | 8.40 |
| IWV | T3 | meta_para | 11.33 | 110.10 | 22.68 | 112.04 | 21.54 | 111.24 | 8.40 |
| IWV | T3 | meta_para | 11.33 | 82.82 | 18.34 | 122.72 | 21.54 | 111.24 | 8.40 |
| IWV | T3 | meta_para | 11.33 | 82.82 | 18.34 | 92.46 | 21.44 | 95.92 | 8.40 |
| IWV | T3 | para_ortho | 41.06 | 105.00 | 20.24 | 103.01 | 52.43 | 83.11 | 46.42 |
| IWV | T3 | para_ortho | 41.06 | 93.58 | 29.09 | 117.47 | 52.43 | 83.11 | 46.42 |
| IWV | T3 | para_ortho | 41.06 | 105.00 | 20.24 | 103.01 | 52.43 | 83.11 | 46.42 |
| IWV | T3 | para_ortho | 41.06 | 93.58 | 29.09 | 117.47 | 52.43 | 83.11 | 46.42 |
| IWV | T3 | para_ortho | 41.06 | 105.00 | 20.24 | 103.01 | 52.43 | 83.11 | 46.42 |
| IWV | T3 | para_ortho | 41.06 | 93.58 | 29.09 | 117.47 | 52.43 | 83.11 | 46.42 |
| IWV | T3 | para_meta | 55.01 | 118.41 | 71.30 | 121.47 | 59.89 | 154.49 | 68.01 |
| IWV | T3 | para_meta | 55.01 | 118.41 | 71.30 | 121.47 | 59.89 | 154.49 | 68.01 |
| IWV | T3 | para_meta | 55.01 | 118.41 | 71.30 | 121.47 | 59.89 | 154.49 | 68.01 |
| IWV | T3 | para_para | 0.00 | 48.88 | 7.81 | 142.56 | 26.79 | 104.57 | 45.89 |
| IWV | T3 | para_para | 0.00 | 115.48 | 31.52 | 111.67 | 26.79 | 104.57 | 45.89 |
| IWV | T3 | para_para | 0.00 | 48.88 | 7.81 | 142.56 | 26.79 | 104.57 | 45.89 |
| IWV | T3 | para_para | 0.00 | 115.48 | 31.52 | 111.67 | 26.79 | 104.57 | 45.89 |
| IWV | T3 | para_para | 0.00 | 48.88 | 7.81 | 142.56 | 26.79 | 104.57 | 45.89 |
| IWV | T3 | para_para | 0.00 | 115.48 | 31.52 | 111.67 | 26.79 | 104.57 | 45.89 |
| IWV | T3 | para_para | 0.00 | 48.88 | 7.81 | 142.56 | 26.79 | 104.57 | 45.89 |
| IWV | T3 | para_para | 0.00 | 115.48 | 31.52 | 111.67 | 26.79 | 104.57 | 45.89 |
| IWV | T7 | ortho_benzene | 21.80 | 85.15 | 32.09 | 164.59 | 98.17 | 134.51 | 28.02 |
| IWV | T7 | ortho_benzene | 21.80 | 85.15 | 32.09 | 86.85 | 39.43 | 111.46 | 28.02 |
| IWV | T7 | ortho_benzene | 21.80 | 85.15 | 32.09 | 164.59 | 98.17 | 134.51 | 28.02 |
| IWV | T7 | ortho_benzene | 21.80 | 85.15 | 32.09 | 86.85 | 39.43 | 111.46 | 28.02 |
| IWV | T7 | ortho_benzene | 21.80 | 85.15 | 32.09 | 164.59 | 98.17 | 134.51 | 28.02 |
| IWV | T7 | ortho_benzene | 21.80 | 85.15 | 32.09 | 86.85 | 39.43 | 111.46 | 28.02 |
| IWV | T7 | ortho_benzene | 21.80 | 85.15 | 32.09 | 164.59 | 98.17 | 134.51 | 28.02 |
| IWV | T7 | ortho_benzene | 21.80 | 85.15 | 32.09 | 86.85 | 39.43 | 111.46 | 28.02 |
| IWV | T7 | meta_benzene | 79.26 | 127.30 | 52.95 | 112.57 | 34.06 | 93.38 | 62.23 |
| IWV | T7 | meta_benzene | 79.26 | 127.30 | 52.95 | 102.19 | 32.17 | 119.32 | 62.23 |
| IWV | T7 | meta_benzene | 79.26 | 127.30 | 52.95 | 112.57 | 34.06 | 93.38 | 62.23 |
| IWV | T7 | meta_benzene | 79.26 | 127.30 | 52.95 | 102.19 | 32.17 | 119.32 | 62.23 |
| IWV | T7 | meta_benzene | 79.26 | 127.30 | 52.95 | 112.57 | 34.06 | 93.38 | 62.23 |
| IWV | T7 | meta_benzene | 79.26 | 127.30 | 52.95 | 102.19 | 32.17 | 119.32 | 62.23 |
| IWV | T7 | meta_benzene | 79.26 | 127.30 | 52.95 | 112.57 | 34.06 | 93.38 | 62.23 |
| IWV | T7 | meta_benzene | 79.26 | 127.30 | 52.95 | 102.19 | 32.17 | 119.32 | 62.23 |
| IWV | T7 | para_benzene | 0.00 | 78.04 | 29.60 | 94.19 | 7.44 | 52.75 | 10.19 |
| IWV | T7 | para_benzene | 0.00 | 78.04 | 29.60 | 94.19 | 7.44 | 52.75 | 10.19 |
| IWV | T7 | para_benzene | 0.00 | 78.04 | 29.60 | 94.19 | 7.44 | 52.75 | 10.19 |
| IWV | T7 | para_benzene | 0.00 | 78.04 | 29.60 | 94.19 | 7.44 | 52.75 | 10.19 |
| IWV | T7 | ortho_ortho | 61.18 | 138.64 | 72.33 | 168.21 | 70.93 | 147.55 | 102.46 |
| IWV | T7 | ortho_ortho | 61.18 | 138.64 | 72.33 | 271.74 | 56.65 | 233.80 | 102.46 |
| IWV | T7 | ortho_ortho | 61.18 | 126.41 | 45.84 | 190.01 | 70.93 | 147.55 | 102.46 |
| IWV | T7 | ortho_ortho | 61.18 | 126.41 | 45.84 | 164.55 | 56.65 | 233.80 | 102.46 |
| IWV | T7 | ortho_ortho | 61.18 | 138.64 | 72.33 | 168.21 | 70.93 | 147.55 | 102.46 |
| IWV | T7 | ortho_ortho | 61.18 | 138.64 | 72.33 | 271.74 | 56.65 | 233.80 | 102.46 |
| IWV | T7 | ortho_ortho | 61.18 | 126.41 | 45.84 | 190.01 | 70.93 | 147.55 | 102.46 |
| IWV | T7 | ortho_ortho | 61.18 | 126.41 | 45.84 | 164.55 | 56.65 | 233.80 | 102.46 |
| IWV | T7 | ortho_ortho | 61.18 | 138.64 | 72.33 | 168.21 | 70.93 | 147.55 | 102.46 |
| IWV | T7 | ortho_ortho | 61.18 | 138.64 | 72.33 | 271.74 | 56.65 | 233.80 | 102.46 |
| IWV | T7 | ortho_ortho | 61.18 | 126.41 | 45.84 | 190.01 | 70.93 | 147.55 | 102.46 |
| IWV | T7 | ortho_ortho | 61.18 | 126.41 | 45.84 | 164.55 | 56.65 | 233.80 | 102.46 |
| IWV | T7 | ortho_ortho | 61.18 | 138.64 | 72.33 | 168.21 | 70.93 | 147.55 | 102.46 |
| IWV | T7 | ortho_ortho | 61.18 | 138.64 | 72.33 | 271.74 | 56.65 | 233.80 | 102.46 |
| IWV | T7 | ortho_ortho | 61.18 | 126.41 | 45.84 | 190.01 | 70.93 | 147.55 | 102.46 |
| IWV | T7 | ortho_ortho | 61.18 | 126.41 | 45.84 | 164.55 | 56.65 | 233.80 | 102.46 |
| IWV | T7 | ortho_meta | 88.58 | 159.69 | 89.08 | 228.60 | 92.85 | 217.24 | 82.49 |
| IWV | T7 | ortho_meta | 88.58 | 159.69 | 89.08 | 156.01 | 101.18 | 189.58 | 82.49 |
| IWV | T7 | ortho_meta | 88.58 | 159.69 | 89.08 | 228.60 | 92.85 | 217.24 | 82.49 |
| IWV | T7 | ortho_meta | 88.58 | 159.69 | 89.08 | 156.01 | 101.18 | 189.58 | 82.49 |
| IWV | T7 | ortho_meta | 88.58 | 159.69 | 89.08 | 228.60 | 92.85 | 217.24 | 82.49 |
| IWV | T7 | ortho_meta | 88.58 | 159.69 | 89.08 | 156.01 | 101.18 | 189.58 | 82.49 |
| IWV | T7 | ortho_para | 66.71 | 163.57 | 79.60 | 139.57 | 71.47 | 144.39 | 83.97 |
| IWV | T7 | ortho_para | 66.71 | 163.57 | 79.60 | 147.47 | 71.47 | 173.32 | 83.97 |
| IWV | T7 | ortho_para | 66.71 | 115.43 | 61.39 | 177.53 | 55.38 | 173.32 | 83.97 |
| IWV | T7 | ortho_para | 66.71 | 115.43 | 61.39 | 130.57 | 55.38 | 144.39 | 83.97 |
| IWV | T7 | ortho_para | 66.71 | 163.57 | 79.60 | 139.57 | 71.47 | 144.39 | 83.97 |
| IWV | T7 | ortho_para | 66.71 | 163.57 | 79.60 | 147.47 | 71.47 | 173.32 | 83.97 |
| IWV | T7 | ortho_para | 66.71 | 115.43 | 61.39 | 177.53 | 55.38 | 173.32 | 83.97 |
| IWV | T7 | ortho_para | 66.71 | 115.43 | 61.39 | 130.57 | 55.38 | 144.39 | 83.97 |
| IWV | T7 | ortho_para | 66.71 | 163.57 | 79.60 | 139.57 | 71.47 | 144.39 | 83.97 |
| IWV | T7 | ortho_para | 66.71 | 163.57 | 79.60 | 147.47 | 71.47 | 173.32 | 83.97 |
| IWV | T7 | ortho_para | 66.71 | 115.43 | 61.39 | 177.53 | 55.38 | 173.32 | 83.97 |
| IWV | T7 | ortho_para | 66.71 | 115.43 | 61.39 | 130.57 | 55.38 | 144.39 | 83.97 |
| IWV | T7 | meta_ortho | 88.45 | 148.15 | 70.91 | 121.15 | 56.59 | 156.28 | 30.53 |
| IWV | T7 | meta_ortho | 88.45 | 148.15 | 70.91 | 117.66 | 54.88 | 223.31 | 30.53 |
| IWV | T7 | meta_ortho | 88.45 | 175.81 | 48.00 | 118.29 | 56.59 | 156.28 | 30.53 |
| IWV | T7 | meta_ortho | 88.45 | 175.81 | 48.00 | 117.25 | 54.88 | 223.31 | 30.53 |
| IWV | T7 | meta_ortho | 88.45 | 148.15 | 70.91 | 121.15 | 56.59 | 156.28 | 30.53 |
| IWV | T7 | meta_ortho | 88.45 | 148.15 | 70.91 | 117.66 | 54.88 | 223.31 | 30.53 |
| IWV | T7 | meta_ortho | 88.45 | 175.81 | 48.00 | 118.29 | 56.59 | 156.28 | 30.53 |
| IWV | T7 | meta_ortho | 88.45 | 175.81 | 48.00 | 117.25 | 54.88 | 223.31 | 30.53 |
| IWV | T7 | meta_ortho | 88.45 | 148.15 | 70.91 | 121.15 | 56.59 | 156.28 | 30.53 |
| IWV | T7 | meta_ortho | 88.45 | 148.15 | 70.91 | 117.66 | 54.88 | 223.31 | 30.53 |
| IWV | T7 | meta_ortho | 88.45 | 175.81 | 48.00 | 118.29 | 56.59 | 156.28 | 30.53 |
| IWV | T7 | meta_ortho | 88.45 | 175.81 | 48.00 | 117.25 | 54.88 | 223.31 | 30.53 |
| IWV | T7 | meta_meta | 55.55 | 119.56 | 66.90 | 156.07 | 66.04 | 124.27 | 68.49 |
| IWV | T7 | meta_meta | 55.55 | 119.56 | 66.90 | 143.88 | 89.21 | 190.69 | 68.49 |
| IWV | T7 | meta_meta | 55.55 | 119.56 | 66.90 | 156.07 | 66.04 | 124.27 | 68.49 |
| IWV | T7 | meta_meta | 55.55 | 119.56 | 66.90 | 143.88 | 89.21 | 190.69 | 68.49 |
| IWV | T7 | meta_meta | 55.55 | 119.56 | 66.90 | 156.07 | 66.04 | 124.27 | 68.49 |
| IWV | T7 | meta_meta | 55.55 | 119.56 | 66.90 | 143.88 | 89.21 | 190.69 | 68.49 |
| IWV | T7 | meta_meta | 55.55 | 119.56 | 66.90 | 156.07 | 66.04 | 124.27 | 68.49 |
| IWV | T7 | meta_meta | 55.55 | 119.56 | 66.90 | 143.88 | 89.21 | 190.69 | 68.49 |
| IWV | T7 | meta_para | 45.51 | 146.56 | 30.68 | 153.00 | 36.39 | 97.87 | 61.05 |
| IWV | T7 | meta_para | 45.51 | 146.56 | 30.68 | 113.25 | 32.73 | 143.19 | 61.05 |
| IWV | T7 | meta_para | 45.51 | 141.16 | 25.37 | 165.39 | 32.73 | 143.19 | 61.05 |
| IWV | T7 | meta_para | 45.51 | 141.16 | 25.37 | 120.06 | 36.39 | 97.87 | 61.05 |
| IWV | T7 | meta_para | 45.51 | 146.56 | 30.68 | 153.00 | 36.39 | 97.87 | 61.05 |
| IWV | T7 | meta_para | 45.51 | 146.56 | 30.68 | 113.25 | 32.73 | 143.19 | 61.05 |
| IWV | T7 | meta_para | 45.51 | 141.16 | 25.37 | 165.39 | 32.73 | 143.19 | 61.05 |
| IWV | T7 | meta_para | 45.51 | 141.16 | 25.37 | 120.06 | 36.39 | 97.87 | 61.05 |
| IWV | T7 | meta_para | 45.51 | 146.56 | 30.68 | 153.00 | 36.39 | 97.87 | 61.05 |
| IWV | T7 | meta_para | 45.51 | 146.56 | 30.68 | 113.25 | 32.73 | 143.19 | 61.05 |
| IWV | T7 | meta_para | 45.51 | 141.16 | 25.37 | 165.39 | 32.73 | 143.19 | 61.05 |
| IWV | T7 | meta_para | 45.51 | 141.16 | 25.37 | 120.06 | 36.39 | 97.87 | 61.05 |
| IWV | T7 | para_ortho | 74.89 | 122.27 | 49.45 | 144.13 | 46.21 | 119.77 | 71.80 |
| IWV | T7 | para_ortho | 74.89 | 105.42 | 49.64 | 182.48 | 46.21 | 119.77 | 71.80 |
| IWV | T7 | para_ortho | 74.89 | 122.27 | 49.45 | 144.13 | 46.21 | 119.77 | 71.80 |
| IWV | T7 | para_ortho | 74.89 | 105.42 | 49.64 | 182.48 | 46.21 | 119.77 | 71.80 |
| IWV | T7 | para_ortho | 74.89 | 122.27 | 49.45 | 144.13 | 46.21 | 119.77 | 71.80 |
| IWV | T7 | para_ortho | 74.89 | 105.42 | 49.64 | 182.48 | 46.21 | 119.77 | 71.80 |
| IWV | T7 | para_meta | 56.68 | 147.20 | 64.90 | 202.71 | 61.69 | 150.24 | 87.50 |
| IWV | T7 | para_meta | 56.68 | 147.20 | 64.90 | 202.71 | 61.69 | 150.24 | 87.50 |
| IWV | T7 | para_meta | 56.68 | 147.20 | 64.90 | 202.71 | 61.69 | 150.24 | 87.50 |
| IWV | T7 | para_para | 37.46 | 81.99 | 39.12 | 123.62 | 49.73 | 133.45 | 65.56 |
| IWV | T7 | para_para | 37.46 | 103.95 | 42.07 | 153.00 | 49.73 | 133.45 | 65.56 |
| IWV | T7 | para_para | 37.46 | 81.99 | 39.12 | 123.62 | 49.73 | 133.45 | 65.56 |
| IWV | T7 | para_para | 37.46 | 103.95 | 42.07 | 153.00 | 49.73 | 133.45 | 65.56 |
| IWV | T7 | para_para | 37.46 | 81.99 | 39.12 | 123.62 | 49.73 | 133.45 | 65.56 |
| IWV | T7 | para_para | 37.46 | 103.95 | 42.07 | 153.00 | 49.73 | 133.45 | 65.56 |
| IWV | T7 | para_para | 37.46 | 81.99 | 39.12 | 123.62 | 49.73 | 133.45 | 65.56 |
| IWV | T7 | para_para | 37.46 | 103.95 | 42.07 | 153.00 | 49.73 | 133.45 | 65.56 |
| UTL | T2 | ortho_benzene | 63.85 | 96.04 | 15.31 | 64.38 | 47.57 | 132.78 | 17.74 |
| UTL | T2 | ortho_benzene | 63.85 | 96.04 | 15.31 | 102.43 | 41.10 | 90.87 | 17.74 |
| UTL | T2 | ortho_benzene | 63.85 | 96.04 | 15.31 | 64.38 | 47.57 | 132.78 | 17.74 |
| UTL | T2 | ortho_benzene | 63.85 | 96.04 | 15.31 | 102.43 | 41.10 | 90.87 | 17.74 |
| UTL | T2 | ortho_benzene | 63.85 | 96.04 | 15.31 | 64.38 | 47.57 | 132.78 | 17.74 |
| UTL | T2 | ortho_benzene | 63.85 | 96.04 | 15.31 | 102.43 | 41.10 | 90.87 | 17.74 |
| UTL | T2 | ortho_benzene | 63.85 | 96.04 | 15.31 | 64.38 | 47.57 | 132.78 | 17.74 |
| UTL | T2 | ortho_benzene | 63.85 | 96.04 | 15.31 | 102.43 | 41.10 | 90.87 | 17.74 |
| UTL | T2 | meta_benzene | 51.60 | 75.09 | 12.68 | 99.43 | -7.05 | 54.62 | 3.71 |
| UTL | T2 | meta_benzene | 51.60 | 75.09 | 12.68 | 70.75 | -13.93 | 52.06 | 3.71 |
| UTL | T2 | meta_benzene | 51.60 | 75.09 | 12.68 | 99.43 | -7.05 | 54.62 | 3.71 |
| UTL | T2 | meta_benzene | 51.60 | 75.09 | 12.68 | 70.75 | -13.93 | 52.06 | 3.71 |
| UTL | T2 | meta_benzene | 51.60 | 75.09 | 12.68 | 99.43 | -7.05 | 54.62 | 3.71 |
| UTL | T2 | meta_benzene | 51.60 | 75.09 | 12.68 | 70.75 | -13.93 | 52.06 | 3.71 |
| UTL | T2 | meta_benzene | 51.60 | 75.09 | 12.68 | 99.43 | -7.05 | 54.62 | 3.71 |
| UTL | T2 | meta_benzene | 51.60 | 75.09 | 12.68 | 70.75 | -13.93 | 52.06 | 3.71 |
| UTL | T2 | para_benzene | 0.00 | 62.79 | -59.22 | 18.60 | 0.67 | 43.15 | 19.81 |
| UTL | T2 | para_benzene | 0.00 | 62.79 | -59.22 | 18.60 | 0.67 | 43.15 | 19.81 |
| UTL | T2 | para_benzene | 0.00 | 62.79 | -59.22 | 18.60 | 0.67 | 43.15 | 19.81 |
| UTL | T2 | para_benzene | 0.00 | 62.79 | -59.22 | 18.60 | 0.67 | 43.15 | 19.81 |
| UTL | T2 | ortho_ortho | 27.56 | 59.04 | 4.55 | 120.65 | 28.78 | 129.37 | 21.25 |
| UTL | T2 | ortho_ortho | 27.56 | 59.04 | 4.55 | 119.65 | 20.13 | 202.11 | 21.25 |
| UTL | T2 | ortho_ortho | 27.56 | 63.07 | -8.90 | 124.37 | 20.13 | 102.11 | 21.25 |
| UTL | T2 | ortho_ortho | 27.56 | 63.07 | -8.90 | 77.84 | 28.78 | 109.37 | 21.25 |
| UTL | T2 | ortho_ortho | 27.56 | 59.04 | 4.55 | 120.65 | 28.78 | 129.37 | 21.25 |
| UTL | T2 | ortho_ortho | 27.56 | 59.04 | 4.55 | 119.65 | 20.13 | 202.11 | 21.25 |
| UTL | T2 | ortho_ortho | 27.56 | 63.07 | -8.90 | 124.37 | 20.13 | 102.11 | 21.25 |
| UTL | T2 | ortho_ortho | 27.56 | 63.07 | -8.90 | 77.84 | 28.78 | 109.37 | 21.25 |
| UTL | T2 | ortho_ortho | 27.56 | 59.04 | 4.55 | 120.65 | 28.78 | 129.37 | 21.25 |
| UTL | T2 | ortho_ortho | 27.56 | 59.04 | 4.55 | 119.65 | 20.13 | 202.11 | 21.25 |
| UTL | T2 | ortho_ortho | 27.56 | 63.07 | -8.90 | 124.37 | 20.13 | 102.11 | 21.25 |
| UTL | T2 | ortho_ortho | 27.56 | 63.07 | -8.90 | 77.84 | 28.78 | 109.37 | 21.25 |
| UTL | T2 | ortho_ortho | 27.56 | 59.04 | 4.55 | 120.65 | 28.78 | 129.37 | 21.25 |
| UTL | T2 | ortho_ortho | 27.56 | 59.04 | 4.55 | 119.65 | 20.13 | 202.11 | 21.25 |
| UTL | T2 | ortho_ortho | 27.56 | 63.07 | -8.90 | 124.37 | 20.13 | 102.11 | 21.25 |
| UTL | T2 | ortho_ortho | 27.56 | 63.07 | -8.90 | 77.84 | 28.78 | 109.37 | 21.25 |
| UTL | T2 | ortho_meta | 14.79 | 86.37 | 50.69 | 151.17 | 53.18 | 109.32 | 56.69 |
| UTL | T2 | ortho_meta | 14.79 | 86.37 | 50.69 | 88.20 | 48.15 | 132.32 | 56.69 |
| UTL | T2 | ortho_meta | 14.79 | 86.37 | 50.69 | 151.17 | 53.18 | 109.32 | 56.69 |
| UTL | T2 | ortho_meta | 14.79 | 86.37 | 50.69 | 88.20 | 48.15 | 132.32 | 56.69 |
| UTL | T2 | ortho_meta | 14.79 | 86.37 | 50.69 | 151.17 | 53.18 | 109.32 | 56.69 |
| UTL | T2 | ortho_meta | 14.79 | 86.37 | 50.69 | 88.20 | 48.15 | 132.32 | 56.69 |
| UTL | T2 | ortho_para | 6.41 | 84.13 | 7.46 | 121.04 | 33.53 | 83.86 | 3.81 |
| UTL | T2 | ortho_para | 6.41 | 84.13 | 7.46 | 69.88 | 33.53 | 139.60 | 3.81 |
| UTL | T2 | ortho_para | 6.41 | 96.83 | -16.74 | 108.16 | -6.26 | 83.86 | 3.81 |
| UTL | T2 | ortho_para | 6.41 | 96.83 | -16.74 | 120.46 | -6.26 | 139.60 | 3.81 |
| UTL | T2 | ortho_para | 6.41 | 84.13 | 7.46 | 121.04 | 33.53 | 83.86 | 3.81 |
| UTL | T2 | ortho_para | 6.41 | 84.13 | 7.46 | 69.88 | 33.53 | 139.60 | 3.81 |
| UTL | T2 | ortho_para | 6.41 | 96.83 | -16.74 | 108.16 | -6.26 | 83.86 | 3.81 |
| UTL | T2 | ortho_para | 6.41 | 96.83 | -16.74 | 120.46 | -6.26 | 139.60 | 3.81 |
| UTL | T2 | ortho_para | 6.41 | 84.13 | 7.46 | 121.04 | 33.53 | 83.86 | 3.81 |
| UTL | T2 | ortho_para | 6.41 | 84.13 | 7.46 | 69.88 | 33.53 | 139.60 | 3.81 |
| UTL | T2 | ortho_para | 6.41 | 96.83 | -16.74 | 108.16 | -6.26 | 83.86 | 3.81 |
| UTL | T2 | ortho_para | 6.41 | 96.83 | -16.74 | 120.46 | -6.26 | 139.60 | 3.81 |
| UTL | T2 | meta_ortho | 17.79 | 85.59 | -4.90 | 67.00 | -20.11 | 140.69 | 14.61 |
| UTL | T2 | meta_ortho | 17.79 | 85.59 | -4.90 | 83.01 | -7.42 | 141.21 | 14.61 |
| UTL | T2 | meta_ortho | 17.79 | 51.73 | 11.56 | 76.90 | -20.11 | 140.69 | 14.61 |
| UTL | T2 | meta_ortho | 17.79 | 51.73 | 11.56 | 87.77 | -7.42 | 141.21 | 14.61 |
| UTL | T2 | meta_ortho | 17.79 | 85.59 | -4.90 | 67.00 | -20.11 | 140.69 | 14.61 |
| UTL | T2 | meta_ortho | 17.79 | 85.59 | -4.90 | 83.01 | -7.42 | 141.21 | 14.61 |
| UTL | T2 | meta_ortho | 17.79 | 51.73 | 11.56 | 76.90 | -20.11 | 140.69 | 14.61 |
| UTL | T2 | meta_ortho | 17.79 | 51.73 | 11.56 | 87.77 | -7.42 | 141.21 | 14.61 |
| UTL | T2 | meta_ortho | 17.79 | 85.59 | -4.90 | 67.00 | -20.11 | 140.69 | 14.61 |
| UTL | T2 | meta_ortho | 17.79 | 85.59 | -4.90 | 83.01 | -7.42 | 141.21 | 14.61 |
| UTL | T2 | meta_ortho | 17.79 | 51.73 | 11.56 | 76.90 | -20.11 | 140.69 | 14.61 |
| UTL | T2 | meta_ortho | 17.79 | 51.73 | 11.56 | 87.77 | -7.42 | 141.21 | 14.61 |
| UTL | T2 | meta_meta | 18.24 | 82.55 | 38.95 | 72.85 | -10.03 | 87.33 | 44.37 |
| UTL | T2 | meta_meta | 18.24 | 82.55 | 38.95 | 70.78 | 14.14 | 102.56 | 44.37 |
| UTL | T2 | meta_meta | 18.24 | 82.55 | 38.95 | 72.85 | -10.03 | 87.33 | 44.37 |
| UTL | T2 | meta_meta | 18.24 | 82.55 | 38.95 | 70.78 | 14.14 | 102.56 | 44.37 |
| UTL | T2 | meta_meta | 18.24 | 82.55 | 38.95 | 72.85 | -10.03 | 87.33 | 44.37 |
| UTL | T2 | meta_meta | 18.24 | 82.55 | 38.95 | 70.78 | 14.14 | 102.56 | 44.37 |
| UTL | T2 | meta_meta | 18.24 | 82.55 | 38.95 | 72.85 | -10.03 | 87.33 | 44.37 |
| UTL | T2 | meta_meta | 18.24 | 82.55 | 38.95 | 70.78 | 14.14 | 102.56 | 44.37 |
| UTL | T2 | meta_para | -11.32 | 69.47 | -1.68 | 60.78 | -25.15 | 119.97 | -20.71 |
| UTL | T2 | meta_para | -11.32 | 69.47 | -1.68 | 70.15 | 7.72 | 79.36 | -20.71 |
| UTL | T2 | meta_para | -11.32 | 56.21 | -6.10 | 48.57 | -25.15 | 119.97 | -20.71 |
| UTL | T2 | meta_para | -11.32 | 56.21 | -6.10 | 78.28 | 7.72 | 79.36 | -20.71 |
| UTL | T2 | meta_para | -11.32 | 69.47 | -1.68 | 60.78 | -25.15 | 119.97 | -20.71 |
| UTL | T2 | meta_para | -11.32 | 69.47 | -1.68 | 70.15 | 7.72 | 79.36 | -20.71 |
| UTL | T2 | meta_para | -11.32 | 56.21 | -6.10 | 48.57 | -25.15 | 119.97 | -20.71 |
| UTL | T2 | meta_para | -11.32 | 56.21 | -6.10 | 78.28 | 7.72 | 79.36 | -20.71 |
| UTL | T2 | meta_para | -11.32 | 69.47 | -1.68 | 60.78 | -25.15 | 119.97 | -20.71 |
| UTL | T2 | meta_para | -11.32 | 69.47 | -1.68 | 70.15 | 7.72 | 79.36 | -20.71 |
| UTL | T2 | meta_para | -11.32 | 56.21 | -6.10 | 48.57 | -25.15 | 119.97 | -20.71 |
| UTL | T2 | meta_para | -11.32 | 56.21 | -6.10 | 78.28 | 7.72 | 79.36 | -20.71 |
| UTL | T2 | para_ortho | 4.39 | 53.30 | -4.31 | 74.01 | 7.54 | 61.75 | -8.88 |
| UTL | T2 | para_ortho | 4.39 | 92.92 | -14.50 | 63.31 | 7.54 | 61.75 | -8.88 |
| UTL | T2 | para_ortho | 4.39 | 53.30 | -4.31 | 74.01 | 7.54 | 61.75 | -8.88 |
| UTL | T2 | para_ortho | 4.39 | 92.92 | -14.50 | 63.31 | 7.54 | 61.75 | -8.88 |
| UTL | T2 | para_ortho | 4.39 | 53.30 | -4.31 | 74.01 | 7.54 | 61.75 | -8.88 |
| UTL | T2 | para_ortho | 4.39 | 92.92 | -14.50 | 63.31 | 7.54 | 61.75 | -8.88 |
| UTL | T2 | para_meta | 2.30 | 82.35 | 41.06 | 152.50 | 16.10 | 106.09 | 13.04 |
| UTL | T2 | para_meta | 2.30 | 82.35 | 41.06 | 152.50 | 16.10 | 106.09 | 13.04 |
| UTL | T2 | para_meta | 2.30 | 82.35 | 41.06 | 152.50 | 16.10 | 106.09 | 13.04 |
| UTL | T2 | para_para | -36.70 | 52.33 | -6.63 | 45.78 | 9.23 | 60.26 | 29.38 |
| UTL | T2 | para_para | -36.70 | 33.37 | -14.63 | 61.59 | 9.23 | 60.26 | 29.38 |
| UTL | T2 | para_para | -36.70 | 52.33 | -6.63 | 45.78 | 9.23 | 60.26 | 29.38 |
| UTL | T2 | para_para | -36.70 | 33.37 | -14.63 | 61.59 | 9.23 | 60.26 | 29.38 |
| UTL | T2 | para_para | -36.70 | 52.33 | -6.63 | 45.78 | 9.23 | 60.26 | 29.38 |
| UTL | T2 | para_para | -36.70 | 33.37 | -14.63 | 61.59 | 9.23 | 60.26 | 29.38 |
| UTL | T2 | para_para | -36.70 | 52.33 | -6.63 | 45.78 | 9.23 | 60.26 | 29.38 |
| UTL | T2 | para_para | -36.70 | 33.37 | -14.63 | 61.59 | 9.23 | 60.26 | 29.38 |
| UTL | T4 | ortho_benzene | 58.57 | 168.18 | 54.70 | 121.92 | 19.39 | 126.69 | 37.14 |
| UTL | T4 | ortho_benzene | 58.57 | 168.18 | 54.70 | 121.06 | 49.26 | 100.17 | 37.14 |
| UTL | T4 | ortho_benzene | 58.57 | 168.18 | 54.70 | 121.92 | 19.39 | 126.69 | 37.14 |
| UTL | T4 | ortho_benzene | 58.57 | 168.18 | 54.70 | 121.06 | 49.26 | 100.17 | 37.14 |
| UTL | T4 | ortho_benzene | 58.57 | 168.18 | 54.70 | 121.92 | 19.39 | 126.69 | 37.14 |
| UTL | T4 | ortho_benzene | 58.57 | 168.18 | 54.70 | 121.06 | 49.26 | 100.17 | 37.14 |
| UTL | T4 | ortho_benzene | 58.57 | 168.18 | 54.70 | 121.92 | 19.39 | 126.69 | 37.14 |
| UTL | T4 | ortho_benzene | 58.57 | 168.18 | 54.70 | 121.06 | 49.26 | 100.17 | 37.14 |
| UTL | T4 | meta_benzene | 51.63 | 96.29 | 46.04 | 84.48 | -11.53 | 112.16 | 32.59 |
| UTL | T4 | meta_benzene | 51.63 | 96.29 | 46.04 | 104.45 | 35.58 | 93.09 | 32.59 |
| UTL | T4 | meta_benzene | 51.63 | 96.29 | 46.04 | 84.48 | -11.53 | 112.16 | 32.59 |
| UTL | T4 | meta_benzene | 51.63 | 96.29 | 46.04 | 104.45 | 35.58 | 93.09 | 32.59 |
| UTL | T4 | meta_benzene | 51.63 | 96.29 | 46.04 | 84.48 | -11.53 | 112.16 | 32.59 |
| UTL | T4 | meta_benzene | 51.63 | 96.29 | 46.04 | 104.45 | 35.58 | 93.09 | 32.59 |
| UTL | T4 | meta_benzene | 51.63 | 96.29 | 46.04 | 84.48 | -11.53 | 112.16 | 32.59 |
| UTL | T4 | meta_benzene | 51.63 | 96.29 | 46.04 | 104.45 | 35.58 | 93.09 | 32.59 |
| UTL | T4 | para_benzene | 53.53 | 82.24 | 49.75 | 95.39 | 11.99 | 126.33 | 34.50 |
| UTL | T4 | para_benzene | 53.53 | 82.24 | 49.75 | 95.39 | 11.99 | 126.33 | 34.50 |
| UTL | T4 | para_benzene | 53.53 | 82.24 | 49.75 | 95.39 | 11.99 | 126.33 | 34.50 |
| UTL | T4 | para_benzene | 53.53 | 82.24 | 49.75 | 95.39 | 11.99 | 126.33 | 34.50 |
| UTL | T4 | ortho_ortho | 37.77 | 139.76 | -8.96 | 128.70 | 52.87 | 158.59 | 82.73 |
| UTL | T4 | ortho_ortho | 37.77 | 139.76 | -8.96 | 100.23 | 15.20 | 143.32 | 82.73 |
| UTL | T4 | ortho_ortho | 37.77 | 109.13 | 18.67 | 133.91 | 15.20 | 143.32 | 82.73 |
| UTL | T4 | ortho_ortho | 37.77 | 109.13 | 18.67 | 117.86 | 52.87 | 158.59 | 82.73 |
| UTL | T4 | ortho_ortho | 37.77 | 139.76 | -8.96 | 128.70 | 52.87 | 158.59 | 82.73 |
| UTL | T4 | ortho_ortho | 37.77 | 139.76 | -8.96 | 100.23 | 15.20 | 143.32 | 82.73 |
| UTL | T4 | ortho_ortho | 37.77 | 109.13 | 18.67 | 133.91 | 15.20 | 143.32 | 82.73 |
| UTL | T4 | ortho_ortho | 37.77 | 109.13 | 18.67 | 117.86 | 52.87 | 158.59 | 82.73 |
| UTL | T4 | ortho_ortho | 37.77 | 139.76 | -8.96 | 128.70 | 52.87 | 158.59 | 82.73 |
| UTL | T4 | ortho_ortho | 37.77 | 139.76 | -8.96 | 100.23 | 15.20 | 143.32 | 82.73 |
| UTL | T4 | ortho_ortho | 37.77 | 109.13 | 18.67 | 133.91 | 15.20 | 143.32 | 82.73 |
| UTL | T4 | ortho_ortho | 37.77 | 109.13 | 18.67 | 117.86 | 52.87 | 158.59 | 82.73 |
| UTL | T4 | ortho_ortho | 37.77 | 139.76 | -8.96 | 128.70 | 52.87 | 158.59 | 82.73 |
| UTL | T4 | ortho_ortho | 37.77 | 139.76 | -8.96 | 100.23 | 15.20 | 143.32 | 82.73 |
| UTL | T4 | ortho_ortho | 37.77 | 109.13 | 18.67 | 133.91 | 15.20 | 143.32 | 82.73 |
| UTL | T4 | ortho_ortho | 37.77 | 109.13 | 18.67 | 117.86 | 52.87 | 158.59 | 82.73 |
| UTL | T4 | ortho_meta | 68.04 | 108.86 | 39.19 | 156.36 | 83.92 | 171.58 | 95.84 |
| UTL | T4 | ortho_meta | 68.04 | 108.86 | 39.19 | 112.05 | 66.44 | 128.37 | 95.84 |
| UTL | T4 | ortho_meta | 68.04 | 108.86 | 39.19 | 156.36 | 83.92 | 171.58 | 95.84 |
| UTL | T4 | ortho_meta | 68.04 | 108.86 | 39.19 | 112.05 | 66.44 | 128.37 | 95.84 |
| UTL | T4 | ortho_meta | 68.04 | 108.86 | 39.19 | 156.36 | 83.92 | 171.58 | 95.84 |
| UTL | T4 | ortho_meta | 68.04 | 108.86 | 39.19 | 112.05 | 66.44 | 128.37 | 95.84 |
| UTL | T4 | ortho_para | 42.71 | 103.63 | 12.88 | 91.22 | 36.47 | 91.97 | 19.38 |
| UTL | T4 | ortho_para | 42.71 | 103.63 | 12.88 | 114.58 | 36.47 | 87.23 | 19.38 |
| UTL | T4 | ortho_para | 42.71 | 102.44 | 0.04 | 79.44 | 26.43 | 87.23 | 19.38 |
| UTL | T4 | ortho_para | 42.71 | 102.44 | 0.04 | 117.11 | 26.43 | 91.97 | 19.38 |
| UTL | T4 | ortho_para | 42.71 | 103.63 | 12.88 | 91.22 | 36.47 | 91.97 | 19.38 |
| UTL | T4 | ortho_para | 42.71 | 103.63 | 12.88 | 114.58 | 36.47 | 87.23 | 19.38 |
| UTL | T4 | ortho_para | 42.71 | 102.44 | 0.04 | 79.44 | 26.43 | 87.23 | 19.38 |
| UTL | T4 | ortho_para | 42.71 | 102.44 | 0.04 | 117.11 | 26.43 | 91.97 | 19.38 |
| UTL | T4 | ortho_para | 42.71 | 103.63 | 12.88 | 91.22 | 36.47 | 91.97 | 19.38 |
| UTL | T4 | ortho_para | 42.71 | 103.63 | 12.88 | 114.58 | 36.47 | 87.23 | 19.38 |
| UTL | T4 | ortho_para | 42.71 | 102.44 | 0.04 | 79.44 | 26.43 | 87.23 | 19.38 |
| UTL | T4 | ortho_para | 42.71 | 102.44 | 0.04 | 117.11 | 26.43 | 91.97 | 19.38 |
| UTL | T4 | meta_ortho | 31.73 | 119.86 | -16.86 | 90.69 | -39.74 | 113.45 | 60.05 |
| UTL | T4 | meta_ortho | 31.73 | 119.86 | -16.86 | 79.78 | 31.91 | 104.64 | 60.05 |
| UTL | T4 | meta_ortho | 31.73 | 98.33 | 14.10 | 104.22 | -39.74 | 113.45 | 60.05 |
| UTL | T4 | meta_ortho | 31.73 | 98.33 | 14.10 | 92.62 | 31.91 | 104.64 | 60.05 |
| UTL | T4 | meta_ortho | 31.73 | 119.86 | -16.86 | 90.69 | -39.74 | 113.45 | 60.05 |
| UTL | T4 | meta_ortho | 31.73 | 119.86 | -16.86 | 79.78 | 31.91 | 104.64 | 60.05 |
| UTL | T4 | meta_ortho | 31.73 | 98.33 | 14.10 | 104.22 | -39.74 | 113.45 | 60.05 |
| UTL | T4 | meta_ortho | 31.73 | 98.33 | 14.10 | 92.62 | 31.91 | 104.64 | 60.05 |
| UTL | T4 | meta_ortho | 31.73 | 119.86 | -16.86 | 90.69 | -39.74 | 113.45 | 60.05 |
| UTL | T4 | meta_ortho | 31.73 | 119.86 | -16.86 | 79.78 | 31.91 | 104.64 | 60.05 |
| UTL | T4 | meta_ortho | 31.73 | 98.33 | 14.10 | 104.22 | -39.74 | 113.45 | 60.05 |
| UTL | T4 | meta_ortho | 31.73 | 98.33 | 14.10 | 92.62 | 31.91 | 104.64 | 60.05 |
| UTL | T4 | meta_meta | 50.03 | 92.25 | 56.50 | 66.96 | 11.06 | 94.19 | 80.43 |
| UTL | T4 | meta_meta | 50.03 | 92.25 | 56.50 | 64.06 | 5.72 | 106.92 | 80.43 |
| UTL | T4 | meta_meta | 50.03 | 92.25 | 56.50 | 66.96 | 11.06 | 94.19 | 80.43 |
| UTL | T4 | meta_meta | 50.03 | 92.25 | 56.50 | 64.06 | 5.72 | 106.92 | 80.43 |
| UTL | T4 | meta_meta | 50.03 | 92.25 | 56.50 | 66.96 | 11.06 | 94.19 | 80.43 |
| UTL | T4 | meta_meta | 50.03 | 92.25 | 56.50 | 64.06 | 5.72 | 106.92 | 80.43 |
| UTL | T4 | meta_meta | 50.03 | 92.25 | 56.50 | 66.96 | 11.06 | 94.19 | 80.43 |
| UTL | T4 | meta_meta | 50.03 | 92.25 | 56.50 | 64.06 | 5.72 | 106.92 | 80.43 |
| UTL | T4 | meta_para | 21.93 | 53.90 | -7.54 | 68.60 | -61.28 | 89.54 | 3.53 |
| UTL | T4 | meta_para | 21.93 | 53.90 | -7.54 | 77.49 | 13.21 | 70.76 | 3.53 |
| UTL | T4 | meta_para | 21.93 | 85.63 | -15.36 | 128.91 | -61.28 | 89.54 | 3.53 |
| UTL | T4 | meta_para | 21.93 | 85.63 | -15.36 | 73.59 | 13.21 | 70.76 | 3.53 |
| UTL | T4 | meta_para | 21.93 | 53.90 | -7.54 | 68.60 | -61.28 | 89.54 | 3.53 |
| UTL | T4 | meta_para | 21.93 | 53.90 | -7.54 | 77.49 | 13.21 | 70.76 | 3.53 |
| UTL | T4 | meta_para | 21.93 | 85.63 | -15.36 | 128.91 | -61.28 | 89.54 | 3.53 |
| UTL | T4 | meta_para | 21.93 | 85.63 | -15.36 | 73.59 | 13.21 | 70.76 | 3.53 |
| UTL | T4 | meta_para | 21.93 | 53.90 | -7.54 | 68.60 | -61.28 | 89.54 | 3.53 |
| UTL | T4 | meta_para | 21.93 | 53.90 | -7.54 | 77.49 | 13.21 | 70.76 | 3.53 |
| UTL | T4 | meta_para | 21.93 | 85.63 | -15.36 | 128.91 | -61.28 | 89.54 | 3.53 |
| UTL | T4 | meta_para | 21.93 | 85.63 | -15.36 | 73.59 | 13.21 | 70.76 | 3.53 |
| UTL | T4 | para_ortho | 9.72 | 55.48 | 15.30 | 85.35 | 32.79 | 109.16 | 22.27 |
| UTL | T4 | para_ortho | 9.72 | 78.84 | -7.85 | 95.25 | 32.79 | 109.16 | 22.27 |
| UTL | T4 | para_ortho | 9.72 | 55.48 | 15.30 | 85.35 | 32.79 | 109.16 | 22.27 |
| UTL | T4 | para_ortho | 9.72 | 78.84 | -7.85 | 95.25 | 32.79 | 109.16 | 22.27 |
| UTL | T4 | para_ortho | 9.72 | 55.48 | 15.30 | 85.35 | 32.79 | 109.16 | 22.27 |
| UTL | T4 | para_ortho | 9.72 | 78.84 | -7.85 | 95.25 | 32.79 | 109.16 | 22.27 |
| UTL | T4 | para_meta | 17.01 | 86.50 | 44.43 | 129.79 | 31.99 | 139.16 | 18.28 |
| UTL | T4 | para_meta | 17.01 | 86.50 | 44.43 | 129.79 | 31.99 | 139.16 | 18.28 |
| UTL | T4 | para_meta | 17.01 | 86.50 | 44.43 | 129.79 | 31.99 | 139.16 | 18.28 |
| UTL | T4 | para_para | 0.00 | 40.43 | -7.57 | 80.82 | 1.83 | 104.91 | 34.65 |
| UTL | T4 | para_para | 0.00 | 70.84 | 7.53 | 92.23 | 1.83 | 104.91 | 34.65 |
| UTL | T4 | para_para | 0.00 | 40.43 | -7.57 | 80.82 | 1.83 | 104.91 | 34.65 |
| UTL | T4 | para_para | 0.00 | 70.84 | 7.53 | 92.23 | 1.83 | 104.91 | 34.65 |
| UTL | T4 | para_para | 0.00 | 40.43 | -7.57 | 80.82 | 1.83 | 104.91 | 34.65 |
| UTL | T4 | para_para | 0.00 | 70.84 | 7.53 | 92.23 | 1.83 | 104.91 | 34.65 |
| UTL | T4 | para_para | 0.00 | 40.43 | -7.57 | 80.82 | 1.83 | 104.91 | 34.65 |
| UTL | T4 | para_para | 0.00 | 70.84 | 7.53 | 92.23 | 1.83 | 104.91 | 34.65 |
| FAU | T1 | ortho_benzene | 0.00 | 94.18 | 6.18 | 105.77 | 36.50 | 102.27 | 18.14 |
| FAU | T1 | ortho_benzene | 0.00 | 94.18 | 6.18 | 96.11 | 8.33 | 48.70 | 18.14 |
| FAU | T1 | ortho_benzene | 0.00 | 94.18 | 6.18 | 105.77 | 36.50 | 102.27 | 18.14 |
| FAU | T1 | ortho_benzene | 0.00 | 94.18 | 6.18 | 96.11 | 8.33 | 48.70 | 18.14 |
| FAU | T1 | ortho_benzene | 0.00 | 94.18 | 6.18 | 105.77 | 36.50 | 102.27 | 18.14 |
| FAU | T1 | ortho_benzene | 0.00 | 94.18 | 6.18 | 96.11 | 8.33 | 48.70 | 18.14 |
| FAU | T1 | ortho_benzene | 0.00 | 94.18 | 6.18 | 105.77 | 36.50 | 102.27 | 18.14 |
| FAU | T1 | ortho_benzene | 0.00 | 94.18 | 6.18 | 96.11 | 8.33 | 48.70 | 18.14 |
| FAU | T1 | meta_benzene | 20.56 | 69.58 | -5.23 | 39.03 | -12.36 | 75.35 | 10.43 |
| FAU | T1 | meta_benzene | 20.56 | 69.58 | -5.23 | 60.37 | -22.44 | 106.61 | 10.43 |
| FAU | T1 | meta_benzene | 20.56 | 69.58 | -5.23 | 39.03 | -12.36 | 75.35 | 10.43 |
| FAU | T1 | meta_benzene | 20.56 | 69.58 | -5.23 | 60.37 | -22.44 | 106.61 | 10.43 |
| FAU | T1 | meta_benzene | 20.56 | 69.58 | -5.23 | 39.03 | -12.36 | 75.35 | 10.43 |
| FAU | T1 | meta_benzene | 20.56 | 69.58 | -5.23 | 60.37 | -22.44 | 106.61 | 10.43 |
| FAU | T1 | meta_benzene | 20.56 | 69.58 | -5.23 | 39.03 | -12.36 | 75.35 | 10.43 |
| FAU | T1 | meta_benzene | 20.56 | 69.58 | -5.23 | 60.37 | -22.44 | 106.61 | 10.43 |
| FAU | T1 | para_benzene | 21.39 | 74.75 | 13.82 | 70.78 | 2.00 | 54.72 | 7.29 |
| FAU | T1 | para_benzene | 21.39 | 74.75 | 13.82 | 70.78 | 2.00 | 54.72 | 7.29 |
| FAU | T1 | para_benzene | 21.39 | 74.75 | 13.82 | 70.78 | 2.00 | 54.72 | 7.29 |
| FAU | T1 | para_benzene | 21.39 | 74.75 | 13.82 | 70.78 | 2.00 | 54.72 | 7.29 |
| FAU | T1 | ortho_ortho | 87.43 | 125.06 | 33.52 | 147.99 | 59.07 | 145.87 | 38.65 |
| FAU | T1 | ortho_ortho | 87.43 | 125.06 | 33.52 | 135.87 | 72.98 | 137.29 | 38.65 |
| FAU | T1 | ortho_ortho | 87.43 | 138.45 | 56.80 | 169.45 | 59.07 | 145.87 | 38.65 |
| FAU | T1 | ortho_ortho | 87.43 | 138.45 | 56.80 | 133.84 | 72.98 | 137.29 | 38.65 |
| FAU | T1 | ortho_ortho | 87.43 | 125.06 | 33.52 | 147.99 | 59.07 | 145.87 | 38.65 |
| FAU | T1 | ortho_ortho | 87.43 | 125.06 | 33.52 | 135.87 | 72.98 | 137.29 | 38.65 |
| FAU | T1 | ortho_ortho | 87.43 | 138.45 | 56.80 | 169.45 | 59.07 | 145.87 | 38.65 |
| FAU | T1 | ortho_ortho | 87.43 | 138.45 | 56.80 | 133.84 | 72.98 | 137.29 | 38.65 |
| FAU | T1 | ortho_ortho | 87.43 | 125.06 | 33.52 | 147.99 | 59.07 | 145.87 | 38.65 |
| FAU | T1 | ortho_ortho | 87.43 | 125.06 | 33.52 | 135.87 | 72.98 | 137.29 | 38.65 |
| FAU | T1 | ortho_ortho | 87.43 | 138.45 | 56.80 | 169.45 | 59.07 | 145.87 | 38.65 |
| FAU | T1 | ortho_ortho | 87.43 | 138.45 | 56.80 | 133.84 | 72.98 | 137.29 | 38.65 |
| FAU | T1 | ortho_ortho | 87.43 | 125.06 | 33.52 | 147.99 | 59.07 | 145.87 | 38.65 |
| FAU | T1 | ortho_ortho | 87.43 | 125.06 | 33.52 | 135.87 | 72.98 | 137.29 | 38.65 |
| FAU | T1 | ortho_ortho | 87.43 | 138.45 | 56.80 | 169.45 | 59.07 | 145.87 | 38.65 |
| FAU | T1 | ortho_ortho | 87.43 | 138.45 | 56.80 | 133.84 | 72.98 | 137.29 | 38.65 |
| FAU | T1 | ortho_meta | 64.58 | 132.18 | 95.71 | 172.89 | 86.43 | 138.25 | 87.15 |
| FAU | T1 | ortho_meta | 64.58 | 132.18 | 95.71 | 161.52 | 74.76 | 155.82 | 87.15 |
| FAU | T1 | ortho_meta | 64.58 | 132.18 | 95.71 | 172.89 | 86.43 | 138.25 | 87.15 |
| FAU | T1 | ortho_meta | 64.58 | 132.18 | 95.71 | 161.52 | 74.76 | 155.82 | 87.15 |
| FAU | T1 | ortho_meta | 64.58 | 132.18 | 95.71 | 172.89 | 86.43 | 138.25 | 87.15 |
| FAU | T1 | ortho_meta | 64.58 | 132.18 | 95.71 | 161.52 | 74.76 | 155.82 | 87.15 |
| FAU | T1 | ortho_para | 46.90 | 134.97 | 53.51 | 124.96 | 56.60 | 106.77 | 51.95 |
| FAU | T1 | ortho_para | 46.90 | 134.97 | 53.51 | 115.67 | 56.60 | 98.78 | 51.95 |
| FAU | T1 | ortho_para | 46.90 | 77.18 | 42.67 | 113.24 | 38.57 | 98.78 | 51.95 |
| FAU | T1 | ortho_para | 46.90 | 77.18 | 42.67 | 149.15 | 38.57 | 106.77 | 51.95 |
| FAU | T1 | ortho_para | 46.90 | 134.97 | 53.51 | 124.96 | 56.60 | 106.77 | 51.95 |
| FAU | T1 | ortho_para | 46.90 | 134.97 | 53.51 | 115.67 | 56.60 | 98.78 | 51.95 |
| FAU | T1 | ortho_para | 46.90 | 77.18 | 42.67 | 113.24 | 38.57 | 98.78 | 51.95 |
| FAU | T1 | ortho_para | 46.90 | 77.18 | 42.67 | 149.15 | 38.57 | 106.77 | 51.95 |
| FAU | T1 | ortho_para | 46.90 | 134.97 | 53.51 | 124.96 | 56.60 | 106.77 | 51.95 |
| FAU | T1 | ortho_para | 46.90 | 134.97 | 53.51 | 115.67 | 56.60 | 98.78 | 51.95 |
| FAU | T1 | ortho_para | 46.90 | 77.18 | 42.67 | 113.24 | 38.57 | 98.78 | 51.95 |
| FAU | T1 | ortho_para | 46.90 | 77.18 | 42.67 | 149.15 | 38.57 | 106.77 | 51.95 |
| FAU | T1 | meta_ortho | 49.11 | 105.62 | 32.37 | 109.98 | 37.31 | 112.79 | 69.39 |
| FAU | T1 | meta_ortho | 49.11 | 105.62 | 32.37 | 100.25 | 34.68 | 114.19 | 69.39 |
| FAU | T1 | meta_ortho | 49.11 | 65.42 | 7.20 | 122.14 | 34.68 | 114.19 | 69.39 |
| FAU | T1 | meta_ortho | 49.11 | 65.42 | 7.20 | 112.60 | 37.31 | 112.79 | 69.39 |
| FAU | T1 | meta_ortho | 49.11 | 105.62 | 32.37 | 109.98 | 37.31 | 112.79 | 69.39 |
| FAU | T1 | meta_ortho | 49.11 | 105.62 | 32.37 | 100.25 | 34.68 | 114.19 | 69.39 |
| FAU | T1 | meta_ortho | 49.11 | 65.42 | 7.20 | 122.14 | 34.68 | 114.19 | 69.39 |
| FAU | T1 | meta_ortho | 49.11 | 65.42 | 7.20 | 112.60 | 37.31 | 112.79 | 69.39 |
| FAU | T1 | meta_ortho | 49.11 | 105.62 | 32.37 | 109.98 | 37.31 | 112.79 | 69.39 |
| FAU | T1 | meta_ortho | 49.11 | 105.62 | 32.37 | 100.25 | 34.68 | 114.19 | 69.39 |
| FAU | T1 | meta_ortho | 49.11 | 65.42 | 7.20 | 122.14 | 34.68 | 114.19 | 69.39 |
| FAU | T1 | meta_ortho | 49.11 | 65.42 | 7.20 | 112.60 | 37.31 | 112.79 | 69.39 |
| FAU | T1 | meta_meta | 48.93 | 124.11 | 79.61 | 108.31 | 55.85 | 111.24 | 100.88 |
| FAU | T1 | meta_meta | 48.93 | 124.11 | 79.61 | 109.17 | 36.73 | 118.15 | 100.88 |
| FAU | T1 | meta_meta | 48.93 | 124.11 | 79.61 | 108.31 | 55.85 | 111.24 | 100.88 |
| FAU | T1 | meta_meta | 48.93 | 124.11 | 79.61 | 109.17 | 36.73 | 118.15 | 100.88 |
| FAU | T1 | meta_meta | 48.93 | 124.11 | 79.61 | 108.31 | 55.85 | 111.24 | 100.88 |
| FAU | T1 | meta_meta | 48.93 | 124.11 | 79.61 | 109.17 | 36.73 | 118.15 | 100.88 |
| FAU | T1 | meta_meta | 48.93 | 124.11 | 79.61 | 108.31 | 55.85 | 111.24 | 100.88 |
| FAU | T1 | meta_meta | 48.93 | 124.11 | 79.61 | 109.17 | 36.73 | 118.15 | 100.88 |
| FAU | T1 | meta_para | 50.07 | 95.49 | 39.22 | 114.30 | 27.31 | 109.76 | 23.20 |
| FAU | T1 | meta_para | 50.07 | 95.49 | 39.22 | 98.74 | 46.36 | 117.28 | 23.20 |
| FAU | T1 | meta_para | 50.07 | 67.88 | 15.37 | 110.17 | 27.31 | 109.76 | 23.20 |
| FAU | T1 | meta_para | 50.07 | 67.88 | 15.37 | 101.99 | 46.36 | 107.28 | 23.20 |
| FAU | T1 | meta_para | 50.07 | 95.49 | 39.22 | 114.30 | 27.31 | 109.76 | 23.20 |
| FAU | T1 | meta_para | 50.07 | 95.49 | 39.22 | 98.74 | 46.36 | 117.28 | 23.20 |
| FAU | T1 | meta_para | 50.07 | 67.88 | 15.37 | 110.17 | 27.31 | 109.76 | 23.20 |
| FAU | T1 | meta_para | 50.07 | 67.88 | 15.37 | 101.99 | 46.36 | 107.28 | 23.20 |
| FAU | T1 | meta_para | 50.07 | 95.49 | 39.22 | 114.30 | 27.31 | 109.76 | 23.20 |
| FAU | T1 | meta_para | 50.07 | 95.49 | 39.22 | 98.74 | 46.36 | 117.28 | 23.20 |
| FAU | T1 | meta_para | 50.07 | 67.88 | 15.37 | 110.17 | 27.31 | 109.76 | 23.20 |
| FAU | T1 | meta_para | 50.07 | 67.88 | 15.37 | 101.99 | 46.36 | 107.28 | 23.20 |
| FAU | T1 | para_ortho | 48.82 | 103.28 | 32.91 | 89.23 | 51.00 | 163.59 | 65.48 |
| FAU | T1 | para_ortho | 48.82 | 81.94 | 33.86 | 142.31 | 51.00 | 163.59 | 65.48 |
| FAU | T1 | para_ortho | 48.82 | 103.28 | 32.91 | 89.23 | 51.00 | 163.59 | 65.48 |
| FAU | T1 | para_ortho | 48.82 | 81.94 | 33.86 | 142.31 | 51.00 | 163.59 | 65.48 |
| FAU | T1 | para_ortho | 48.82 | 103.28 | 32.91 | 89.23 | 51.00 | 163.59 | 65.48 |
| FAU | T1 | para_ortho | 48.82 | 81.94 | 33.86 | 142.31 | 51.00 | 163.59 | 65.48 |
| FAU | T1 | para_meta | 39.00 | 113.60 | 86.72 | 141.20 | 47.56 | 118.14 | 76.18 |
| FAU | T1 | para_meta | 39.00 | 113.60 | 86.72 | 141.20 | 47.56 | 118.14 | 76.18 |
| FAU | T1 | para_meta | 39.00 | 113.60 | 86.72 | 141.20 | 47.56 | 118.14 | 76.18 |
| FAU | T1 | para_para | 0.00 | 110.37 | 18.89 | 143.34 | 43.89 | 116.33 | 37.76 |
| FAU | T1 | para_para | 0.00 | 38.80 | 17.73 | 158.05 | 43.89 | 116.33 | 37.76 |
| FAU | T1 | para_para | 0.00 | 110.37 | 18.89 | 143.34 | 43.89 | 116.33 | 37.76 |
| FAU | T1 | para_para | 0.00 | 38.80 | 17.73 | 158.05 | 43.89 | 116.33 | 37.76 |
| FAU | T1 | para_para | 0.00 | 110.37 | 18.89 | 143.34 | 43.89 | 116.33 | 37.76 |
| FAU | T1 | para_para | 0.00 | 38.80 | 17.73 | 158.05 | 43.89 | 116.33 | 37.76 |
| FAU | T1 | para_para | 0.00 | 110.37 | 18.89 | 143.34 | 43.89 | 116.33 | 37.76 |
| FAU | T1 | para_para | 0.00 | 38.80 | 17.73 | 158.05 | 43.89 | 116.33 | 37.76 |

**Table S23**. Zeolite rescaled Direct Path free energy barriers in kJ/mol. Transalkylation paths have been rescaled relative to the minimum I0 energy identified among the BOG (T1: meta-benzene), IWV (T7: ortho-benzene), and UTL (T2: para-benzene). Scaled disproportionation pathways have been referenced to the minimum I0 energy found among BOG (T1: meta–para), IWV (T7: meta–para), and UTL (T2: para–para).

| Framework | Aluminum | Isomer | I0 | TS1 | I4 |
|---|---|---|---|---|---|
| BOG | T1 | ortho_benzene | 40.93 | 129.96 | 29.87 |
| BOG | T1 | ortho_benzene | 40.93 | 129.96 | 29.87 |
| BOG | T1 | ortho_benzene | 40.93 | 129.96 | 29.87 |
| BOG | T1 | ortho_benzene | 40.93 | 129.96 | 29.87 |
| BOG | T1 | meta_benzene | 16.36 | 117.16 | 22.51 |
| BOG | T1 | meta_benzene | 16.36 | 117.16 | 22.51 |
| BOG | T1 | meta_benzene | 16.36 | 117.16 | 22.51 |
| BOG | T1 | meta_benzene | 16.36 | 117.16 | 22.51 |
| BOG | T1 | para_benzene | 0.00 | 141.63 | 42.80 |
| BOG | T1 | para_benzene | 0.00 | 141.63 | 42.80 |
| BOG | T1 | para_benzene | 0.00 | 141.63 | 42.80 |
| BOG | T1 | para_benzene | 0.00 | 141.63 | 42.80 |
| BOG | T1 | ortho_ortho | 98.75 | 187.45 | 164.12 |
| BOG | T1 | ortho_ortho | 98.75 | 187.45 | 164.12 |
| BOG | T1 | ortho_ortho | 98.75 | 187.45 | 164.12 |
| BOG | T1 | ortho_ortho | 98.75 | 187.45 | 164.12 |
| BOG | T1 | ortho_meta | 92.40 | 173.19 | 43.18 |
| BOG | T1 | ortho_meta | 92.40 | 173.19 | 43.18 |
| BOG | T1 | ortho_meta | 92.40 | 173.19 | 43.18 |
| BOG | T1 | ortho_para | 46.07 | 166.82 | 76.65 |
| BOG | T1 | ortho_para | 46.07 | 166.82 | 76.65 |
| BOG | T1 | ortho_para | 46.07 | 166.82 | 76.65 |
| BOG | T1 | meta_ortho | 47.41 | 157.60 | 84.18 |
| BOG | T1 | meta_ortho | 47.41 | 157.60 | 84.18 |
| BOG | T1 | meta_ortho | 47.41 | 157.60 | 84.18 |
| BOG | T1 | meta_meta | 95.75 | 251.53 | 86.26 |
| BOG | T1 | meta_meta | 95.75 | 251.53 | 86.26 |
| BOG | T1 | meta_meta | 95.75 | 251.53 | 86.26 |
| BOG | T1 | meta_meta | 95.75 | 251.53 | 86.26 |
| BOG | T1 | meta_para | 27.06 | 127.55 | 57.06 |
| BOG | T1 | meta_para | 27.06 | 127.55 | 57.06 |
| BOG | T1 | meta_para | 27.06 | 127.55 | 57.06 |
| BOG | T1 | para_ortho | 60.12 | 149.30 | 72.11 |
| BOG | T1 | para_ortho | 60.12 | 149.30 | 72.11 |
| BOG | T1 | para_ortho | 60.12 | 149.30 | 72.11 |
| BOG | T1 | para_meta | 25.91 | 289.90 | 95.25 |
| BOG | T1 | para_meta | 25.91 | 289.90 | 95.25 |
| BOG | T1 | para_meta | 25.91 | 289.90 | 95.25 |
| BOG | T1 | para_para | 0.00 | 133.20 | 48.43 |
| BOG | T1 | para_para | 0.00 | 133.20 | 48.43 |
| BOG | T1 | para_para | 0.00 | 133.20 | 48.43 |
| BOG | T1 | para_para | 0.00 | 133.20 | 48.43 |
| BOG | T2 | ortho_benzene | 48.67 | 133.77 | 49.81 |
| BOG | T2 | ortho_benzene | 48.67 | 133.77 | 49.81 |
| BOG | T2 | ortho_benzene | 48.67 | 133.77 | 49.81 |
| BOG | T2 | ortho_benzene | 48.67 | 133.77 | 49.81 |
| BOG | T2 | meta_benzene | 16.12 | 119.36 | 25.92 |
| BOG | T2 | meta_benzene | 16.12 | 119.36 | 25.92 |
| BOG | T2 | meta_benzene | 16.12 | 119.36 | 25.92 |
| BOG | T2 | meta_benzene | 16.12 | 119.36 | 25.92 |
| BOG | T2 | para_benzene | 19.82 | 106.28 | 28.03 |
| BOG | T2 | para_benzene | 19.82 | 106.28 | 28.03 |
| BOG | T2 | para_benzene | 19.82 | 106.28 | 28.03 |
| BOG | T2 | para_benzene | 19.82 | 106.28 | 28.03 |
| BOG | T2 | ortho_ortho | 74.55 | 165.04 | 101.76 |
| BOG | T2 | ortho_ortho | 74.55 | 165.04 | 101.76 |
| BOG | T2 | ortho_ortho | 74.55 | 165.04 | 101.76 |
| BOG | T2 | ortho_ortho | 74.55 | 165.04 | 101.76 |
| BOG | T2 | ortho_meta | 150.65 | 231.79 | 136.12 |
| BOG | T2 | ortho_meta | 150.65 | 231.79 | 136.12 |
| BOG | T2 | ortho_meta | 150.65 | 231.79 | 136.12 |
| BOG | T2 | ortho_para | 66.74 | 170.14 | 87.44 |
| BOG | T2 | ortho_para | 66.74 | 170.14 | 87.44 |
| BOG | T2 | ortho_para | 66.74 | 170.14 | 87.44 |
| BOG | T2 | meta_ortho | 79.14 | 194.54 | 104.86 |
| BOG | T2 | meta_ortho | 79.14 | 194.54 | 104.86 |
| BOG | T2 | meta_ortho | 79.14 | 194.54 | 104.86 |
| BOG | T2 | meta_meta | 50.64 | 239.86 | 93.66 |
| BOG | T2 | meta_meta | 50.64 | 239.86 | 93.66 |
| BOG | T2 | meta_meta | 50.64 | 239.86 | 93.66 |
| BOG | T2 | meta_meta | 50.64 | 239.86 | 93.66 |
| BOG | T2 | meta_para | 19.18 | 176.18 | 63.81 |
| BOG | T2 | meta_para | 19.18 | 176.18 | 63.81 |
| BOG | T2 | meta_para | 19.18 | 176.18 | 63.81 |
| BOG | T2 | para_ortho | 51.20 | 197.64 | 123.55 |
| BOG | T2 | para_ortho | 51.20 | 197.64 | 123.55 |
| BOG | T2 | para_ortho | 51.20 | 197.64 | 123.55 |
| BOG | T2 | para_meta | 61.72 | 170.77 | 72.43 |
| BOG | T2 | para_meta | 61.72 | 170.77 | 72.43 |
| BOG | T2 | para_meta | 61.72 | 170.77 | 72.43 |
| BOG | T2 | para_para | 25.31 | 111.39 | 73.98 |
| BOG | T2 | para_para | 25.31 | 111.39 | 73.98 |
| BOG | T2 | para_para | 25.31 | 111.39 | 73.98 |
| BOG | T2 | para_para | 25.31 | 111.39 | 73.98 |
| IWV | T3 | ortho_benzene | 4.80 | 110.04 | 20.89 |
| IWV | T3 | ortho_benzene | 4.80 | 110.04 | 20.89 |
| IWV | T3 | ortho_benzene | 4.80 | 110.04 | 20.89 |
| IWV | T3 | ortho_benzene | 4.80 | 110.04 | 20.89 |
| IWV | T3 | meta_benzene | 10.32 | 85.97 | 26.31 |
| IWV | T3 | meta_benzene | 10.32 | 85.97 | 26.31 |
| IWV | T3 | meta_benzene | 10.32 | 85.97 | 26.31 |
| IWV | T3 | meta_benzene | 10.32 | 85.97 | 26.31 |
| IWV | T3 | para_benzene | 28.80 | 85.36 | -2.82 |
| IWV | T3 | para_benzene | 28.80 | 85.36 | -2.82 |
| IWV | T3 | para_benzene | 28.80 | 85.36 | -2.82 |
| IWV | T3 | para_benzene | 28.80 | 85.36 | -2.82 |
| IWV | T3 | ortho_ortho | 53.36 | 182.74 | 41.23 |
| IWV | T3 | ortho_ortho | 53.36 | 182.74 | 41.23 |
| IWV | T3 | ortho_ortho | 53.36 | 182.74 | 41.23 |
| IWV | T3 | ortho_ortho | 53.36 | 182.74 | 41.23 |
| IWV | T3 | ortho_meta | 51.58 | 213.61 | 113.33 |
| IWV | T3 | ortho_meta | 51.58 | 213.61 | 113.33 |
| IWV | T3 | ortho_meta | 51.58 | 213.61 | 113.33 |
| IWV | T3 | ortho_para | 53.40 | 171.46 | 63.38 |
| IWV | T3 | ortho_para | 53.40 | 171.46 | 63.38 |
| IWV | T3 | ortho_para | 53.40 | 171.46 | 63.38 |
| IWV | T3 | meta_ortho | 70.37 | 158.47 | 57.23 |
| IWV | T3 | meta_ortho | 70.37 | 158.47 | 57.23 |
| IWV | T3 | meta_ortho | 70.37 | 158.47 | 57.23 |
| IWV | T3 | meta_meta | 36.23 | 206.64 | 68.12 |
| IWV | T3 | meta_meta | 36.23 | 206.64 | 68.12 |
| IWV | T3 | meta_meta | 36.23 | 206.64 | 68.12 |
| IWV | T3 | meta_meta | 36.23 | 206.64 | 68.12 |
| IWV | T3 | meta_para | 11.33 | 214.10 | 8.40 |
| IWV | T3 | meta_para | 11.33 | 214.10 | 8.40 |
| IWV | T3 | meta_para | 11.33 | 214.10 | 8.40 |
| IWV | T3 | para_ortho | 41.06 | 167.75 | 46.42 |
| IWV | T3 | para_ortho | 41.06 | 167.75 | 46.42 |
| IWV | T3 | para_ortho | 41.06 | 167.75 | 46.42 |
| IWV | T3 | para_meta | 55.01 | 162.26 | 68.01 |
| IWV | T3 | para_meta | 55.01 | 162.26 | 68.01 |
| IWV | T3 | para_meta | 55.01 | 162.26 | 68.01 |
| IWV | T3 | para_para | 0.00 | 111.07 | 45.89 |
| IWV | T3 | para_para | 0.00 | 111.07 | 45.89 |
| IWV | T3 | para_para | 0.00 | 111.07 | 45.89 |
| IWV | T3 | para_para | 0.00 | 111.07 | 45.89 |
| IWV | T7 | ortho_benzene | 21.80 | 96.39 | 28.02 |
| IWV | T7 | ortho_benzene | 21.80 | 96.39 | 28.02 |
| IWV | T7 | ortho_benzene | 21.80 | 96.39 | 28.02 |
| IWV | T7 | ortho_benzene | 21.80 | 96.39 | 28.02 |
| IWV | T7 | meta_benzene | 79.26 | 151.23 | 62.23 |
| IWV | T7 | meta_benzene | 79.26 | 151.23 | 62.23 |
| IWV | T7 | meta_benzene | 79.26 | 151.23 | 62.23 |
| IWV | T7 | meta_benzene | 79.26 | 151.23 | 62.23 |
| IWV | T7 | para_benzene | 0.00 | 130.03 | 10.19 |
| IWV | T7 | para_benzene | 0.00 | 130.03 | 10.19 |
| IWV | T7 | para_benzene | 0.00 | 130.03 | 10.19 |
| IWV | T7 | para_benzene | 0.00 | 130.03 | 10.19 |
| IWV | T7 | ortho_ortho | 61.18 | 166.18 | 102.46 |
| IWV | T7 | ortho_ortho | 61.18 | 166.18 | 102.46 |
| IWV | T7 | ortho_ortho | 61.18 | 166.18 | 102.46 |
| IWV | T7 | ortho_ortho | 61.18 | 166.18 | 102.46 |
| IWV | T7 | ortho_meta | 88.58 | 221.92 | 82.49 |
| IWV | T7 | ortho_meta | 88.58 | 221.92 | 82.49 |
| IWV | T7 | ortho_meta | 88.58 | 221.92 | 82.49 |
| IWV | T7 | ortho_para | 66.71 | 164.96 | 83.97 |
| IWV | T7 | ortho_para | 66.71 | 164.96 | 83.97 |
| IWV | T7 | ortho_para | 66.71 | 164.96 | 83.97 |
| IWV | T7 | meta_ortho | 88.45 | 190.16 | 30.53 |
| IWV | T7 | meta_ortho | 88.45 | 190.16 | 30.53 |
| IWV | T7 | meta_ortho | 88.45 | 190.16 | 30.53 |
| IWV | T7 | meta_meta | 55.55 | 177.46 | 68.49 |
| IWV | T7 | meta_meta | 55.55 | 177.46 | 68.49 |
| IWV | T7 | meta_meta | 55.55 | 177.46 | 68.49 |
| IWV | T7 | meta_meta | 55.55 | 177.46 | 68.49 |
| IWV | T7 | meta_para | 45.51 | 146.61 | 61.05 |
| IWV | T7 | meta_para | 45.51 | 146.61 | 61.05 |
| IWV | T7 | meta_para | 45.51 | 146.61 | 61.05 |
| IWV | T7 | para_ortho | 74.89 | 160.18 | 71.80 |
| IWV | T7 | para_ortho | 74.89 | 160.18 | 71.80 |
| IWV | T7 | para_ortho | 74.89 | 160.18 | 71.80 |
| IWV | T7 | para_meta | 56.68 | 213.14 | 87.50 |
| IWV | T7 | para_meta | 56.68 | 213.14 | 87.50 |
| IWV | T7 | para_meta | 56.68 | 213.14 | 87.50 |
| IWV | T7 | para_para | 37.46 | 228.13 | 65.56 |
| IWV | T7 | para_para | 37.46 | 228.13 | 65.56 |
| IWV | T7 | para_para | 37.46 | 228.13 | 65.56 |
| IWV | T7 | para_para | 37.46 | 228.13 | 65.56 |
| UTL | T2 | ortho_benzene | 63.85 | 136.78 | 17.74 |
| UTL | T2 | ortho_benzene | 63.85 | 136.78 | 17.74 |
| UTL | T2 | ortho_benzene | 63.85 | 136.78 | 17.74 |
| UTL | T2 | ortho_benzene | 63.85 | 136.78 | 17.74 |
| UTL | T2 | meta_benzene | 51.60 | 125.69 | 3.71 |
| UTL | T2 | meta_benzene | 51.60 | 125.69 | 3.71 |
| UTL | T2 | meta_benzene | 51.60 | 125.69 | 3.71 |
| UTL | T2 | meta_benzene | 51.60 | 125.69 | 3.71 |
| UTL | T2 | para_benzene | 0.00 | 100.35 | 19.81 |
| UTL | T2 | para_benzene | 0.00 | 100.35 | 19.81 |
| UTL | T2 | para_benzene | 0.00 | 100.35 | 19.81 |
| UTL | T2 | para_benzene | 0.00 | 100.35 | 19.81 |
| UTL | T2 | ortho_ortho | 27.56 | 147.46 | 21.25 |
| UTL | T2 | ortho_ortho | 27.56 | 147.46 | 21.25 |
| UTL | T2 | ortho_ortho | 27.56 | 147.46 | 21.25 |
| UTL | T2 | ortho_ortho | 27.56 | 147.46 | 21.25 |
| UTL | T2 | ortho_meta | 14.79 | 156.01 | 56.69 |
| UTL | T2 | ortho_meta | 14.79 | 156.01 | 56.69 |
| UTL | T2 | ortho_meta | 14.79 | 156.01 | 56.69 |
| UTL | T2 | ortho_para | 6.41 | 87.62 | 3.81 |
| UTL | T2 | ortho_para | 6.41 | 87.62 | 3.81 |
| UTL | T2 | ortho_para | 6.41 | 87.62 | 3.81 |
| UTL | T2 | meta_ortho | 17.79 | 96.72 | 14.61 |
| UTL | T2 | meta_ortho | 17.79 | 96.72 | 14.61 |
| UTL | T2 | meta_ortho | 17.79 | 96.72 | 14.61 |
| UTL | T2 | meta_meta | 18.24 | 179.27 | 44.37 |
| UTL | T2 | meta_meta | 18.24 | 179.27 | 44.37 |
| UTL | T2 | meta_meta | 18.24 | 179.27 | 44.37 |
| UTL | T2 | meta_meta | 18.24 | 179.27 | 44.37 |
| UTL | T2 | meta_para | -11.32 | 88.79 | -20.71 |
| UTL | T2 | meta_para | -11.32 | 88.79 | -20.71 |
| UTL | T2 | meta_para | -11.32 | 88.79 | -20.71 |
| UTL | T2 | para_ortho | 4.39 | 93.18 | -8.88 |
| UTL | T2 | para_ortho | 4.39 | 93.18 | -8.88 |
| UTL | T2 | para_ortho | 4.39 | 93.18 | -8.88 |
| UTL | T2 | para_meta | 2.30 | 138.63 | 13.04 |
| UTL | T2 | para_meta | 2.30 | 138.63 | 13.04 |
| UTL | T2 | para_meta | 2.30 | 138.63 | 13.04 |
| UTL | T2 | para_para | -36.70 | 68.41 | 29.38 |
| UTL | T2 | para_para | -36.70 | 68.41 | 29.38 |
| UTL | T2 | para_para | -36.70 | 68.41 | 29.38 |
| UTL | T2 | para_para | -36.70 | 68.41 | 29.38 |
| UTL | T4 | ortho_benzene | 58.57 | 139.33 | 37.14 |
| UTL | T4 | ortho_benzene | 58.57 | 139.33 | 37.14 |
| UTL | T4 | ortho_benzene | 58.57 | 139.33 | 37.14 |
| UTL | T4 | ortho_benzene | 58.57 | 139.33 | 37.14 |
| UTL | T4 | meta_benzene | 51.63 | 122.65 | 32.59 |
| UTL | T4 | meta_benzene | 51.63 | 122.65 | 32.59 |
| UTL | T4 | meta_benzene | 51.63 | 122.65 | 32.59 |
| UTL | T4 | meta_benzene | 51.63 | 122.65 | 32.59 |
| UTL | T4 | para_benzene | 53.53 | 109.26 | 34.50 |
| UTL | T4 | para_benzene | 53.53 | 109.26 | 34.50 |
| UTL | T4 | para_benzene | 53.53 | 109.26 | 34.50 |
| UTL | T4 | para_benzene | 53.53 | 109.26 | 34.50 |
| UTL | T4 | ortho_ortho | 37.77 | 110.75 | 82.73 |
| UTL | T4 | ortho_ortho | 37.77 | 110.75 | 82.73 |
| UTL | T4 | ortho_ortho | 37.77 | 110.75 | 82.73 |
| UTL | T4 | ortho_ortho | 37.77 | 110.75 | 82.73 |
| UTL | T4 | ortho_meta | 68.04 | 204.49 | 95.84 |
| UTL | T4 | ortho_meta | 68.04 | 204.49 | 95.84 |
| UTL | T4 | ortho_meta | 68.04 | 204.49 | 95.84 |
| UTL | T4 | ortho_para | 42.71 | 128.86 | 19.38 |
| UTL | T4 | ortho_para | 42.71 | 128.86 | 19.38 |
| UTL | T4 | ortho_para | 42.71 | 128.86 | 19.38 |
| UTL | T4 | meta_ortho | 31.73 | 106.49 | 60.05 |
| UTL | T4 | meta_ortho | 31.73 | 106.49 | 60.05 |
| UTL | T4 | meta_ortho | 31.73 | 106.49 | 60.05 |
| UTL | T4 | meta_meta | 50.03 | 144.25 | 80.43 |
| UTL | T4 | meta_meta | 50.03 | 144.25 | 80.43 |
| UTL | T4 | meta_meta | 50.03 | 144.25 | 80.43 |
| UTL | T4 | meta_meta | 50.03 | 144.25 | 80.43 |
| UTL | T4 | meta_para | 21.93 | 113.25 | 3.53 |
| UTL | T4 | meta_para | 21.93 | 113.25 | 3.53 |
| UTL | T4 | meta_para | 21.93 | 113.25 | 3.53 |
| UTL | T4 | para_ortho | 9.72 | 95.37 | 22.27 |
| UTL | T4 | para_ortho | 9.72 | 95.37 | 22.27 |
| UTL | T4 | para_ortho | 9.72 | 95.37 | 22.27 |
| UTL | T4 | para_meta | 17.01 | 97.67 | 18.28 |
| UTL | T4 | para_meta | 17.01 | 97.67 | 18.28 |
| UTL | T4 | para_meta | 17.01 | 97.67 | 18.28 |
| UTL | T4 | para_para | 0.00 | 182.19 | 34.65 |
| UTL | T4 | para_para | 0.00 | 182.19 | 34.65 |
| UTL | T4 | para_para | 0.00 | 182.19 | 34.65 |
| UTL | T4 | para_para | 0.00 | 182.19 | 34.65 |
| FAU | T1 | ortho_benzene | 0.00 | 131.09 | 18.14 |
| FAU | T1 | ortho_benzene | 0.00 | 131.09 | 18.14 |
| FAU | T1 | ortho_benzene | 0.00 | 131.09 | 18.14 |
| FAU | T1 | ortho_benzene | 0.00 | 131.09 | 18.14 |
| FAU | T1 | meta_benzene | 20.56 | 95.31 | 10.43 |
| FAU | T1 | meta_benzene | 20.56 | 95.31 | 10.43 |
| FAU | T1 | meta_benzene | 20.56 | 95.31 | 10.43 |
| FAU | T1 | meta_benzene | 20.56 | 95.31 | 10.43 |
| FAU | T1 | para_benzene | 21.39 | 104.56 | 7.29 |
| FAU | T1 | para_benzene | 21.39 | 104.56 | 7.29 |
| FAU | T1 | para_benzene | 21.39 | 104.56 | 7.29 |
| FAU | T1 | para_benzene | 21.39 | 104.56 | 7.29 |
| FAU | T1 | ortho_ortho | 87.43 | 177.53 | 38.65 |
| FAU | T1 | ortho_ortho | 87.43 | 177.53 | 38.65 |
| FAU | T1 | ortho_ortho | 87.43 | 177.53 | 38.65 |
| FAU | T1 | ortho_ortho | 87.43 | 177.53 | 38.65 |
| FAU | T1 | ortho_meta | 64.58 | 167.44 | 87.15 |
| FAU | T1 | ortho_meta | 64.58 | 167.44 | 87.15 |
| FAU | T1 | ortho_meta | 64.58 | 167.44 | 87.15 |
| FAU | T1 | ortho_para | 46.90 | 132.26 | 51.95 |
| FAU | T1 | ortho_para | 46.90 | 132.26 | 51.95 |
| FAU | T1 | ortho_para | 46.90 | 132.26 | 51.95 |
| FAU | T1 | meta_ortho | 49.11 | 163.72 | 69.39 |
| FAU | T1 | meta_ortho | 49.11 | 163.72 | 69.39 |
| FAU | T1 | meta_ortho | 49.11 | 163.72 | 69.39 |
| FAU | T1 | meta_meta | 48.93 | 156.90 | 100.88 |
| FAU | T1 | meta_meta | 48.93 | 156.90 | 100.88 |
| FAU | T1 | meta_meta | 48.93 | 156.90 | 100.88 |
| FAU | T1 | meta_meta | 48.93 | 156.90 | 100.88 |
| FAU | T1 | meta_para | 50.07 | 123.24 | 23.20 |
| FAU | T1 | meta_para | 50.07 | 123.24 | 23.20 |
| FAU | T1 | meta_para | 50.07 | 123.24 | 23.20 |
| FAU | T1 | para_ortho | 48.82 | 155.32 | 65.48 |
| FAU | T1 | para_ortho | 48.82 | 155.32 | 65.48 |
| FAU | T1 | para_ortho | 48.82 | 155.32 | 65.48 |
| FAU | T1 | para_meta | 39.00 | 150.94 | 76.18 |
| FAU | T1 | para_meta | 39.00 | 150.94 | 76.18 |
| FAU | T1 | para_meta | 39.00 | 150.94 | 76.18 |
| FAU | T1 | para_para | 0.00 | 157.62 | 37.76 |
| FAU | T1 | para_para | 0.00 | 157.62 | 37.76 |
| FAU | T1 | para_para | 0.00 | 157.62 | 37.76 |
| FAU | T1 | para_para | 0.00 | 157.62 | 37.76 |

**Table S24**. Relative free energy stabilities, $G_{TA} - G_{TB}$, for the different BOG, IWV and UTL models studied in kJ/mol. Intermediates from I column, Transition States from TS column.

| Isomer | I | BOG (T2-T1) | IWV (T7-T3) | UTL (T4-T2) | TS | BOG (T2-T1) | IWV (T7-T3) | UTL (T4-T2) |
|---|---|---|---|---|---|---|---|---|
| ortho_benzene | I1 | 5.00 | 16.27 | 7.55 | TS1 | -5.33 | -15.01 | 10.97 |
| meta_benzene | I1 | -1.20 | 39.04 | 1.84 | TS1 | -8.76 | 41.91 | -22.09 |
| para_benzene | I1 | 8.20 | -24.85 | 38.28 | TS1 | -37.34 | 34.27 | 11.34 |
| ortho_ortho | I1 | -24.32 | 37.35 | 35.56 | TS1 | -35.97 | -20.43 | -36.38 |
| ortho_meta | I1 | 43.00 | 19.86 | 43.28 | TS1 | 46.19 | -2.48 | 54.62 |
| ortho_para | I1 | 16.37 | 10.26 | 34.00 | TS1 | -0.52 | -48.75 | 19.44 |
| meta_ortho | I1 | 24.80 | 6.05 | 12.07 | TS1 | 25.08 | -0.91 | 25.64 |
| meta_meta | I1 | -37.01 | 8.04 | 23.77 | TS1 | -38.00 | -32.21 | -43.08 |
| meta_para | I1 | 4.64 | 27.93 | 33.24 | TS1 | 37.33 | -75.45 | 34.08 |
| para_ortho | I1 | -13.01 | 20.36 | 9.29 | TS1 | 23.53 | -7.22 | -6.56 |
| para_meta | I1 | 24.10 | -6.13 | 10.85 | TS1 | -99.45 | 23.43 | -33.38 |
| para_para | I1 | 8.38 | 34.97 | 15.56 | TS1 | -22.03 | 72.64 | 106.97 |
| ortho_benzene | I2 | 75.32 | 4.14 | 39.46 | TS2 | -60.73 | -7.30 | 61.92 |
| meta_benzene | I2 | -30.17 | 15.58 | 35.21 | TS2 | 7.82 | 49.85 | 14.04 |
| para_benzene | I2 | -21.46 | 29.07 | 98.17 | TS2 | -0.26 | -0.69 | 13.58 |
| ortho_ortho | I2 | -30.10 | 6.93 | 1.44 | TS2 | 25.50 | 22.34 | 48.47 |
| ortho_ortho | I2 | 8.10 | 7.39 | -9.95 | TS2 | -18.59 | -26.18 | 68.19 |
| ortho_meta | I2 | 28.06 | -10.74 | -6.72 | TS2 | -52.78 | -3.18 | 13.37 |
| ortho_para | I2 | -26.11 | 13.68 | 8.23 | TS2 | -18.31 | 38.58 | 10.47 |
| ortho_para | I2 | -4.49 | 36.91 | 17.40 | TS2 | 9.63 | -11.27 | 5.40 |
| meta_ortho | I2 | 48.93 | -3.25 | -9.85 | TS2 | -55.07 | 1.86 | 28.60 |
| meta_ortho | I2 | -5.50 | 33.43 | 10.51 | TS2 | 27.15 | 13.63 | 32.71 |
| meta_meta | I2 | -11.10 | -2.55 | 22.55 | TS2 | 22.97 | 8.89 | -36.54 |
| meta_para | I2 | -25.03 | 2.12 | -12.26 | TS2 | 44.28 | 56.79 | 2.06 |
| meta_para | I2 | 10.08 | 6.19 | -5.13 | TS2 | -65.22 | 29.14 | -5.16 |
| para_ortho | I2 | 26.02 | 14.16 | 12.89 | TS2 | 28.32 | 0.32 | -4.38 |
| para_ortho | I2 | -9.47 | 12.15 | 9.57 | TS2 | -46.05 | 17.48 | -15.28 |
| para_meta | I2 | -9.28 | -1.70 | 3.03 | TS2 | -33.67 | 27.39 | -0.63 |
| para_para | I2 | 0.29 | 31.57 | 21.06 | TS2 | -44.75 | 31.55 | 20.86 |
| para_para | I2 | 4.06 | 10.33 | 10.94 | TS2 | -5.21 | -6.45 | -0.98 |
| ortho_benzene | I3 | 5.54 | 14.42 | -22.94 | TS3 | -40.09 | -19.40 | 52.62 |
| ortho_benzene | I3 | -3.58 | 48.92 | 12.50 | TS3 | 72.13 | 47.45 | 12.21 |
| meta_benzene | I3 | 4.83 | 0.82 | 47.76 | TS3 | -20.74 | 22.74 | 9.81 |
| meta_benzene | I3 | -10.99 | 18.13 | 1.10 | TS3 | -50.10 | 11.80 | 3.89 |
| para_benzene | I3 | 7.95 | 0.56 | 12.18 | TS3 | -8.03 | 20.32 | 24.61 |
| ortho_ortho | I3 | 37.92 | 6.12 | -12.95 | TS3 | 18.03 | 33.84 | 10.34 |
| ortho_ortho | I3 | 6.81 | 6.12 | 17.36 | TS3 | 48.74 | 74.44 | 3.76 |
| ortho_ortho | I3 | 6.81 | -6.95 | -12.95 | TS3 | 58.83 | -19.59 | -8.07 |
| ortho_ortho | I3 | 37.92 | -6.95 | 17.36 | TS3 | 57.13 | 37.74 | 26.93 |
| ortho_meta | I3 | 6.35 | -7.17 | 17.25 | TS3 | 47.84 | -36.75 | 13.44 |
| ortho_meta | I3 | 9.27 | -20.03 | 17.29 | TS3 | 28.93 | -17.86 | 24.16 |
| ortho_para | I3 | -29.83 | -1.40 | 6.80 | TS3 | -37.14 | 8.54 | -11.86 |
| ortho_para | I3 | -29.83 | -9.75 | 25.49 | TS3 | 19.82 | -27.62 | -25.21 |
| ortho_para | I3 | 23.13 | -9.75 | 25.49 | TS3 | 34.94 | -19.67 | 4.59 |
| ortho_para | I3 | 23.13 | -1.40 | 6.80 | TS3 | 31.41 | 2.20 | 23.60 |
| meta_ortho | I3 | -21.26 | 11.64 | 52.00 | TS3 | 26.00 | -1.50 | 6.07 |
| meta_ortho | I3 | 0.16 | 11.64 | 52.00 | TS3 | 98.04 | -0.25 | 24.13 |
| meta_ortho | I3 | 0.16 | 32.46 | -37.44 | TS3 | 89.38 | -13.56 | 10.65 |
| meta_ortho | I3 | -21.26 | 32.46 | -37.44 | TS3 | 3.40 | -3.81 | 9.47 |
| meta_meta | I3 | 50.70 | 43.78 | 4.96 | TS3 | -12.55 | -43.60 | -6.71 |
| meta_meta | I3 | -5.96 | -14.65 | 22.79 | TS3 | 54.46 | -13.34 | -3.38 |
| meta_para | I3 | 3.12 | 7.93 | 33.10 | TS3 | 14.75 | 38.66 | 19.25 |
| meta_para | I3 | 3.12 | 11.24 | 33.10 | TS3 | -30.83 | 17.20 | 27.96 |
| meta_para | I3 | -14.95 | 11.24 | -70.81 | TS3 | 14.64 | 6.84 | -2.47 |
| meta_para | I3 | -14.95 | 7.93 | -70.81 | TS3 | 51.62 | 21.13 | 65.45 |
| para_ortho | I3 | 5.12 | -8.31 | 8.29 | TS3 | 18.54 | 50.98 | 31.33 |
| para_ortho | I3 | 5.12 | -8.31 | 8.29 | TS3 | -19.79 | 31.46 | 27.40 |
| para_meta | I3 | -61.86 | 10.97 | 13.83 | TS3 | -33.07 | 60.35 | -15.79 |
| para_para | I3 | 11.49 | 13.93 | -4.05 | TS3 | 12.01 | 32.82 | 11.22 |
| para_para | I3 | 11.49 | 13.93 | -4.05 | TS3 | 29.75 | -22.42 | 34.03 |
| ortho_benzene | I4 | 13.32 | -1.67 | 10.77 | TS4 | -50.02 | 36.51 | 3.78 |
| ortho_benzene | I4 | 13.32 | -1.67 | 10.77 | TS4 | 42.83 | 45.56 | -8.13 |
| meta_benzene | I4 | 0.22 | -6.20 | 20.80 | TS4 | 29.69 | 9.17 | 37.78 |
| meta_benzene | I4 | 0.22 | -6.20 | 20.80 | TS4 | -47.81 | 25.72 | 55.59 |
| para_benzene | I4 | -22.55 | 10.75 | 12.18 | TS4 | -13.01 | 6.71 | 71.80 |
| ortho_ortho | I4 | -68.90 | 41.34 | 37.70 | TS4 | -29.27 | 92.88 | -39.28 |
| ortho_ortho | I4 | -68.90 | 41.34 | 37.70 | TS4 | 35.70 | 23.76 | 25.75 |
| ortho_meta | I4 | 69.53 | -18.21 | 24.40 | TS4 | -6.25 | -13.26 | 52.96 |
| ortho_meta | I4 | 69.53 | -18.21 | 24.40 | TS4 | 36.94 | -2.58 | -0.24 |
| ortho_para | I4 | 5.08 | 20.17 | 10.67 | TS4 | 21.18 | -15.03 | -36.23 |
| ortho_para | I4 | 5.08 | 20.17 | 10.67 | TS4 | -0.11 | -3.08 | 6.66 |
| meta_ortho | I4 | 9.09 | -15.17 | 28.44 | TS4 | 45.90 | 39.94 | -29.98 |
| meta_ortho | I4 | 9.09 | -15.17 | 28.44 | TS4 | -28.41 | 41.88 | -15.19 |
| meta_meta | I4 | 2.27 | 4.66 | 31.61 | TS4 | -27.09 | 3.83 | 11.54 |
| meta_meta | I4 | 2.27 | 4.66 | 31.61 | TS4 | -51.95 | 37.92 | -47.85 |
| meta_para | I4 | -6.90 | 39.13 | 23.91 | TS4 | 32.38 | 41.65 | 17.28 |
| meta_para | I4 | -6.90 | 39.13 | 23.91 | TS4 | -48.24 | 0.97 | -33.43 |
| para_ortho | I4 | 44.51 | 16.99 | 28.11 | TS4 | 14.48 | 42.79 | 47.74 |
| para_meta | I4 | -22.03 | 7.82 | 11.51 | TS4 | -47.42 | -7.75 | 30.95 |
| para_para | I4 | 4.29 | 4.35 | 14.49 | TS4 | -47.93 | 12.71 | 39.56 |